\documentclass[12pt]{article}

\usepackage{jheppub, bm, wrapfig,float,array}
\usepackage[utf8]{inputenc}
\numberwithin{equation}{section}
\renewcommand\afterTocRuleSpace{\bigskip}

\usepackage{amsfonts}
\usepackage{amsmath}
\usepackage{color}
\usepackage{graphicx}
\usepackage{float}
\usepackage[T1]{fontenc}
\usepackage[utf8]{inputenc}
\usepackage{alphabeta}
\usepackage{fancyvrb}
\usepackage[dvipsnames]{xcolor}
\usepackage{bold-extra}
\usepackage{lmodern}
\usepackage{graphicx}
\usepackage{lmodern}
\usepackage{rotating}
\usepackage{tikz}
\usetikzlibrary{arrows}
\usetikzlibrary{shapes.geometric,calc,arrows, positioning,shapes.misc,decorations.markings}
\tikzset{
  big arrow/.style={
    decoration={markings,mark=at position 1 with {\arrow[scale=2,#1]{>}}},
    postaction={decorate},
    shorten >=0.4pt},
  big arrow/.default=black}

\newcommand{\bea}{\begin{eqnarray}}
\newcommand{\eea}{\end{eqnarray}}
\newcommand{\be}{\begin{equation}}
\newcommand{\ee}{\end{equation}}
\newcommand{\bit}{\begin{itemize}}
\newcommand{\eit}{\end{itemize}}
\newcommand{\ben}{\begin{enumerate}}
\newcommand{\een}{\end{enumerate}}
\newcommand{\nn}{\nonumber}
\renewcommand{\ni}{\noindent}

\newcommand{\wh}{\widehat}

\newcommand{\half}{\frac{1}{2}}

\newcommand{\Z}{{\mathbb Z}}
\newcommand{\R}{{\mathbb R}}
\newcommand{\C}{{\mathbb C}}

\newcommand{\bF}{{\mathbb F}}
\renewcommand{\P}{{\mathbb P}}

\newcommand{\cF}{\mathcal{F}}

\newcommand{\cL}{\mathcal{L}}

\newcommand{\cN}{\mathcal{N}}
\newcommand{\cO}{\mathcal{O}}

\newcommand{\cR}{\mathcal{R}}
\newcommand{\cS}{\mathcal{S}}
\newcommand{\cT}{\mathcal{T}}

\newcommand{\F}{\mathsf{F}}
\renewcommand{\S}{\mathsf{S}}
\newcommand{\Asym}{\mathsf{\Lambda}^2}

\newcommand{\tAsym}{\mathsf{\Lambda}^3}

\newcommand{\nAsym}{\mathsf{\Lambda}^n}
\newcommand{\bnAsym}{\bar{\mathsf{\Lambda}}^n}

\newcommand{\Sym}{\mathsf{S}^2}
\newcommand{\bSym}{\bar{\mathsf{S}}^2}
\renewcommand{\C}{\mathsf{C}}

\renewcommand{\C}{\mathsf{C}}

\newcommand{\fT}{\mathfrak{T}}

\newcommand{\fe}{\mathfrak{e}}
\newcommand{\ff}{\mathfrak{f}}
\newcommand{\fg}{\mathfrak{g}}
\newcommand{\fh}{\mathfrak{h}}
\newcommand{\su}{\mathfrak{su}}
\renewcommand{\sp}{\mathfrak{sp}}
\newcommand{\so}{\mathfrak{so}}
\renewcommand{\u}{\mathfrak{u}}

\newcommand{\ubf}[1]{\underline{\bf #1}}

\def\tr{\mathop{\mathrm{tr}}\nolimits}

\newcommand{\lra}{\leftrightarrow}
\newcommand{\llra}{\longleftrightarrow}
\newcommand{\Lra}{\longrightarrow}

\title{Twisted Circle Compactifications of $6d$ SCFTs}

\author{Lakshya Bhardwaj$\,^a$, Patrick Jefferson$\,^b$, Hee-Cheol Kim$\,^c$, Houri-Christina Tarazi$\,^a$, Cumrun Vafa$\,^a$}

\affiliation{$^a$ Department of Physics, Harvard University, Cambridge, MA 02138, USA\\
$^b$ Center for Theoretical Physics, Department of Physics, Massachusetts Institute of Technology, 77 Massachusetts Avenue, Cambridge, MA 02139, USA\\$^c$ Department of Physics, POSTECH, Pohang 790-784, Korea}

\abstract{We study $6d$ superconformal field theories (SCFTs) compactified on a circle with arbitrary twists. The theories obtained after compactification, often referred to as $5d$ Kaluza-Klein (KK) theories, can be viewed as starting points for RG flows to $5d$ SCFTs. According to a conjecture, all $5d$ SCFTs can be obtained in this fashion. We compute the Coulomb branch prepotential for all $5d$ KK theories obtainable in this manner and associate to these theories a smooth local genus one fibered Calabi-Yau threefold in which is encoded information about all possible RG flows to $5d$ SCFTs. These Calabi-Yau threefolds provide hitherto unknown M-theory duals of F-theory configurations compactified on a circle with twists. For certain exceptional KK theories that do not admit a standard geometric description we propose an algebraic description that appears to retain the properties of the local Calabi-Yau threefolds necessary to determine RG flows to $5d$ SCFTs, along with other relevant physical data.
}

\begin{document}

\maketitle

\section{Introduction} \label{intro}
Recently, there has been a resurgence of interest in the problem of classifying $5d$ superconformal field theories (SCFTs), with a particular emphasis on exploring the relationship between $5d$ UV fixed points and $6d$ UV fixed points \cite{Jefferson:2017ahm,Jefferson:2018irk,Bhardwaj:2018yhy,Bhardwaj:2018vuu,DelZotto:2017pti,Apruzzi:2018nre,Apruzzi:2019vpe,Apruzzi:2019opn,Apruzzi:2019enx,Bhardwaj:2019jtr,DelZotto:2018tcj}. The motivation for studying this relationship is the observation that all known $5d$ SCFTs can be organized into families of theories (connected to one another by RG flows) whose ``progenitors'' are $6d$ SCFTs compactified on a circle \cite{Jefferson:2017ahm,Jefferson:2018irk}, and hence every $6d$ SCFT compactified on a circle provides a natural starting point for the systematic identification of a large family of $5d$ SCFTs. 

While it has been appreciated in the literature for some time that circle compactifications of $6d$ SCFTs can flow to $5d$ SCFTs, only recently has the existence of a $6d$ UV fixed point been understood in an intrinsically $5d$ setting. To understand this point, let us recall that the most widely used method for identifying $5d$ SCFTs is to construct a candidate effective field theory assumed to be a relevant deformation of a $5d$ UV fixed point, and to verify the effective theory passes a number of consistency checks which are believed to be sufficient to guarantee the existence of a such a non-trivial UV fixed point. This method, which has been used to construct numerous examples of UV complete minimally supersymmetric $5d$ QFTs---both by means of standard gauge theoretic methods \cite{Seiberg:1996bd,Intriligator:1997pq,Jefferson:2017ahm}, as well as string theory constructions such as $(p,q)$ 5-brane configurations in type IIB string theory \cite{Aharony:1997bh,Aharony:1997ju,Bergman:2015dpa,Zafrir:2015ftn,Hayashi:2019yxj,Hayashi:2018lyv,Hayashi:2018bkd} and M-theory compactifications on local Calabi-Yau threefolds \cite{Morrison:1996xf,Douglas:1996xp,Jefferson:2018irk,Closset:2018bjz}---has also led to the identification of numerous examples of theories that despite not satisfying the criteria necessary for the existence of a non-trivial 5d UV completion, nonetheless exhibit certain features that suggest they can be UV completed in 6d. All known examples of such theories are characterized by the emergence of an intrinsic length scale that is interpreted as the size of a compactification circle, and it has been argued that each of these theories is a circle compactification of a $6d$ SCFT possibly twisted by the action of a discrete global symmetry\footnote{Twisting the theory around the circle means that we introduce a holonomy for the background gauge fields associated to discrete global symmetries of the theory.}; see for example \cite{Jefferson:2018irk,Hayashi:2015vhy,Hayashi:2015zka,Zafrir:2015rga,Hayashi:2019yxj,Kim:2015jba,Bhardwaj:2018yhy,Bhardwaj:2018vuu,Hayashi:2015fsa}. These observations have led to the identification of a set of criteria believed sufficient to imply the existence of a $6d$ UV completion for certain 5d theories, and this introduces the possibility of also classifying circle compactifications of $6d$ SCFTs using 5d physics.

It was recently conjectured \cite{Jefferson:2018irk} that all $5d$ SCFTs can be obtained via RG flows starting from $5d$ Kaluza-Klein (KK) theories. The latter are defined as $6d$ SCFTs compactified on a circle (of finite radius) possibly with discrete twists around the circle. Given a 5d KK theory, the RG flows of interest correspond to integrating out BPS particles from the 5d KK theory---thus, if the full BPS spectrum is known then according to the conjecture of \cite{Jefferson:2018irk} it is possible to classify all $5d$ SCFTs by systematically studying all possible RG flows from the $5d$ KK theory. 

 In this paper, we focus on the geometric approach in which one realizes a $5d$ KK theory via a compactification of M-theory on a genus one fibered Calabi-Yau threefold. The set of holomorphic curves in the threefold completely encode the information about the spectrum of BPS particles required to track all RG flows down to $5d$ SCFTs. Therefore, a precursor to classifying RG flows from 5d KK theories to 5d SCFTs is to geometrically classify all 5d KK theories themselves in terms of Calabi-Yau threefolds. See \cite{Bhardwaj:2019jtr} (also \cite{Jefferson:2018irk}) for explicit application of this geometric procedure to the classification of $5d$ SCFTs up to rank three.

It is believed that all $6d$ SCFTs can be constructed by compactifying F-theory on singular elliptically fibered Calabi-Yau threefolds admitting certain singular limits characterized by the contraction of holomorphic curves in the base of the fibration. Here we should distinguish between two different kinds of compactifications of F-theory depending on whether or not they contain O7$^+$ plane from the point of view of type IIB string theory. If there is no O7$^+$, then the compactification is said to lie in the \emph{unfrozen} phase of F-theory; otherwise it is said to lie in the \emph{frozen} phase \cite{Witten:1997bs,deBoer:2001wca,Tachikawa:2015wka} of F-theory. These two phases are qualitatively different in the following sense: The rules for converting geometry in the unfrozen phase to the corresponding $6d$ physics are far more straightforward than the rules for converting geometry in the frozen phase to the corresponding $6d$ physics \cite{Bhardwaj:2018jgp}. See \cite{Heckman:2015bfa,Heckman:2013pva} (see also \cite{Bhardwaj:2015xxa}) for the classification of $6d$ SCFTs arising from the unfrozen phase of F-theory, and \cite{Bhardwaj:2019hhd} for the classification of $6d$ SCFTs arising from the frozen phase of F-theory.

A $5d$ KK theory corresponding to the untwisted compactification of a $6d$ SCFT arising in the unfrozen phase can be constructed by compactifying M-theory on a Calabi-Yau threefold which is a resolution of the Calabi-Yau threefold arising in the F-theory construction. This fact is a special case of the duality between M-theory and (unfrozen phase of) F-theory compactified on a circle (without any twist). Explicit resolution of all Calabi-Yau threefolds associated to $6d$ SCFTs was performed by \cite{Bhardwaj:2018vuu,Bhardwaj:2018yhy}, and hence the Calabi-Yau threefolds associated to corresponding $5d$ KK theories was determined. These threefolds are elliptically fibered since the threefolds associated to $6d$ SCFTs are elliptically fibered to begin with.

In this paper, we extend the work of \cite{Bhardwaj:2018vuu,Bhardwaj:2018yhy} and determine a resolved local Calabi-Yau threefold describing every $5d$ KK theory, with the exception of certain examples which do not appear to admit a conventional geometric description\footnote{For these examples, we propose an algebraic description which mimics certain properties of the Calabi-Yau threefolds associated to other KK theories. This algebraic description can be used to compute RG flows starting from these KK theories to $5d$ SCFTs. In the paper we sometimes abuse terminology and use the word `geometry' to refer to both theories that admit a conventional geometric description along with those (i.e. ``non-geometric'' theories) for which only an algebraic description is available.}. Not only do we include twisted compactifications of $6d$ SCFTs arising in the unfrozen phase, but also the untwisted and twisted compactifications of $6d$ SCFTs arising in the frozen phase. We find that these Calabi-Yau threefolds are in general only genus one fibered and may not be elliptically fibered, which means that the fibration may not admit a zero section.

Our analysis can be divided into two parts. In the first part of the analysis, which is purely field theoretic, we determine the prepotential for each $5d$ KK theory by using the following observations: Each $6d$ SCFT admits a $6d$ gauge theory description which can be reduced on a circle with an appropriate twist to obtain a canonical $5d$ gauge theory description of the associated $5d$ KK theory. The Green-Schwarz term in $6d$ reduces to a Chern-Simons term in the $5d$ gauge theory, which induces a tree-level contribution to the prepotential. Combining this contribution with the one-loop contribution coming from the $5d$ gauge theory produces the full prepotential for the $5d$ KK theory. In the second part of the analysis, we interpret the prepotential as describing the triple intersection numbers of 4-cycles inside a yet to be determined Calabi-Yau threefold. Using the data of these triple intersection numbers, along with some other consistency conditions, we are able to determine a description of the Calabi-Yau threefold as a neighborhood of intersecting K\"ahler surfaces along the lines of the discussion in \cite{Jefferson:2018irk,Bhardwaj:2018yhy,Bhardwaj:2018vuu}, and we verify that each threefold admits the structure of genus one fibration\footnote{See for example \cite{Morrison:2014era} for a discussion of F-theory compactifications on genus one fibered, in contrast to elliptically fibered, Calabi-Yau varieties.}. By construction, compactifying M-theory on this Calabi-Yau threefold leads to the $5d$ KK theory whose prepotential we computed in the first part of the analysis.

One can view these Calabi-Yau threefolds as providing hitherto unknown M-theory duals of general unfrozen and frozen F-theory configurations compactified on a circle possibly with a discrete twist. Even though we have provided explicit results only for F-theory configurations realizing $6d$ SCFTs, our methods should in principle apply to any general F-theory configuration.

Notice that at no step in our analysis do we distinguish between $6d$ SCFTs arising from the unfrozen phase and $6d$ SCFTs arising from the frozen phase. Thus, according to our analysis, the rules for converting geometry into the corresponding $5d$ physics are uniform irrespective of whether the $5d$ KK theory arises from the compactification of a $6d$ SCFT lying in the frozen or the unfrozen phase. In other words, the frozen and unfrozen six-dimensional compactifications of F-theory are given a unified geometric description\footnote{Some of the frozen theories belong to the class of exceptional KK theories which do not admit a conventional geometric description, and thus to which we only associate an algebraic description.} in M-theory.

We close the introduction with a brief overview of the structure of Calabi-Yau threefolds that we associate to $5d$ KK theories. By construction, the structure of these threefolds descends from the structure of $6d$ SCFTs. Recall that an important object characterizing a $6d$ SCFT is the matrix of Dirac pairings of ``fundamental'' BPS strings visible on the tensor branch of the $6d$ SCFT. The matrix of Dirac pairings is a symmetric, positive definite, integer matrix with positive entries on the diagonal and non-positive off-diagonal entries. Thus, the Dirac pairing matrix is analogous to the Cartan matrix of a simply laced Lie algebra, and we can associate to this matrix a graph analogous to a Dynkin graph for a simply laced Lie algebra.

As discussed in more detail later in the paper, the matrix of Dirac pairings descends to a matrix of Chern-Simons terms in the canonical gauge theory associated to the $5d$ KK theory, where the precise map between the two matrices depends on the choice of twist. We find that $5d$ KK theories end up organizing themselves according to this matrix of Chern-Simons terms. Like the matrix of Dirac pairings, the matrix of Chern-Simons terms is in general a positive definite, integer matrix with positive entries on the diagonal and non-positive off-diagonal entries, where off-diagonal entries can only be zero if their transposes are also zero. But, unlike the matrix of Dirac pairings, the matrix of Chern-Simons terms is not necessarily a symmetric matrix. Thus, the matrix of Chern-Simons couplings is analogous to the Cartan matrix of a general (simply or non-simply laced) Lie algebra, and we associate to it a graph analogous to a Dynkin graph for a general Lie algebra.

In this way, $5d$ KK theories are characterized by graphs that generalize Dynkin graphs. The associated Calabi-Yau geometry is assembled according to the structure of this graph:
\bit
\item To each node in the graph, we associate a collection of Hirzebruch surfaces intersecting with each other. In fact, we associate a family of such collections parametrized by an integer $\nu$, where the collections labeled by different values of $\nu$ are related to one another by flop transitions. A key point is that a certain linear combination of the $\P^1$ fibers of these Hirzebruch surfaces has genus one, and an appropriate multiple of the genus one fiber is identified physically with the KK mode of momentum one around the circle.
\item To a pair of nodes connected to each other by some edges, we associate certain gluing\footnote{When two K\"ahler surfaces intersect transversely along a common holomorphic curve inside of a Calabi-Yau threefold, the intersection implies that a holomorphic curve inside one of the two surfaces is identified with a holomorphic curve inside of the other surface. We refer to this identification as a \emph{gluing} together of the two surfaces.} rules. These gluing rules describe how to glue the collection of surfaces associated to a node to the collection of surfaces associated to another node. These gluing rules capture the data of intersections between the two collections of surfaces. In general, the gluing rules provided in this paper work only for a subset of the values of $\nu$ parametrizing the two collections of surfaces being glued together. Our claim is that given a $5d$ KK theory, we can always find at least one value of $\nu$ for each node in the associated graph such that the gluing rules for each edge work. 

By applying these gluing rules, it can be checked that a multiple of the genus one fiber in one collection of surfaces is glued to a multiple of the genus fiber in the other collection of surfaces. These multiples are such that the KK mode associated to one collection is identified with the KK mode associated to the other collection. This must be so since there is only a single KK mode associated to the full KK theory and the genus one fibers inside each collection are merely different geometric manifestations of the same mode.
\item Once we are done gluing all the collections of surfaces according to the gluing rules associated to each edge, we obtain a larger collection of surfaces intersecting with each other. The Calabi-Yau threefold associated to the KK theory is by definition a local neighborhood of this larger collection of surfaces. As we have described above, this Calabi-Yau threefold is canonically genus one fibered.
\eit

The rest of the paper is organized as follows. In Section \ref{re}, we review how all $6d$ SCFTs can be neatly encapsulated in terms of graphs that capture the data of the tensor branch of the corresponding $6d$ SCFTs. We list all the possible vertices and edges appearing in such graphs. Our presentation treats unfrozen and frozen cases on an equal footing. Another distinguishing feature of our presentation is that we carefully distinguish different theories having the same gauge algebra content and same Dirac pairing. This includes the theta angle for $\sp(n)$, different distributions of hypers between the spinor and cospinor representations of $\so(12)$, as well as some frozen cases.

In Section \ref{SKK}, we study all the possible twists of $6d$ SCFTs once they are compactified on  a circle. Each twist leads to a different $5d$ KK theory. The different twists of a $6d$ SCFT $\fT$ are characterized by equivalence classes in the group of discrete global symmetries of $\fT$. We show that these equivalence classes can be described by foldings of the graphs $\Sigma_\fT$ associated to $\fT$ along with choice of an outer automorphism for each gauge algebra appearing in the low energy theory on the tensor branch of $\fT$. Thus, different $5d$ KK theories are also classified by graphs that generalize the graphs classifying $6d$ SCFTs. We provide a list of all the possible vertices and edges that can appear in the graphs associated to $5d$ KK theories.

In Section \ref{PPKK}, we provide a prescription to obtain the prepotential of any $5d$ KK theory. This is done by compactifying the low energy gauge theory appearing on the tensor branch of the corresponding $6d$ SCFT on a circle with the corresponding twist. This leads to a $5d$ gauge theory whose prepotential, along with a shift, is identified as the prepotential for the $5d$ KK theory.

In Section \ref{GKK}, we associate a genus-one fibered Calabi-Yau threefold to each $5d$ KK theory, except for a few exceptional cases, for which we provide an algebraic description mimicking the essential properties of genus one fibered Calabi-Yau threefolds. The chief ingredient in the determination of the threefold is the prepotential determined in Section \ref{PPKK}. The prepotential captures the data of the triple intersection numbers of surfaces inside the threefold. Once a description of the threefold as a local neighborhood of a collection of surfaces glued to each other is presented, these triple intersections can be computed in a multitude of different ways. Demanding all of these different computations to give the same result leads to strong consistency constraints on such a description and often uniquely fixes the description (up to isomorphisms). Other consistency conditions playing a crucial role are also discussed in Section \ref{GF}. 

\ni The description of the geometry is provided in two different steps according to the structure of the graph associated to the $5d$ KK theory under study. First, a part of the geometry is assigned to each vertex in the graph according to results presented in Section \ref{RO}. Then, depending on the configuration of edges in the graph, different parts of the geometry corresponding to different vertices in the graph are glued to each other via the gluing rules presented in Sections \ref{AR} and \ref{ARng}.

In Section \ref{conc}, we present our conclusions. In Appendix \ref{back}, we review some geometric background relevant for this paper. In Appendix \ref{Exc}, we address certain exceptional examples of geometries and gluing rules that do not admit a straightforward analysis following the main methods described in this paper. In Appendix \ref{concrete}, we provide a concrete and non-trivial check of our proposal for computing the prepotential and geometries associated to $5d$ KK theories. We demonstrate that a $5d$ KK theory arising from a non-trivial twist (involving a permutation of tensor multiplets) of a $6d$ SCFT has a $5d$ gauge theory description found in earlier studies by using brane constructions. In Appendix \ref{comparison}, we provide some more checks of our proposal. Finally, in Appendix \ref{mathematica} we provide instructions for using the Mathematica notebook submitted as an ancillary file along with this paper. The Mathematica notebook allows one to compute the prepotential for $5d$ KK theories involving one or two nodes. Combining these results, one can obtain the prepotential for any $5d$ KK theory. The notebook also converts the prepotential into triple intersection numbers for the associated geometry and displays these intersection numbers in a graphical form. 

\section{Structure of $6d$ SCFTs}\label{re}

\begin{table}[htbp]
\begin{center}
\begin{tabular}{|c|c|l|} \hline
\raisebox{-.4\height}{\begin{tikzpicture}
\node at (-0.5,0.5) {$\Omega^{ii}$};
\node at (-0.55,0.9) {$\fg_i$};
\end{tikzpicture}} &Comments& Hypermultiplet content 
\\ \hline
 \hline
\raisebox{-.4\height}{ \begin{tikzpicture}
\node at (-0.5,0.45) {1};
\node at (-0.45,0.9) {$\sp(n)_\theta$};
\end{tikzpicture}}&$\theta=0,\pi$&$(2n+8)\F$
\\ \hline
\raisebox{-.4\height}{ \begin{tikzpicture}
\node at (-0.5,0.45) {1};
\node at (-0.45,0.9) {$\su(n)$};
\end{tikzpicture}}&$n\ge3$&$(n+8)\F+\Asym$
\\ \hline
\raisebox{-.4\height}{ \begin{tikzpicture}
\node at (-0.5,0.45) {$1$};
\node at (-0.45,0.9) {$\su(\wh n)$};
\end{tikzpicture}}&$n\ge8$; frozen; non-geometric&$(n-8)\F+\Sym$
\\ \hline
\raisebox{-.4\height}{ \begin{tikzpicture}
\node at (-0.5,0.45) {1};
\node at (-0.45,0.9) {$\su(\tilde 6)$};
\end{tikzpicture}}&&$15\F+\half\tAsym$
\\ \hline
\raisebox{-.4\height}{ \begin{tikzpicture}
\node at (-0.5,0.45) {2};
\node at (-0.45,0.9) {$\su(n)$};
\end{tikzpicture}}&&$2n\F$
\\ \hline
\raisebox{-.4\height}{ \begin{tikzpicture}
\node at (-0.5,0.45) {3};
\node at (-0.45,0.9) {$\su(3)$};
\end{tikzpicture}}&&
\\ \hline
\raisebox{-.4\height}{ \begin{tikzpicture}
\node at (-0.5,0.45) {4};
\node at (-0.45,0.9) {$\so(n)$};
\end{tikzpicture}}&$n\ge8$&$(n-8)\F$
\\ \hline
\raisebox{-.4\height}{ \begin{tikzpicture}
\node at (-0.5,0.45) {$k$};
\node at (-0.45,0.9) {$\so(8)$};
\end{tikzpicture}}&$1\le k\le3$&$(4-k)\F+(4-k)\S+(4-k)\C$
\\ \hline
\raisebox{-.4\height}{ \begin{tikzpicture}
\node at (-0.5,0.45) {$k$};
\node at (-0.45,0.9) {$\so(n)$};
\end{tikzpicture}}&$1\le k\le3$; $7\le n\le12,n\neq8$&$(n-4-k)\F+2^{\lceil\frac{9-n}{2}\rceil}(4-k)\S$
\\ \hline
\raisebox{-.4\height}{ \begin{tikzpicture}
\node at (-0.5,0.45) {$k$};
\node at (-0.45,0.9) {$\so(\wh{12})$};
\end{tikzpicture}}&$k=1,2$&$(8-k)\F+\half(3-k)\S+\half\C$
\\ \hline
\raisebox{-.4\height}{ \begin{tikzpicture}
\node at (-0.5,0.45) {2};
\node at (-0.45,0.9) {$\so(13)$};
\end{tikzpicture}}&&$7\F+\half\S$
\\ \hline
\raisebox{-.4\height}{ \begin{tikzpicture}
\node at (-0.5,0.45) {$k$};
\node at (-0.45,0.9) {$\fg_2$};
\end{tikzpicture}}&$1\le k\le3$&$(10-3k)\F$
\\ \hline
\raisebox{-.4\height}{ \begin{tikzpicture}
\node at (-0.5,0.45) {$k$};
\node at (-0.45,0.9) {$\ff_4$};
\end{tikzpicture}}&$1\le k\le5$&$(5-k)\F$
\\ \hline
\raisebox{-.4\height}{ \begin{tikzpicture}
\node at (-0.5,0.45) {$k$};
\node at (-0.45,0.9) {$\fe_6$};
\end{tikzpicture}}&$1\le k\le6$&$(6-k)\F$
\\ \hline
\raisebox{-.4\height}{ \begin{tikzpicture}
\node at (-0.5,0.45) {$k$};
\node at (-0.45,0.9) {$\fe_7$};
\end{tikzpicture}}&$1\le k\le8$&$\half(8-k)\F$
\\ \hline
\raisebox{-.4\height}{ \begin{tikzpicture}
\node at (-0.5,0.45) {$12$};
\node at (-0.5,0.9) {$\fe_8$};
\end{tikzpicture}}&&
\\ \hline
\end{tabular}
\end{center}
\caption{List of all the possible nodes with non-trivial $\fg_i$ appearing in graphs associated to $6d$ SCFTs. A hat or a tilde distinguishes different nodes having same values of $\Omega^{ii}$ and $\fg_i$.}	\label{TR1}
\end{table}

In this section, we review the fact that $6d$ SCFTs are characterized by graphs that are analogous to Dynkin graphs associated to simply laced Lie algebras. In the next section, we will show that $5d$ KK theories are also characterized by similar graphs that are instead analogous to Dynkin graphs associated to general (i.e. both simply laced and non-simply laced) Lie algebras.

\begin{table}[htbp]
\begin{center}
\begin{tabular}{|c|c|c|} \hline
\raisebox{-.4\height}{\begin{tikzpicture}
\node at (-0.5,0.5) {$\Omega^{ii}$};
\node at (-0.55,0.9) {$\fg_i$};
\end{tikzpicture}}& Comments&Flavor symmetry algebra, $\ff$\\ \hline
 \hline
\raisebox{-.4\height}{ \begin{tikzpicture}
\node at (-0.5,0.45) {1};
\node at (-0.45,0.9) {$\sp(0)_\theta$};
\end{tikzpicture}}&$\theta=0,\pi$&$\fe_8$\\ \hline
\raisebox{-.4\height}{ \begin{tikzpicture}
\node at (-0.5,0.45) {2};
\node at (-0.45,0.9) {$\su(1)$};
\end{tikzpicture}}&&$\su(2)$\\ \hline
\end{tabular}
\end{center}
\caption{List of all the possible nodes with trivial $\fg_i$ that can appear in graphs associated to $6d$ SCFTs. If $\Omega^{ii}=2$, we refer to the trivial gauge algebra as $\su(1)$ and if $\Omega^{ii}=1$, we refer to the trivial gauge algebra as $\sp(0)$. In the latter case, sometimes a $\Z_2$ valued theta angle is physically relevant. We also list the flavor symmetry algebra $\ff$ for each case. The sum of gauge algebras neighboring each such node must be contained inside the corresponding $\ff$.}	\label{R1N}
\end{table}

The low-energy theory on the tensor branch of a $6d$ SCFT $\fT$ can be organized in terms of tensor multiplets $B_i$. There is a gauge algebra $\fg_i$ associated to each $i$ where $\fg_i$ can either be a simple or a trivial algebra. Each tensor multiplet $B_i$ is also associated to a ``fundamental'' BPS string excitation $S^i$ such that the charge of $S^i$ under $B_j$ is the Kronecker delta $\delta^i_j$. The Dirac pairing $\Omega^{ij}$ between $S^i$ and $S^j$ appears in the Green-Schwarz term in the Lagrangian
\be \label{GS}
\Omega^{ij}B_i\wedge\tr(F_j^2)
\ee
where $F_j$ is the field strength for $\fg_j$ if $\fg_j$ is simple and $F_j=0$ if $\fg_j$ is trivial.

\begin{table}[htbp]
\begin{center}
\begin{tabular}{|c|c|l|} \hline
\raisebox{-.4\height}{\begin{tikzpicture}
\node (v1) at (-0.5,0.5) {$\Omega^{ii}$};
\node at (-0.55,0.9) {$\fg_i$};
\begin{scope}[shift={(1.5,0)}]
\node (v2) at (-0.5,0.5) {$\Omega^{jj}$};
\node at (-0.55,0.9) {$\fg_j$};
\end{scope}
\draw  (v1) edge (v2);
\end{tikzpicture}} &Comments& Mixed hyper content
\\ \hline
\hline
 \raisebox{-.4\height}{\begin{tikzpicture}
\node (v1) at (-0.5,0.45) {1};
\node at (-0.45,0.9) {$\sp(n_i)$};
\begin{scope}[shift={(1.5,0)}]
\node (v2) at (-0.5,0.45) {2};
\node at (-0.45,0.9) {$\su(n_j)$};
\end{scope}
\draw  (v1) edge (v2);
\end{tikzpicture}}&$n_i\le n_j$; $n_j\le2n_i+7$&$\F\otimes\F$
\\ 
 \raisebox{-.4\height}{\begin{tikzpicture}
\node (v1) at (-0.5,0.45) {1};
\node at (-0.45,0.9) {$\sp(n_i)_\theta$};
\begin{scope}[shift={(1.5,0)}]
\node (v2) at (-0.5,0.45) {2};
\node at (-0.45,0.9) {$\su(n_j)$};
\end{scope}
\draw  (v1) edge (v2);
\end{tikzpicture}}&$n_j=2n_i+8$; $\theta=0,\pi$&$\F\otimes\F$
\\ 
\raisebox{-.4\height}{ \begin{tikzpicture}
\node (v1) at (-0.5,0.45) {1};
\node at (-0.45,0.9) {$\sp(n_i)$};
\begin{scope}[shift={(1.5,0)}]
\node (v2) at (-0.5,0.45) {$k$};
\node at (-0.45,0.9) {$\so(n_j)$};
\end{scope}
\draw  (v1) edge (v2);
\end{tikzpicture}}&$n_i\le n_j-4-k$; $n_j\le4n_i+16$; $2\le k\le4$&$\half(\F\otimes\F)$
\\ 
\raisebox{-.4\height}{ \begin{tikzpicture}
\node (v1) at (-0.5,0.45) {1};
\node at (-0.45,0.9) {$\sp(n_i)$};
\begin{scope}[shift={(1.5,0)}]
\node (v2) at (-0.5,0.45) {$2$};
\node at (-0.45,0.9) {$\so(\wh{12})$};
\end{scope}
\draw  (v1) edge (v2);
\end{tikzpicture}}&$n_i\le 6$&$\half(\F\otimes\F)$
\\ 
\raisebox{-.4\height}{ \begin{tikzpicture}
\node (v1) at (-0.5,0.45) {1};
\node at (-0.45,0.9) {$\sp(n_i)$};
\begin{scope}[shift={(1.5,0)}]
\node (v2) at (-0.5,0.45) {$k$};
\node at (-0.45,0.9) {$\so(8)$};
\end{scope}
\draw  [dashed](v1) -- (v2);
\end{tikzpicture}}&$n_i\le 4-k$; $k\le3$&$\half(\F\otimes\S)$
\\ 
\raisebox{-.4\height}{ \begin{tikzpicture}
\node (v1) at (-0.5,0.45) {1};
\node at (-0.45,0.9) {$\sp(n_i)$};
\begin{scope}[shift={(1.5,0)}]
\node (v2) at (-0.5,0.45) {$k$};
\node at (-0.45,0.9) {$\so(7)$};
\end{scope}
\draw  [dashed](v1) -- (v2);
\end{tikzpicture}}&$n_i\le 8-2k$; $k=2,3$&$\half(\F\otimes\S)$
\\ 
\raisebox{-.4\height}{ \begin{tikzpicture}
\node (v1) at (-0.5,0.45) {1};
\node at (-0.45,0.9) {$\sp(n_i)$};
\begin{scope}[shift={(1.5,0)}]
\node (v2) at (-0.5,0.45) {$k$};
\node at (-0.45,0.9) {$\fg_2$};
\end{scope}
\draw  (v1) edge (v2);
\end{tikzpicture}\:\:\:\:}&$n_i\le 10-3k$; $k=2,3$&$\half(\F\otimes\F)$
\\ 
\raisebox{-.4\height}{ \begin{tikzpicture}
\node (v1) at (-0.5,0.45) {1};
\node at (-0.45,0.9) {$\su(n_i)$};
\begin{scope}[shift={(1.5,0)}]
\node (v2) at (-0.5,0.45) {2};
\node at (-0.45,0.9) {$\su(n_j)$};
\end{scope}
\draw  (v1) edge (v2);
\end{tikzpicture}}&$n_i\le 2n_j$; $n_j\le n_i+8$&$\F\otimes\F$
\\ 
\raisebox{-.4\height}{ \begin{tikzpicture}
\node (v1) at (-0.5,0.45) {1};
\node at (-0.45,0.9) {$\su(\wh{n}_i)$};
\begin{scope}[shift={(1.5,0)}]
\node (v2) at (-0.5,0.45) {2};
\node at (-0.45,0.9) {$\su(n_j)$};
\end{scope}
\draw  (v1) edge (v2);
\end{tikzpicture}}&$n_i\le 2n_j$; $n_j\le n_i-8$&$\F\otimes\F$
\\ 
\raisebox{-.4\height}{ \begin{tikzpicture}
\node (v1) at (-0.5,0.45) {1};
\node at (-0.45,0.9) {$\su(\tilde{6})$};
\begin{scope}[shift={(1.5,0)}]
\node (v2) at (-0.5,0.45) {2};
\node at (-0.45,0.9) {$\su(n_j)$};
\end{scope}
\draw  (v1) edge (v2);
\end{tikzpicture}}&$3\le n_j\le 15$&$\F\otimes\F$
\\ 
\raisebox{-.4\height}{ \begin{tikzpicture}
\node (v1) at (-0.5,0.45) {2};
\node at (-0.45,0.9) {$\su(n_i)$};
\begin{scope}[shift={(1.5,0)}]
\node (v2) at (-0.5,0.45) {2};
\node at (-0.45,0.9) {$\su(n_j)$};
\end{scope}
\draw  (v1) edge (v2);
\end{tikzpicture}}&$n_i\le 2n_j$; $n_j\le 2n_i$&$\F\otimes\F$
\\ 
\raisebox{-.4\height}{ \begin{tikzpicture}
\node (v1) at (-0.5,0.45) {2};
\node at (-0.45,0.9) {$\su(n_i)$};
\begin{scope}[shift={(1.5,0)}]
\node (v2) at (-0.5,0.45) {4};
\node at (-0.45,0.9) {$\so(n_j)$};
\end{scope}
\node (v3) at (0.25,0.45) {\tiny{2}};
\draw (v1)--(v3);
\draw (v2)--(v3);
\end{tikzpicture}}&$n_i\le n_j-8$; $n_j\le 2n_i$; frozen&$\F\otimes\F$
\\ 
\raisebox{-.4\height}{ \begin{tikzpicture}
\node (v1) at (-0.5,0.45) {2};
\node at (-0.45,0.9) {$\su(2)$};
\begin{scope}[shift={(1.5,0)}]
\node (v2) at (-0.5,0.45) {$k$};
\node at (-0.45,0.9) {$\so(7)$};
\end{scope}
\draw[dashed]  (v1) edge (v2);
\end{tikzpicture}}&$1\le k\le3$&$\half(\F\otimes\S)$
\\ 
\raisebox{-.4\height}{ \begin{tikzpicture}
\node (v1) at (-0.5,0.45) {2};
\node at (-0.45,0.9) {$\su(2)$};
\begin{scope}[shift={(1.5,0)}]
\node (v2) at (-0.5,0.45) {$k$};
\node at (-0.45,0.9) {$\fg_2$};
\end{scope}
\draw  (v1) edge (v2);
\end{tikzpicture}\:\:\:\:}&$1\le k\le3$&$\half(\F\otimes\F)$
\\ \hline
\end{tabular}
\end{center}
\caption{List of all the possible edges between two gauge-theoretic nodes that can appear in graphs characterizing $6d$ SCFTs. An edge with $2$ in the middle of it denotes the fact that there are two edges between the two asocciated nodes. Solid edges denote matter in bifundamental and dashed edges denote matter in $\F\otimes \S$. The theta angle of $\sp(n)$ is only displayed when it is physically relevant.}\label{TR2}
\end{table}

\begin{table}[htbp]
\begin{center}
\begin{tabular}{|c|c|l|} \hline
\raisebox{-.15\height}{\begin{tikzpicture}
\node (v1) at (-0.5,0.5) {$\Omega^{ii}$};
\node at (-0.55,0.9) {$\fg_i$};
\begin{scope}[shift={(1.5,0)}]
\node (v2) at (-0.5,0.5) {$\Omega^{jj}$};
\node at (-0.55,0.9) {$\fg_j$};
\end{scope}
\draw  (v1) edge (v2);
\end{tikzpicture}} &Comments& Mixed hyper content
\\ \hline
\hline
\raisebox{-.15\height}{\begin{tikzpicture}
\node (v1) at (-0.5,0.45) {1};
\node at (-0.45,0.9) {$\sp(0)$};
\begin{scope}[shift={(1.5,0)}]
\node (v2) at (-0.5,0.45) {2};
\node at (-0.45,0.9) {$\su(n)$};
\end{scope}
\draw  (v1) edge (v2);
\end{tikzpicture}}&$n\le 9, n\neq8$&
\\ 
\raisebox{-.15\height}{\begin{tikzpicture}
\node (v1) at (-0.5,0.45) {1};
\node at (-0.45,0.9) {$\sp(0)_\theta$};
\begin{scope}[shift={(1.5,0)}]
\node (v2) at (-0.5,0.45) {2};
\node at (-0.45,0.9) {$\su(n)$};
\end{scope}
\draw  (v1) edge (v2);
\end{tikzpicture}}&$n=8$; $\theta=0,\pi$&
\\ 
\raisebox{-.15\height}{\begin{tikzpicture}
\node (v1) at (-0.5,0.45) {1};
\node at (-0.45,0.9) {$\sp(0)$};
\begin{scope}[shift={(1.5,0)}]
\node (v2) at (-0.5,0.45) {3};
\node at (-0.45,0.9) {$\su(3)$};
\end{scope}
\draw  (v1) edge (v2);
\end{tikzpicture}}&&
\\ 
\raisebox{-.15\height}{ \begin{tikzpicture}
\node (v1) at (-0.5,0.45) {1};
\node at (-0.45,0.9) {$\sp(0)$};
\begin{scope}[shift={(1.5,0)}]
\node (v2) at (-0.5,0.45) {$k$};
\node at (-0.45,0.9) {$\so(n)$};
\end{scope}
\draw  (v1) edge (v2);
\end{tikzpicture}}&$n\le16$; $2\le k\le 4$&
\\ 
\raisebox{-.15\height}{ \begin{tikzpicture}
\node (v1) at (-0.5,0.45) {1};
\node at (-0.45,0.9) {$\sp(0)$};
\begin{scope}[shift={(1.5,0)}]
\node (v2) at (-0.5,0.45) {$2$};
\node at (-0.45,0.9) {$\so(\wh{12})$};
\end{scope}
\draw  (v1) edge (v2);
\end{tikzpicture}}&&
\\ 
\raisebox{-.15\height}{ \begin{tikzpicture}
\node (v1) at (-0.5,0.45) {1};
\node at (-0.45,0.9) {$\sp(0)$};
\begin{scope}[shift={(1.5,0)}]
\node (v2) at (-0.5,0.45) {$k$};
\node at (-0.45,0.9) {$\fg_2$};
\end{scope}
\draw  (v1) edge (v2);
\end{tikzpicture}\:\:\:\:}&$k=2,3$&
\\ 
\raisebox{-.15\height}{ \begin{tikzpicture}
\node (v1) at (-0.5,0.45) {1};
\node at (-0.45,0.9) {$\sp(0)$};
\begin{scope}[shift={(1.5,0)}]
\node (v2) at (-0.5,0.45) {$k$};
\node at (-0.45,0.9) {$\ff_4$};
\end{scope}
\draw  (v1) edge (v2);
\end{tikzpicture}\:\:\:\:}&$2\le k\le 5$&
\\ 
\raisebox{-.15\height}{ \begin{tikzpicture}
\node (v1) at (-0.5,0.45) {1};
\node at (-0.45,0.9) {$\sp(0)$};
\begin{scope}[shift={(1.5,0)}]
\node (v2) at (-0.5,0.45) {$k$};
\node at (-0.45,0.9) {$\fe_6$};
\end{scope}
\draw  (v1) edge (v2);
\end{tikzpicture}\:\:\:\:}&$2\le k\le 6$&
\\ 
\raisebox{-.15\height}{ \begin{tikzpicture}
\node (v1) at (-0.5,0.45) {1};
\node at (-0.45,0.9) {$\sp(0)$};
\begin{scope}[shift={(1.5,0)}]
\node (v2) at (-0.5,0.45) {$k$};
\node at (-0.45,0.9) {$\fe_7$};
\end{scope}
\draw  (v1) edge (v2);
\end{tikzpicture}\:\:\:\:}&$2\le k\le 8$&
\\ 
\raisebox{-.15\height}{ \begin{tikzpicture}
\node (v1) at (-0.5,0.45) {1};
\node at (-0.45,0.9) {$\sp(0)$};
\begin{scope}[shift={(1.5,0)}]
\node (v2) at (-0.5,0.45) {$12$};
\node at (-0.45,0.9) {$\fe_8$};
\end{scope}
\draw  (v1) edge (v2);
\end{tikzpicture}\:\:\:\:}&&
\\ \hline
\raisebox{-.15\height}{\begin{tikzpicture}
\node (v1) at (-0.5,0.45) {2};
\node at (-0.45,0.9) {$\su(1)$};
\begin{scope}[shift={(1.5,0)}]
\node (v2) at (-0.5,0.45) {1};
\node at (-0.45,0.9) {$\sp(1)$};
\end{scope}
\draw  (v1) edge (v2);
\end{tikzpicture}}&&$\half\F$ in $\fg_j=\sp(1)$
\\ 
\raisebox{-.15\height}{\begin{tikzpicture}
\node (v1) at (-0.5,0.45) {2};
\node at (-0.45,0.9) {$\su(1)$};
\begin{scope}[shift={(1.5,0)}]
\node (v2) at (-0.5,0.45) {2};
\node at (-0.45,0.9) {$\su(2)$};
\end{scope}
\draw  (v1) edge (v2);
\end{tikzpicture}}&&$\half\F$ in $\fg_j=\su(2)$
\\ \hline
\end{tabular}
\end{center}
\caption{List of all the possible edges between a gauge-theoretic and a non-gauge-theoretic node that can appear in graphs characterizing $6d$ SCFTs. The theta angle of $\sp(0)$ is only displayed when it is physically relevant.}\label{R2P}
\end{table}

\begin{table}[htbp]
\begin{center}
\begin{tabular}{|c|} \hline
\raisebox{-.15\height}{\begin{tikzpicture}
\node (v1) at (-0.5,0.5) {$\Omega^{ii}$};
\node at (-0.55,0.9) {$\fg_i$};
\begin{scope}[shift={(1.5,0)}]
\node (v2) at (-0.5,0.5) {$\Omega^{jj}$};
\node at (-0.55,0.9) {$\fg_j$};
\end{scope}
\draw  (v1) edge (v2);
\end{tikzpicture}}
\\ \hline
\hline
\raisebox{-.15\height}{\begin{tikzpicture}
\node (v1) at (-0.5,0.45) {1};
\node at (-0.45,0.9) {$\sp(0)$};
\begin{scope}[shift={(1.5,0)}]
\node (v2) at (-0.5,0.45) {2};
\node at (-0.45,0.9) {$\su(1)$};
\end{scope}
\draw  (v1) edge (v2);
\end{tikzpicture}}
\\ \hline
\raisebox{-.15\height}{\begin{tikzpicture}
\node (v1) at (-0.5,0.45) {2};
\node at (-0.45,0.9) {$\su(1)$};
\begin{scope}[shift={(1.5,0)}]
\node (v2) at (-0.5,0.45) {2};
\node at (-0.45,0.9) {$\su(1)$};
\end{scope}
\draw  (v1) edge (v2);
\end{tikzpicture}}
\\ \hline
\end{tabular}
\end{center}
\caption{List of all the possible edges between two non-gauge-theoretic nodes that can appear in graphs characterizing $6d$ SCFTs. The theta angle of $\sp(0)$ is not displayed since it is not physically relevant.}\label{R2N}
\end{table}

\begin{table}[htbp]
\begin{center}
\begin{tabular}{|c|l|} \hline
\raisebox{-.15\height}{\begin{tikzpicture}
\node (v1) at (-0.5,0.5) {$\Omega^{ii}$};
\node at (-0.55,0.9) {$\fg_i$};
\begin{scope}[shift={(1.5,0)}]
\node (v2) at (-0.5,0.5) {$1$};
\node at (-0.55,0.9) {$\sp(0)$};
\end{scope}
\begin{scope}[shift={(3,0)}]
\node (v3) at (-0.5,0.5) {$\Omega^{kk}$};
\node at (-0.55,0.9) {$\fg_k$};
\end{scope}
\draw  (v1) edge (v2);
\draw  (v2) edge (v3);
\end{tikzpicture}}&Comments
\\ \hline
\hline
\raisebox{-.15\height}{\begin{tikzpicture}
\node (v1) at (-0.5,0.5) {$2$};
\node at (-0.55,0.9) {$\su(2)$};
\begin{scope}[shift={(1.5,0)}]
\node (v2) at (-0.5,0.5) {$1$};
\node at (-0.55,0.9) {$\sp(0)$};
\end{scope}
\begin{scope}[shift={(3,0)}]
\node (v3) at (-0.5,0.5) {$k$};
\node at (-0.55,0.9) {$\fg$};
\end{scope}
\draw  (v1) edge (v2);
\draw  (v2) edge (v3);
\end{tikzpicture}\:\:\:\:}&$k\ge3$; $\fg=\fe_7,\fe_6,\ff_4,\fg_2,\so(n\le12)$
\\ 
\raisebox{-.15\height}{\begin{tikzpicture}
\node (v1) at (-0.5,0.5) {$k$};
\node at (-0.55,0.9) {$\su(3)$};
\begin{scope}[shift={(1.5,0)}]
\node (v2) at (-0.5,0.5) {$1$};
\node at (-0.55,0.9) {$\sp(0)$};
\end{scope}
\begin{scope}[shift={(3,0)}]
\node (v3) at (-0.5,0.5) {$l$};
\node at (-0.55,0.9) {$\fg$};
\end{scope}
\draw  (v1) edge (v2);
\draw  (v2) edge (v3);
\end{tikzpicture}\:\:\:\:}&$k,l\ge2$; $k+l\ge5$; $\fg=\fe_6,\ff_4,\fg_2,\so(n\le10),\su(n\le6)$
\\ 
\raisebox{-.15\height}{\begin{tikzpicture}
\node (v1) at (-0.5,0.5) {$2$};
\node at (-0.55,0.9) {$\su(4)$};
\begin{scope}[shift={(1.5,0)}]
\node (v2) at (-0.5,0.5) {$1$};
\node at (-0.55,0.9) {$\sp(0)$};
\end{scope}
\begin{scope}[shift={(3,0)}]
\node (v3) at (-0.5,0.5) {$k$};
\node at (-0.55,0.9) {$\fg$};
\end{scope}
\draw  (v1) edge (v2);
\draw  (v2) edge (v3);
\end{tikzpicture}\:\:\:\:}&$k=3,4$; $\fg=\fg_2,\so(n\le10)$
\\ 
\raisebox{-.15\height}{\begin{tikzpicture}
\node (v1) at (-0.5,0.5) {$k$};
\node at (-0.55,0.9) {$\so(7)$};
\begin{scope}[shift={(1.5,0)}]
\node (v2) at (-0.5,0.5) {$1$};
\node at (-0.55,0.9) {$\sp(0)$};
\end{scope}
\begin{scope}[shift={(3,0)}]
\node (v3) at (-0.5,0.5) {$l$};
\node at (-0.55,0.9) {$\fg$};
\end{scope}
\draw  (v1) edge (v2);
\draw  (v2) edge (v3);
\end{tikzpicture}\:\:\:\:}&$k,l\ge2$; $k+l\ge5$; $\fg=\fg_2,\so(n\le9)$
\\ 
\raisebox{-.15\height}{\begin{tikzpicture}
\node (v1) at (-0.5,0.5) {$k$};
\node at (-0.55,0.9) {$\so(8)$};
\begin{scope}[shift={(1.5,0)}]
\node (v2) at (-0.5,0.5) {$1$};
\node at (-0.55,0.9) {$\sp(0)$};
\end{scope}
\begin{scope}[shift={(3,0)}]
\node (v3) at (-0.5,0.5) {$l$};
\node at (-0.55,0.9) {$\fg$};
\end{scope}
\draw  (v1) edge (v2);
\draw  (v2) edge (v3);
\end{tikzpicture}\:\:\:\:}&$k,l\ge2$; $k+l\ge5$; $\fg=\fg_2,\so(8)$
\\ 
\raisebox{-.15\height}{\begin{tikzpicture}
\node (v1) at (-0.5,0.5) {$k$};
\node at (-0.55,0.9) {$\so(9)$};
\begin{scope}[shift={(1.5,0)}]
\node (v2) at (-0.5,0.5) {$1$};
\node at (-0.55,0.9) {$\sp(0)$};
\end{scope}
\begin{scope}[shift={(3,0)}]
\node (v3) at (-0.5,0.5) {$l$};
\node at (-0.55,0.9) {$\fg$};
\end{scope}
\draw  (v1) edge (v2);
\draw  (v2) edge (v3);
\end{tikzpicture}\:\:\:\:}&$k,l\ge2$; $k+l\ge5$; $\fg=\fg_2$
\\ 
\raisebox{-.15\height}{\begin{tikzpicture}
\node (v1) at (-0.5,0.5) {$k$};
\node at (-0.55,0.9) {$\fg_2$};
\begin{scope}[shift={(1.5,0)}]
\node (v2) at (-0.5,0.5) {$1$};
\node at (-0.55,0.9) {$\sp(0)$};
\end{scope}
\begin{scope}[shift={(3,0)}]
\node (v3) at (-0.5,0.5) {$l$};
\node at (-0.55,0.9) {$\fg$};
\end{scope}
\draw  (v1) edge (v2);
\draw  (v2) edge (v3);
\end{tikzpicture}\:\:\:\:}&$k,l\ge2$; $k+l\ge5$; $\fg=\ff_4,\fg_2$
\\ \hline
\end{tabular}
\end{center}
\caption{List of all the possibilities for multiple neighbors of $\sp(0)$.}\label{R3E}
\end{table}

$[\Omega^{ij}]$ is a symmetric, positive definite matrix with all of its entries valued in integers. Thus, it is analogous to the Cartan matrix for a simply laced Lie algebra. The only possible values for off-diagonal entries are $\Omega^{ij}=0,-1,-2$. We note that $\Omega^{ij}=-2$ is only possible for $6d$ SCFTs arising from the frozen phase of F-theory \cite{Bhardwaj:2019hhd,Bhardwaj:2018jgp}. 

We can thus display the data of a $6d$ SCFT in terms of an associated graph $\Sigma_\fT$ that is constructed as follows:
\bit
\item \ubf{Nodes}: For each tensor multiplet $B_i$, we place a node $i$ with value \raisebox{-.125\height}{ \begin{tikzpicture}
\node at (-0.5,0.45) {$\Omega^{ii}$};
\node at (-0.45,0.9) {$\fg_i$};
\end{tikzpicture}}. All such possibilities are listed in Table \ref{TR1} when $\fg_i$ is non-trivial, and in Table \ref{R1N} when $\fg_i$ is trivial. In the former case, each node contributes hypers charged under a representation $\cR_i$ of $\fg_i$ where $\cR_i$ is shown in Table \ref{TR1}. In the latter case, for the node with $\fg_i=\sp(0)$, an important role is played by the adjoint representation of $\fe_8$, which is formed by the BPS string excitations associated to this node.

We note that the node \raisebox{-.125\height}{ \begin{tikzpicture}
\node at (-0.5,0.45) {$1$};
\node at (-0.45,0.9) {$\su(\wh n)$};
\end{tikzpicture}} only arises in the frozen phase of F-theory.

In the case of $\Omega^{ii}=1$ and $\fg_i=\sp(n)$, there is a possibility of a $\Z_2$ valued $6d$ theta angle which is physically relevant (in the context of $6d$ SCFTs) only when the $2n+8$ hypers in fundamental are gauged by a neighboring $\su(2n+8)$ gauge algebra. For $\fg_i=\sp(0)$, the theta angle is physically relevant (in the context of $6d$ SCFTs) only if there is a neighboring $\su(8)$ gauge algebra \cite{Mekareeya:2017jgc}. This can be understood in terms of two different embeddings of $\su(8)$ into $\fe_8$ (both having embedding index one), so that the adjoint of $\fe_8$ decomposes differently in the two cases, leading to different spectrum of string excitations.

In the case of $\Omega^{ii}=1$ and $\fg_i=\su(6)$, there are two possible choices of matter content. We distinguish the non-standard choice of matter content by denoting the corresponding $\fg_i$ as $\su(\tilde 6)$. 

In the case of $\fg_i=\so(12)$, the two spinor representations $\S$ and $\C$ are not conjugate to each other but have same contributions to the anomaly polynomial. The total number of hypers in the two spinor representations is fixed by the value of $\Omega^{ii}$. But since the two spinor representations are not conjugate, the relative distribution of hypers between the two makes a difference. For $\Omega^{ii}=1,2$, we can obtain two inequivalent theories in this way (note that the existence of two inequivalent theories with $\mathfrak{so}(12)$ gauge symmetry was pointed out in \cite{DelZotto:2018tcj}.) The version containing both $\S$ and $\C$ is distinguished from the one contataining only $\S$ by denoting its $\fg_i$ as $\so(\wh{12})$.
\item \ubf{Edges}: Consider two nodes $i$ and $j$ whose values are \raisebox{-.125\height}{ \begin{tikzpicture}
\node at (-0.5,0.45) {$\Omega^{ii}$};
\node at (-0.45,0.9) {$\fg_i$};
\end{tikzpicture}} and \raisebox{-.125\height}{ \begin{tikzpicture}
\node at (-0.5,0.45) {$\Omega^{jj}$};
\node at (-0.45,0.9) {$\fg_j$};
\end{tikzpicture}} respectively. We place $-\Omega^{ij}$ number of edges between $i$ and $j$. For instance, if $\Omega^{ij}=-1$, then we display this as
\be
\begin{tikzpicture}
\node (v1) at (-0.5,0.45) {$\Omega^{ii}$};
\node at (-0.45,0.9) {$\fg_i$};
\begin{scope}[shift={(2,0)}]
\node (v2) at (-0.5,0.45) {$\Omega^{jj}$};
\node at (-0.45,0.9) {$\fg_j$};
\end{scope}
\draw  (v1) -- (v2);
\end{tikzpicture}
\ee
and, if $\Omega^{ij}=-2$, then we display this as
\be
\begin{tikzpicture}
\node (v1) at (-0.5,0.45) {$\Omega^{ii}$};
\node at (-0.45,0.9) {$\fg_i$};
\begin{scope}[shift={(2,0)}]
\node (v2) at (-0.5,0.45) {$\Omega^{jj}$};
\node at (-0.45,0.9) {$\fg_j$};
\end{scope}
\node (v3) at (0.5,0.45) {\tiny{2}};
\draw (v1)--(v3);
\draw (v2)--(v3);
\end{tikzpicture}
\ee
There are no edges between nodes $i$ and $j$ if $\Omega^{ij}=0$. All the possible edges are listed in Table \ref{TR2} when both $\fg_i$ and $\fg_j$ are non-trivial, in Table \ref{R2P} when only one of $\fg_i$ and $\fg_j$ is non-trivial, and in Table \ref{R2N} when both $\fg_i$ and $\fg_j$ are trivial. 

Each edge corresponds to a hyper transforming in a mixed representation $\cR_{ij}=\cR_{ij,i}\otimes\cR_{ij,j}$ of $\fg_i\oplus\fg_j$ where $\cR_{ij,i}$ is a representation of $\fg_i$ and $\cR_{ij,j}$ is a representation of $\fg_j$. The possible $\cR_{ij}$ are shown in the third column of Table \ref{TR2}. Note that we must have $\oplus_j\cR_{ij,i}^{\oplus \text{dim}(\cR_{ij,j})}\subseteq\cR_i$ as representations of $\fg_i$ for each node $i$.

In the case of $\Omega^{ii}=1$, $\fg_i=\sp(n_i)$, $\Omega^{jj}=k$, $\fg_j=\so(7,8)$ and $\Omega^{ij}=-1$, there are two possible mixed representations $\half(\F\otimes\F)$ or $\half(\F\otimes\S)$. We distinguish the case $\half(\F\otimes\S)$ by denoting the corresponding edge as a dashed line. Notice that when $\fg_j=\so(8)$, the dashed edge is only physically relevant when it is a part of a configuration of form
\be
\begin{tikzpicture}
\node (v1) at (-0.5,0.45) {1};
\node at (-0.45,0.9) {$\sp(n_i)$};
\begin{scope}[shift={(1.5,0)}]
\node (v2) at (-0.5,0.45) {$k$};
\node at (-0.45,0.9) {$\so(8)$};
\end{scope}
\begin{scope}[shift={(3,0)}]
\node (v3) at (-0.5,0.45) {$1$};
\node at (-0.45,0.9) {$\sp(n_k)$};
\end{scope}
\draw [dashed]  (v1) -- (v2);
\draw  (v2) -- (v3);
\end{tikzpicture}
\ee
Otherwise, the dashed edge can be converted to the non-dashed edge by applying an outer-automorphism of $\so(8)$.

\item \ubf{Multiple neighbors of $\sp(0)$}: Consider a node $i$ with value \raisebox{-.125\height}{ \begin{tikzpicture}
\node at (-0.5,0.45) {1};
\node at (-0.45,0.9) {$\sp(0)$};
\end{tikzpicture}}. Related to the fact that the flavor symmetry algebra associated to this node is $\fe_8$, it can be shown that its neighbors must satisfy $\oplus_j \fg_j\subseteq \fe_8$ where only those $j$ are included in the sum for which $\Omega^{ij}=-1$. In fact all such subalgebras are realized except\footnote{It can be shown that the embedding index of each neighboring $\fg_j$ inside $\fe_8$ must be one. The only possible embedding of $\so(13)\oplus\su(2)$ into $\fe_8$ follows from first embedding $\so(13)\oplus\su(2)$ into $\so(16)$ as a special maximal subalgebra and then embedding $\so(16)$ into $\fe_8$ as a regular maximal subalgebra. The embedding index of the $\su(2)$ factor under this embedding is two rather than one, thus $\so(13)\oplus\su(2)$ cannot be realized as a neighbor of $\sp(0)$. The absence of $\so(13)\oplus\su(2)$ neighbor was first noticed in \cite{Frey:2018vpw}.} for $\so(13)\oplus\su(2)$.

In the context of $6d$ SCFTs, it is not possible for $\sp(0)$ to have more than two neighbors. We collect all the possibilities for multiple neighbors of $\sp(0)$ in Table \ref{R3E}.
\eit
Notice that the relationship between $\Sigma_\fT$ and $[\Omega^{ij}]$ is analogous to the relationship between Dynkin graph and Cartan matrix of a simply laced Lie algebra.

\section{Structure of $5d$ KK theories}\label{SKK}
\subsection{Twists}
Consider a QFT $\fT$ that admits a discrete global symmetry group $\Gamma$. When we compactify $\fT$ on a circle, we have the option of ``twisting'' $\fT$ around the circle. This means that we introduce a holonomy $\gamma\in\Gamma$ for the background gauge field corresponding to $\Gamma$. Note that the number of distinct twists is not given by the number of elements in $\Gamma$, but rather by the number of conjugacy classes in $\Gamma$. This is because two holonomies that are conjugate in $\Gamma$ are physically equivalent and thus lead to the same twist.

In this section, we will explore all the possible twists for $6d$ SCFTs. Each twist leads to a different $5d$ KK theory.

\subsection{Discrete symmetries from outer automorphisms}\label{DS1}
A general discrete symmetry of a $6d$ SCFT $\fT$ is generated by combining two kinds of basic discrete symmetries. We start by discussing the first kind of basic discrete symmetries. These arise from outer automorphisms of gauge algebras $\fg_i$.

$\su(n)$ for $n\ge3$, $\so(2m)$ for $m\ge4$ and $\fe_6$ admit an order two outer automorphism that we call $\cO^{(2)}$. It exchanges the roots in the following fashion
\be\nn
\begin{tikzpicture}
\begin{scope}[shift={(0,7)}]
\draw[fill=black]
(0,0) node (v1) {} circle [radius=0.1]
(1,0) node (v3) {} circle [radius=0.1]
(3,0) node (v5) {} circle [radius=0.1]
(4,0) node (v6) {} circle [radius=0.1]
(5,0) node (v7) {} circle [radius=0.1]
(7,0) node (v10) {} circle [radius=0.1]
(8,0) node (v11) {} circle [radius=0.1];
\node (v2) at (2,0) {$\cdots$};
\node (v8) at (6,0) {$\cdots$};
\draw  (v1) edge (v3);
\draw  (v3) edge (v2);
\draw  (v2) edge (v5);
\draw  (v5) edge (v6);
\draw  (v6) edge (v7);
\draw  (v7) edge (v8);
\draw  (v8) edge (v10);
\draw  (v10) edge (v11);
\draw[<->,red,thick] (v3) .. controls (1.4,-1.6) and (6.6,-1.6) .. (v10);
\draw[<->,red,thick] (v5) .. controls (3.2,-0.8) and (4.8,-0.8) .. (v7);
\draw[<->,red,thick] (v1) .. controls (0.4,-2.4) and (7.6,-2.4) .. (v11);
\node at (-2,0) {\ubf{$\su(2n)$, $\cO^{(2)}$}:};
\end{scope}
\end{tikzpicture}
\ee
\be\nn
\begin{tikzpicture}
\begin{scope}[shift={(0,7)}]
\draw[fill=black]
(0,0) node (v1) {} circle [radius=0.1]
(1,0) node (v3) {} circle [radius=0.1]
(3,0) node (v5) {} circle [radius=0.1]
(4,0) node (v7) {} circle [radius=0.1]
(6,0) node (v10) {} circle [radius=0.1]
(7,0) node (v11) {} circle [radius=0.1];
\node (v2) at (2,0) {$\cdots$};
\node (v8) at (5,0) {$\cdots$};
\draw  (v1) edge (v3);
\draw  (v3) edge (v2);
\draw  (v2) edge (v5);
\draw  (v5) edge (v7);
\draw  (v7) edge (v8);
\draw  (v8) edge (v10);
\draw  (v10) edge (v11);
\draw[<->,red,thick] (v3) .. controls (1.4,-1.2) and (5.6,-1.2) .. (v10);
\draw[<->,red,thick] (v5) .. controls (3,-0.6) and (4,-0.6) .. (v7);
\draw[<->,red,thick] (v1) .. controls (0.4,-1.8) and (6.6,-1.8) .. (v11);
\node at (-2,0) {\ubf{$\su(2n+1)$, $\cO^{(2)}$}:};
\end{scope}
\end{tikzpicture}
\ee
\be\nn
\begin{tikzpicture}
\begin{scope}[shift={(0,3)}]
\draw[fill=black]
(0,0) node (v1) {} circle [radius=0.1]
(1,0) node (v3) {} circle [radius=0.1]
(2,0) node (v4) {} circle [radius=0.1]
(4,0) node (v5) {} circle [radius=0.1]
(5,0) node (v6) {} circle [radius=0.1]
(6,-1) node (v7) {} circle [radius=0.1]
(6,1) node (v8) {} circle [radius=0.1];
\node (v2) at (3,0) {$\cdots$};
\draw  (v1) edge (v3);
\draw  (v3) edge (v4);
\draw  (v4) edge (v2);
\draw  (v2) edge (v5);
\draw  (v5) edge (v6);
\draw  (v6) edge (v7);
\draw  (v8) edge (v6);
\draw[<->,red,thick] (v8) .. controls (6.8,0.4) and (6.8,-0.4) .. (v7);
\node at (-2,0) {\ubf{$\so(2n)$, $\cO^{(2)}$}:};
\end{scope}
\end{tikzpicture}
\ee
\be\nn
\begin{tikzpicture}
\begin{scope}[shift={(0,-0.5)}]
\draw[fill=black]
(0,0) node (v1) {} circle [radius=0.1]
(1,0) node (v3) {} circle [radius=0.1]
(2,0) node (v4) {} circle [radius=0.1]
(4,0) node (v5) {} circle [radius=0.1]
(3,0) node (v7) {} circle [radius=0.1]
(2,1) node (v8) {} circle [radius=0.1];
\node (v2) at (3,0) {};
\draw  (v1) edge (v3);
\draw  (v3) edge (v4);
\draw  (v4) edge (v2);
\draw  (v2) edge (v5);
\draw  (v4) edge (v7);
\draw  (v4) edge (v8);
\draw[<->,red,thick] (v3) .. controls (1.2,-0.8) and (2.8,-0.8) .. (v2);
\draw[<->,red,thick] (v1) .. controls (0.4,-1.6) and (3.6,-1.6) .. (v5);
\node at (-2,0) {\ubf{$\fe_6$, $\cO^{(2)}$}:};
\end{scope}
\end{tikzpicture}
\ee
$\so(8)$ also admits an order three outer automorphism which we call $\cO^{(3)}$. It cyclically permutes the roots as shown below
\be\nn
\begin{tikzpicture}
\begin{scope}[]
\draw[fill=black]
(-1.5,0) node (v1) {} circle [radius=.1]
(0,0) node (v3) {} circle [radius=.1]
(60:1.5) node (v7) {} circle [radius=.1]
(-60:1.5) node (v8) {} circle [radius=.1];
\draw  (v1) edge (v3);
\draw  (v3) edge (v7);
\draw  (v8) edge (v3);
\draw[<-,red,thick] (v8) .. controls (-20:1.8) and (20:1.8) .. (v7);
\draw[<-,red,thick] (v7) .. controls (100:1.8) and (140:1.8) .. (v1);
\draw[<-,red,thick] (v1) .. controls (220:1.8) and (260:1.8) .. (v8);
\end{scope}
\node at (-3.5,0) {\ubf{$\so(8)$, $\cO^{(3)}$}:};
\end{tikzpicture}
\ee
The full group of outer automorphisms of $\so(8)$ is the symmetric group $S_3$ which can be generated by combining $\cO^{(2)}$ and $\cO^{(3)}$. Note that $\cO^{(2)}$ and $\cO^{(3)}$ are not conjugate to each other (since they have different orders) and hence we need to consider both of them.

The above action of an outer automorphism $\cO^{(q)}$ (for $q=2,3$) on the roots of $\fg$ translates to an action on the Dynkin coefficients of the weights for representations of $\fg$. In other words, the action of $\cO^{(q)}$ can be viewed as an action on representations of $\fg$---see Table \ref{OOA}.

\begin{table}[htbp]
\begin{center}
\begin{tabular}{|c|c|l|} \hline
$\fg$&\raisebox{-.15\height}{$\cO^{(q)}$}&\raisebox{-.15\height}{$\cO^{(q)}\cdot\cR_\fg$}
\\ \hline
 \hline
$\su(m)$&\raisebox{-.15\height}{$\cO^{(2)}$}&\raisebox{-.15\height}{$\F\llra\bar{\F}$, $\nAsym\llra\bnAsym$, $\Sym\llra\bSym$}
\\ \hline
$\so(2m)$&\raisebox{-.15\height}{$\cO^{(2)}$}&\raisebox{-.15\height}{$\F\Lra\F$, $\S\llra\C$}
\\ \hline
$\fe_6$&\raisebox{-.15\height}{$\cO^{(2)}$}&\raisebox{-.15\height}{$\F\llra\bar{\F}$}
\\ \hline
$\so(8)$&\raisebox{-.15\height}{$\cO^{(3)}$}&\raisebox{-.15\height}{$\F\Lra\S$, $\S\Lra\C$, $\C\Lra\F$}
\\ \hline
\end{tabular}
\end{center}
\caption{List of non-trivial outer automorphisms $\cO^{(q)}$ of $\fg$ and their actions $\cO^{(q)}\cdot\cR_\fg$ on various irreducible representations $\cR_\fg$ of $\fg$. $\F$ denotes fundamental representation, $\nAsym$ denotes the irreducible $n$-index antisymmetric representation, $\Sym$ denotes the irreducible 2-index symmetric representation, and $\S$ and $\C$ denote irreducible spinor and cospinor representations. Bar on top of a representation denotes the complex conjugate of that representation. $\F$ of $\so(2m)$ is left invariant by the action of $\cO^{(2)}$.}	\label{OOA}
\end{table}

An outer automorphism $\cO^{(q_i)}$ of a gauge algebra $\fg_i\in\fT$ is a symmetry of $\fT$ if
\begin{align}
\cO^{(q_i)}\cdot\cR_i&=\cR_i\\
\cO^{(q_i)}\cdot\cR_{ij,i}&=\cR_{ij,i}\:\:\:\:\:\:\forall j
\end{align}
where $\cO^{(q_i)}\cdot\cR$ denotes the action of $\cO^{(q_i)}$ on $\cR$. We should keep in mind that a hyper in a representation $\cR$ is the same as a hyper in representation $\bar{\cR}$. So, $\cR_i$ and $\cR_{ij,i}$ are only defined up to complex conjugation on constituent irreps. Thus, whenever $\cR\lra\bar{\cR}$ in Table \ref{OOA}, it means that two distinct hypers in $\cR$ are interchanged with each other under the action of the outer automorphism.

As an example consider the $6d$ theory given by
\be\label{ex0}
\begin{tikzpicture}
\node at (-0.5,0.45) {2};
\node at (-0.45,0.9) {$\su(n)$};
\end{tikzpicture}
\ee
The theory includes $2n$ hypers in $\F$. The outer automorphism $\cO^{(2)}$ of $\su(n)$ descends to a discrete symmetry of the theory whose action on the hypermultiplets can be manifested as follows. We divide the $2n$ hypers into two ordered sets such that each set contains $n$ hypers. Then we exchange these two sets with each other.

\subsection{Discrete symmetries from permutation of tensor multiplets}\label{DS2}
Now we turn to a discussion of the second kind of basic discrete symmetries. These arise from permutation of tensor multiplets $i\to S(i)$ such that 
\begin{align}
\fg_{S(i)}&=\fg_{i} \label{S1}\\
\Omega^{S(i)S(j)}&=\Omega^{ij} \label{S2}
\end{align}
for all $i,j$. This is a symmetry of $\fT$ if 
\begin{align}
\cR_{S(i)}&\simeq\cR_{i} \label{R1}\\
\cR_{S(i)S(j)}&\simeq\cR_{ij} \label{R2}
\end{align}
for all $i,j$.

As an example, consider the $6d$ theory given by
\be\label{ex1}
\begin{tikzpicture}
\node (v1) at (-0.5,0.45) {1};
\node at (-0.45,0.9) {$\sp(n)$};
\begin{scope}[shift={(1.5,0)}]
\node (v2) at (-0.5,0.45) {4};
\node at (-0.45,0.9) {$\so(m)$};
\end{scope}
\begin{scope}[shift={(3,0)}]
\node (v3) at (-0.5,0.45) {$1$};
\node at (-0.45,0.9) {$\sp(n)$};
\end{scope}
\begin{scope}[shift={(4.5,0)}]
\node (v4) at (-0.5,0.45) {$4$};
\node at (-0.45,0.9) {$\so(p)$};
\end{scope}
\begin{scope}[shift={(-1.5,0)}]
\node (v0) at (-0.5,0.45) {$4$};
\node at (-0.45,0.9) {$\so(p)$};
\end{scope}
\draw  (v1) -- (v0);
\draw  (v1) -- (v2);
\draw  (v2) -- (v3);
\draw  (v4) -- (v3);
\end{tikzpicture}
\ee
The permutation
\be
\begin{tikzpicture}
\node (v1) at (-0.5,0.45) {1};
\node at (-0.45,0.9) {$\sp(n)$};
\begin{scope}[shift={(1.5,0)}]
\node (v2) at (-0.5,0.45) {4};
\node at (-0.45,0.9) {$\so(m)$};
\end{scope}
\begin{scope}[shift={(3,0)}]
\node (v3) at (-0.5,0.45) {$1$};
\node at (-0.45,0.9) {$\sp(n)$};
\end{scope}
\begin{scope}[shift={(4.5,0)}]
\node (v4) at (-0.5,0.45) {$4$};
\node at (-0.45,0.9) {$\so(p)$};
\end{scope}
\begin{scope}[shift={(-1.5,0)}]
\node (v0) at (-0.5,0.45) {$4$};
\node at (-0.45,0.9) {$\so(p)$};
\end{scope}
\draw  (v1) -- (v0);
\draw  (v1) -- (v2);
\draw  (v2) -- (v3);
\draw  (v4) -- (v3);
\draw[<->,red,thick] (v0) .. controls (-1,-2) and (3,-2) .. (v4);
\draw[<->,red,thick] (v1) .. controls (0,-1) and (2,-1) .. (v3);
\end{tikzpicture}
\ee
is a symmetry of the theory.

As another example, consider the $6d$ theory given by
\be\label{ex2}
\begin{tikzpicture}
\node (v1) at (-0.5,0.45) {2};
\node at (-0.45,0.9) {$\su(n)$};
\begin{scope}[shift={(1.5,0)}]
\node (v2) at (-0.5,0.45) {2};
\node at (-0.45,0.9) {$\su(n)$};
\end{scope}
\begin{scope}[shift={(-1.5,0)}]
\node (v0) at (-0.5,0.45) {2};
\node at (-0.45,0.9) {$\su(m)$};
\end{scope}
\begin{scope}[shift={(3,0)}]
\node (v3) at (-0.5,0.45) {2};
\node at (-0.45,0.9) {$\su(m)$};
\end{scope}
\draw  (v1) -- (v2);
\draw  (v0) edge (v1);
\draw  (v2) edge (v3);
\end{tikzpicture}
\ee
The permutation
\be\label{eg2}
\begin{tikzpicture}
\node (v1) at (-0.5,0.45) {2};
\node at (-0.45,0.9) {$\su(n)$};
\begin{scope}[shift={(1.5,0)}]
\node (v2) at (-0.5,0.45) {2};
\node at (-0.45,0.9) {$\su(n)$};
\end{scope}
\begin{scope}[shift={(-1.5,0)}]
\node (v0) at (-0.5,0.45) {2};
\node at (-0.45,0.9) {$\su(m)$};
\end{scope}
\begin{scope}[shift={(3,0)}]
\node (v3) at (-0.5,0.45) {2};
\node at (-0.45,0.9) {$\su(m)$};
\end{scope}
\draw  (v1) -- (v2);
\draw  (v0) edge (v1);
\draw  (v2) edge (v3);
\draw[<->,red,thick] (v0) .. controls (-1.5,-1.5) and (2,-1.5) .. (v3);
\draw[<->,red,thick] (v1) .. controls (-0.5,-0.5) and (1,-0.5) .. (v2);
\end{tikzpicture}
\ee
is a symmetry of the theory.

Now, consider a permutation $S$ that is a symmetry of $\fT$. We can use the data of $S$ to convert $[\Omega^{ij}]$ into another matrix $[\Omega_S^{\alpha\beta}]$. Here $\alpha$, $\beta$ etc. parametrize orbits of nodes $i$ under the iterative action of $S$. To define a particular entry $\Omega_S^{\alpha\beta}$, we pick a node $i$ lying in the orbit $\alpha$ and let
\be\label{fmatrix}
\Omega_S^{\alpha\beta}=\sum_{j\in\beta}\Omega^{ij}
\ee
where the sum is over all nodes $j$ lying in the orbit $\beta$. Notice that the resulting matrix $[\Omega_S^{\alpha\beta}]$ need not be symmetric but must be positive definite. It turns out for $S$ associated to $6d$ SCFTs that whenever $\Omega_S^{\alpha\beta}\neq \Omega_S^{\beta\alpha}$, then the smaller of the two entries is $-1$. Thus, $[\Omega_S^{\alpha\beta}]$ is analogous to the Cartan matrix for a general (i.e. either simply laced or non-simply laced) Lie algebra.

Let us compute the matrix $[\Omega_S^{\alpha\beta}]$ for the above example (\ref{ex1}). To start with, $[\Omega^{ij}]$ is
\[\begin{pmatrix}
4&-1&0&0&0\\
-1&1&-1&0&0\\
0&-1&4&-1&0\\
0&0&-1&1&-1\\
0&0&0&-1&4
\end{pmatrix}\]
There are three orbits. The third node lies in the first orbit, the second and fourth nodes lie in the second orbit, and the first and fifth nodes lie in the third orbit. Applying our prescription (\ref{fmatrix}), we find that $[\Omega_S^{\alpha\beta}]$ is
\[\begin{pmatrix}
4&-2&0\\
-1&1&-1\\
0&-1&4
\end{pmatrix}\]

Similarly, we can compute the matrix $[\Omega_S^{\alpha\beta}]$ for the above example (\ref{ex2}). $[\Omega^{ij}]$ is
\[\begin{pmatrix}
2&-1&0&0\\
-1&2&-1&0\\
0&-1&2&-1\\
0&0&-1&2
\end{pmatrix}\]
and $[\Omega_S^{\alpha\beta}]$ is
\[\begin{pmatrix}
1&-1\\
-1&2
\end{pmatrix}\].

Now, we define a graph $\Sigma^S_\fT$ associated to $[\Omega_S^{\alpha\beta}]$:
\bit
\item \ubf{Nodes}: The nodes of $\Sigma^S_\fT$ are in one-to-one correspondence with the set of orbits $\alpha$. The value of node $\alpha$ is \raisebox{-.125\height}{ \begin{tikzpicture}
\node at (-0.5,0.45) {$\Omega^{ii}$};
\node at (-0.45,0.9) {$\fg_i$};
\end{tikzpicture}} where $i$ is a node of $\Sigma_\fT$ lying in the orbit $\alpha$.
\item \ubf{Edges}: Let $\alpha\neq\beta$ and let $\Omega_S^{\alpha\beta}\geq\Omega_S^{\beta\alpha}$. Then we place $-\Omega_S^{\alpha\beta}$ number of edges between nodes $\alpha$ and $\beta$. If $\Omega_S^{\alpha\beta}=\Omega_S^{\beta\alpha}$, then the edges are undirected. If $\Omega_S^{\alpha\beta}>\Omega_S^{\beta\alpha}$, then all the edges are directed from $\alpha$ to $\beta$.
\item \ubf{Self-edges}: Let $l_\alpha=\Omega^{ii}-\Omega_S^{\alpha\alpha}$ where $i$ is a node of $\Sigma_\fT$ lying in the orbit $\alpha$. Then, we introduce $l_\alpha$ edges such that the source and target of each edge is the same node $\alpha$.
\eit
$\Sigma^S_\fT$ can be understood as a folding\footnote{Notice that, unlike the foldings of Dynkin diagrams, the foldings of graphs $\Sigma_\fT$ can lead to self-edges.} of $\Sigma_\fT$ by the action of $S$. Observe that the relationship between $\Sigma^S_\fT$ and $[\Omega_S^{\alpha\beta}]$ is analogous to the relationship between the Dynkin graph and Cartan matrix for a general (i.e. either simply laced or non-simply laced) Lie algebra.

For our example (\ref{ex1}), the folded graph $\Sigma^S_\fT$ is
\be\label{eg1}
\begin{tikzpicture}
\node (v1) at (-0.5,0.45) {$4$};
\node at (-0.45,0.9) {$\so(m)$};
\begin{scope}[shift={(2,0)}]
\node (v2) at (-0.5,0.45) {$1$};
\node at (-0.45,0.9) {$\sp(n)$};
\end{scope}
\begin{scope}[shift={(4,0)}]
\node (v4) at (-0.5,0.45) {$4$};
\node at (-0.45,0.9) {$\so(p)$};
\end{scope}
\node (v3) at (0.5,0.45) {\tiny{2}};
\draw (v1)--(v3);
\draw[<-] (v2)--(v3);
\draw  (v2) edge (v4);
\end{tikzpicture}
\ee
and for (\ref{ex2}), the folded graph $\Sigma^S_\fT$ is
\be\label{eg2}
\begin{tikzpicture}
\node (v1) at (-0.5,0.4) {2};
\node at (-0.45,0.9) {$\su(n)$};
\begin{scope}[shift={(2,0)}]
\node (v4) at (-0.5,0.4) {$2$};
\node at (-0.45,0.9) {$\su(m)$};
\end{scope}
\draw (v1) .. controls (-1.5,-0.5) and (0.5,-0.5) .. (v1);
\draw  (v1) edge (v4);
\end{tikzpicture}
\ee

We note that, starting from the data of $\Sigma^S_\fT$, we can only reconstruct $S$ up to conjugation. But this is enough to keep track of the twist associated to $S$. Thus, throughout this paper, we will specify twists via folded graphs $\Sigma^S_\fT$ and will not refer to an explicit $S$ inducing the folding.

\subsection{General discrete symmetries}

We now discuss twists associated to general discrete symmetries that combine the basic discrete symmetries discussed in Sections \ref{DS1} and \ref{DS2}. That is, we consider actions of the form
\be\label{gt}
\left(\prod_{i}\cO^{(q_i)}\right)S
\ee
where $S$ is a permutation of the tensor multiplets and $\cO^{(q_i)}$ is an outer automorphism of order $q_i$ of gauge algebra $\fg_i$, where each $q_i\in\{1,2,3\}$ and $q_i=1$ denotes the identity automorphism. (\ref{gt}) is a symmetry of the $6d$ theory $\fT$ only if
\begin{align}
\fg_{S(i)}&\simeq\fg_{i}\\
\Omega^{S(i)S(j)}&=\Omega^{ij}
\end{align}
and
\begin{align}
\cO^{(q_{S(i)})}\cdot\cR_i&=\cR_{S(i)}\\
\cO^{(q_{S(i)})}\cdot\cR_{S(i)S(j),S(i)}&=\cR_{ij,i}
\end{align}
As in Section \ref{DS2}, we associate the matrix $[\Omega_S^{\alpha\beta}]$ to the twist generated by the action of (\ref{gt}).

As an example, consider the $6d$ SCFT
\be\label{ex3}
\begin{tikzpicture}
\node (v1) at (-0.5,0.45) {2};
\node at (-0.45,0.9) {$\su(n)$};
\begin{scope}[shift={(1.5,0)}]
\node (v2) at (-0.5,0.45) {2};
\node at (-0.45,0.9) {$\su(m)$};
\end{scope}
\begin{scope}[shift={(-1.5,0)}]
\node (v0) at (-0.5,0.45) {2};
\node at (-0.45,0.9) {$\su(m)$};
\end{scope}
\draw  (v1) -- (v2);
\draw  (v0) edge (v1);
\end{tikzpicture}
\ee
Suppose we want to perform the outer-automorphism $\cO^{(2)}$ for the middle $\su(n)$ node. Recall from the discussion around (\ref{ex0}) that the outer automorphism of $\su(n)$ exchanges the fundamental hypers in pairs. However, the graph in (\ref{ex3}) indicates that the fundamental hypers of the middle $\mathfrak{su}(n)$ algebra are part of bifundamental representations formed by taking the tensor product with the fundamental representations of the neighboring $\mathfrak{su}(m)$ algebras. Therefore, if we want $\cO^{(2)}$ to be a symmetry of the theory, we must permute the two neighboring $\su(m)$ as well. Thus, $\cO^{(2)}$ by itself is not a symmetry of the theory, but its combination with the permutation
\be

\ee
is a symmetry of the theory. Thus, we see that in general it is not possible to decompose a general symmetry of the form (\ref{gt}) into more basic symmetries discussed earlier.

As another illustrative example, consider
\be
%
\ee
Consider sending the left $\so(2m)$ to the right $\so(2m)$ with an outer automorphism $\cO^{(2)}$, and sending the right $\so(2m)$ to the left $\so(2m)$ without any outer automorphism. We can represent this action as
\be\label{ex4}
%
\ee
This action is a symmetry of the theory and is represented as
\be
\cO^{(1)}_1\cO^{(2)}_3S
\ee
in the notation of (\ref{gt}). Here we have labeled the nodes as $1,2,3$ from left to right and the subscript of $\cO$ denotes the node it is acting at. We can also consider the action
\be\label{ex5}
%
\ee
which is also a symmetry of the theory and is represented as
\be\label{exx5}
\cO^{(2)}_1\cO^{(2)}_3S
\ee
in the notation of (\ref{gt}).

\begin{table}[htbp]
\begin{center}
\begin{tabular}{|c|c|l|} \hline
\raisebox{-.4\height}{\begin{tikzpicture}
\node at (-0.5,0.5) {$\Omega^{\alpha\alpha}$};
\node at (-0.55,0.9) {$\fg_\alpha$};
\end{tikzpicture}} &Comments
\\ \hline
 \hline
 \raisebox{-.4\height}{ \begin{tikzpicture}
\node at (-0.5,0.4) {1};
\node at (-0.45,0.9) {$\su(n)^{(2)}$};
\end{tikzpicture}}&$n=3,4$
\\ \hline
\raisebox{-.4\height}{ \begin{tikzpicture}
\node at (-0.5,0.4) {2};
\node at (-0.45,0.9) {$\su(n)^{(2)}$};
\end{tikzpicture}}&$n\ge3$
\\ \hline
\raisebox{-.4\height}{ \begin{tikzpicture}
\node at (-0.5,0.4) {$3$};
\node at (-0.45,0.9) {$\su(3)^{(2)}$};
\end{tikzpicture}}&
\\ \hline
\raisebox{-.4\height}{ \begin{tikzpicture}
\node at (-0.5,0.4) {4};
\node at (-0.45,0.9) {$\so(2n)^{(2)}$};
\end{tikzpicture}}&$n\ge5$
\\ \hline
\raisebox{-.4\height}{ \begin{tikzpicture}
\node at (-0.5,0.4) {$k$};
\node at (-0.45,0.9) {$\so(8)^{(q)}$};
\end{tikzpicture}}&$1\le k\le 4$; $q=2,3$
\\ \hline
\raisebox{-.4\height}{ \begin{tikzpicture}
\node at (-0.5,0.4) {$2$};
\node at (-0.45,0.9) {$\so(10)^{(2)}$};
\end{tikzpicture}}&
\\ \hline
\raisebox{-.4\height}{ \begin{tikzpicture}
\node at (-0.5,0.4) {2};
\node at (-0.45,0.9) {$\so(\wh{12})^{(2)}$};
\end{tikzpicture}}&
\\ \hline
\raisebox{-.4\height}{ \begin{tikzpicture}
\node at (-0.5,0.4) {$k$};
\node at (-0.45,0.9) {$\fe_6^{(2)}$};
\end{tikzpicture}}&$k=2,4,6$
\\ \hline
\raisebox{-.4\height}{\begin{tikzpicture}
\node (v1) at (-0.5,0.4) {2};
\node at (-0.45,0.9) {$\su(n)^{(1)}$};
\draw (v1) .. controls (-1.5,-0.5) and (0.5,-0.5) .. (v1);
\end{tikzpicture}}&$n\ge1; \text{non-geometric}$
\\ \hline
\end{tabular}
\end{center}
\caption{List of all the new nodes that can appear in graphs associated to $5d$ KK theories. We also list all the possibilities where an edge starts and ends on the same node. The comment ``non-geometric'' for the last entry refers to the fact that there is no completely geometric description of this node. See also a node appearing in Table \ref{TR1}. If a KK theory involves either of these two kinds of nodes, then it does not admit a conventional geometric description.}	\label{KR1}
\end{table}

\begin{table}[htbp]
\begin{center}
\begin{tabular}{|c|c|l|} \hline
\raisebox{-.4\height}{\begin{tikzpicture}
\node (v1) at (-0.5,0.4) {$\Omega^{\alpha\alpha}$};
\node at (-0.55,0.9) {$\fg_\alpha^{(q_\alpha)}$};
\begin{scope}[shift={(2,0)}]
\node (v2) at (-0.5,0.4) {$\Omega^{\beta\beta}$};
\node at (-0.55,0.9) {$\fg_\beta^{(q_\beta)}$};
\end{scope}
\draw  (v1) edge (v2);
\end{tikzpicture}} &Comments
\\ \hline
\hline
 \raisebox{-.4\height}{\begin{tikzpicture}
\node (v1) at (-0.5,0.4) {1};
\node at (-0.45,0.9) {$\sp(n_\alpha)^{(1)}$};
\begin{scope}[shift={(2,0)}]
\node (v2) at (-0.5,0.4) {$k$};
\node at (-0.45,0.9) {$\so(n_\beta)^{(2)}$};
\end{scope}
\draw  (v1) edge (v2);
\end{tikzpicture}}&$n_\alpha\le n_\beta-4-k$; $n_\beta\le4n_\alpha+14$; $2\le k\le4$
\\ 
 \raisebox{-.4\height}{\begin{tikzpicture}
\node (v1) at (-0.5,0.4) {1};
\node at (-0.45,0.9) {$\sp(n_\alpha)^{(1)}$};
\begin{scope}[shift={(2,0)}]
\node (v2) at (-0.5,0.4) {$2$};
\node at (-0.45,0.9) {$\so(\wh{12})^{(2)}$};
\end{scope}
\draw  (v1) edge (v2);
\end{tikzpicture}}&$n_\alpha\le 6$
\\ 
 \raisebox{-.4\height}{\begin{tikzpicture}
\node (v1) at (-0.5,0.4) {2};
\node at (-0.45,0.9) {$\su(n_\alpha)^{(1)}$};
\begin{scope}[shift={(2,0)}]
\node (v2) at (-0.5,0.4) {$4$};
\node at (-0.45,0.9) {$\so(n_\beta)^{(2)}$};
\end{scope}
\node (v3) at (0.5,0.4) {\tiny{2}};
\draw  (v1) -- (v3);
\draw  (v2) -- (v3);
\end{tikzpicture}}&$n_\alpha\le n_\beta-8$; $n_\beta\le2n_\alpha$
\\ 
 \raisebox{-.6\height}{\begin{tikzpicture}
\node (v1) at (-0.5,0.4) {2};
\node at (-0.45,0.9) {$\su(n_\alpha)^{(1)}$};
\begin{scope}[shift={(2,0)}]
\node (v2) at (-0.5,0.4) {$2$};
\node at (-0.45,0.9) {$\su(n_\beta)^{(1)}$};
\end{scope}
\draw  (v1) -- (v2);
\draw (v1) .. controls (-1.5,-0.5) and (0.5,-0.5) .. (v1);
\end{tikzpicture}}&$n_\alpha\le 2n_\beta$; $n_\beta\le n_\alpha$
\\
\hline
\end{tabular}
\end{center}
\caption{List of all the new undirected edges that can appear in graphs characterizing $5d$ KK theories.}\label{FTR2}
\end{table}

\begin{table}[htbp]
\begin{center}
\begin{tabular}{|c|c|l|} \hline
\raisebox{-.4\height}{\begin{tikzpicture}
\node (v1) at (-0.5,0.4) {$\Omega^{\alpha\alpha}$};
\node at (-0.55,0.9) {$\fg_\alpha^{(q_\alpha)}$};
\begin{scope}[shift={(2,0)}]
\node (v2) at (-0.5,0.4) {$\Omega^{\beta\beta}$};
\node at (-0.55,0.9) {$\fg_\beta^{(q_\beta)}$};
\end{scope}
\node (v3) at (0.5,0.4) {\tiny{$e$}};
\draw  (v1) edge (v3);
\draw  [->](v3) -- (v2);
\end{tikzpicture}} &Comments
\\ \hline
\hline
 \raisebox{-.4\height}{\begin{tikzpicture}
\node (v1) at (-0.5,0.4) {1};
\node at (-0.45,0.9) {$\sp(n_\alpha)^{(1)}$};
\begin{scope}[shift={(2,0)}]
\node (v2) at (-0.5,0.4) {$k$};
\node at (-0.45,0.9) {$\so(n_\beta)^{(q_\beta)}$};
\end{scope}
\node (v3) at (0.5,0.4) {\tiny{$2$}};
\draw  (v1) edge (v3);
\draw [->] (v3) edge (v2);
\end{tikzpicture}}&$n_\alpha\le n_\beta-4-k$; $n_\beta\le2n_\alpha+10-2q_\beta$; $k=3,4$; $q_\beta=1,2$
\\ 
 \raisebox{-.4\height}{\begin{tikzpicture}
\node (v1) at (-0.5,0.4) {1};
\node at (-0.45,0.9) {$\sp(n_\alpha)^{(1)}$};
\begin{scope}[shift={(2,0)}]
\node (v2) at (-0.5,0.4) {$3$};
\node at (-0.45,0.9) {$\so(7)^{(1)}$};
\end{scope}
\node (v3) at (0.5,0.4) {\tiny{$2$}};
\draw [dashed] (v1) -- (v3);
\draw [dashed,->] (v3) -- (v2);
\end{tikzpicture}\:\:}&$n_\alpha=1,2$
\\ 
 \raisebox{-.4\height}{\begin{tikzpicture}
\node (v1) at (-0.5,0.4) {1};
\node at (-0.45,0.9) {$\sp(1)^{(1)}$};
\begin{scope}[shift={(2,0)}]
\node (v2) at (-0.5,0.4) {$3$};
\node at (-0.45,0.9) {$\fg_2^{(1)}$};
\end{scope}
\node (v3) at (0.5,0.4) {\tiny{$2$}};
\draw  (v1) -- (v3);
\draw [->] (v3) -- (v2);
\end{tikzpicture}\:\:\:\:}&
\\
\hline
\end{tabular}
\end{center}
\caption{List of all the possible directed edges between two gauge-theoretic nodes that can appear in graphs characterizing $5d$ KK theories. An arrow with $e$ in the middle of it denotes $e$ edges directed in the direction of arrow. Solid edges arise from foldings of solid edges and dashed edges arise from foldings of dashed edges. A partially dashed and partially solid edge with 2 in the middle of it arises from a folding together of a dashed edge and a solid edge.}\label{BTR2}
\end{table}

\begin{table}[htbp]
\begin{center}
\begin{tabular}{|c|c|l|} \hline
\raisebox{-.4\height}{\begin{tikzpicture}
\node (v1) at (-0.5,0.4) {$\Omega^{\alpha\alpha}$};
\node at (-0.55,0.9) {$\fg_\alpha^{(q_\alpha)}$};
\begin{scope}[shift={(2,0)}]
\node (v2) at (-0.5,0.4) {$\Omega^{\beta\beta}$};
\node at (-0.55,0.9) {$\fg_\beta^{(q_\beta)}$};
\end{scope}
\node (v3) at (0.5,0.4) {\tiny{$e$}};
\draw  (v1) edge (v3);
\draw  [->](v3) -- (v2);
\end{tikzpicture}} &Comments
\\ \hline
\hline
 \raisebox{-.4\height}{\begin{tikzpicture}
\node (v1) at (-0.5,0.4) {1};
\node at (-0.45,0.9) {$\sp(0)^{(1)}$};
\begin{scope}[shift={(2,0)}]
\node (v2) at (-0.5,0.4) {$3$};
\node at (-0.45,0.9) {$\su(3)^{(q)}$};
\end{scope}
\node (v3) at (0.5,0.4) {\tiny{$2$}};
\draw  (v1) edge (v3);
\draw [->] (v3) edge (v2);
\end{tikzpicture}}&$q=1,2$
\\ 
 \raisebox{-.4\height}{\begin{tikzpicture}
\node (v1) at (-0.5,0.4) {1};
\node at (-0.45,0.9) {$\sp(0)^{(1)}$};
\begin{scope}[shift={(2,0)}]
\node (v2) at (-0.5,0.4) {$3$};
\node at (-0.45,0.9) {$\so(7)^{(1)}$};
\end{scope}
\node (v3) at (0.5,0.4) {\tiny{$2$}};
\draw  (v1) edge (v3);
\draw [->] (v3) edge (v2);
\end{tikzpicture}}&
\\ 
 \raisebox{-.4\height}{\begin{tikzpicture}
\node (v1) at (-0.5,0.4) {1};
\node at (-0.45,0.9) {$\sp(0)^{(1)}$};
\begin{scope}[shift={(2,0)}]
\node (v2) at (-0.5,0.4) {$k$};
\node at (-0.45,0.9) {$\so(8)^{(1)}$};
\end{scope}
\node (v3) at (0.5,0.4) {\tiny{$2$}};
\draw  (v1) edge (v3);
\draw [->] (v3) edge (v2);
\end{tikzpicture}}&$k=3,4$
\\ 
 \raisebox{-.4\height}{\begin{tikzpicture}
\node (v1) at (-0.5,0.4) {1};
\node at (-0.45,0.9) {$\sp(0)^{(1)}$};
\begin{scope}[shift={(2,0)}]
\node (v2) at (-0.5,0.4) {$3$};
\node at (-0.45,0.9) {$\fg_2^{(1)}$};
\end{scope}
\node (v3) at (0.5,0.4) {\tiny{$2$}};
\draw  (v1) edge (v3);
\draw [->] (v3) edge (v2);
\end{tikzpicture}\:\:\:\:}&
\\
\hline
\end{tabular}
\end{center}
\caption{List of all the possible directed edges involving at least one non-gauge-theoretic node that can appear in graphs characterizing $5d$ KK theories.}\label{BTR2N}
\end{table}

\begin{table}[htbp]
\begin{center}
\begin{tabular}{|c|l|} \hline
\raisebox{-.15\height}{\begin{tikzpicture}
\node (v1) at (-0.5,0.4) {$\Omega^{ii}$};
\node at (-0.45,0.9) {$\fg^{(q_i)}_i$};
\begin{scope}[shift={(1.75,0)}]
\node (v2) at (-0.5,0.4) {$1$};
\node at (-0.45,0.9) {$\sp(0)^{(1)}$};
\end{scope}
\begin{scope}[shift={(3.5,0)}]
\node (v3) at (-0.5,0.4) {$\Omega^{kk}$};
\node at (-0.45,0.9) {$\fg^{(q_k)}_k$};
\end{scope}
\draw  (v1) edge (v2);
\draw  (v2) edge (v3);
\end{tikzpicture}}&Comments
\\ \hline
\hline
\raisebox{-.15\height}{\begin{tikzpicture}
\node (v1) at (-0.5,0.4) {$2$};
\node at (-0.45,0.9) {$\su(2)^{(1)}$};
\begin{scope}[shift={(1.75,0)}]
\node (v2) at (-0.5,0.4) {$1$};
\node at (-0.45,0.9) {$\sp(0)^{(1)}$};
\end{scope}
\begin{scope}[shift={(3.5,0)}]
\node (v3) at (-0.5,0.4) {$k$};
\node at (-0.45,0.9) {$\fg^{(q)}$};
\end{scope}
\draw  (v1) edge (v2);
\draw  (v2) edge (v3);
\end{tikzpicture}\:\:\:\:}&$k\ge3$; $\fg^{(q)}=\fe_6^{(2)},\so(8,10)^{(2)},\so(8)^{(3)}$
\\ 
\raisebox{-.15\height}{\begin{tikzpicture}
\node (v1) at (-0.5,0.4) {$k$};
\node at (-0.45,0.9) {$\su(3)^{(1)}$};
\begin{scope}[shift={(1.75,0)}]
\node (v2) at (-0.5,0.4) {$1$};
\node at (-0.45,0.9) {$\sp(0)^{(1)}$};
\end{scope}
\begin{scope}[shift={(3.5,0)}]
\node (v3) at (-0.5,0.4) {$l$};
\node at (-0.45,0.9) {$\fg^{(q)}$};
\end{scope}
\draw  (v1) edge (v2);
\draw  (v2) edge (v3);
\end{tikzpicture}\:\:\:\:}&$k,l\ge3$; $k+l\ge5$; $\fg^{(q)}=\so(8)^{(2)},\so(8)^{(3)},\su(n\le4)^{(2)}$
\\ 
\raisebox{-.15\height}{\begin{tikzpicture}
\node (v1) at (-0.5,0.4) {$k$};
\node at (-0.45,0.9) {$\su(3)^{(2)}$};
\begin{scope}[shift={(1.75,0)}]
\node (v2) at (-0.5,0.4) {$1$};
\node at (-0.45,0.9) {$\sp(0)^{(1)}$};
\end{scope}
\begin{scope}[shift={(3.5,0)}]
\node (v3) at (-0.5,0.4) {$l$};
\node at (-0.45,0.9) {$\fg^{(1)}$};
\end{scope}
\draw  (v1) edge (v2);
\draw  (v2) edge (v3);
\end{tikzpicture}\:\:\:\:}&$k,l\ge2$; $k+l\ge5$; $\fg=\ff_4,\fg_2,\so(n\le9),\su(4)$
\\
\hline
\end{tabular}
\end{center}
\caption{List of all the new possibilities for multiple neighbors of $\sp(0)^{(1)}$ connected to it by undirected edges.}\label{FR3E}
\end{table}

Now, let $\fg_\alpha=\fg_i$ and $\Omega^{\alpha\alpha}=\Omega^{ii}$ where $i$ is a node of $\Sigma_\fT$ lying in the orbit $\alpha$ of $S$. Then $\cO^{(q_i)}$ can be viewed as an outer automorphism of $\fg_\alpha$. Let us define an outer automorphism $\cO^{(q_\alpha)}$ of $\fg_\alpha$ by
\be
\cO^{(q_\alpha)}=\prod_{i\in\alpha}\cO^{(q_i)}
\ee
where each $\cO^{(q_i)}$ on the right hand side is viewed as an outer automorphism of $\fg_\alpha$ and the $\cO^{(q_i)}$ for all $i$ lying in the orbit $\alpha$ are then multiplied with each other to produce the outer automorphism $\cO^{(q_\alpha)}$ of $\fg_\alpha$. Notice that we have chosen some ordering of various $i$ while evaluating the product $\prod_{i\in\alpha}\cO^{(q_i)}$. Different orderings produce different but conjugate $\cO^{(q_\alpha)}$. Thus, we leave the ordering unspecified since we are only interested in the conjugacy class of $\cO^{(q_\alpha)}$.

We can now associate a graph $\Sigma^{S,\{q_\alpha\}}_\fT$ to the action of (\ref{gt}). We start from the graph $\Sigma^S_\fT$ defined in Section \ref{DS2} and modify the values of the node $\alpha$ to \raisebox{-.125\height}{ \begin{tikzpicture}
\node at (-0.5,0.4) {$\Omega_S^{\alpha\alpha}$};
\node at (-0.45,0.95) {$\fg_\alpha^{(q_\alpha)}$};
\end{tikzpicture}}
where $i$ is a node of $\Sigma_\fT$ lying in the orbit $\alpha$. The graph obtained after this simple modification is what we refer to as $\Sigma^{S,\{q_\alpha\}}_\fT$.

Note that the data of $\Sigma^{S,\{q_\alpha\}}_\fT$ is enough to reconstruct the action (\ref{gt}) up to conjugation. Thus, we will capture the twist associated to the action (\ref{gt}) by the graph $\Sigma^{S,\{q_\alpha\}}_\fT$ and call the resulting $5d$ KK theory as $\fT^{KK}_{S,\{q_\alpha\}}$.

For the example discussed around (\ref{ex3}), $\Sigma^{S,\{q_\alpha\}}_\fT$ is
\be\label{exa3}
\begin{tikzpicture}
\node (v1) at (-0.5,0.45) {$2$};
\node at (-0.45,0.9) {$\su(n)^{(2)}$};
\begin{scope}[shift={(2,0)}]
\node (v2) at (-0.5,0.45) {$2$};
\node at (-0.45,0.9) {$\su(m)^{(1)}$};
\end{scope}
\node (v3) at (0.5,0.45) {\tiny{2}};
\draw (v1)--(v3);
\draw[<-] (v2)--(v3);
\end{tikzpicture}
\ee
Similarly, for (\ref{ex4}), $\Sigma^{S,\{q_\alpha\}}_\fT$ is
\be
\begin{tikzpicture}
\node (v1) at (-0.5,0.45) {$1$};
\node at (-0.45,0.9) {$\sp(n)^{(1)}$};
\begin{scope}[shift={(2,0)}]
\node (v2) at (-0.5,0.45) {$4$};
\node at (-0.45,0.9) {$\so(2m)^{(2)}$};
\end{scope}
\node (v3) at (0.5,0.45) {\tiny{2}};
\draw (v1)--(v3);
\draw[<-] (v2)--(v3);
\end{tikzpicture}
\ee
However, for (\ref{ex5}), $\Sigma^{S,\{q_\alpha\}}_\fT$ is
\be\label{fg1}
\begin{tikzpicture}
\node (v1) at (-0.5,0.45) {$1$};
\node at (-0.45,0.9) {$\sp(n)^{(1)}$};
\begin{scope}[shift={(2,0)}]
\node (v2) at (-0.5,0.45) {$4$};
\node at (-0.45,0.9) {$\so(2m)^{(1)}$};
\end{scope}
\node (v3) at (0.5,0.45) {\tiny{2}};
\draw (v1)--(v3);
\draw[<-] (v2)--(v3);
\end{tikzpicture}
\ee
which is the same as $\Sigma^{S,\{q_\alpha\}}_\fT$ for the symmetry
\be\label{ex6}
\begin{tikzpicture}
\node (v1) at (-0.5,0.45) {1};
\node at (-0.45,0.9) {$\sp(n)$};
\begin{scope}[shift={(1.5,0)}]
\node (v2) at (-0.5,0.45) {4};
\node(w2) at (-0.45,0.9) {$\so(2m)$};
\end{scope}
\begin{scope}[shift={(-1.5,0)}]
\node (v0) at (-0.5,0.45) {4};
\node(w0) at (-0.45,0.9) {$\so(2m)$};
\end{scope}
\draw  (v1) -- (v2);
\draw  (v0) edge (v1);
\draw[<->,red,thick] (v0) .. controls (-1.5,-1) and (0.5,-1) .. (v2);
\end{tikzpicture}
\ee
which does not involve any outer automorphisms. Thus, according to our claim, (\ref{ex5}) and \ref{ex6}) must be in the same conjugacy class. Let us demonstrate it explicitly. Conjugating (\ref{exx5}) by $\cO^{(2)}_1$, we get
\begin{align}
&\cO^{(2)}_1(\cO^{(2)}_1\cO^{(2)}_3S)\cO^{(2)}_1\\
=&\cO^{(2)}_3S\cO^{(2)}_1\\
=&\cO^{(2)}_3\cO^{(2)}_3S\\
=&S
\end{align}
Thus, the KK theories corresponding to (\ref{ex5}) and (\ref{ex6}) must be the same, and we denote it by the folded graph (\ref{fg1}).

In a similar fashion, by studying various $6d$ SCFTs and their symmetries, we can isolate all the possible ingredients that can appear in graphs of the form $\Sigma^{S,\{q_\alpha\}}_\fT$ associated to $5d$ KK theories:
\bit
\item First of all, the nodes listed in Tables \ref{TR1} and \ref{R1N} are all allowed. We simply write each gauge algebra $\fg$ appearing in Table \ref{TR1} as $\fg^{(1)}$.
\item Similarly, the edges appearing in Tables \ref{TR2}, \ref{R2P} and \ref{R2N} are all allowed with each gauge algebra being written as $\fg^{(1)}$.
\item The new nodes that can appear in graphs associated to $5d$ KK theories but do not appear in graphs associated to $6d$ SCFTs are listed in Table \ref{KR1}.
\item The new undirected edges appearing for graphs associated to $5d$ KK theories are listed in Table \ref{FTR2}.

The configuration
\be\label{bound1}
\begin{tikzpicture}
\node (v1) at (-0.5,0.4) {1};
\node at (-0.45,0.9) {$\sp(n_\alpha)^{(1)}$};
\begin{scope}[shift={(2,0)}]
\node (v2) at (-0.5,0.4) {$k$};
\node at (-0.45,0.9) {$\so(n_\beta)^{(2)}$};
\end{scope}
\draw  (v1) edge (v2);
\end{tikzpicture}
\ee
for $n_\beta=4n_\alpha+16$ and $n_\alpha\ge0$ is not allowed since the choice of theta angle for $\sp(n_\alpha)$ in the associated $6d$ theory is correlated to the choice of a spinor representation of the neighboring $\so(4n_\alpha+16)$. Thus, the outer automorphism $\cO^{(2)}$ of $\so(4n_\alpha+16)$ is not a symmetry of the theory\footnote{The authors thank Gabi Zafrir for a discussion on this point}.
\item The directed edges between two nodes both carrying a non-trivial gauge algebra are listed in Table \ref{BTR2}.

The configuration
\be\label{bound2}
\begin{tikzpicture}
\node (v1) at (-0.5,0.4) {1};
\node at (-0.45,0.9) {$\sp(n_\alpha)^{(1)}$};
\begin{scope}[shift={(2,0)}]
\node (v2) at (-0.5,0.4) {$k$};
\node at (-0.45,0.9) {$\so(n_\beta)^{(2)}$};
\end{scope}
\node (v3) at (0.5,0.4) {\tiny{$2$}};
\draw  (v1) edge (v3);
\draw [->] (v3) edge (v2);
\end{tikzpicture}
\ee
with $n_\beta=2n_\alpha+8$ is not allowed. This configuration descends from (\ref{ex4}) with $n=n_\alpha$ and $m=n_\alpha+4$. Recall that the choice of theta angle of the gauge algebra $\sp(n_\alpha)$ is equivalent to the choice of a spinor representation of its flavor symmetry algebra $\so(4n_\alpha+16)$. But $\so(2n_\alpha+8)\oplus\so(2n_\alpha+8)$ subalgebra of $\so(4n_\alpha+16)$ is gauged. The $\S$ of $\so(4n_\alpha+16)$ decomposes as $(\S\otimes\C)\oplus(\C\otimes\S)$ of $\so(2n_\alpha+8)\oplus\so(2n_\alpha+8)$ which is sent to $(\C\otimes\C)\oplus(\S\otimes\S)$ of $\so(2n_\alpha+8)\oplus\so(2n_\alpha+8)$ by the action depicted in (\ref{ex4}). Thus, (\ref{ex4}) is not a symmetry when $n=n_\alpha$ and $m=n_\alpha+4$.

For similar reasons, the configuration
\be\label{bound3}
\begin{tikzpicture}
\node (v1) at (-0.5,0.4) {1};
\node at (-0.45,0.9) {$\sp(n_\alpha)^{(1)}$};
\begin{scope}[shift={(2,0)}]
\node (v2) at (-0.5,0.4) {$k$};
\node at (-0.45,0.9) {$\so(n_\beta)^{(2)}$};
\end{scope}
\node (v3) at (0.5,0.4) {\tiny{$3$}};
\draw  (v1) edge (v3);
\draw [->] (v3) edge (v2);
\end{tikzpicture}
\ee
with $3n_\beta=4n_\alpha+16$ is not allowed.

The KK theory
\be
\begin{tikzpicture}
\node (v1) at (-0.5,0.4) {3};
\node at (-0.45,0.9) {$\so(8)^{(2)}$};
\begin{scope}[shift={(2,0)}]
\node (v2) at (-0.5,0.4) {1};
\node at (-0.45,0.9) {$\sp(1)^{(1)}$};
\end{scope}
\node (v3) at (0.5,0.4) {\tiny{$2$}};
\draw [dashed] (v1) edge (v3);
\draw  [->](v3) -- (v2);
\end{tikzpicture}
\ee
arises from the $6d$ SCFT
\be
\begin{tikzpicture}
\node (v1) at (-0.5,0.4) {3};
\node at (-0.45,0.9) {$\so(8)$};
\begin{scope}[shift={(2,0)}]
\node (v2) at (-0.5,0.4) {1};
\node at (-0.45,0.9) {$\sp(1)$};
\end{scope}
\begin{scope}[shift={(-2,0)}]
\node (v0) at (-0.5,0.4) {1};
\node at (-0.45,0.9) {$\sp(1)$};
\end{scope}
\draw  (v1) edge (v2);
\draw [dashed] (v0) -- (v1);
\end{tikzpicture}
\ee
by performing the outer automorphism $\cO^{(2)}$ of $\so(8)$ which permutes $\F$ and $\S$, and hence induces the exchange of the two $\sp(1)$.
\item Other kinds of directed edges are listed in Table \ref{BTR2N}.

Due to similar reasons as explained above, the configuration
\be
\begin{tikzpicture}
\node (v1) at (-0.5,0.4) {1};
\node at (-0.45,0.9) {$\sp(0)^{(1)}$};
\begin{scope}[shift={(2,0)}]
\node (v2) at (-0.5,0.4) {$k$};
\node at (-0.45,0.9) {$\so(8)^{(2)}$};
\end{scope}
\node (v3) at (0.5,0.4) {\tiny{$2$}};
\draw  (v1) edge (v3);
\draw [->] (v3) edge (v2);
\end{tikzpicture}
\ee
is not allowed.
\item There are various kinds of possibilities for multiple neighbors of $\sp(0)^{(1)}$. All of the possibilities listed in Table \ref{R3E} are allowed with the substitution of $\fg^{(1)}$ in place of every trivial or non-trivial algebra $\fg$ appearing in that table. New possibilities involving undirected edges are listed in Table \ref{FR3E}. These are obtained by performing outer automorphisms on the possibilities listed in Table \ref{R3E}. However, some of the outer automorphisms do not yield a symmetry of the the theory. 

For example, consider the decomposition of the adjoint $\mathbf{248}$ of $\fe_8$ under $\su(3)\oplus\fe_6$
\be\label{e8su3e6}
\mathbf{248}\to\mathbf{(8,1)}\oplus\mathbf{(1,78)}\oplus\mathbf{(3,27)}\oplus\mathbf{(3',27')}
\ee
It can be seen from the above decomposition that neither the outer automorphism of $\su(3)$ nor the outer automorphism of $\fe_6$ is a symmetry of the decomposition, implying that neither the configuration
\be\label{su3Te6}
\begin{tikzpicture}
\node (v1) at (-0.5,0.4) {$k$};
\node at (-0.45,0.9) {$\su(3)^{(2)}$};
\begin{scope}[shift={(2,0)}]
\node (v2) at (-0.5,0.4) {$1$};
\node at (-0.45,0.9) {$\sp(0)^{(1)}$};
\end{scope}
\begin{scope}[shift={(4,0)}]
\node (v3) at (-0.5,0.4) {$l$};
\node at (-0.45,0.9) {$\fe_6^{(1)}$};
\end{scope}
\draw  (v1) edge (v2);
\draw  (v2) edge (v3);
\end{tikzpicture}
\ee
nor the configuration
\be\label{su3e6T}
\begin{tikzpicture}
\node (v1) at (-0.5,0.4) {$k$};
\node at (-0.45,0.9) {$\su(3)^{(1)}$};
\begin{scope}[shift={(2,0)}]
\node (v2) at (-0.5,0.4) {$1$};
\node at (-0.45,0.9) {$\sp(0)^{(1)}$};
\end{scope}
\begin{scope}[shift={(4,0)}]
\node (v3) at (-0.5,0.4) {$l$};
\node at (-0.45,0.9) {$\fe_6^{(2)}$};
\end{scope}
\draw  (v1) edge (v2);
\draw  (v2) edge (v3);
\end{tikzpicture}
\ee
is an allowed KK theory. However, the configuration
\be\label{su3Te6T}
\begin{tikzpicture}
\node (v1) at (-0.5,0.4) {$k$};
\node at (-0.45,0.9) {$\su(3)^{(2)}$};
\begin{scope}[shift={(2,0)}]
\node (v2) at (-0.5,0.4) {$1$};
\node at (-0.45,0.9) {$\sp(0)^{(1)}$};
\end{scope}
\begin{scope}[shift={(4,0)}]
\node (v3) at (-0.5,0.4) {$l$};
\node at (-0.45,0.9) {$\fe_6^{(2)}$};
\end{scope}
\draw  (v1) edge (v2);
\draw  (v2) edge (v3);
\end{tikzpicture}
\ee
is an allowed KK theory since the combined outer automorphism of $\su(3)$ and $\fe_6$ is indeed a symmetry of the decomposition (\ref{e8su3e6}). Correspondingly, neither (\ref{su3Te6}) nor (\ref{su3e6T}) appears in the Table \ref{FR3E}, while (\ref{su3Te6T}) does appear in Table \ref{FR3E}.

Similarly, the reader can check that the following configurations do not give rise to allowed KK theories:
\be
\begin{tikzpicture}
\node (v1) at (-0.5,0.4) {$k$};
\node at (-0.45,0.9) {$\so(8)^{(2)}$};
\begin{scope}[shift={(2,0)}]
\node (v2) at (-0.5,0.4) {$1$};
\node at (-0.45,0.9) {$\sp(0)^{(1)}$};
\end{scope}
\begin{scope}[shift={(4,0)}]
\node (v3) at (-0.5,0.4) {$l$};
\node at (-0.45,0.9) {$\so(8)^{(q)}$};
\end{scope}
\draw  (v1) edge (v2);
\draw  (v2) edge (v3);
\end{tikzpicture}
\ee
for $q=1,3$. However, $q=2$ is allowed.
\be
\begin{tikzpicture}
\node (v1) at (-0.5,0.4) {$k$};
\node at (-0.45,0.9) {$\su(4)^{(p)}$};
\begin{scope}[shift={(2,0)}]
\node (v2) at (-0.5,0.4) {$1$};
\node at (-0.45,0.9) {$\sp(0)^{(1)}$};
\end{scope}
\begin{scope}[shift={(4,0)}]
\node (v3) at (-0.5,0.4) {$l$};
\node at (-0.45,0.9) {$\so(10)^{(q)}$};
\end{scope}
\draw  (v1) edge (v2);
\draw  (v2) edge (v3);
\end{tikzpicture}
\ee
for $(p,q)$ equal to $(1,2)$ and $(2,1)$. However, $(1,1)$ and $(2,2)$ are allowed.
\be
\begin{tikzpicture}
\node (v1) at (-0.5,0.4) {$k$};
\node at (-0.45,0.9) {$\su(3)^{(p)}$};
\begin{scope}[shift={(2,0)}]
\node (v2) at (-0.5,0.4) {$1$};
\node at (-0.45,0.9) {$\sp(0)^{(1)}$};
\end{scope}
\begin{scope}[shift={(4,0)}]
\node (v3) at (-0.5,0.4) {$l$};
\node at (-0.45,0.9) {$\so(10)^{(q)}$};
\end{scope}
\draw  (v1) edge (v2);
\draw  (v2) edge (v3);
\end{tikzpicture}
\ee
for $(p,q)$ equal to $(1,2)$ and $(2,1)$.
\be
\begin{tikzpicture}
\node (v1) at (-0.5,0.4) {$k$};
\node at (-0.45,0.9) {$\su(3)^{(p)}$};
\begin{scope}[shift={(2,0)}]
\node (v2) at (-0.5,0.4) {$1$};
\node at (-0.45,0.9) {$\sp(0)^{(1)}$};
\end{scope}
\begin{scope}[shift={(4,0)}]
\node (v3) at (-0.5,0.4) {$l$};
\node at (-0.45,0.9) {$\su(5,6)^{(q)}$};
\end{scope}
\draw  (v1) edge (v2);
\draw  (v2) edge (v3);
\end{tikzpicture}
\ee
for $(p,q)$ equal to $(1,2)$ and $(2,1)$.
\be
\begin{tikzpicture}
\node (v1) at (-0.5,0.4) {$k$};
\node at (-0.45,0.9) {$\su(2)^{(1)}$};
\begin{scope}[shift={(2,0)}]
\node (v2) at (-0.5,0.4) {$1$};
\node at (-0.45,0.9) {$\sp(0)^{(1)}$};
\end{scope}
\begin{scope}[shift={(4,0)}]
\node (v3) at (-0.5,0.4) {$l$};
\node at (-0.45,0.9) {$\so(12)^{(2)}$};
\end{scope}
\draw  (v1) edge (v2);
\draw  (v2) edge (v3);
\end{tikzpicture}
\ee
\be
\begin{tikzpicture}
\node (v1) at (-0.5,0.4) {$k$};
\node at (-0.45,0.9) {$\su(4)^{(p)}$};
\begin{scope}[shift={(2,0)}]
\node (v2) at (-0.5,0.4) {$1$};
\node at (-0.45,0.9) {$\sp(0)^{(1)}$};
\end{scope}
\begin{scope}[shift={(4,0)}]
\node (v3) at (-0.5,0.4) {$l$};
\node at (-0.45,0.9) {$\so(8)^{(3)}$};
\end{scope}
\draw  (v1) edge (v2);
\draw  (v2) edge (v3);
\end{tikzpicture}
\ee
for $p=1,2$.
\be
\begin{tikzpicture}
\node (v1) at (-0.5,0.4) {$k$};
\node at (-0.45,0.9) {$\so(7)^{(1)}$};
\begin{scope}[shift={(2,0)}]
\node (v2) at (-0.5,0.4) {$1$};
\node at (-0.45,0.9) {$\sp(0)^{(1)}$};
\end{scope}
\begin{scope}[shift={(4,0)}]
\node (v3) at (-0.5,0.4) {$l$};
\node at (-0.45,0.9) {$\so(8)^{(3)}$};
\end{scope}
\draw  (v1) edge (v2);
\draw  (v2) edge (v3);
\end{tikzpicture}
\ee
\item It is not possible for $\sp(0)^{(1)}$ to have multiple neighbors when one of the neighbors is connected to it by a directed edge going outwards from $\sp(0)^{(1)}$. This is simply a consequence of the fact that $\sp(0)$ cannot have three neighbors in the context of $6d$ SCFTs.

However, it is possible for $\sp(0)^{(1)}$ to have multiple neighbors with some neighbors having directed edges pointing inwards towards $\sp(0)^{(1)}$. These possibilities can be simply obtained by replacing one or more undirected edges appearing in Tables \ref{R3E} and \ref{FR3E} by suitable directed edges (pointing inwards) taken from Table \ref{BTR2N}. One has to ensure that the matrix associated to the resulting configuration is positive definite, which disallows some substitutions. We do not pursue a full classification of such cases since they won't be useful in this paper. Later on, in Section \ref{DE}, we will provide a general prescription to obtain the gluing rules associated to such directed edges from the gluing rules associated to their ``parent'' undirected edges.
\eit

\section{Prepotential for $5d$ KK theories}\label{PPKK}
The goal of this section is to propose a formula for the prepotential of a $5d$ KK theory $\fT^{KK}_{S,\{q_\alpha\}}$ starting from the tensor branch description of the corresponding $6d$ SCFT $\fT$.

\subsection{Prepotential}

\begin{table}[htbp]
\begin{center}
\begin{tabular}{|c|c|c|l|} \hline
$\fg$&$\cO^{(q)}$&$\fh$& $\cR_\fg\to\cR_\fh$
\\ \hline
 \hline
$\su(2m)$&$\cO^{(2)}$&$\sp(m)$&\raisebox{-.15\height}{$\F\to\F$, $\bar\F\to\F$, $\Asym\to\Asym\oplus\mathsf{1}$}
\\ \hline
$\su(2m+1)$&$\cO^{(2)}$&$\sp(m)$&\raisebox{-.15\height}{$\F\to\F\oplus\mathsf{1}$, $\bar\F\to\F\oplus\mathsf{1}$}
\\ \hline
$\so(2m)$&$\cO^{(2)}$&$\so(2m-1)$&\raisebox{-.15\height}{$\F\to\F\oplus\mathsf{1}$, $\S\to\S$, $\C\to\S$}
\\ \hline
$\fe_6$&$\cO^{(2)}$&$\ff_4$&\raisebox{-.15\height}{$\F\to\F\oplus\mathsf{1}$, $\bar\F\to\F\oplus\mathsf{1}$}
\\ \hline
$\so(8)$&$\cO^{(3)}$&$\fg_2$&\raisebox{-.15\height}{$\F\to\F\oplus\mathsf{1}$, $\S\to\F\oplus\mathsf{1}$, $\C\to\F\oplus\mathsf{1}$}
\\ \hline
\end{tabular}
\end{center}
\caption{The table displays the invariant algebra $\fh$ when $\fg$ is quotiented by $\cO^{(q)}$. An irrep $\cR_\fg$ of $\fg$ decomposes to an irrep $\cR_\fh$ of $\fh$ and this decomposition is displayed (for representations relevant in this paper) in the column labeled $\cR_\fg\to\cR_\fh$. $\mathsf{1}$ denotes the singlet representation.}\label{Inv}
\end{table}
Compactify a $6d$ SCFT $\fT$ on a circle with a twist $S,\{q_\alpha\}$ around the circle. Let us analyze the low energy theory. Every node $\alpha$ in $\Sigma^{S,\{q_\alpha\}}_\fT$ gives rise to a low energy $5d$ gauge algebra $\fh_\alpha=\fg_\alpha/\cO^{(q_\alpha)}$ which is the subalgebra of $\fg_\alpha$ left invariant by the action of outer automorphism $\cO^{(q_\alpha)}$. In this paper, our choice of outer automorphisms is such that the invariant subalgebras are those listed in Table \ref{Inv}. For each node $\alpha$, we obtain an additional $\u(1)_\alpha$ gauge algebra in the low energy $5d$ theory coming from the reduction of a tensor multiplet $B_i$ on the circle where $i$ lies in the orbit $\alpha$.

Now we determine the spectrum of hypermultiplets charged under $\oplus_\alpha\fh_\alpha$ under the low energy $5d$ theory. First of all, for every node $i$ in $6d$ theory, we define $\cT_i=\oplus_j\cR_{ij,i}^{\oplus \text{dim}(\cR_{ij,j})}$. Recall that $\cT_i\subseteq\cR_i$ and hence the $6d$ theory contains hypermultiplets charged under representation $\cS_i$ of $\fg_i$ where $\cS_i$ is defined such that $\cS_i\oplus\cT_i=\cR_i$. $\cS_i$ is the representation formed by those hypers that are only charged $\fg_i$ and not under any other gauge algebra $\fg_j$ with $j\neq i$.

As detailed in Table \ref{Inv}, irreducible representations $\cR_{\fg_\alpha}$ of $\fg_\alpha$ can be viewed as irreducible representations of $\cR_{\fh_\alpha}$. We can thus view hypers transforming in representation $\cS_i$ of $\fg_i$ as transforming in a representation of $\fh_\alpha$. Let us denote this representation of $\fh_\alpha$ by $\tilde\cS_\alpha$. The outer automorphism $\cO^{(q_\alpha)}$ then permutes constituent irreps inside $\tilde\cS_\alpha$ and thus acts on $\tilde\cS_\alpha$ as an automorphism. The low energy $5d$ theory then contains hypers transforming in the representation
\be
\cS_\alpha:=\tilde\cS_\alpha/\cO^{(q_\alpha)}
\ee
These hypers are only charged under $\fh_\alpha$ and not under any other gauge algebra $\fh_\beta$ with $\beta\neq\alpha$.

Now consider other hypermultiplets that are charged under multiple gauge algebras in the $6d$ theory. These descend to hypermultiplets charged under multiple gauge algebras in the low energy $5d$ theory plus some hypers only charged under the individual algebras. Consider the mixed representation $\cR_{ij}=\cR_{ij,i}\otimes\cR_{ij,j}$ of $\fg_i\oplus\fg_j$ in the $6d$ theory. Let $i$ and $j$ lie in orbits $\alpha$ and $\beta$ respectively. Let $\cR_{ij,i}$ decompose as $\cR_{\alpha\beta,\alpha}\oplus n_{\alpha\beta,\alpha}\mathsf{1}$ when viewed as a representation of $\fh_\alpha$, where $\cR_{\alpha\beta,\alpha}$ is the full subrepresentation that is charged non-trivially under $\fh_\alpha$. Similarly, let $\cR_{ij,j}$ decompose as $\cR_{\alpha\beta,\beta}\oplus n_{\alpha\beta,\beta}\mathsf{1}$ when viewed as a representation of $\fh_\beta$, where $\cR_{\alpha\beta,\beta}$ is the full subrepresentation that is charged non-trivially under $\fh_\beta$. Then, under the twist, $\cR_{ij}$ descends to a mixed representation $\cR_{\alpha\beta}$ of $\fh_\alpha\oplus\fh_\beta$ plus representations $\cS_{\alpha\beta,\alpha}$ and $\cS_{\alpha\beta,\beta}$ of $\fh_\alpha$ and $\fh_\beta$ respectively. Here $\cR_{\alpha\beta}=\cR_{\alpha\beta,\alpha}\otimes\cR_{\alpha\beta,\beta}$, $\S_{\alpha\beta,\alpha}=n_{\alpha\beta,\beta}\cR_{\alpha\beta,\alpha}$, and  $\S_{\alpha\beta,\beta}=n_{\alpha\beta,\alpha}\cR_{\alpha\beta,\beta}$.

In addition to the above, we also obtain hypers in the symmetric product $\text{Sym}^2(\cR_{ij,i})$ for all $j\neq i$ such that both $j$ and $i$ are in the same orbit $\alpha$. Thus, the full representation $\cR_\alpha$ formed by hypers under $\fh_\alpha$ is
\be
\cR_\alpha=\oplus_{j\in\alpha}\text{Sym}^2(\cR_{ij,i})|_{\fh_\alpha}\oplus\cS_\alpha\oplus_\beta\left(\cR_{\alpha\beta,\alpha}^{\oplus \text{dim}(\cR_{\alpha\beta,\beta})}\oplus\cS_{\alpha\beta,\alpha}\right)
\ee
where $\text{Sym}^2(\cR_{ij,i})|_{\fh_\alpha}$ means that we view $\text{Sym}^2(\cR_{ij,i})$ as a representation of $\fh_\alpha$. Note that in the above expression, $i$ is a fixed node in the orbit $\alpha$, $j$ cannot equal $i$, and $\beta$ cannot equal $\alpha$. There are no hypers charged under $\u(1)_\alpha$. Just as the representations $\cR_i$ and $\cR_{ij}$ for all $i$ and $j$ determine the full matter content for $6d$ SCFTs, the representations $\cR_\alpha$ and $\cR_{\alpha\beta}$ for all $\alpha$ and $\beta$ determine the full matter content for $5d$ KK theories.

As an example, let us determine the low energy $5d$ theory for (\ref{ex4}). The $5d$ gauge algebra is $\fh=\sp(n)\oplus\so(2m-1)$. A half-bifundamental of $\sp(n)\oplus\so(2m)$ decomposes as a half-bifundamental of $\sp(n)\oplus\so(2m-1)$ plus a half-fundamental of $\sp(n)$. Thus, the two half-bifundamentals between the $\sp(n)$ and the two $\so(2m)$ in (\ref{ex4}) descend to a half-bifundamental of $\fh$ plus a half-fundamental of $\sp(n)$ in the $5d$ theory. There are $2m-8-n$ extra fundamentals of the left $\so(2m)$ in (\ref{ex4}) not charged under any other gauge algebra. Similarly, there are $2m-8-n$ extra fundamentals of the right $\so(2m)$ in (\ref{ex4}) not charged under any other gauge algebra. These two sets of fundamentals descend to $2m-8-n$ fundamentals of $\so(2m-1)$ in the $5d$ theory. We also obtain $2m-8-n$ singlets that decouple and so we ignore them. Finally, there are $2n+8-2m$ extra fundamentals of $\sp(n)$ in (\ref{ex4}) not charged under any other gauge algebra. These hypers descend to $2n+8-2m$ extra fundamentals of $\sp(n)$ in the low energy $5d$ theory that are not charged under $\so(2m-1)$. To recap, the low energy $5d$ theory is an $\sp(n)\oplus\so(2m-1)$ gauge theory with a half-bifundamental plus $4n+17-4m$ half-fundamentals of $\sp(n)$ plus $2m-8-n$ fundamentals of $\so(2m-1)$.

As another example, let us determine the low energy $5d$ theory for (\ref{eg2}). The two $\su(m)$ get identified to a single $\su(m)$ algebra. Similarly, the two $\su(n)$ get identified to a single $\su(n)$ algebra. Thus the $5d$ gauge algebra is $\fh=\su(n)\oplus\su(m)$. The bifundamentals of $\su(m)\oplus\su(n)$ descend to a single bifundamental of $\fh$. The bifundamental of $\su(n)\oplus\su(n)$ descends to $\Sym$ of $\su(n)$. Furthermore, we obtain $n-m$ extra fundamentals of $\su(n)$ and $2m-n$ extra fundamentals of $\su(m)$. Thus, the low energy $5d$ theory is an $\su(n)\oplus\su(m)$ gauge theory with a bifundamental plus $(2m-n)\F$ of $\su(m)$ plus $(n-m)\F\oplus\Sym$ of $\su(n)$.

The low energy $5d$ gauge theory also contains tree-level Chern-Simons terms that arise from the reduction of (\ref{GS}) on the circle. These can be written as
\be \label{CST}
\Omega_S^{\alpha\beta}A_{0,\alpha}\wedge\tr(F_\beta^2)
\ee
where $A_{0,\alpha}$ is the gauge field corresponding to the $\u(1)_\alpha$ obtained by reducing $B_\alpha$ on the circle and $F_\beta$ is the gauge field strength for $\fh_\beta$. In writing (\ref{CST}), we have used the fact that the index of $\fh_\beta$ in $\fg_\beta$ is one which is true for our choice of $\fh$ listed in Table \ref{OOA}. (\ref{CST}) contributes the following tree-level term to the prepotential
\be \label{PPT}
6\mathcal{F}^{\text{tree}}_{S,\{q_\alpha\}}=6\sum_{\alpha,\beta}\half\Omega_S^{\alpha\beta}\phi_{0,\alpha}\left(K_\beta^{ab}\phi_{a,\beta}\phi_{b,\beta}\right)
\ee
where $\phi_{0,\alpha}$ is the scalar living in the vector multiplet corresponding to $\u(1)_\alpha$ and $\phi_{a,\beta}$ are scalars living in the vector multiplets corresponding to $\u(1)_{a,\beta}$ which parametrize the Cartan of $\fh_\beta$. Here $K_\beta^{ab}$ is the Killing form on $\fh_\beta$ normalized such that its diagonal entries are minimum positive integers while keeping all the other entries integer valued.

Let $\fh=\oplus_\alpha \fh_\alpha$ be the total gauge algebra visible at low energies. The low energy hypermultiplets form some representation $\cR$ of $\fh$ which decomposes into irreducible representations of $\fh$ as $\cR=\oplus_f\cR_f$. Note that it is possible to have $f\neq f'$ such that $\cR_f=\cR_{f'}$. In other words, the index $f$ distinguishes multiple copies of representation $\cR_f$. Now we can add the one-loop contribution to the prepotential (\ref{PPT}) to obtain
\be \label{PP}
6\mathcal{F}_{S,\{q_\alpha\}}=\sum_{\alpha,\beta}3\Omega_S^{\alpha\beta}\phi_{0,\alpha}\left(K_\beta^{ab}\phi_{a,\beta}\phi_{b,\beta}\right)+\frac{1}{2}\left(\sum_{r}|r \cdot \phi|^3- \sum_f \sum_{w(\cR_f)}|w(\cR_f) \cdot \phi + m_f|^3\right)
\ee
where $r$ are the roots of $\fh=\oplus_\alpha \fh_\alpha$, $w(\cR_f)$ parametrize weights of $\cR_f$ and $m_f\in\R$ is a mass term for each full\footnote{Half-hypermultiplets do not admit mass parameters unless completed into a full hypermultiplet.} hypermultiplet $f$. The notation $w\cdot\phi$ denotes the scalar product of the Dynkin coefficients of the weight $w$ with Coulomb branch parameters. Note that similar approaches for computing prepotentials of 5d theories have appeared in the literature---see for example \cite{Bonetti:2011mw,Bonetti:2013cza,Grimm:2015zea}.

In (\ref{PP}) we must impose that mass terms for hypers belonging to $\cS_{\alpha\beta,\alpha}$ and $\cS_{\alpha\beta,\beta}$ equal the mass term for hypers belonging to $\cR_{\alpha\beta}$. This is because $\cR_{\alpha\beta}$, $\cS_{\alpha\beta,\alpha}$ and $\cS_{\alpha\beta,\beta}$ all descend from the same $6d$ representation $\cR_{ij}$ which has only a single $\u(1)$ symmetry rotating it. The Wilson lines for this $\u(1)$ around the compactification circle gives rise to the mass terms for $\cR_{\alpha\beta}$, $\cS_{\alpha\beta,\alpha}$ and $\cS_{\alpha\beta,\beta}$, and hence all these mass terms must be equal.

We propose that (\ref{PP}) is the full exact prepotential for $\fT^{KK}_{S,\{q_\alpha\}}$ where we have ignored the terms involving the mass parameter $\frac{1}{R}$ where $R$ is the radius of compactification. We are justified in doing so since these terms do not play any role in this paper. Moreover, only the part of $6\mathcal{F}_{S,\{q_\alpha\}}$ that is cubic in Coulomb branch parameters $\phi_{a,\alpha}$ is relevant to the discussion in this paper; so, for convenience, we denote the part of the prepotential cubic in Coulomb branch parameters by $6\mathcal{F}^\phi_{S,\{q_\alpha\}}$.

Notice that fixing the relative values of $\phi_{a,\alpha}$ and $m_f$ fixes the signs of the terms inside absolute values in (\ref{PP}). As the relative values of $\phi_{a,\alpha}$ and $m_f$ are changed, the sign of some of the terms in (\ref{PP}) changes. This leads to jumps in the coefficients of various terms in the resulting $6\mathcal{F}^\phi_{S,\{q_\alpha\}}$. This means that different relative values of $\phi_{a,\alpha}$ and $m_f$ lead to different phases inside the Coulomb branch of the $5d$ KK theory.

Let us illustrate through a simple example of the KK theory specified by the graph
\be\label{ex8}
\begin{tikzpicture}
\node at (-0.5,0.45) {2};
\node at (-0.45,0.9) {$\su(3)^{(1)}$};
\end{tikzpicture}
\ee
This theory has six hypers in fundamental of $\su(3)$. The Dynkin coefficients of the positive roots of $\su(3)$ are $(2,-1)$, $(1,1)$ and $(-1,2)$.  The Dynkin coefficients for the weights of fundamental are $(1,0)$, $(-1,1)$ and $(0,-1)$. The Killing form is
\[\begin{pmatrix}
2&-1\\
-1&2
\end{pmatrix}\]
and $\Omega_S^{\alpha\beta}$ is a $1\times1$ matrix which equals $2$. Without loss of generality, we can take $r\cdot\phi$ for positive roots to be positive. This implies that $r\cdot\phi$ for negative roots is negative. 

Let us first fix all the mass terms to be zero. Then the first weight $(1,0)$ contributes with a positive sign since the positivity of $r\cdot\phi$ for positive roots implies that $\phi_1$ is positive. Similarly, the third weight $(0,-1)$ contributes with a negative sign to the prepotential. However, the sign of second weight $(-1,1)$ cannot be determined uniquely, and hence the theory has two phases when all mass parameters vanish. These two phases are distinguished by the sign $s$ of the contribution due to the weight $(-1,1)$. The prepotential can be written as
\begin{align}\label{ex7}
6\cF^\phi=6\cF=&12\phi_0\left(\phi_1^2+\phi_2^2-\phi_1\phi_2\right)+\left(\left(2 \phi _1-\phi _2\right)^3+\left(\phi _1+\phi _2\right)^3+\left(2 \phi _2-\phi _1\right)^3\right)\nn\\&-3 \left(s\left(\phi _2-\phi _1\right)^3+\phi _1^3+\phi _2^3\right)
\end{align}
Here $12\phi_0\left(\phi_1^2+\phi_2^2-\phi_1\phi_2\right)$ is the contribution coming from the Green-Schwarz term in $6d$, $\left(2 \phi _1-\phi _2\right)^3+\left(\phi _1+\phi _2\right)^3+\left(2 \phi _2-\phi _1\right)^3$ is the contribution coming from the positive and negative roots, and $-3 \left(s\left(\phi _2-\phi _1\right)^3+\phi _1^3+\phi _2^3\right)$ is the contribution coming from the weights of six hypers in fundamental.

When we turn on mass parameters, the sign of the weights corresponding to different hypers can be changed. For example, consider turning on a mass parameter for one of the fundamentals $m_1$ while keeping the mass parameters for the other five fundamentals zero. Now we obtain contributions from terms of the form $|m_1+\phi_1|$, $|m_1-\phi_1+\phi_2|$ and $|m_1-\phi_2|$. Depending on the value of $m_1$, we go through various new phases of the theory which are parametrized by choices of signs of these three terms. For example, suppose that $m_1$ is positive and very large, so that all the three terms are positive. Moreover, assume that $\phi_2-\phi_1$ is positive, so that $s=+1$. Then the resulting phase is governed by the following prepotential
\begin{align}
6\cF=&12\phi_0\left(\phi_1^2+\phi_2^2-\phi_1\phi_2\right)+\left(\left(2 \phi _1-\phi _2\right)^3+\left(\phi _1+\phi _2\right)^3+\left(2 \phi _2-\phi _1\right)^3\right)\nn\\&-\frac52 \left(\left(\phi _2-\phi _1\right)^3+\phi _1^3+\phi _2^3\right)-\half\left(\left(\phi _2-\phi _1+m_1\right)^3+(\phi _1+m_1)^3+(-\phi _2+m_1)^3\right)
\end{align}
which implies that the truncated prepotential is
\begin{align}\label{eg7}
6\cF^\phi=&12\phi_0\left(\phi_1^2+\phi_2^2-\phi_1\phi_2\right)+\left(\left(2 \phi _1-\phi _2\right)^3+\left(\phi _1+\phi _2\right)^3+\left(2 \phi _2-\phi _1\right)^3\right)\nn\\&-3 \left(\left(\phi _2-\phi _1\right)^3+\phi _1^3\right)-2\phi_2^3
\end{align}

We caution the reader that there can be phases of the KK theory which cannot be traversed by changing the signs of various contributions to the prepotential. In other words, they are not visible to the canonical low energy gauge theory that we associated to the KK theory in the beginning of this subsection. We will refer to such phases as non-gauge theoretic. This terminology does not mean that the low energy theory governing such phases cannot be understood as Coulomb branch of a gauge theory. Rather it simply means that low energy theory governing such phases cannot be understood as part of Coulomb branch of the canonical gauge theory associated to the corresponding KK theory.

\subsection{Shifting the prepotential}\label{Sshift}

\begin{table}[htbp]
\begin{center}
\begin{tabular}{cc}
\ubf{$\su(n)^{(1)}$}:
\vspace{2mm}
&
\ubf{$\fe_6^{(1)}$}:
\vspace{2mm}\\
\begin{tikzpicture}[scale=1]
\draw[fill=black]
(0,0) node (v1) {} circle [radius=.1]
(1,0) node (v3) {} circle [radius=.1]
(2,0) node (v4) {} circle [radius=.1]
(4,0) node (v5) {} circle [radius=.1]
(5,0) node (v6) {} circle [radius=.1]
(6,0) node (v7) {} circle [radius=.1];
\draw[fill=white]
(3,1.5) node (v8) {} circle [radius=.1];
\node (v2) at (3,0) {$\cdots$};
\draw  (v1) edge (v3);
\draw  (v3) edge (v4);
\draw  (v4) edge (v2);
\draw  (v2) edge (v5);
\draw  (v5) edge (v6);
\draw  (v6) edge (v7);
\draw  (v1) edge (v8);
\draw  (v7) edge (v8);
\node at (0,-0.3) {\scriptsize{$1$}};
\node at (3,1.8) {\scriptsize{$1$}};
\node at (1,-0.3) {\scriptsize{$1$}};
\node at (2,-0.3) {\scriptsize{$1$}};
\node at (4,-0.3) {\scriptsize{$1$}};
\node at (5,-0.3) {\scriptsize{$1$}};
\node at (6,-0.3) {\scriptsize{$1$}};
\node at (0,-0.6) {\scriptsize{{\color{red} 1}}};
\node at (1,-0.6) {\scriptsize{{\color{red} 1}}};
\node at (2,-0.6) {\scriptsize{{\color{red} 1}}};
\node at (4,-0.6) {\scriptsize{{\color{red} 1}}};
\node at (5,-0.6) {\scriptsize{{\color{red} 1}}};
\node at (6,-0.6) {\scriptsize{{\color{red} 1}}};
\node at (3,2.1) {\scriptsize{{\color{red} 1}}};
\end{tikzpicture}
\vspace{2mm}
&
\begin{tikzpicture} [scale=1]
\draw[fill=black]
(0,0) node (v1) {} circle [radius=.1]
(1,0) node (v3) {} circle [radius=.1]
(2,0) node (v4){} circle [radius=.1]
(4,0) node (v5) {} circle [radius=.1]
(3,0) node (v7) {} circle [radius=.1]
(2,1) node (v8) {} circle [radius=.1];
\draw[fill=white]
(2,2) node (v9) {} circle [radius=.1];
\node (v2) at (3,0) {};
\draw  (v1) edge (v3);
\draw  (v3) edge (v4);
\draw  (v4) edge (v2);
\draw  (v2) edge (v5);
\draw  (v4) edge (v7);
\draw  (v4) edge (v8);
\draw  (v9) edge (v8);
\node at (2.3,1) {\scriptsize{$2$}};
\node at (0,-0.3) {\scriptsize{$1$}};
\node at (1,-0.3) {\scriptsize{$2$}};
\node at (2,-0.3) {\scriptsize{$3$}};
\node at (3,-0.3) {\scriptsize{$2$}};
\node at (4,-0.3) {\scriptsize{$1$}};
\node at (2.3,2) {\scriptsize{$1$}};
\node at (0,-0.6) {\scriptsize{{\color{red}$1$}}};
\node at (1,-0.6) {\scriptsize{{\color{red}$2$}}};
\node at (2,-0.6) {\scriptsize{{\color{red}$3$}}};
\node at (3,-0.6) {\scriptsize{{\color{red}$2$}}};
\node at (4,-0.6) {\scriptsize{{\color{red}$1$}}};
\node at (2.6,1) {\scriptsize{{\color{red}$2$}}};
\node at (2.6,2) {\scriptsize{{\color{red}$1$}}};
\end{tikzpicture} 
\vspace{2mm}\\
\ubf{$\so(2n+1)^{(1)}$}: 
\vspace{2mm}
&
\ubf{$\fe_7^{(1)}$}:
\vspace{2mm}\\
\begin{tikzpicture}[scale=1]
\draw[fill=black]
(0,0) node (v1) {} circle [radius=.1]
(1,0) node (v3) {} circle [radius=.1]
(2,0) node (v4) {} circle [radius=.1]
(4,0) node (v5) {} circle [radius=.1]
(5,0) node (v6) {} circle [radius=.1]
(6,0) node (v7) {} circle [radius=.1];
\draw[fill=white]
(1,1) node (v8) {} circle [radius=.1];
\node (v2) at (3,0) {$\cdots$};
\node at (0,-0.3) {\scriptsize{$1$}};
\node at (1,-0.3) {\scriptsize{$2$}};
\node at (2,-0.3) {\scriptsize{$2$}};
\node at (4,-0.3) {\scriptsize{$2$}};
\node at (5,-0.3) {\scriptsize{$2$}};
\node at (6,-0.3) {\scriptsize{$1$}};
\node at (1.3,1) {\scriptsize{$1$}};
\draw  (v1) edge (v3);
\draw  (v3) edge (v4);
\draw  (v4) edge (v2);
\draw  (v2) edge (v5);
\draw  (v5) edge (v6);
\draw  (v3) edge (v8);
\node at (0,-0.6) {\scriptsize{\color{red}$1$}};
\node at (1,-0.6) {\scriptsize{\color{red}$2$}};
\node at (2,-0.6) {\scriptsize{\color{red}$2$}};
\node at (4,-0.6) {\scriptsize{\color{red}$2$}};
\node at (5,-0.6) {\scriptsize{\color{red}$2$}};
\node at (6,-0.6) {\scriptsize{\color{red}$2$}};
\node at (1.6,1) {\scriptsize{\color{red}$1$}};
\begin{scope}[shift={(-0,0)}]
\draw (5.15,0.025) -- (5.825,0.025) (5.825,-0.025) -- (5.15,-0.025);
\draw (5.425,0.15) -- (5.575,0) -- (5.425,-0.15);
\end{scope}
\end{tikzpicture}
\vspace{2mm}
&
 \begin{tikzpicture}[scale=1]
\draw[fill=black]
(0,0) node (v1) {} circle [radius=.1]
(1,0) node (v3) {} circle [radius=.1]
(2,0) node (v4){} circle [radius=.1]
(4,0) node (v5) {} circle [radius=.1]
(3,0) node (v7) {} circle [radius=.1]
(5,0) node (v10) {} circle [radius=.1]
(2,1) node (v8) {} circle [radius=.1];
\draw[fill=white]
(-1,0) node (v9) {} circle [radius=.1];
\node (v2) at (3,0) {};
\draw  (v1) edge (v3);
\draw  (v3) edge (v4);
\draw  (v4) edge (v2);
\draw  (v2) edge (v5);
\draw  (v4) edge (v7);
\draw  (v4) edge (v8);
\draw  (v5) edge (v10);
\draw  (v9) edge (v1);
\node at (2.3,1) {\scriptsize{$2$}};
\node at (-1,-0.3) {\scriptsize{$1$}};
\node at (0,-0.3) {\scriptsize{$2$}};
\node at (1,-0.3) {\scriptsize{$3$}};
\node at (2,-0.3) {\scriptsize{$4$}};
\node at (3,-0.3) {\scriptsize{$3$}};
\node at (4,-0.3) {\scriptsize{$2$}};
\node at (5,-0.3) {\scriptsize{$1$}};
\node at (-1,-0.6) {\scriptsize{\color{red}$1$}};
\node at (0,-0.6) {\scriptsize{\color{red}$2$}};
\node at (1,-0.6) {\scriptsize{\color{red}$3$}};
\node at (2,-0.6) {\scriptsize{\color{red}$4$}};
\node at (3,-0.6) {\scriptsize{\color{red}$3$}};
\node at (4,-0.6) {\scriptsize{\color{red}$2$}};
\node at (5,-0.6) {\scriptsize{\color{red}$1$}};
\node at (2.6,1) {\scriptsize{\color{red}$2$}};
\end{tikzpicture}  
\vspace{2mm}\\
\ubf{$\sp(n)^{(1)}$}:
\vspace{2mm}
& 
\ubf{$\fe_8^{(1)}$}:
\vspace{2mm}\\ 
\begin{tikzpicture}[scale=1]
\draw[fill=black]
(1,0) node (v3) {} circle [radius=.1]
(2,0) node (v4) {} circle [radius=.1]
(4,0) node (v5) {} circle [radius=.1]
(5,0) node (v6) {} circle [radius=.1]
(6,0) node (v7) {} circle [radius=.1];
\draw[fill=white] (0,0) node (v1) {} circle [radius=.1];
\node (v2) at (3,0) {$\cdots$};
\node at (0,-0.3) {\scriptsize{$1$}};
\node at (1,-0.3) {\scriptsize{$1$}};
\node at (2,-0.3) {\scriptsize{$1$}};
\node at (4,-0.3) {\scriptsize{$1$}};
\node at (5,-0.3) {\scriptsize{$1$}};
\node at (6,-0.3) {\scriptsize{$1$}};
\node at (0,-0.6) {\scriptsize{\color{red}$1$}};
\node at (1,-0.6) {\scriptsize{\color{red}$2$}};
\node at (2,-0.6) {\scriptsize{\color{red}$2$}};
\node at (4,-0.6) {\scriptsize{\color{red}$2$}};
\node at (5,-0.6) {\scriptsize{\color{red}$2$}};
\node at (6,-0.6) {\scriptsize{\color{red}$1$}};
\draw  (v3) edge (v4);
\draw  (v4) edge (v2);
\draw  (v2) edge (v5);
\draw  (v5) edge (v6);
\begin{scope}[]
\draw (5.15,0.025) -- (5.825,0.025) (5.825,-0.025) -- (5.15,-0.025);
\draw (5.6,0.15) -- (5.45,0) -- (5.6,-0.15);
\end{scope}
\begin{scope}[shift={(-5,0)}]
\draw (5.15,0.025) -- (5.825,0.025) (5.825,-0.025) -- (5.15,-0.025);
\draw (5.4,0.15) -- (5.55,0) -- (5.4,-0.15);
\end{scope}
\end{tikzpicture}
\vspace{2mm}
&
\begin{tikzpicture}[scale=1]
\draw[fill=black]
(0,0) node (v1) {} circle [radius=.1]
(1,0) node (v3) {} circle [radius=.1]
(2,0) node (v4){} circle [radius=.1]
(4,0) node (v5) {} circle [radius=.1]
(3,0) node (v7) {} circle [radius=.1]
(5,0) node (v10) {} circle [radius=.1]
(6,0) node (v11) {} circle [radius=.1]
(2,1) node (v8) {} circle [radius=.1];
\draw[fill=white]
(7,0) node (v9) {} circle [radius=.1];
\node (v2) at (3,0) {} ;
\draw  (v1) edge (v3);
\draw  (v3) edge (v4);
\draw  (v4) edge (v2);
\draw  (v2) edge (v5);
\draw  (v4) edge (v7);
\draw  (v4) edge (v8);
\draw  (v5) edge (v10);
\draw  (v11) edge (v10);
\draw  (v9) edge (v11);
\node at (2.3,1) {\scriptsize{$3$}};
\node at (6,-0.3) {\scriptsize{$2$}};
\node at (7,-0.3) {\scriptsize{$1$}};
\node at (0,-0.3) {\scriptsize{$2$}};
\node at (1,-0.3) {\scriptsize{$4$}};
\node at (2,-0.3) {\scriptsize{$6$}};
\node at (3,-0.3) {\scriptsize{$5$}};
\node at (4,-0.3) {\scriptsize{$4$}};
\node at (5,-0.3) {\scriptsize{$3$}};
\node at (2.6,1) {\scriptsize{\color{red}$3$}};
\node at (6,-0.6) {\scriptsize{\color{red}$2$}};
\node at (7,-0.6) {\scriptsize{\color{red}$1$}};
\node at (0,-0.6) {\scriptsize{\color{red}$2$}};
\node at (1,-0.6) {\scriptsize{\color{red}$4$}};
\node at (2,-0.6) {\scriptsize{\color{red}$6$}};
\node at (3,-0.6) {\scriptsize{\color{red}$5$}};
\node at (4,-0.6) {\scriptsize{\color{red}$4$}};
\node at (5,-0.6) {\scriptsize{\color{red}$3$}};
\end{tikzpicture} 
\vspace{2mm}\\
\ubf{$\so(2n)^{(1)} $}:
\vspace{2mm}
&
\ubf{$\ff_4^{(1)}$}:
\vspace{2mm}\\
\begin{tikzpicture}[scale=1]
\draw[fill=black]
(0,-1) node (v1) {} circle [radius=.1]
(1,0) node (v3) {} circle [radius=.1]
(2,0) node (v4) {} circle [radius=.1]
(4,0) node (v5) {} circle [radius=.1]
(5,0) node (v6) {} circle [radius=.1]
(6,-1) node (v7) {} circle [radius=.1]
(6,1) node (v8) {} circle [radius=.1];
\draw[fill=white]
(0,1) node (v9) {} circle [radius=.1];
\node (v2) at (3,0) {$\cdots$};
\node at (0,-1.3) {\scriptsize{$1$}};
\node at (1,-0.3) {\scriptsize{$2$}};
\node at (2,-0.3) {\scriptsize{$2$}};
\node at (4,-0.3) {\scriptsize{$2$}};
\node at (5,-0.3) {\scriptsize{$2$}};
\node at (6,-1.3) {\scriptsize{$1$}};
\node at (6,1.3) {\scriptsize{$1$}};
\node at (0,1.3) {\scriptsize{$1$}};
\node at (0,-1.6) {\scriptsize{\color{red}$1$}};
\node at (1,-0.6) {\scriptsize{\color{red}$2$}};
\node at (2,-0.6) {\scriptsize{\color{red}$2$}};
\node at (4,-0.6) {\scriptsize{\color{red}$2$}};
\node at (5,-0.6) {\scriptsize{\color{red}$2$}};
\node at (6,-1.6) {\scriptsize{\color{red}$1$}};
\node at (6,1.6) {\scriptsize{\color{red}$1$}};
\node at (0,1.6) {\scriptsize{\color{red}$1$}};
\draw  (v1) edge (v3);
\draw  (v9) edge (v3);
\draw  (v3) edge (v4);
\draw  (v4) edge (v2);
\draw  (v2) edge (v5);
\draw  (v5) edge (v6);
\draw  (v6) edge (v7);
\draw  (v8) edge (v6);
\end{tikzpicture}
&
 \begin{tikzpicture}[scale=1]
\draw[fill=black]
(0,0) node (v1) {} circle [radius=.1]
(1,0) node (v3) {} circle [radius=.1]
(2,0) node (v4){} circle [radius=.1]
(3,0) node (v7) {} circle [radius=.1];
\draw[fill=white]
(-1,0) node (v8) {} circle [radius=.1];
\node (v2) at (3,0){} ;
\draw  (v1) edge (v3);
\draw  (v4) edge (v2);
\draw  (v4) edge (v7);
\draw  (v1) edge (v8);
\node at (-1,-0.3) {\scriptsize{$1$}};
\node at (0,-0.3) {\scriptsize{$2$}};
\node at (1,-0.3) {\scriptsize{$3$}};
\node at (2,-0.3) {\scriptsize{$2$}};
\node at (3,-0.3) {\scriptsize{$1$}}; 
\node at (-1,-0.6) {\scriptsize{\color{red}$1$}};
\node at (0,-0.6) {\scriptsize{\color{red}$2$}};
\node at (1,-0.6) {\scriptsize{\color{red}$3$}};
\node at (2,-0.6) {\scriptsize{\color{red}$4$}};
\node at (3,-0.6) {\scriptsize{\color{red}$2$}};
\begin{scope}[shift={(-4,0)}]
\draw (5.15,0.025) -- (5.825,0.025) (5.825,-0.025) -- (5.15,-0.025);
\draw (5.425,0.15) -- (5.575,0) -- (5.425,-0.15);
\end{scope}
\begin{scope}[shift={(0,-2.4)}]
\draw[fill=black]
(0,0) node (v1) {} circle [radius=0.1]
(1,0) node (v3) {} circle [radius=0.1];
\draw[fill=white]
(2,0) node (v8) {} circle [radius=0.1];
\node (v4) at (0,0.03) {};
\node (v5) at (1,0.03) {} ;
\node (v6) at (0,-0.03) {} ;
\node (v7) at (1,-0.03) {} ;
\draw (v3) -- (v8);
\node at (2,-0.3) {\scriptsize{$1$}};
\node at (0,-0.3) {\scriptsize{$1$}};
\node at (1,-0.3) {\scriptsize{$2$}};
\node at (2,-0.6) {\scriptsize{\color{red}$1$}};
\node at (0,-0.6) {\scriptsize{\color{red}$3$}};
\node at (1,-0.6) {\scriptsize{\color{red}$2$}};
\begin{scope}[shift={(-5,0)}]
\draw (5.15,0.05) -- (5.825,0.05) (5.825,0) -- (5.15,0) (5.825,-0.05) -- (5.15,-0.05);
\draw (5.575,0.175) -- (5.4,0) -- (5.575,-0.175);
\end{scope}
\end{scope}
\node at (1,-1.6) {{\bf\underline{$\mathfrak{g}^{(1)}_2$}}:};
\end{tikzpicture}
\end{tabular}
\end{center}
\caption{Untwisted affine Lie algebras. The affine node is shown as a hollow circle. The numbers in black $d_a^\vee$ denote the column null vector for the Cartan matrix, popularly known as dual Coxeter labels. The numbers in red $d_a$ denote the row null vector for the Cartan matrix, popularly known as Coxeter labels.}\label{UA}
\end{table}

\begin{table}[htbp]
\begin{center}
\begin{tabular}{cc}
\ubf{$\su(2n+1)^{(2)}$}:
\vspace{2mm}
&
\ubf{$\fe_6^{(2)}$}:
\vspace{2mm}\\
\begin{tikzpicture}[scale=1]
\draw[fill=black]
(1,0) node (v3) {} circle [radius=.1]
(2,0) node (v4) {} circle [radius=.1]
(4,0) node (v5) {} circle [radius=.1]
(5,0) node (v6) {} circle [radius=.1]
(6,0) node (v7) {} circle [radius=.1];
\draw[fill=white]
(0,0) node (v1) {} circle [radius=.1];
\node (v2) at (3,0) {$\cdots$};
\node at (0,-0.3) {\scriptsize{$1$}};
\node at (1,-0.3) {\scriptsize{$2$}};
\node at (2,-0.3) {\scriptsize{$2$}};
\node at (4,-0.3) {\scriptsize{$2$}};
\node at (5,-0.3) {\scriptsize{$2$}};
\node at (6,-0.3) {\scriptsize{$2$}};
\node at (0,-0.6) {\scriptsize{\color{red}$2$}};
\node at (1,-0.6) {\scriptsize{\color{red}$2$}};
\node at (2,-0.6) {\scriptsize{\color{red}$2$}};
\node at (4,-0.6) {\scriptsize{\color{red}$2$}};
\node at (5,-0.6) {\scriptsize{\color{red}$2$}};
\node at (6,-0.6) {\scriptsize{\color{red}$1$}};
\draw  (v3) edge (v4);
\draw  (v4) edge (v2);
\draw  (v2) edge (v5);
\draw  (v5) edge (v6);
\begin{scope}[shift={(-5,0)}]
\draw (5.15,0.025) -- (5.825,0.025) (5.825,-0.025) -- (5.15,-0.025);
\draw (5.6,0.15) -- (5.45,0) -- (5.6,-0.15);
\end{scope}
\begin{scope}[shift={(-0,0)}]
\draw (5.15,0.025) -- (5.825,0.025) (5.825,-0.025) -- (5.15,-0.025);
\draw (5.6,0.15) -- (5.45,0) -- (5.6,-0.15);
\end{scope}
\end{tikzpicture}
\vspace{2mm}
&
\begin{tikzpicture}[scale=1]
\draw[fill=black]
(0,0) node (v1) {} circle [radius=.1]
(1,0) node (v3) {} circle [radius=.1]
(2,0) node (v4){} circle [radius=.1]
(3,0) node (v7) {} circle [radius=.1];
\draw[fill=white]
(-1,0) node (v8) {} circle [radius=.1];
\node (v2) at (3,0) {};
\draw  (v1) edge (v3);
\draw  (v4) edge (v2);
\draw  (v4) edge (v7);
\draw  (v1) edge (v8);
\node at (-1,-0.3) {\scriptsize{$1$}};
\node at (0,-0.3) {\scriptsize{$2$}};
\node at (1,-0.3) {\scriptsize{$3$}};
\node at (2,-0.3) {\scriptsize{$4$}};
\node at (3,-0.3) {\scriptsize{$2$}};
\node at (-1,-0.6) {\scriptsize{\color{red}$1$}};
\node at (0,-0.6) {\scriptsize{\color{red}$2$}};
\node at (1,-0.6) {\scriptsize{\color{red}$3$}};
\node at (2,-0.6) {\scriptsize{\color{red}$2$}};
\node at (3,-0.6) {\scriptsize{\color{red}$1$}};
\begin{scope}[shift={(-4,0)}]
\draw (5.15,0.025) -- (5.825,0.025) (5.825,-0.025) -- (5.15,-0.025);
\draw (5.6,0.15) -- (5.45,0) -- (5.6,-0.15);
\end{scope}
\end{tikzpicture}
\vspace{2mm}\\
\ubf{$\so(2n)^{(2)}$}:  
\vspace{2mm}
&
\ubf{$\so(8)^{(3)}$}:    
\vspace{2mm}\\
\begin{tikzpicture}[scale=1]
\draw[fill=black]
(1,0) node (v3) {} circle [radius=.1]
(2,0) node (v4) {} circle [radius=.1]
(4,0) node (v5) {} circle [radius=.1]
(5,0) node (v6) {} circle [radius=.1]
(6,0) node (v7) {} circle [radius=.1];
\draw[fill=white] (0,0) node (v1) {} circle [radius=.1];
\node (v2) at (3,0) {$\cdots$};
\node at (0,-0.3) {\scriptsize{$1$}};
\node at (1,-0.3) {\scriptsize{$2$}};
\node at (2,-0.3) {\scriptsize{$2$}};
\node at (4,-0.3) {\scriptsize{$2$}};
\node at (5,-0.3) {\scriptsize{$2$}};
\node at (6,-0.3) {\scriptsize{$1$}};
\node at (0,-0.6) {\scriptsize{\color{red}$1$}};
\node at (1,-0.6) {\scriptsize{\color{red}$1$}};
\node at (2,-0.6) {\scriptsize{\color{red}$1$}};
\node at (4,-0.6) {\scriptsize{\color{red}$1$}};
\node at (5,-0.6) {\scriptsize{\color{red}$1$}};
\node at (6,-0.6) {\scriptsize{\color{red}$1$}};
\draw  (v3) edge (v4);
\draw  (v4) edge (v2);
\draw  (v2) edge (v5);
\draw  (v5) edge (v6);
\begin{scope}[shift={(-5,0)}]
\draw (5.15,0.025) -- (5.825,0.025) (5.825,-0.025) -- (5.15,-0.025);
\draw (5.6,0.15) -- (5.45,0) -- (5.6,-0.15);
\end{scope}
\begin{scope}[shift={(-0,0)}]
\draw (5.15,0.025) -- (5.825,0.025) (5.825,-0.025) -- (5.15,-0.025);
\draw (5.4,0.15) -- (5.55,0) -- (5.4,-0.15);
\end{scope}
\end{tikzpicture}
\vspace{2mm}
&
\begin{tikzpicture}[scale=1]
\draw[fill=black]
(0,0) node (v1) {} circle [radius=.1]
(1,0) node (v3) {} circle [radius=.1];
\draw[fill=white]
(2,0) node (v8) {} circle [radius=.1];
\node at (0,-0.3) {\scriptsize{$3$}};
\node at (1,-0.3) {\scriptsize{$2$}};
\node at (2,-0.3) {\scriptsize{$1$}};
\node at (0,-0.6) {\scriptsize{\color{red}$1$}};
\node at (1,-0.6) {\scriptsize{\color{red}$2$}};
\node at (2,-0.6) {\scriptsize{\color{red}$1$}};
\node (v4) at (0,0.05) {};
\node (v5) at (1,0.05) {};
\node (v6) at (0,-0.05) {};
\node (v7) at (1,-0.05) {};
\draw (v3) -- (v8);
\begin{scope}[shift={(-5,0)}]
\draw (5.15,0.05) -- (5.825,0.05) (5.825,0) -- (5.15,0) (5.825,-0.05) -- (5.15,-0.05);
\draw (5.4,0.175) -- (5.575,0) -- (5.4,-0.175);
\end{scope}
\end{tikzpicture}
\vspace{2mm}\\
\ubf{$\su(2n)^{(2)}$}:
\vspace{0mm}
&
\ubf{$\su(4)^{(2)}$}:
\vspace{2mm}\\
\begin{tikzpicture}[scale=1]
\draw[fill=black]
(0,0) node (v1) {} circle [radius=.1]
(1,0) node (v3) {} circle [radius=.1]
(2,0) node (v4) {} circle [radius=.1]
(4,0) node (v5) {} circle [radius=.1]
(5,0) node (v6) {} circle [radius=.1]
(6,0) node (v7) {} circle [radius=.1];
\draw[fill=white]
(1,1) node (v8) {} circle [radius=.1];
\node (v2) at (3,0) {$\cdots$};
\node at (0,-0.3) {\scriptsize{$1$}};
\node at (1,-0.3) {\scriptsize{$2$}};
\node at (2,-0.3) {\scriptsize{$2$}};
\node at (4,-0.3) {\scriptsize{$2$}};
\node at (5,-0.3) {\scriptsize{$2$}};
\node at (6,-0.3) {\scriptsize{$2$}};
\node at (1,1.3) {\scriptsize{$1$}};
\node at (0,-0.6) {\scriptsize{\color{red}$1$}};
\node at (1,-0.6) {\scriptsize{\color{red}$2$}};
\node at (2,-0.6) {\scriptsize{\color{red}$2$}};
\node at (4,-0.6) {\scriptsize{\color{red}$2$}};
\node at (5,-0.6) {\scriptsize{\color{red}$2$}};
\node at (6,-0.6) {\scriptsize{\color{red}$1$}};
\node at (1,1.6) {\scriptsize{\color{red}$1$}};
\draw  (v1) edge (v3);
\draw  (v3) edge (v4);
\draw  (v4) edge (v2);
\draw  (v2) edge (v5);
\draw  (v5) edge (v6);
\draw  (v3) edge (v8);
\begin{scope}[shift={(-0,0)}]
\draw (5.15,0.025) -- (5.825,0.025) (5.825,-0.025) -- (5.15,-0.025);
\draw (5.6,0.15) -- (5.45,0) -- (5.6,-0.15);
\end{scope}
\end{tikzpicture}
&
\begin{tikzpicture}[scale=1]
\draw[fill=black]
(1,0) node (v3) {} circle [radius=0.1]
(2,0) node (v4) {} circle [radius=0.1];
\draw[fill=white] 
(0,0) node (v1) {} circle [radius=0.1];
\node at (0,-0.3) {\scriptsize{$1$}};
\node at (1,-0.3) {\scriptsize{$2$}};
\node at (2,-0.3) {\scriptsize{$1$}};
\node at (0,-0.6) {\scriptsize{\color{red}$1$}};
\node at (1,-0.6) {\scriptsize{\color{red}$1$}};
\node at (2,-0.6) {\scriptsize{\color{red}$1$}};
\begin{scope}[shift={(-5,0)}]
\draw (5.15,0.025) -- (5.825,0.025) (5.825,-0.025) -- (5.15,-0.025);
\draw (5.6,0.15) -- (5.45,0) -- (5.6,-0.15);
\end{scope}
\begin{scope}[shift={(-4,0)}]
\draw (5.15,0.025) -- (5.825,0.025) (5.825,-0.025) -- (5.15,-0.025);
\draw (5.4,0.15) -- (5.55,0) -- (5.4,-0.15);
\end{scope}
\begin{scope}[shift={(0.5,-2.4)}]
\draw[fill=black]
(1,0) node (v3) {} circle [radius=.1];
\draw[fill=white]
(0,0) node (v8) {} circle [radius=.1];
\node (v4) at (0,0.05) {};
\node (v5) at (1,0.05) {};
\node (v6) at (0,-0.05) {};
\node (v7) at (1,-0.05) {};
\node at (0,-0.3) {\scriptsize{$1$}};
\node at (1,-0.3) {\scriptsize{$2$}};
\node at (0,-0.6) {\scriptsize{\color{red}$2$}};
\node at (1,-0.6) {\scriptsize{\color{red}$1$}};
\begin{scope}[shift={(-5,0)}]
\draw (5.15,0.075) -- (5.825,0.075) (5.15,0.025) -- (5.825,0.025) (5.825,-0.025) -- (5.15,-0.025) (5.825,-0.075) -- (5.15,-0.075);
\draw (5.6,0.2) -- (5.4,0) -- (5.6,-0.2);
\end{scope}
\end{scope}
\node at (1,-1.6) {{\bf\underline{$\mathfrak{su}(3)^{(2)}$}}:};
\end{tikzpicture}
\end{tabular}
\end{center}
\caption{Twisted affine Lie algebras. The affine node is shown as a hollow circle. The numbers in black $d^\vee_a$ denote the column null vector for the Cartan matrix, popularly known as dual Coxeter labels. The numbers in red $d_a$ denote the row null vector for the Cartan matrix, popularly known as Coxeter labels. The total number of nodes for $\su(2n+1)^{(2)}$ is $n+1$, for $\so(2n)^{(2)}$ is $n$, and for $\su(2n)^{(2)}$ is $n+1$.}\label{TA}
\end{table}

Consider a $6d$ theory $\fT$ with gauge algebras $\fg_i$ on its tensor branch. Consider further compactifying $\fT$ on a circle of finite size without a twist. On a generic point of the resulting $5d$ Coulomb branch, the massive BPS spectrum includes W-bosons for the corresponding untwisted affine gauge algebras $\fg_i^{(1)}$. In other words, the abelian gauge algebra visible at low energies on the Coulomb branch is $\oplus_a\u(1)_{a,i}$ parametrizing the Cartan of $\fg_i$ plus a $\u(1)_{0,i}$ responsible for affinization. The $\u(1)_{i}$ arising from the reduction of tensor multiplet $B_i$ is central to $\oplus_a\u(1)_{a,i}\oplus\u(1)_{0,i}$. The untwisted Lie algebras are listed in Table \ref{UA} along with their Coxeter and dual Coxeter labels.

We now generalize the above statements to the twisted case. Consider compactifying $\fT$ on a circle of finite size with a twist $S,\{q_\alpha\}$. On a generic point of the resulting $5d$ Coulomb branch, the massive BPS spectrum includes W-bosons for the corresponding twisted/untwisted affine gauge algebras $\fg_\alpha^{(q_\alpha)}$. In other words, the abelian gauge algebra visible at low energies on the Coulomb branch is $\oplus_a\u(1)_{a,\alpha}$ parametrizing the Cartan of $\fh_\alpha$ plus a $\u(1)_{0,\alpha}$ responsible for affinization. The $\u(1)_{\alpha}$ arising from the reduction of tensor multiplet $B_i$ (with $i$ in orbit of $\alpha$) is central to $\oplus_a\u(1)_{a,\alpha}\oplus\u(1)_{0,\alpha}$. The twisted Lie algebras are listed in Table \ref{TA} along with their Coxeter and dual Coxeter labels.

The charge under $\u(1)_{b,\alpha}$ (corresponding to a simple co-root $e_b^\vee$) of a W-boson $W_a$ (corresponding to simple root $e_a$ of $\fg_\alpha^{(q_\alpha)}$) is given by the element $A_{ab}$ of the Cartan matrix. Now consider the $\u(1)$ embedding into $\oplus_{b=0}^{r_\alpha}\u(1)_{b,\alpha}$ by the map $e^{i\theta}\to\oplus_{b=0}^{r_\alpha}\left(e^{id^\vee_b\theta}\right)_b$ where $\left(e^{id^\vee_b\theta}\right)_b$ is the element $e^{id^\vee_b\theta}$ of $\u(1)_{b,\alpha}$ and $d^\vee_b$ are dual Coxeter labels of $\fg_\alpha^{(q_\alpha)}$ listed in Tables \ref{UA} and \ref{TA}. Since all the W-bosons $W_a$ are uncharged under this $\u(1)$, it follows that this $\u(1)$ can be identified with the central $\u(1)_\alpha$. The charge of a particle $n_\alpha$ under $\u(1)_\alpha$ can be written as $\sum_{b=0}^{r_\alpha}d^\vee_b n_{b,\alpha}$ where $n_{b,\alpha}$ is the charge of the particle under $\u(1)_{b,\alpha}$. 

The truncated prepotential $6\mathcal{F}^\phi_{S,\{q_\alpha\}}$ is written in terms of Coulomb branch parameters $\phi_{b,\alpha}$ (with $1\le b\le r_\alpha$) corresponding to $\u(1)_{b,\alpha}$ and $\phi_{0,\alpha}$ corresponding to $\u(1)_\alpha$. To facilitate comparison with geometry, we wish to write the prepotential in terms of Coulomb branch parameters corresponding to $\u(1)_{b,\alpha}$ for $0\le b\le r_\alpha$. This is achieved by performing the following replacement in $6\mathcal{F}^\phi_{S,\{q_\alpha\}}$
\be\label{shift}
\phi_{b,\alpha}\to\phi_{b,\alpha}-d^\vee_b\phi_{0,\alpha}
\ee
for all $1\le b\le r_\alpha$ and for all $\alpha$.\footnote{Note that the shift (\ref{shift}) has been studied before the in the literature in relation to resolutions of elliptically fibered Calabi-Yau threefolds; in these examples, the effect of the shift is to expand the K\"ahler form $J$ in basis of primitive divisors---see for example \cite{Grimm:2010ks}.} We will call the prepotential obtained after this shift as $\tilde\cF_{S,\{q_\alpha\}}$. The Coulomb branch parameter $\phi_{0,\alpha}$ in $\tilde\cF_{S,\{q_\alpha\}}$ corresponds to $\u(1)_{0,\alpha}$ rather than $\u(1)_\alpha$.

For illustrative purposes, we note that the shift for our example (\ref{ex8}) is
\begin{align}
\phi_1&\to\phi_1-\phi_0\nn\\
\phi_2&\to\phi_2-\phi_0\nn
\end{align}
which means that the shifted prepotential corresponding (\ref{ex7}) is
\be\label{eq8}
6\tilde\cF=8\phi_0^3+8\phi_1^3+2\phi_2^3-6\phi_1\phi_0^2+6\phi_1\phi_2^2-6\phi_2\phi_0^2-12\phi_2\phi_1^2
\ee
where we have chosen the phase $s=+1$.

The shifted prepotential for (\ref{eg7}) is
\be\label{eg8}
6\tilde\cF=7\phi_0^3+8\phi_1^3+3\phi_2^3-6\phi_1\phi_0^2+6\phi_1\phi_2^2-3\phi_2\phi_0^2-3\phi_0\phi_2^2-12\phi_2\phi_1^2
\ee

A Mathematica notebook accompanying the submission of this paper can be used to compute the contribution to $6\tilde\cF$ (in any gauge-theoretic phase) from a single node or two nodes connected by an edge. 
Using these two results, one can write the contribution to $6\tilde\cF$ from two nodes connected by an edge as contributions from the two nodes alone and a contribution from the edge. Thus, we can figure out what is the contribution to $6\tilde\cF$ by each possible edge. Combining the contributions from the nodes and the edges, one can obtain $6\tilde\cF_{S,\{q_\alpha\}}$ for any arbitrary graph $\Sigma^{S,\{q_\alpha\}}_\fT$. More details and the instructions for using the notebook can be found in Appendix \ref{mathematica}.

\section{Geometries associated to $5d$ KK theories}\label{GKK}
In this section, we will show that we can associate (at least one) genus-one fibered Calabi-Yau threefold $X_{S,\{q_\alpha\}}$ to every $5d$ KK theory\footnote{We remind the reader that this statement is not completely true for KK theories involving the last node in Table \ref{KR1}. For such KK theories, we only propose an algebraic description whose structure closely mimics the structure of genus-one fibered Calabi-Yau threefolds to be discussed in the next subsection \ref{GF}.} $\fT^{KK}_{S,\{q_\alpha\}}$. Compactifying M-theory on $X_{S,\{q_\alpha\}}$ produces the Coulomb branch of $\fT^{KK}_{S,\{q_\alpha\}}$. Some of the results appearing below also appeared in \cite{Esole:2017kyr,Esole:2015xfa,Esole:2018mqb,Esole:2017qeh,Esole:2017rgz,Esole:2011sm,Esole:2014bka,Esole:2014hya,DelZotto:2017pti,Bhardwaj:2018yhy,Bhardwaj:2018vuu}

\subsection{General features}\label{GF}
In this subsection, we start with a description of general features of the geometric structure of $X_{S,\{q_\alpha\}}$ and the relationship between this geometry and the low energy effective theory governing the Coulomb branch of the KK theory $\fT^{KK}_{S,\{q_\alpha\}}$.

We will show that $X_{S,\{q_\alpha\}}$ can be realized as a local neighborhood of a collection of irreducible compact holomorphic surfaces intersecting with each other pairwise transversely. As we will see, the surfaces fall into families indexed by $\alpha$. We denote the irreducible surfaces in each family $\alpha$ as $S_{a,\alpha}$ where $0\le a\le r_\alpha$ (where $r_\alpha$ is the rank of $\fh_\alpha$). The Kahler parameters associated to $S_{a,\alpha}$ are identified as the Coulomb branch parameters $\phi_{a,\alpha}$ of the corresponding $5d$ KK theory discussed in the previous section. Whenever $\fh_\alpha$ is trivial, the rank of $\fh_\alpha$ is zero and hence there is only a single surface $S_{0,\alpha}$ associated to the node $\alpha$  in that case.

\subsubsection{Triple intersection numbers and the prepotential}\label{tripre}
A key role in the relationship between $X_{S,\{q_\alpha\}}$ and $\fT^{KK}_{S,\{q_\alpha\}}$ is played by the shifted prepotential $6\tilde\cF_{S,\{q_\alpha\}}$. The coefficients $c_{a\alpha,b\beta,c\gamma}$ of $\phi_{a,\alpha}\phi_{b,\beta}\phi_{c,\gamma}$ in $6\tilde\cF_{S,\{q_\alpha\}}$ capture the triple intersection numbers of surfaces in $X_{S,\{q_\alpha\}}$ as follows:
\begin{align}
c_{a\alpha,a\alpha,a\alpha}&=S_{a,\alpha}\cdot S_{a,\alpha}\cdot S_{a,\alpha}\\
c_{a\alpha,a\alpha,b\beta}&=3S_{a,\alpha}\cdot S_{a,\alpha}\cdot S_{b,\beta}\\
c_{a\alpha,b\beta,c\gamma}&=6S_{a,\alpha}\cdot S_{b,\beta}\cdot S_{c,\gamma}
\end{align}
where $(a,\alpha),(b,\beta),(c,\gamma)$ denote distinct non-equal indices. 

A triple intersection product of three surfaces can be computed via intersection numbers inside any one of the three surfaces. To explain it, let us first first define the notion of ``gluing curves''. Consider the intersection locus $\cL_{a\alpha,b\beta}$ between two distinct surfaces $S_{a,\alpha}$ and $S_{b,\beta}$ in $X_{S,\{q_\alpha\}}$. $\cL_{a\alpha,b\beta}$ splits into geometrically irreducible components as $\sum _i\cL^i_{a\alpha,b\beta}$. Each $\cL^i_{a\alpha,b\beta}$ appears as an irreducible curve $C^i_{a,\alpha;b,\beta}$ in $S_{a,\alpha}$ and an irreducible curve $C^i_{b,\beta;a,\alpha}$ in $S_{b,\beta}$. In other words, we can manufacture the intersection of $S_{a,\alpha}$ and $S_{b,\beta}$ by identifying the curves
\be\label{iden}
C^i_{a,\alpha;b,\beta}\sim C^i_{b,\beta;a,\alpha}
\ee
with each other for all $i$. Identifying pairs of curves in the above fashion can be thought of as ``gluing together'' two surfaces along those curves\footnote{On multiple occasions throughout this paper, we abuse the language and denote the identification of two curves as ``gluing'' of the two curves.}. The reducible curve $C_{a,\alpha;b,\beta}:=\sum_iC^i_{a,\alpha;b,\beta}$ is called the ``total gluing curve'' in $S_{a,\alpha}$ for the intersection of $S_{a,\alpha}$ and $S_{b,\beta}$. Similarly, $C_{b,\beta;a,\alpha}:=\sum_iC^i_{b,\beta;a,\alpha}$ is called the total gluing curve in $S_{b,\beta}$  for the intersection of $S_{a,\alpha}$ and $S_{b,\beta}$.

As two distinct surfaces $S_{a,\alpha}$ and $S_{b,\beta}$ can intersect each other, so can a single surface $S_{a,\alpha}$ intersect itself. Much as above for the intersection of two distinct surfaces, the self-intersection of $S_{a,\alpha}$ can be captured in terms of gluings
\be
C^i_{a,\alpha}\sim D^i_{a,\alpha}
\ee
where $C^i_{a,\alpha}$ and $D^i_{a,\alpha}$ are irreducible curves in $S_{a,\alpha}$.

Then the triple intersection numbers can be expressed as:
\begin{align}
S_{a,\alpha}\cdot S_{a,\alpha}\cdot S_{a,\alpha}&=K'_{a,\alpha}\cdot K'_{a,\alpha}\\
S_{a,\alpha}\cdot S_{a,\alpha}\cdot S_{b,\beta}&=K'_{a,\alpha}\cdot C_{a,\alpha;b,\beta}=C_{b,\beta;a,\alpha}^2\\
S_{a,\alpha}\cdot S_{b,\beta}\cdot S_{c,\gamma}&=C_{a,\alpha;b,\beta}\cdot C_{a,\alpha;c,\gamma}=C_{b,\beta;c,\gamma}\cdot C_{b,\beta;a,\alpha}=C_{c,\gamma;a,\alpha}\cdot C_{c,\gamma;b,\beta}
\end{align}
where
\be
K'_{{a,\alpha}}:=K_{a,\alpha}+\sum_i\left(C^i_{a,\alpha}+D^i_{a,\alpha}\right)
\ee
and $K_{{a,\alpha}}$ denotes the canonical class of $S_{a,\alpha}$.

As an illustrative example consider the KK theory (\ref{ex8}) for which the shifted prepotential in a particular phase is displayed in (\ref{eq8}). We propose that the associated geometry is as follows. Since there is a single node, we drop the index $\alpha$ and only display the index $a$. The surfaces are $S_0=\bF_0$, $S_1=\bF_2$, $S_2=\bF_4^6$. The gluing curves between $S_0$ and $S_1$ are $C_{0;1}=e,C_{1;0}=e$. The gluing curves between $S_1$ and $S_2$ are $C_{1;2}=h,C_{2;1}=e$. The gluing curves between $S_2$ and $S_0$ are $C_{2;0}=h-\sum x_i,C_{0;2}=e$. 

Now we can check that the intersections of these curves indeed give rise to the various coefficients in (\ref{eq8}):
\bit
\item First of all, recall from (\ref{K2}) that $K^2=8-b$ for $\bF_n^b$. Indeed, the coefficients of $\phi_a^3$ in (\ref{eq8}) equal $K_a^2$. 
\item One third the coefficient of $\phi_0\phi_1^2$ is zero which matches $C_{0;1}^2=(e^2)_{S_0}$ where $(e^2)_{S_0}$ denotes that the intersection number $e^2$ is computed inside $S_0$ and that in particular the curve $e$ is inside $S_0$. The coefficient also matches $K_1\cdot C_{1;0}=(K\cdot e)_{S_1}=0$. One third of the coefficient of $\phi_2\phi_0^2$ is $-2$ which indeed matches $C_{2;0}^2=\left((h-\sum x_i)^2\right)_{S_2}=\left(h^2-\sum x_i^2\right)_{S_2}=4-6=-2$ and $K_0\cdot C_{0;2}=(K\cdot e)_{S_0}=-2$. Similarly, we can check the matching of such intersection numbers with one third the coefficients of other terms of the form $\phi_a\phi_b^2$.
\item  One sixth the coefficient of $\phi_0\phi_1\phi_2$ is zero which matches $C_{0;1}\cdot C_{0;2}=(e^2)_{S_0}=0$, $C_{1;2}\cdot C_{1;0}=(h\cdot e)_{S_1}=0$, and $C_{2,0}\cdot C_{2;1}=\left((h-\sum x_i)\cdot e\right)_{S_2}=0$.
\eit

On the other hand, the geometry associated to (\ref{eg8}) has $S_0=\bF_0^1$, $S_1=\bF_2$ and $S_2=\bF_4^5$. The gluing curves between $S_0$ and $S_1$ are $C_{0;1}=e,C_{1;0}=e$. The gluing curves between $S_1$ and $S_2$ are $C_{1;2}=h,C_{2;1}=e$. The gluing curves between $S_2$ and $S_0$ are $C_{2;0}=h-\sum x_i,C_{0;2}=e-x$. Here $x$ denotes the exceptional curve of the blowup of $S_0$ and $x_i$ denote the exceptional curves of the blowups of $S_2$. One can check that the intersections of these curves indeed give rise to the various coefficients in (\ref{eg8}).

\subsubsection{Consistency of gluings: volume matching, the Calabi-Yau condition, and irreducibility} \label{consistency}
Not every pair of curves can be identified with one another to form a consistent gluing. First of all, the topology of the two curves must be identical. This implies that a geometrically irreducible curve in one surface can only be identified with a geometrically irreducible curve in another surface, and furthermore that the genera (as defined in Appendix \ref{genus}) of the two curves must be identical and non-negative. If $C \subset S$ is an irreducible curve, then a necessary condition that must be satisfied by $C$ is that for any other irreducible curve $C' \subset S$ such that $C \ne C$, the intersection product must be non-negative:
	\begin{align}
		 C \cdot C' \geq 0. 	
	\end{align}
In this paper, some of the algebraic examples are non-geometric (i.e. do not admit a conventional geometric description satisfying these consistency conditions) because they involve gluings which identify a geometrically \emph{reducible} curve in one surface with a geometrically irreducible curve in another surface. Despite this apparent pathology, these examples nevertheless satisfy the remaining conditions described below.

In addition to the above topological constraints, the volumes of a pair of gluing curves must be the same. The volume of a curve $C$ is computed by intersecting the curve with the Kahler class $J$ via
\be
\text{vol}(C)=-J\cdot C
\ee
where
\be
J=\sum_{a,\alpha}\phi_{a,\alpha}S_{a,\alpha}+\sum_f m_f N_f
\ee
where $m_f$ are mass parameters and $N_f$ are non-compact surfaces corresponding to those mass parameters. The contribution of mass parameters to the volume will not play a prominent role in this paper, so we define a truncated Kahler class $J^\phi$ which only keep track of the contribution of Coulomb branch parameters to the volume
\be
J^\phi=\sum_{a,\alpha}\phi_{a,\alpha}S_{a,\alpha}
\ee
The volume of $C$ equals the mass of the BPS state obtained by wrapping an M2 brane on $C$ because the intersection number
\be\label{SC}
-S_{a,\alpha}\cdot C
\ee
captures the charge under $\u(1)_{a,\alpha}$ of the BPS state arising from M2 brane wrapping $C$. If $C$ lies in $S_{a,\alpha}$, then the intersection (\ref{SC}) is computed via
\be\label{SC1}
S_{a,\alpha}\cdot C=K'_{a,\alpha}\cdot C
\ee
If $C$ lies in some other surface $S_{b,\beta}$, then (\ref{SC}) is computed via
\be\label{SC2}
S_{a,\alpha}\cdot C= C_{b,\beta;a,\alpha}\cdot C
\ee
Now, for (\ref{iden}) to be consistent we must have
\be\label{CYP}
J^\phi\cdot C^i_{a,\alpha;b,\beta}=J^\phi\cdot C^i_{b,\beta;a,\alpha}
\ee
which is an important consistency condition for constructing $X_{S,\{q_\alpha\}}$. We have checked that (\ref{CYP}) is satisfied for all the geometries presented in this paper.

Finally, the gluing curves also have to satisfy the \emph{Calabi-Yau condition} which states that
\be\label{CY}
\left(C^i_{a,\alpha;b,\beta}\right)^2+\left(C^i_{b,\beta;a,\alpha}\right)^2=2g-2
\ee
where $g$ is the genus of $C^i_{a,\alpha;b,\beta}$. See \cite{Jefferson:2018irk,Bhardwaj:2018vuu} for more details.

Notice that in special situations the Calabi-Yau condition (\ref{CY}) is automatically satisfied as long as we satisfy (\ref{CYP}). This is the situation when there is a single gluing curve $C_{a,\alpha;b,\beta}\sim C_{b,\beta;a,\alpha}$ between two surfaces $S_{a,\alpha}$ and $S_{b,\beta}$ such that neither of them is a self-glued surface. Then, (\ref{CYP}) implies
\be
K\cdot C_{a,\alpha;b,\beta}=C_{b,\beta;a,\alpha}^2
\ee
Adding $C_{a,\alpha;b,\beta}^2$ to both sides of the above equation we get
\be
C_{a,\alpha;b,\beta}^2+C_{b,\beta;a,\alpha}^2=2g-2
\ee

As an example, in what preceded above we discussed the geometry associated to (\ref{eq8}). We can check that (\ref{CYP}) is satisfied for all the gluing curves in the geometry. For instance,
\begin{align}
J^\phi\cdot C_{0;1}&=\phi_{0}\left(K_0\cdot C_{0;1}\right)+\phi_1 C_{0;1}^2+\phi_2 \left(C_{0;2}\cdot C_{0;1}\right)\\
&=\phi_0\left(K\cdot e\right)_{S_0}+\phi_1 \left(e^2\right)_{S_0}+\phi_2 \left(e^2\right)_{S_0}\\
&=-2\phi_0
\end{align}
and comparing it with
\begin{align}
J^\phi\cdot C_{1;0}&=\phi_0 C_{1;0}^2+\phi_{1}\left(K_1\cdot C_{1;0}\right)+\phi_2 \left(C_{1;2}\cdot C_{1;0}\right)\\
&=\phi_0 \left(e^2\right)_{S_1}+\phi_1\left(K\cdot e\right)_{S_1}+\phi_2 \left(e\cdot h\right)_{S_1}\\
&=-2\phi_0
\end{align}
we find that indeed the gluing $C_{0;1}\sim C_{1;0}$ is consistent. Similarly, it can be checked that all the other gluings are consistent as well. In a similar fashion, one can also check that all of the gluings in the geometry associated to (\ref{eg8}) discussed above satisfy (\ref{CYP}).

\subsubsection{Weights, phase transitions and flops}\label{flops}
A hypermultiplet transforming in a representation $\cR_f$ of the $5d$ gauge algebra $\fh=\oplus_\alpha\fh_\alpha$ appears as a collection of curves inside $X_{S,\{q_\alpha\}}$. These curves are characterized as follows. Let $m_f$ be the mass parameter corresponding to $\cR_f$. For each weight $w(\cR_f)$ of $\cR_f$, define a quantity $\text{vol}\left(w(\cR_f)\right)$, which we call the \emph{virtual volume}, by shifting the quantity 
\be\label{weight}
w(\cR_f)\cdot \phi+m_f
\ee
by the shift (\ref{shift}) for all $\alpha$. Then, one can find a holomorphic curve $C_{w(\cR_f)}$ in $X_{S,\{q_\alpha\}}$ such that 
\be
\text{vol}\left(C_{w(\cR_f)}\right)=|\text{vol}\left(w(\cR_f)\right)|
\ee

In general, the curve $C_{w(\cR_f)}$ can be a positive linear combination of curves living inside various irreducible surfaces. However, some of the curves $C_{w(\cR_f)}$ turn out to be living purely inside a single irreducible surface $S_{a,\alpha}$. If such a curve $C_{w}$ has genus zero and self-intersection $-1$ inside $S_{a,\alpha}$, then one can perform a \emph{flop transition}\footnote{This transition corresponds to blowing down $C$ inside $S_{a,\alpha}$ and performing a blow-up in the neighboring surfaces intersecting $C$ transversally. We will explain such transitions via various illustrations throughout this paper. More detailed background can be found in Section 2 of \cite{Bhardwaj:2018vuu}.} on $X_{S,\{q_\alpha\}}$ by flopping $C$, which corresponds to a phase transition in the Coulomb branch of the $5d$ gauge theory described in previous section. We refer to such a flop transition as a ``gauge-theoretic flop transition'' to distinguish it from the flop transitions associated to more general $-1$ curves not associated to any hypermultiplet.

Let the geometry obtained after the flop transition associated to $C_w$ be $X'_{S,\{q_\alpha\}}$. As for $X_{S,\{q_\alpha\}}$, there exist curves $C'_{w(\cR_f)}$ in $X'_{S,\{q_\alpha\}}$ associated to weights $w(\cR_f)$ such that
\be
\text{vol}\left(C'_{w(\cR_f)}\right)=|\text{vol}'\left(w(\cR_f)\right)|
\ee
where $\text{vol}'\left(w(\cR_f)\right)$ is the shift of the quantity (\ref{weight}) computed in the new phase. The relationship between the two virtual volumes $\text{vol}'\left(w(\cR_f)\right)$ and $\text{vol}\left(w(\cR_f)\right)$ is
\be
\text{vol}'\left(w(\cR_f)\right)=\text{vol}\left(w(\cR_f)\right)
\ee
for all $w(\cR_f)\neq w$, and 
\be\label{volflop}
\text{vol}'\left(w\right)=-\text{vol}\left(w\right)
\ee
with a minus sign.

We know from the analysis presented in the last section that the canonical $5d$ gauge theory associated to (\ref{ex8}) is an $\su(3)$ gauge theory with six fundamental hypers. The Dynkin coefficients of the weights of fundamental are $(1,0)$, $(-1,1)$ and $(0,-1)$. We call these weights $w_1$, $w_2$ and $w_3$ respectively. We can compute
\begin{align}
\text{vol}(w_1)&=-\phi_0+\phi_1\\
\text{vol}(w_2)&=-\phi_1+\phi_2\\
\text{vol}(w_3)&=\phi_0-\phi_2
\end{align}
Recall that the phase (\ref{eq8}) corresponds to $\text{vol}(w_1)$ and $\text{vol}(w_2)$ being positive and $\text{vol}(w_3)$ being negative for all the six fundamentals. Now compute the volume of one of the blowups $x_i$ living in the surface $S_2$ in the geometry corresponding to (\ref{eq8}):
\be
\text{vol}(x_i)=-\phi_0+\phi_2
\ee
Thus we see that $C_{w_3}$ for each fundamental is $x_i$. The reader can check that $C_{w_2}=f_2+x_i$ and $C_{w_1}=f_1+f_2+x_i$ where $f_a$ denotes the fiber of the Hirzebruch surface $S_a$.

In fact, the geometries corresponding to (\ref{eq8}) and (\ref{eg8}) are related by a flop transition. We first blow down one of the blowups, say $x_6$, inside $S_2$. Under this blowdown the identity of $S_2$ changes from $\bF_4^6$ to $\bF_4^5$. Since $x_6$ intersects the gluing curve $h-\sum_{i=1}^6 x_i$ at one point, the gluing curve after the blowdown becomes $h-\sum_{i=1}^6 x_i+x_6=h-\sum_{i=1}^5 x_i$. The other gluing curve inside $S_2$ is unaffected since $x_6$ does not intersect with it. Correspondingly, since the gluing curve for $S_1$ in $S_2$ does not intersect $x_6$, the surface $S_1$ is unaffected by the flop transition. However, since the gluing curve for $S_0$ in $S_2$ intersects $x_6$, we have to blowup $S_0$ at a point lying on the gluing curve for $S_2$ inside $S_0$. Under the blowup the identity of $S_0$ changes from $\bF_0$ to $\bF_0^1$. The gluing curve for $S_2$ inside $S_1$ is changed to $e-x$. 

Recall that the phase (\ref{eg8}) corresponds to turning on a large mass $m$ for one of the fundamentals such that 
\be\label{w3flip}
\text{vol}(w_3)=\phi_0-\phi_2+m
\ee
for this fundamental is positive. Correspondingly, we can compute that
\be
\text{vol}(x)=\phi_0-\phi_2
\ee
which indeed matches (\ref{w3flip}) up to the contribution from mass parameter, thus verifying (\ref{volflop}). We are not keeping track of non-compact surfaces in this paper, so we are only able to verify (\ref{volflop}) up to the contribution from $m$.

\subsubsection{Affine Cartan matrices and intersections of fibers}
For each surface $S_{a,\alpha}$ in $X_{S,\{q_\alpha\}}$, we define a canonical fiber $f_{a,\alpha}$ inside it:
\bit
\item If $\fg_\alpha$ is non-trivial, then $S_{a,\alpha}$ will always be a Hirzebruch surface\footnote{In this paper, by a ``Hirzebruch surface'', we refer to a Hirzebruch surface possibly with blowups at generic or non-generic locations. Some background on Hirzebruch surfaces can be found in Appendix \ref{back}.} whose fiber class is the canonical fiber $f_{a,\alpha}$. An M2 brane wrapping this curve gives rise to the W-boson $W_{a,\alpha}$ discussed in last section. 
\item If the node $\alpha$ is
\be
\begin{tikzpicture}
\node at (-0.5,0.45) {2};
\node at (-0.45,0.9) {$\su(1)^{(1)}$};
\end{tikzpicture}
\ee
then it turns out that there is a single corresponding surface $S_{0,\alpha}=F_0^2$ which is self-glued since $e-x$ and $e-y$ are identified with each other where $x$ and $y$ are the exceptional curves corresponding to the two blowups. Due to the self-gluing, the fiber class of $S_{0,\alpha}$ intersects itself inside the threefold $X_{S,\{q_\alpha\}}$ and appears as an elliptic curve with a nodal singularity. It is this fiber class that we refer to as the canonical fiber $f_{0,\alpha}$ in this case.
\item If the node $\alpha$ is
\be
\begin{tikzpicture}
\node at (-0.5,0.45) {1};
\node at (-0.45,0.9) {$\sp(0)^{(1)}_\theta$};
\end{tikzpicture}
\ee
then it turns out that there is a single corresponding surface $S_{0,\alpha}=dP_9$. The del Pezzo surface\footnote{In this paper, by a ``del Pezzo surface $dP_n$'', we refer to a surface which is an $n$ point blowup of $\P^2$ but the blowups can be at non-generic locations. Some background on del Pezzo surfaces can be found in Appendix \ref{back}.} $dP_9$ admits a unique elliptic fiber class $3l-\sum x_i$ which we refer to as the canonical fiber $f_{0,\alpha}$ in this case.
\item If the node $\alpha$ is
\be
\begin{tikzpicture}
\node (v1) at (-0.5,0.4) {2};
\node at (-0.45,0.9) {$\su(1)^{(1)}$};
\draw (v1) .. controls (-1.5,-0.5) and (0.5,-0.5) .. (v1);
\end{tikzpicture}
\ee
then it turns out that there is no completely geometric description. We provide an algebraic description in terms of algebraic properties of the curves inside the surface $S_{0,\alpha}=F_1^2$ which is self-glued since $x$ and $y$ are identified with each other. The canonical fiber in this case is $f_{0,\alpha}=2h+f-2x-2y$ which is a genus one curve of self-intersection zero.
\eit

For each $\alpha$ we find that
\be\label{GCM}
f_{a,\alpha}\cdot S_{b,\alpha}=-A_{ab}
\ee
where $A_{ab}$ is the Cartan matrix of $\fg_\alpha^{(q_\alpha)}$ and $A_{ab}\equiv A_{00}=0$ whenever $\fg_\alpha$ is trivial. This means that the fibers of Hirzebruch surfaces $S_{a,\alpha}$ for a fixed $\alpha$ intersect in the fashion of Dynkin diagram associated to affine Lie algebra $\fg_\alpha^{(q_\alpha)}$. 

Intersection (\ref{GCM}) is of the form $C\cdot S_{a,\alpha}$ where $C$ is some curve in the threefold $X_{S,\{q_\alpha\}}$ and $S_{a,\alpha}$ is a surface inside the threefold. Like the triple intersection numbers of surfaces inside a threefold, such intersections can also be computed in terms of intersection numbers inside a surface. If $C$ is a curve inside $S_{a,\alpha}$, then
\be
C\cdot S_{a,\alpha}=C\cdot K'_{a,\alpha}
\ee
and if $C$ is a curve inside a surface $S_{b,\beta}$ that is distinct from $S_{a,\alpha}$, then
\be
C\cdot S_{a,\alpha}=C\cdot C_{b,\beta;a,\alpha}
\ee

Consider the example of (\ref{eq8}) whose associated geometry was described towards the end of Section \ref{tripre}. We can compute that
\begin{align}
f_0\cdot S_0&=(K\cdot f)_{S_0}=-2\\
f_1\cdot S_1&=(K\cdot f)_{S_1}=-2\\
f_2\cdot S_2&=(K\cdot f)_{S_2}=-2\\
f_0\cdot S_1&=C_{0;1}\cdot f_0=(e\cdot f)_{S_0}=1\\
f_1\cdot S_2&=C_{1;2}\cdot f_1=(h\cdot f)_{S_0}=1\\
f_2\cdot S_0&=C_{2;0}\cdot f_2=\left((h-\sum x_i)\cdot f\right)_{S_0}=1\\
f_1\cdot S_0&=C_{1;0}\cdot f_1=(e\cdot f)_{S_1}=1\\
f_2\cdot S_1&=C_{2;1}\cdot f_2=(e\cdot f)_{S_2}=1\\
f_0\cdot S_2&=C_{0;2}\cdot f_0=(e\cdot f)_{S_0}=1
\end{align}
Thus we see that $f_a\cdot S_b$ indeed reproduces the negative of Cartan matrix of affine Lie algebra $\su(3)^{(1)}$. We can similarly check that the geometry associated to (\ref{eg8}) also leads to the Cartan matrix of $\su(3)^{(1)}$.

\subsubsection{The genus one fibration}
For each $\alpha$, combining the fibers $f_{a,\alpha}$, let us define a fiber $f_\alpha$ via
\be
f_\alpha=d_af_{a,\alpha}
\ee
where $d_a$ are Coxeter labels for $\fg_\alpha^{(q_\alpha)}$ listed (in red color) in Tables \ref{UA} and \ref{TA}. If $\fg_\alpha$ is trivial, then $d_0:=1$. 

We claim that $f_\alpha$ is a genus one fiber. This means that $f_\alpha$ can be obtained by a degeneration of a torus. It is well-known that torus fibers can degenerate into Kodaira fibers, which are collections of rational curves\footnote{This means they have genus zero.} intersecting in the pattern of untwisted affine Dynkin diagrams of type $\su(n)^{(1)}$, $\so(2n)^{(1)}$ and $\fe_n^{(1)}$. The multiplicity of each rational component curve is given by the Coxeter label for the corresponding node in the affine Dynkin diagram. The fiber $f_\alpha$, on the other hand, is composed of rational curves $f_{a,\alpha}$ with their multiplicity given by the Coxeter labels for affine Dynkin diagram $\fg_\alpha^{(q_\alpha)}$. Now, one can notice that every affine Dynkin diagram can be obtained by folding affine Dynkin diagrams of type $\su(n)^{(1)}$, $\so(2n)^{(1)}$ and $\fe_n^{(1)}$ as follows:
\begin{align}
\so(2n)^{(1)}&\to\so(2n-1)^{(1)}\to \so(2n-2)^{(2)}\\
\fe_6^{(1)}&\to\ff_4^{(1)}\to\so(8)^{(3)}\\
\so(8)^{(1)}&\to \so(7)^{(1)}\to\fg_2^{(1)}\\
\so(4n)^{(1)}&\to\su(2n)^{(2)}\to\su(2n-1)^{(2)}\\
\so(8)^{(1)}&\to\so(7)^{(1)}\to\su(4)^{(2)}\to\su(3)^{(2)}\\
\fe_7^{(1)}&\to\fe_6^{(2)}
\end{align}
Moreover, observe that the Coxeter numbers of two nodes are added if they are identified under gluing. This means that $f_\alpha$ can be obtained by identifying the rational components of the Kodaira fibers according to the above folding rules. This explicitly shows that $f_\alpha$ is a genus one fiber.

Moreover, we find that due to the virtue of gluing rules, $f_\alpha$ is glued to $f_\beta$ as
\be\label{tfgr}
q_\alpha (-\Omega_{\beta\alpha}) f_\alpha\sim q_\beta (-\Omega_{\alpha\beta}) f_\beta
\ee
This generalizes the condition in the untwisted unfrozen case \cite{Bhardwaj:2018vuu} where $f_i\sim f_j$ whenever there is an edge between $i$ and $j$ in $\Sigma_\fT$. This shows that certain multiples of genus one fibers are identified with each other as one passes over from one collection of surfaces to another, allowing us to extend the fibration structure consistently throughout the threefold.

More formally, according to a theorem due to Oguiso and Wilson \cite{Oguiso,Wilson}, a threefold $X$ admits an genus one fibration structure if and only if there exists an effective divisor $S_{T^2}$ satisfying
	\begin{align}\label{OW}
		S_{T^2} \cdot S_{T^2} \cdot S_{T^2} = 0, ~~ S_{T^2} \cdot S_{T^2} \ne 0
	\end{align}
where $S_{T^2}$ lives in the extended K\"ahler cone, possibly on the boundary. The extended K\"ahler cone is parameterized by all the Coulomb branch and mass parameters satisfying
\be
J\cdot C\ge0
\ee
for all holomorphic curves $C$ in $X$. Physically, the extended K\"ahler cone corresponds to the Coulomb branch of the (possibly mass deformed) $5d$ theory corresponding to $X$. 

In all of geometries associated to $5d$ KK theories, we can find an $S_{T^2}$ which lies in the extended K\"ahler cone satisfies (\ref{OW}). Pick any node $\alpha$ and define
\be\label{ST2}
S_{T^2}:=\sum_{a=0}^{r_\alpha} d^\vee_a S_{a,\alpha}
\ee
where $d^\vee_a$ are dual Coxeter labels for the associated affine algebra $\fg_\alpha^{(q_\alpha)}$ (see Tables \ref{UA} and \ref{TA}) and $r_\alpha$ is the rank of invariant subalgebra $\fh_\alpha$. If the node $\alpha$ carries a trivial gauge algebra, then we define $d^\vee_0=1$ and take (\ref{ST2}) to be the definition of $S_{T^2}$.

In the gauge theoretic case, the direction parametrized by (\ref{ST2}) is special since all the fibers $f_{a,\alpha}$ have zero volume along this direction\footnote{In fact, non-negativity of the volumes of fibers implies that the only directions in the Coulomb branch when mass parameters are turned off are given by $\sum_a d^\vee_aS_{a,\alpha}$ for various $\alpha$.}
\be
-S_{T^2}\cdot f_{a,\alpha}=\sum_b A_{ab} d^\vee_b=0
\ee
Similarly, in the non-gauge theoretic case 
\be
-S_{T^2}\cdot f_{0,\alpha}=-K'_{0,\alpha}\cdot f_{0,\alpha}=0
\ee
where the last equality can be checked to be true for every non-gauge theoretic case. Moreover, the reader can check using the explicit description of geometries presented in this paper that
\be
S_{T^2}\cdot C\ge0
\ee
for all other holomorphic $C$ in the threefold $X_{S,\{q_\alpha\}}$. So, $S_{T^2}$ as defiend in (\ref{ST2}) lies in the extended K\"ahler cone of $X_{S,\{q_\alpha\}}$.

Now it can be easily checked for all the geometries presented in this paper that
\be
S_{T^2}\cdot S_{T^2}=-q_\alpha\Omega^{\alpha\alpha}\sum_{a=0}^{r_\alpha}(d_af_{a,\alpha})\neq0
\ee
where $d_a$ are the Coxeter labels for $\fg_\alpha^{(q_\alpha)}$ with $d_0:=1$ if $\alpha$ is a non-gauge theoretic node. We can now compute
\be
S_{T^2}\cdot S_{T^2}\cdot S_{T^2}\propto \sum_{a=0}^{r_\alpha}(d_af_{a,\alpha})\cdot (\sum_{b=0}^{r_\alpha}d^\vee_b S_{b,\alpha})=-\sum_{a,b=0}^{r_\alpha}d_a A_{ab} d^\vee_b=0
\ee
thus verifying both the conditions in (\ref{OW}) and establishing the presence of a genus one fibration in $X_{S,\{q_\alpha\}}$.

Let us now discuss the relationship between fibers $f_\alpha$ and the radius of compactification circle $R$. In general, we can find at least one node $\mu$ such that 
\be
n_\mu f_\mu\sim n_{\mu,\alpha} f_\alpha
\ee
with $n_{\mu,\alpha}\ge n_\mu\ge1$ for all $\alpha$ . Then the curve
\be
f:= l_\mu n_\mu f_\mu
\ee
with $l_\mu$ defined in Section \ref{DS2} can be identified with the KK mode of unit momentum in $\fT^{KK}_{S,\{q_\alpha\}}$ and has mass $\frac{1}{R}$ where $R$ is the radius of the circle on which the $6d$ theory $\fT$ has been compactified. Thus, all the $f_\alpha$ can be identified as fractional KK modes with mass $\frac{1}{n_\alpha R}$ where $n_\alpha=l_\mu n_{\mu,\alpha}$. This generalizes the condition in the untwisted unfrozen case where the KK mode is identified with
\be
f:=f_i
\ee
for any $i$, which is consistent since $f_i\sim f_j$ for all $i,j$.

Let us now discuss some examples. For the KK theory
\be
\begin{tikzpicture}
\node (v1) at (-0.5,0.45) {$1$};
\node at (-0.45,0.9) {$\sp(n)^{(1)}$};
\begin{scope}[shift={(2,0)}]
\node (v2) at (-0.5,0.45) {$4$};
\node at (-0.45,0.9) {$\so(2m)^{(2)}$};
\end{scope}
\draw (v1)--(v2);
\end{tikzpicture}
\ee
we find that 
\be
f_{\sp(n)^{(1)}}\sim 2f_{\so(2m)^{(2)}}
\ee
and the KK mode is
\be
f=f_{\sp(n)^{(1)}}
\ee

For the KK theory (\ref{fg1}), we find that
\be
f_{\sp(n)^{(1)}}\sim 2f_{\so(2m)^{(1)}}
\ee
and the KK mode is
\be
f=f_{\sp(n)^{(1)}}
\ee

For the KK theory (\ref{exa3}), our gluing rules say that
\be
2f_{\su(n)^{(2)}}\sim 2f_{\su(m)^{(1)}}
\ee
and the KK mode is
\be
f=2f_{\su(n)^{(2)}}
\ee

For the KK theory (\ref{eg2}), our gluing rules say that
\be
f_{\su(n)^{(1)}}\sim f_{\su(m)^{(1)}}
\ee
and the KK mode is
\be
f=2f_{\su(n)^{(1)}}
\ee

An interesting example to consider is the KK theory defined by the untwisted compactification of the $6d$ SCFT
\be\label{fr}
\begin{tikzpicture}
\node (v1) at (-0.5,0.45) {2};
\node at (-0.45,0.9) {$\su(p)$};
\begin{scope}[shift={(1.5,0)}]
\node (v2) at (-0.5,0.45) {4};
\node at (-0.45,0.9) {$\so(m)$};
\end{scope}
\begin{scope}[shift={(3,0)}]
\node (v4) at (-0.5,0.45) {1};
\node at (-0.45,0.9) {$\sp(n)$};
\end{scope}
\node (v3) at (0.25,0.45) {\tiny{2}};
\draw (v1)--(v3);
\draw (v2)--(v3);
\draw (v4)--(v2);
\end{tikzpicture}
\ee
which arises only in the frozen phase. We find that
\begin{align}
2f_{\su(p)^{(1)}}&\sim 2f_{\so(m)^{(1)}}\\
f_{\so(m)^{(1)}}&\sim f_{\sp(n)^{(1)}}
\end{align}
and the KK mode is
\be\label{frns}
f=2f_{\su(p)^{(1)}}\sim2f_{\so(m)^{(1)}}\sim2f_{\sp(n)^{(1)}}
\ee
If (\ref{fr}) arose in the unfrozen phase of F-theory, then we would have obtained
\be
f=f_{\su(p)^{(1)}}\sim f_{\so(m)^{(1)}}\sim f_{\sp(n)^{(1)}}
\ee
Thus equation (\ref{frns}) is a way to see that (\ref{fr}) cannot arise in the unfrozen phase of F-theory.

\subsection{Geometry for each node}\label{RO}
In this section we will describe the surfaces $S_{a,\alpha}$ along with their intersections associated to a single node $\alpha$.

\subsubsection{Graphical notation}
We will capture the data of the surfaces and their intersections by using a graphical notation that would be a simpler version of the graphical notation used in \cite{Bhardwaj:2018vuu,Bhardwaj:2018yhy}. This subsection is devoted to the explanation of this notation. We find it best to explain the notation with the following example:
\be\label{GN}
\begin{tikzpicture} [scale=1.9]
\node (v1) at (-5,-0.5) {$\mathbf{0_{8}^{2+2}}$};
\node (v2) at (-2.5,-0.5) {$\mathbf{1_{6}^{2+2}}$};
\node (v3) at (-0.2,-0.5) {$\mathbf{2_{0}}$};
\draw  (v2) edge (v3);
\node at (-4.4,-0.4) {\scriptsize{$e$-$\sum y_i$}};
\node at (-3.2,-0.4) {\scriptsize{$h$+$\sum(f$-$y_i)$}};
\node at (-1.8,-0.4) {\scriptsize{$e$-$\sum x_i$-$\sum y_i$}};
\node at (-0.7,-0.4) {\scriptsize{$3e+2f$}};
\node at (-3.2,-0.2) {\scriptsize{$f$-$x_i$,}};
\node at (-4.4,-0.2) {\scriptsize{$f$-$x_i$,}};
\node (w1) at (-3.9,-0.5) {\scriptsize{3}};
\draw  (v1) edge (w1);
\draw  (w1) edge (v2);
\node at (-5.3,-0.7) {\scriptsize{$x_i$}};
\node at (-4.7,-0.7) {\scriptsize{$y_i$}};
\node at (-2.8,-0.7) {\scriptsize{$x_i$}};
\node at (-2.2,-0.7) {\scriptsize{$y_i$}};
\node (v4) at (-5,-1) {\scriptsize{2}};
\node (v5) at (-2.5,-1) {\scriptsize{2}};
\draw (v1) .. controls (-5.1,-0.7) and (-5.4,-0.9) .. (v4);
\draw (v1) .. controls (-4.9,-0.7) and (-4.6,-0.9) .. (v4);
\draw (v2) .. controls (-2.6,-0.7) and (-2.9,-0.9) .. (v5);
\draw (v2) .. controls (-2.4,-0.7) and (-2.1,-0.9) .. (v5);
\end{tikzpicture}
\ee
which is a particular phase of the KK theory
\be
\begin{tikzpicture}
\node at (-0.5,0.45) {2};
\node at (-0.45,0.9) {$\so(8)^{(3)}$};
\end{tikzpicture}
\ee
Since the rank of invariant subalgebra $\fh=\fg_2$ is two, we should have three surfaces in this case labeled by $S_a$ where $0\le a\le 2$. The middle number in the label for each node denotes the index $a$. Thus the node labeled $\mathbf{0_8^{2+2}}$ denotes the surface $S_0$, the node labeled $\mathbf{1_6^{2+2}}$ denotes the surface $S_1$, and the node labeled $\mathbf{2_0}$ denotes the surface $S_2$. 

Every surface $S_a$ is a Hirzebruch surface. The subscript in the label for each node denotes the degree of the corresponding Hirzebruch surface. Thus, $S_0$ has degree 8, $S_1$ has degree 6, and $S_2$ has degree 0. The superscript in the label for each node denotes the number of blowups on the corresponding Hirzebruch surface. Thus, $S_0$ carries $2+2=4$ blowups and hence $S_0=\bF_8^4$, $S_1$ carries $2+2=4$ blowups and hence $S_1=\bF_6^4$, and $S_2$ carries no blowups and hence $S_2=\bF_0$. 

The fact that the four blowups on $S_0$ are displayed as $2+2$ denotes that the four blowups are divided into two sets, with each set containing two blowups. We denote the blowups in the first set as $x_i$ and the blowups in the second set as $y_i$. The same is true for $S_1$. In a general graph, the blowups on a surface can be divided into more than two sets, and the number of blowups inside each set can be different. Whatever may be the case, we adopt the notation of denoting the blowups inside the first set as $x_i$, the blowups inside the second set as $y_i$, the blowups inside the third set as $z_i$ etc.

The label in the middle of an edge between two nodes denotes the number of irreducible components of the intersection locus between the two surfaces corresponding to the two nodes. As already discussed above, each component of the intersection locus can be viewed as an irreducible gluing curve inside each of the surfaces participating in the intersection. Thus, there is a single gluing curve between $S_1$ and $S_2$ in the graph (\ref{GN}), but there are three gluing curves between $S_0$ and $S_1$. The graph also tells us that the surface $S_0$ is a self-glued Hirzebruch surface since there are edges which start and end at $S_0$. Similarly, $S_1$ is also a self-glued surface. We can see that the number of self-gluings in $S_0$ are two, and the number of self-gluings in $S_1$ are also two.

The curves displayed at the ends of edges tell us the identities of various gluing curves. The left end of the edge between $\mathbf{1_6^{2+2}}$ and $\mathbf{2_0}$ reads $e-\sum x_i-\sum y_i$, which means that the corresponding gluing curve inside $S_1$ is $e-\sum x_i-\sum y_i$. The right end of the edge between $\mathbf{1_6^{2+2}}$ and $\mathbf{2_0}$ reads $3e+2f$, which means that the corresponding gluing curve inside $S_2$ is $3e+2f$. We note that whenever we write $\sum x_i$ or $\sum y_i$, we mean a sum of all the blowups in the set of blowups denoted by $x_i$ or $y_i$ respectively. 

In the above graph, the two self-gluings of $S_0$ are displayed by writing $x_i$ at one end and $y_i$ at the other end. This tells us that $x_i$ in $S_0$ is glued to $y_i$ in $S_0$. Since there is no sum over $i$, this gluing is supposed to be true for each valued of $i$. Hence, the two self-gluings are $x_1\sim y_1$ and $x_2\sim y_2$. The same is true for self-gluings of $S_1$.

The gluing curves for the three gluings between $S_0$ and $S_1$ are displayed as $f-x_i,e-\sum y_i$ inside $S_0$ and as $f-x_i,h+\sum(f-y_i)$ inside $S_1$. These are supposed to be read in the order they are written. Thus, unpacking the notation we learn that the three gluings are
\begin{align}
(f-x_1)_{S_0}&\sim (f-x_1)_{S_1}\\
(f-x_2)_{S_0}&\sim (f-x_2)_{S_1}\\
(e-y_1-y_2)_{S_0}&\sim (h+2f-y_1-y_2)_{S_1}
\end{align}

We also sometimes suppress multiplicity of a gluing curve. For example, in the geometry
\be
\begin{tikzpicture} [scale=1.9]
\node (v1) at (-2.5,1.1) {$\mathbf{3^{2+2}_6}$};
\node (v2) at (-2.5,-0.5) {$\mathbf{2_{3}}$};
\node (v3) at (-0.5,-0.5) {$\mathbf{1_{5}^{6+6}}$};
\draw  (v1) edge (v2);
\draw  (v2) edge (v3);
\node at (-2,0.9) {\scriptsize{$f$}};
\node at (-3,0.8) {\scriptsize{$e$-$\sum x_i$-$\sum y_i$}};
\node at (-2.7,-0.2) {\scriptsize{$2h$}};
\node at (-1,-0.6) {\scriptsize{$e$}};
\node at (-2.1,-0.6) {\scriptsize{$h$}};
\node at (-0.6,-0.1) {\scriptsize{$f$-$ x_i$-$y_i$}};
\node (v4) at (-4.4,-0.5) {$\mathbf{0_{1}}$};
\draw  (v4) edge (v2);
\node at (-4,-0.6) {\scriptsize{$h$}};
\node at (-2.9,-0.6) {\scriptsize{$e$}};
\node (v5) at (-1.5,0.3) {\scriptsize{$6$}};
\draw  (v1) edge (v5);
\draw  (v5) edge (v3);
\node (w1) at (-2.5,1.6) {\scriptsize{2}};
\draw (v1) .. controls (-2.6,1.2) and (-3,1.5) .. (w1);
\draw (v1) .. controls (-2.4,1.2) and (-2,1.5) .. (w1);
\node at (-2.9,1.4) {\scriptsize{$x_i$}};
\node at (-2.1,1.4) {\scriptsize{$y_i$}};
\end{tikzpicture}
\ee
the gluing curve for $S_2$ in $S_3$ is displayed simply as $f$. But the edge between $S_2$ and $S_3$ shows that there are six gluing curves involved. This means that the true gluing curve for $S_2$ in $S_3$ is actually six copies of the fiber $f$ of $S_3$.

Now, let us extract the prepotential $6\tilde\cF$ from the graph (\ref{GN}). The coefficient of $\phi_0^3$ is
\be\label{cf1}
(K'^2)_{S_0}=\left((K+\sum x_i+\sum y_i)^2\right)_{S_0}=(K^2+\sum x_i^2+\sum y_i^2+2\sum K\cdot x_i+2\sum K\cdot y_i)_{S_0}
\ee
We have $K^2=8-4=4$ and $K\cdot x_i=K\cdot y_i=-1$, using which (\ref{cf1}) reduces to
\be
(K'^2)_{S_0}=4-2-2-4-4=-8
\ee
Similarly the coefficient of $\phi_1^3$ is $-8$. The coefficient of $\phi_2^3$ is $8$. The coefficient of $\phi_2\phi_1^2$ can be computed as
\be
\left((3e+2f)^2\right)_{S_2}=12
\ee
which coincides with
\be
\left(K'\cdot(e-\sum x_i-\sum y_i)\right)_{S_1}=\left((K+\sum x_i+\sum y_i)\cdot(e-\sum x_i-\sum y_i)\right)_{S_1}=12
\ee
as it should for consistency. We can compute the coefficient of $\phi_0\phi_1^2$ to be
\be
\left(\left((e-\sum y_i)+(f-x_1)+(f-x_2)\right)^2\right)_{S_0}=-8
\ee
which indeed coincides with
\be
\left(K'\cdot\left((h+2f-y_1-y_2)+(f-x_1)+(f-x_2)\right)\right)_{S_1}=-8
\ee
Similarly, we can compute coefficients for other terms of the form $\phi_a\phi_b^2$. Finally, the coefficient of $\phi_0\phi_1\phi_2$ must be 0 since there is no edge between $S_0$ and $S_2$. But this coefficient can also be computed as an intersection number of gluing curves inside $S_1$. Thus, the corresponding intersection number better be zero for consistency. Indeed we find that
\be
\left((e-\sum x_i-\sum y_i)\cdot\left((h+2f-y_1-y_2)+(f-x_1)+(f-x_2)\right)\right)_{S_1}=0
\ee

\subsubsection{Untwisted}
In this subsection, we collect our results for nodes of the form
\be\label{nofu}
\begin{tikzpicture}
\node at (-0.5,0.45) {$k$};
\node at (-0.45,0.9) {$\fg^{(1)}$};
\end{tikzpicture}
\ee
That is, we restrict ourselves to the case where the associated affine Lie algebra is untwisted. All such nodes are displayed in Table \ref{TR1} and Table \ref{R1N}. Most such cases were first studied in \cite{Bhardwaj:2018vuu,Bhardwaj:2018yhy}. We will be able to recover their results. We will associate a collection of geometries parametrized by $\nu$ to each node of the form (\ref{nofu}). Geometries for different values of $\nu$ are flop equivalent as long as there are no neighboring nodes, but might cease to be flop equivalent in the presence of neighboring nodes. The geometries associated to (\ref{nofu}) in \cite{Bhardwaj:2018vuu} are obtained as $\nu=0,1$ versions of the geometries associated in this paper. 

The geometries associated to nodes of the form (\ref{nofu}) are presented below. We will display the corresponding node inside a circle placed at top of the geometry:
\be\label{spt0}
\begin{tikzpicture} [scale=1.9]
\node (v1) at (-2.4,-2) {$\mathbf{0_1^{2n+8-\nu}}$};
\node (v10) at (-0.5,-2) {$\mathbf{1_{2n+2-\nu}}$};
\node (v9) at (0.4,-2) {$\cdots$};
\node (v8) at (1.5,-2) {$\mathbf{(n-2)_{8-\nu}}$};
\node (v5) at (5.2,-2) {$\mathbf{n^\nu_0}$};
\node (v7) at (3.25,-2) {$\mathbf{(n-1)_{6-\nu}}$};
\draw  (v7) edge (v8);
\draw  (v8) edge (v9);
\draw  (v9) edge (v10);
\draw  (v10) edge (v1);
\node at (0,-1.9) {\scriptsize{$e$}};
\node at (-1,-1.9) {\scriptsize{$h$}};
\node at (4.6,-1.9) {\scriptsize{$2e$+$f$-$\sum x_i$}};
\node at (3.9,-1.9) {\scriptsize{$e$}};
\node at (2.6,-1.9) {\scriptsize{$h$}};
\node at (0.9,-1.9) {\scriptsize{$h$}};
\node at (-1.6,-1.9) {\scriptsize{$2h$- $\sum x_i$}};
\node at (2.1,-1.9) {\scriptsize{$e$}};
\draw  (v5) edge (v7);
\begin{scope}[shift={(1.7,-1.9)}]
\node (v1_1) at (-0.5,0.7) {1};
\node at (-0.45,1) {$\sp(n)_{(n+1)\pi}^{(1)}$};
\draw  (-0.5,0.9) ellipse (0.7 and 0.5);
\end{scope}
\end{tikzpicture}
\ee
where $0\le\nu\le 2n+8$, $n\ge1$ and the theta angle should be viewed modulo $2\pi$. We can see that $f_a\cdot S_b$ reproduces the negative of Cartan matrix for untwisted affine Lie algebra $\sp(n)^{(1)}$, where $f_a$ is the canonical fiber of Hirzebruch surface $S_a$. The same hold true for all the examples discussed below in this subsection. One can check in each example below that $f_a\cdot S_b$ reproduces the negative of Cartan matrix for the associated untwisted affine Lie algebra $\fg^{(1)}$.

\be\label{spt1}
\begin{tikzpicture} [scale=1.9]
\node (v1) at (-2.5,-2) {$\mathbf{0_1^{2n+8-\nu}}$};
\node (v10) at (-0.5,-2) {$\mathbf{1_{2n+2-\nu}}$};
\node (v9) at (0.4,-2) {$\cdots$};
\node (v8) at (1.5,-2) {$\mathbf{(n-2)_{8-\nu}}$};
\node (v5) at (5.1,-2) {$\mathbf{n^\nu_1}$};
\node (v7) at (3.25,-2) {$\mathbf{(n-1)_{6-\nu}}$};
\draw  (v7) edge (v8);
\draw  (v8) edge (v9);
\draw  (v9) edge (v10);
\draw  (v10) edge (v1);
\node at (0,-1.9) {\scriptsize{$e$}};
\node at (-1,-1.9) {\scriptsize{$h$}};
\node at (4.6,-1.9) {\scriptsize{$2h$-$\sum x_i$}};
\node at (3.9,-1.9) {\scriptsize{$e$}};
\node at (2.6,-1.9) {\scriptsize{$h$}};
\node at (0.9,-1.9) {\scriptsize{$h$}};
\node at (-1.7,-1.9) {\scriptsize{$2h$- $\sum x_i$}};
\node at (2.1,-1.9) {\scriptsize{$e$}};
\draw  (v5) edge (v7);
\begin{scope}[shift={(1.6,-1.9)}]
\node (v1_1) at (-0.5,0.7) {1};
\node at (-0.45,1) {$\sp(n)_{n\pi}^{(1)}$};
\draw  (-0.5,0.9) ellipse (0.5 and 0.5);
\end{scope}
\end{tikzpicture}
\ee
where $0\le\nu\le 2n+8$, $n\ge1$ and the theta angle should be viewed modulo $2\pi$. See Appendix (\ref{theta}) for more discussion on the relationship between theta angle and geometry. 

Notice that the two geometries (\ref{spt0}) and (\ref{spt1}) are isomorphic by virtue of the isomorphism between $\bF_0^1$ and $\bF_1^1$ discussed in Appendix \ref{Hirz}. Suppose first that $\nu>0$. Then, the isomorphism applied to $S_n$ sends $2h-x_1$ in $\bF_1^{\nu}$ to $2e+f-x_1$ in $\bF_0^{\nu}$, thus mapping the gluing curve for $S_{n-1}$ in $S_n$ in (\ref{spt1}) to the gluing curve for $S_{n-1}$ in $S_n$ in (\ref{spt0}). Thus the whole geometry (\ref{spt1}) is mapped to the geometry (\ref{spt0}) by this isomorphism. For $\nu=0$, the two geometries (\ref{spt1}) and (\ref{spt0}) are flop equivalent due to this isomorphism. This is because they are flop equivalent to $\nu>0$ versions of the geometries (\ref{spt1}) and (\ref{spt0}), and we have already established an isomorphism between the latter geometries.

However, it is possible for this isomorphism to not extend to the full Calabi-Yau threefold when $\sp(n)$ has other neighbors. The gluing curves inside $S_0$ and $S_n$ for the surfaces corresponding to these neighbors might not map to each other under the above isomorphism plus flops. Whenever the isomorphism extends to the full threefold, the $\sp(n)$ theta angle is physically irrelevant. Whenever the isomorphism does not extend to the full threefold, the $\sp(n)$ theta angle is physically relevant. We will see examples of both situations later when we discuss gluing rules for $\sp(n)$.

For $n=0$, we claim that the associated geometry is
\be\label{Gsp0}
\begin{tikzpicture} [scale=1.9]
\node (v2) at (-2.2,-0.5) {$\mathbf{0^{8}_{1}}$};
\begin{scope}[shift={(-1.7,-0.5)}]
\node (v1_1) at (-0.5,0.7) {1};
\node at (-0.45,1) {$\sp(0)_\theta^{(1)}$};
\draw  (-0.5,0.9) ellipse (0.5 and 0.5);
\end{scope}
\end{tikzpicture}
\ee
One way to see this is to notice that both the geometries (\ref{spt0}) and (\ref{spt1}) reduce to (\ref{Gsp0}) in the limit $n=0$. For a more precise way to see that (\ref{Gsp0}) is the correct geometry, see the discussion around (\ref{sp05d}).

\ni When $\sp(0)_\theta^{(1)}$ has no other neighbors, then all the blowups are generic and we can write $S_0=dP_9$. When $\sp(0)_\theta^{(1)}$ has neighbors, it turns out that $S_0=dP_9$ with 9 non-generic blowups is the correct answer, instead of $S_0=\bF_1^8$ with eight non-generic blowups. This is because when the 9 blowups are non-generic, it is not always possible to represent $dP_9$ as $\bF_1^8$ with 8 non-generic blowups. So, $S_0=\bF_1^8$ is not quite the correct answer. See \cite{Bhardwaj:2018vuu} for more discussion on this point. Thus, in this paper, from this point on, we will represent the geometry associated to $\sp(0)_\theta^{(1)}$ by $dP_9$.

\be

\ee
For this geometry, we do not define multiple versions distinguished by the parameter $\nu$. Nevertheless, for uniformity of notation, we denote this geometry with $\nu=0$. Similarly, we will denote all the following geometries having a single unique version with $\nu=0$.

\ni For $n=2$, we have
\be
%
\ee

The above two examples are not completely geometric. See the discussion after equation (\ref{Gsu1l}).

\be
%
\ee
where $0\le\nu\le 4n+2$ and $n\ge1$.

\ni For $n=0$, we claim that the geometry is 
\be\label{Gsu1}
%
\ee
which can be recognized as a limit of $\nu=1$ phase of (\ref{Gsuo2}). See Appendix \ref{ng} for a derivation that this is the correct answer.

\be
%
\ee
where $1\le k\le 3$ and we have divided the $16-4k$ blowups into four sets of $4-k$ blowups each. We label blowups in the four sets by $x_i$, $y_i$, $z_i$ and $w_i$ respectively.

\be
%

\subsubsection{Twisted}
In this subsection, we will generalize our results to nodes of the form
\be\label{nof}
%
\ee
All such nodes are listed in Table \ref{KR1}.
\be
%
\ee
where $m\ge 3$. Notice that the Cartan matrix associated to this geometry is precisely that of $\su(2m)^{(2)}$. Similar comments hold for all the geometries discussed below in this subsection. For each example below, one can check that $f_a\cdot S_b$ reproduces negative of Cartan matrix of the associated twisted affine algebra $\fg^{(q)}$.

\be
%
\ee

Now we discuss some examples which are not completely geometric:

\be\label{GsuoS}
%
\ee
Let us now discuss the reasons why the above five examples are not completely geometric. Let us start with (\ref{Gsu1l}). The geometry for this example contains the $-1$ curve $h-x-y$ and hence an M2 brane wrapping this curve should give rise to a BPS particle. However, this BPS particle cannot appear in the associated $5d$ KK theory for the following reason. The existence of a particle associated to $h-x-y$ implies that the KK mode, which is associated to the elliptic curve $2h+f-2x-2y$, decomposes as a bound state of $h-x-y$ and $h+f-x-y$ but this is a contradiction since these two curves do not meet each other and hence there cannot be such a bound state.

Another reasoning is as follows. The volume of $f$ is $2\phi$ where $\phi$ is the Coulomb branch parameter associated to the above surface. On the other hand, the volume of $h-x-y$ is $-\phi$. Requiring non-negative volumes for both curves implies that $\phi$ must be zero. In other words, there is no direction in the Coulomb branch where all BPS particles have non-negative mass. Thus, this geometry is not \emph{marginal}, in the sense defined by \cite{Jefferson:2018irk}, which is a condition that must be satisfied by geometries associated to KK theories.

The precise sense in which the above self-glued $\bF_1$ surface is associated to the KK theory
\be\label{nong}
\begin{tikzpicture}
\node (v1) at (-0.5,0.4) {2};
\node at (-0.45,0.9) {$\su(1)^{(1)}$};
\draw (v1) .. controls (-1.5,-0.5) and (0.5,-0.5) .. (v1);
\end{tikzpicture}
\ee
is as follows. The Mori cone of the surface is generated by $h-x-y,f-x,x,e$. However, since the curve $h-x-y$ does not correspond to a BPS particle, the generators of the Mori cone thus do not correspond to the fundamental BPS particles\footnote{We define a fundamental BPS particle to be a BPS particle that cannot arise as a bound state of other BPS particles.} in the associated KK theory (\ref{nong}). We propose that the fundamental BPS particles instead correspond to the curves $2h-x-2y,f-x,x,e$. This set of curves satisfies all the properties that must be satisfied by the generators of the Mori cone of a surface. Thus, it is a complete set which can be consistently associated to fundamental BPS particles. The KK mode can be found as a bound state of $2h-x-2y$ and $f-x$. One can check that this set of proposed BPS particles is marginal in the sense that it allows a direction in Coulomb branch with all BPS particles having non-negative volumes. See also Appendix \ref{ng} where we verify that this description of the KK theory allows the existence of an RG flow to an $\cN=2$ $5d$ SCFT, which is a fact well-known in the literature.

There are two viewpoints one can take on the relationship between self-glued $\bF_1$ and the KK theory (\ref{nong}). The first is that indeed compactifying M-theory on this surface leads to the KK theory (\ref{nong}), but the compactification has some extra ingredients which account for the mismatch between the set of Mori cone generators and the set of fundamental BPS particles\footnote{A similar situation occurs in the frozen phase of F-theory \cite{Bhardwaj:2018jgp}, where the set of generators of the Mori cone of the base of a threefold used for compactifying F-theory does not match the set of fundamental BPS strings arising in the associated $6d$ theory.}. The other viewpoint is that the relationship with self-glued $\bF_1$ has no deep meaning and is probably a red herring. At the time of writing of this paper, we do not know which of these two viewpoints, or if either of these two viewpoints, is the correct one. We leave this issue for future exploration, and only use the relationship between the two as an algebraic tool to build a formalism for KK theories from which one can explicitly perform RG flows to $5d$ SCFTs.

Now let us discuss the non-geometric nature of the KK theories 
\be
\begin{tikzpicture}
\node (v1) at (-0.5,0.4) {2};
\node at (-0.45,0.9) {$\su(m)^{(1)}$};
\draw (v1) .. controls (-1.5,-0.5) and (0.5,-0.5) .. (v1);
\end{tikzpicture}
\ee
with $m>1$. Consider as an example the case of $m=3$. The surface $S_2$ contains a gluing curve $e+f-x-2y$ and hence there must be a BPS particle associated to it. However, notice that it decomposes as $e+f-x-2y=(e-x-y)+(f-y)$ such that the components $e-x-y$ and $f-y$ do not intersect each other. This leads to the same problem as discussed above, and we are forced to hypothesize that the fundamental BPS particles are distinct from the generators of Mori cone due to some non-geometric feature in the M-theory compactification. It is also evident that some of the components of the gluing curves in certain surfaces  (which are identified with irreducible curves in adjacent surfaces as part of the gluing construction) fail to satisfy the necessary properties of irreducible curves that are described at the beginning of Section \ref{consistency}.\footnote{For example, in the case $m=3$, one can see that the surface $\textbf{2}_{\textbf{0}}^{\textbf{1} + \textbf{1} + \textbf{1}}$ contains a curve class $e+f- x -2 y$, which is identified with the curve class $h$ in the surface $\textbf{1}_{\textbf{3}}$. Since $h$ is irreducible, this implies that $e+f-x-2y$ must also be irreducible, but this leads to a contradiction (with smoothness) if the usual class $f-y$ remains among the generators of the Mori cone of $\textbf{2}_{\textbf{0}}^{\textbf{1} + \textbf{1} + \textbf{1}}$.} Similar comments apply to each of the $m>1$ models presented above should be regarded as an algebraic proposal which retains many of the features of the local threefolds that seem to be necessary to compute RG flows to 5d SCFTs. 

Similar comments apply to (\ref{GsueF}) and (\ref{GsuoF}), and they are also not conventionally geometric.

\subsection{Gluing rules between two gauge theoretic nodes}\label{AR}
In this section we will describe how to glue the surfaces $S_{a,\alpha}$ corresponding to a node $\alpha$ to the surfaces $S_{b,\beta}$ corresponding to another node $\beta$ if there is an edge between $\alpha$ and $\beta$. The gluing rules are different for different kinds of edges between the two nodes. It turns out that the gluing rules between $\alpha$ and $\beta$ are insensitive to the values of $\Omega^{\alpha\alpha}$ and $\Omega^{\beta\beta}$. This was also true for all of the cases studied in \cite{Bhardwaj:2018vuu}. For this reason, we will often suppress the data of $\Omega^{\alpha\alpha}$ and $\Omega^{\beta\beta}$ in this subsection.

As a preface to the following subsections, we re-emphasize that the gluing rules must be compatible with the general consistency conditions described in Section \ref{consistency}, and those that do not must again be regarded, most conservatively, as an algebraic proposal that retains certain salient features of conventional smooth threefold geometries. The basic, underlying hypothesis of the gluing rules is that, given a pair of geometries corresponding to circle compactifications of 6d SCFTs, if there exists a consistent gluing of these two nodes along their respective genus one fibers, then there must also exist a mutual gauging of the respective global symmetries of the parent 6d SCFTs that allows the two theories to be coupled together in the sense described in Section \ref{re}.

\subsubsection{Undirected edges between untwisted algebras}\label{GR1}
Such edges are displayed in Table \ref{TR2}. The gluing rules for all of these cases except for \raisebox{-.25\height}{\begin{tikzpicture}
\node (w1) at (-0.5,0.9) {$\su(n_\alpha)^{(1)}$};
\begin{scope}[shift={(3,0)}]
\node (w2) at (-0.5,0.9) {$\so(n_\beta)^{(1)}$};
\end{scope}
\node (v3) at (1,0.9) {\tiny{2}};
\draw (w1)--(v3);
\draw (w2)--(v3);
\end{tikzpicture}} were first studied in \cite{Bhardwaj:2018vuu}. We are able to reproduce their results using our methods.

\noindent\ubf{Gluing rules for \raisebox{-.25\height}{\begin{tikzpicture}
\node (w1) at (-0.5,0.9) {$\sp(n_\alpha)_\theta^{(1)}$};
\begin{scope}[shift={(3,0)}]
\node (w2) at (-0.5,0.9) {$\su(n_\beta)^{(1)}$};
\end{scope}
\draw (w1)--(w2);
\end{tikzpicture}}}:
We can take any geometry with $0\le\nu\le 2n_\alpha+8-n_\beta$ for $\sp(n_\alpha)_\theta^{(1)}$, and any geometry with $0\le\nu\le 2n_\beta-2n_\alpha$ for $\su(n_\beta)^{(1)}$. The gluing rules below work irrespective of the value of $\theta$. The gluing rules are:
\bit
\item $f-x_1,x_{n_\beta}$ in $S_{0,\alpha}$ are glued to $f-x_1,x_{2n_\alpha}$ in $S_{0,\beta}$.
\item $x_i-x_{i+1}$ in $S_{0,\alpha}$ is glued to $f$ in $S_{i,\beta}$ for $i=1,\cdots,n_\beta-1$.
\item $x_{i}-x_{i+1},x_{2n_\alpha-i}-x_{2n_\alpha-i+1}$ in $S_{0,\beta}$ are glued to $f,f$ in $S_{i,\alpha}$ for $i=1,\cdots,n_\alpha-1$.
\item $x_{n_\alpha}-x_{n_\alpha+1}$ in $S_{0,\beta}$ is glued to $f$ in $S_{n_\alpha,\alpha}$.
\eit
By convention, the first item in the above list of gluing rules displays the gluings in an order. That is, $f-x_1$ in $S_{0,\alpha}$ is glued to $f-x_1$ in $S_{0,\beta}$ and $x_{n_\beta}$ in $S_{0,\alpha}$ is glued to $x_{2n_\alpha}$ in $S_{0,\beta}$. We will adopt this convention in what follows. All the gluings should be read in the order in which they are written.

Let us label the fiber of the Hirzebruch surface $S_{a,\alpha}$ as $f_{a,\alpha}$ and the fiber of the Hirzebruch surface $S_{b,\beta}$ as $f_{b,\beta}$. According the above gluing rules, $f_{0,\alpha}$ is glued to $f_{0,\beta}-x_1+x_{2n_\alpha}+\sum_{i=1}^{n_\beta-1}f_{i,\beta}$ where $x_1$ and $x_{2n_\alpha}$ are blowups in $S_{0,\beta}$, and $2\sum_{i=1}^{n_\alpha-1}f_{i,\alpha}+f_{n_\alpha,\alpha}$ is glued to $x_1-x_{2n_\alpha}$ in $S_{0,\beta}$. Combining these two we see that 
\be
f_{0,\alpha}+2\sum_{i=1}^{n_\alpha-1}f_{i,\alpha}+f_{n_\alpha,\alpha}\:\:\sim\:\: \sum_{i=0}^{n_\beta-1} f_{i,\beta}
\ee
thus confirming the gluing rule (\ref{tfgr}) for the torus fibers. In a similar fashion, the reader can verify that (\ref{tfgr}) is satisfied for all the gluing rules that follow.

The theta angle of $\sp(n_\alpha)$ is physically irrelevant if $n_\beta<2n_\alpha+8$ and physically relevant if $n_\beta=2n_\alpha+8$. Thus the above gluing rules should allow the isomorphism between (\ref{spt0}) and (\ref{spt1}) to extend to the combined geometry for 
\be\label{spsu}
\begin{tikzpicture}
\node (w1) at (-0.5,0.9) {$\sp(n_\alpha)_\theta^{(1)}$};
\begin{scope}[shift={(3,0)}]
\node (w2) at (-0.5,0.9) {$\su(n_\beta)^{(1)}$};
\end{scope}
\draw (w1)--(w2);
\end{tikzpicture}
\ee
in the case $n_\beta<2n_\alpha+8$, but not in the case of $n_\beta=2n_\alpha+8$.

\ni To see this for $n_\beta<2n_\alpha+8$, we can go to the flop frame $\nu=1$ for $\sp(n_\alpha)_\theta^{(1)}$ without changing the above gluing rules. Then we can implement the map that formed the isomorphism between (\ref{spt0}) and (\ref{spt1}). Since the above gluing rules do not interact with blowups living on $S_{n_\alpha,\alpha}$, the map trivially extends to an isomorphism of the combined geometry associated to (\ref{spsu}). For $n_\beta=2n_\alpha+8$, we cannot reach $\nu>0$ frame without changing the above gluing rules. Thus the map implementing isomorphism between (\ref{spt0}) and (\ref{spt1}) does not extend to an isomorphism of the combined geometry associated to (\ref{spsu}).

\noindent\ubf{Gluing rules for \raisebox{-.25\height}{\begin{tikzpicture}
\node (w1) at (-0.5,0.9) {$\sp(n_\alpha)_\theta^{(1)}$};
\begin{scope}[shift={(3,0)}]
\node (w2) at (-0.5,0.9) {$\so(2n_\beta)^{(1)}$};
\end{scope}
\draw (w1)--(w2);
\end{tikzpicture}}}:
Here we allow $2n_\beta=\wh{12}$. We can take any geometry with $0\le\nu\le 2n_\alpha+8-n_\beta$ for $\sp(n_\alpha)_\theta^{(1)}$, and any geometry with $0\le\nu\le 2n_\beta-4-\Omega^{\beta\beta}-n_\alpha$ for $\so(2n_\beta)^{(1)}$. The gluing rules below work for both values of $\theta$. In the future, if the value of $\theta$ is unspecified, then the gluing rules work for both the values. In our present case, the gluing rules are:
\bit
\item $f-x_1-x_2$ in $S_{0,\alpha}$ is glued to $f$ in $S_{0,\beta}$.
\item $x_i-x_{i+1}$ in $S_{0,\alpha}$ is glued to $f$ in $S_{i,\beta}$ for $i=1,\cdots,n_\beta-1$.
\item $x_{n_{\beta-1}},x_{n_\beta}$ in $S_{0,\alpha}$ are glued to $f-x_1,y_1$ in $S_{n_\beta,\beta}$.
\item $x_{i}-x_{i+1},y_{i+1}-y_{i}$ in $S_{n_\beta,\beta}$ are glued to $f,f$ in $S_{i,\alpha}$ for $i=1,\cdots,n_\alpha-1$.
\item $x_{n_\alpha}-y_{n_\alpha}$ in $S_{n_\beta,\beta}$ is glued to $f$ in $S_{n_\alpha,\alpha}$.
\eit
To show that the theta angle is irrelevant for $n_\beta<2n_\alpha+8$, we first notice that we can go to the flop frame $\nu=1$ for $\sp(n_\alpha)^{(1)}_\theta$ without changing the above gluing rules. Then the isomorphism between (\ref{spt0}) and (\ref{spt1}) extends to an isomorphism of the combined geometry for
\be\label{spsoe}
\begin{tikzpicture}
\node (w1) at (-0.5,0.9) {$\sp(n_\alpha)_\theta^{(1)}$};
\begin{scope}[shift={(3,0)}]
\node (w2) at (-0.5,0.9) {$\so(2n_\beta)^{(1)}$};
\end{scope}
\draw (w1)--(w2);
\end{tikzpicture}
\ee
For $n_\beta=2n_\alpha+8$, the above argument does not work since going to $\nu=1$ frame changes the gluing rules. However, it turns out that the combined geometries for different $\theta$ are flop equivalent up to an outer automorphism of $\so(2n_\beta)$. To see this, notice that the combined geometry for (\ref{spsoe}) is flop equivalent to the following geometry. We pick the frame $\nu=2n_\alpha+8$ for $\sp(n_\alpha)^{(1)}_\theta$ and $\nu=2n_\beta-8$ for $\so(2n_\beta)^{(1)}$ with the gluing rules being:
\bit
\item $f-x_1-x_2$ in $S_{n_\alpha,\alpha}$ is glued to $f$ in $S_{n_\beta,\beta}$.
\item $x_i-x_{i+1}$ in $S_{n_\alpha,\alpha}$ is glued to $f$ in $S_{n_\beta-i,\beta}$ for $i=1,\cdots,n_\beta-1$.
\item $x_{n_{\beta-1}},x_{n_\beta}$ in $S_{n_\alpha,\alpha}$ are glued to $f-x_1,y_1$ in $S_{0,\beta}$.
\item $x_{i}-x_{i+1},y_{i+1}-y_{i}$ in $S_{0,\beta}$ are glued to $f,f$ in $S_{n_\alpha-i,\alpha}$ for $i=1,\cdots,n_\alpha-1$.
\item $x_{n_\alpha}-y_{n_\alpha}$ in $S_{0,\beta}$ is glued to $f$ in $S_{0,\alpha}$.
\eit
Now it is clear that exchanging $f-x_1$ and $x_1$ interchanges $S_{n_\beta,\beta}$ and $S_{n_\beta-1,\beta}$. Thus the choice of theta angle for $\sp(n_\alpha)^{(1)}$ is correlated to the choice of an outer automorphism frame of $\so(2n_\beta)^{(1)}$ for $n_\beta=2n_\alpha+8$.

The gluing rules for a configuration having multiple edges are simply obtained by combining the gluing rules mentioned above. We have to just make sure that we never use the same blowup twice. For example, consider the configuration
\be\label{suspso}
\begin{tikzpicture}
\node (w1) at (-0.5,0.9) {$\sp(n_\alpha)_\theta^{(1)}$};
\begin{scope}[shift={(3,0)}]
\node (w2) at (-0.5,0.9) {$\so(2n_\beta)^{(1)}$};
\end{scope}
\begin{scope}[shift={(-3,0)}]
\node (w3) at (-0.5,0.9) {$\su(n_\gamma)^{(1)}$};
\end{scope}
\draw (w1)--(w2);
\draw  (w3) edge (w1);
\end{tikzpicture}
\ee
Then we can use any geometry with $0\le\nu\le 2n_\alpha+8-n_\beta-n_\gamma$ for $\sp(n_\alpha)_\theta^{(1)}$, any geometry with $0\le\nu\le 2n_\beta-4-\Omega^{\beta\beta}-n_\alpha$ for $\so(2n_\beta)^{(1)}$, and any geometry with $0\le\nu\le 2n_\gamma-2n_\alpha$ for $\su(n_\gamma)^{(1)}$. The gluing rules for the sub-configuration
\be
\begin{tikzpicture}
\node (w1) at (-0.5,0.9) {$\sp(n_\alpha)_\theta^{(1)}$};
\begin{scope}[shift={(3,0)}]
\node (w2) at (-0.5,0.9) {$\so(2n_\beta)^{(1)}$};
\end{scope}
\draw (w1)--(w2);
\end{tikzpicture}
\ee
are the same as the ones listed above, while the gluing rules for the sub-configuration
\be
\begin{tikzpicture}
\node (w1) at (-0.5,0.9) {$\sp(n_\alpha)_\theta^{(1)}$};
\begin{scope}[shift={(3,0)}]
\node (w2) at (-0.5,0.9) {$\su(n_\gamma)^{(1)}$};
\end{scope}
\draw (w1)--(w2);
\end{tikzpicture}
\ee
are as follows:
\bit
\item $f-x_{n_\beta+1},x_{n_\beta+n_\gamma}$ in $S_{0,\alpha}$ are glued to $f-x_1,x_{2n_\alpha}$ in $S_{0,\gamma}$.
\item $x_{n_\beta+i}-x_{n_\beta+i+1}$ in $S_{0,\alpha}$ is glued to $f$ in $S_{i,\gamma}$ for $i=1,\cdots,n_\gamma-1$.
\item $x_{i}-x_{i+1},x_{2n_\alpha-i}-x_{2n_\alpha-i+1}$ in $S_{0,\gamma}$ are glued to $f,f$ in $S_{i,\alpha}$ for $i=1,\cdots,n_\alpha-1$.
\item $x_{n_\alpha}-x_{n_\alpha+1}$ in $S_{0,\gamma}$ is glued to $f$ in $S_{n_\alpha,\alpha}$.
\eit
In a similar way, by choosing mutually exclusive sets of blowups, we can combine the gluing rules to obtain geometries for graphs with multiple algebras and edges between them. Sometimes some of the blowups are allowed to appear in more than one gluing rules. In such cases, we will explicitly mention such blowups and the configurations in which they can appear in multiple gluing rules.

\noindent\ubf{Gluing rules for \raisebox{-.25\height}{\begin{tikzpicture}
\node (w1) at (-0.5,0.9) {$\sp(n_\alpha)_\theta^{(1)}$};
\begin{scope}[shift={(3.3,0)}]
\node (w2) at (-0.5,0.9) {$\so(2n_\beta+1)^{(1)}$};
\end{scope}
\draw (w1)--(w2);
\end{tikzpicture}}}:
We can take any geometry with $1\le\nu\le 2n_\alpha+8-n_\beta$ for $\sp(n_\alpha)_\theta^{(1)}$, and any geometry with $0\le\nu\le 2n_\beta-3-\Omega^{\beta\beta}-n_\alpha$ for $\so(2n_\beta+1)^{(1)}$. The gluing rules are:
\bit
\item $f-x_1-x_2$ in $S_{0,\alpha}$ is glued to $f$ in $S_{0,\beta}$.
\item $x_i-x_{i+1}$ in $S_{0,\alpha}$ is glued to $f$ in $S_{i,\beta}$ for $i=1,\cdots,n_\beta-1$.
\item $x_{n_{\beta}},x_{n_\beta}$ in $S_{0,\alpha}$ are glued to $x_1,y_1$ in $S_{n_\beta,\beta}$.
\item $x_{i+1}-x_{i},y_{i+1}-y_{i}$ in $S_{n_\beta,\beta}$ are glued to $f,f$ in $S_{i,\alpha}$ for $i=1,\cdots,n_\alpha-1$.
\item $f-x_{n_\alpha},f-y_{n_\alpha}$ in $S_{n_\beta,\beta}$ are glued to $f-x_1,x_1$ in $S_{n_\alpha,\alpha}$.
\eit
To show that the theta angle is irrelevant, use the map that exchanges $x_1$ and $f-x_1$ in $S_{n_\alpha,\alpha}$. If this is accompanied by $x_i\lra y_i$ in $S_{n_\beta,\beta}$, then the gluing rules remain unchanged.

Consider a configuration of the form
\be\label{soospsoo}
\begin{tikzpicture}
\node (w1) at (-0.5,0.9) {$\sp(n_\alpha)_\theta^{(1)}$};
\begin{scope}[shift={(3,0)}]
\node (w2) at (-0.5,0.9) {$\so(2n_\beta+1)^{(1)}$};
\end{scope}
\begin{scope}[shift={(-3,0)}]
\node (w3) at (-0.5,0.9) {$\so(2n_\gamma+1)^{(1)}$};
\end{scope}
\draw (w1)--(w2);
\draw  (w3) edge (w1);
\end{tikzpicture}
\ee
We wish to emphasize that we use the same blowup $x_1$ on $S_{n_\alpha,\alpha}$ in the gluing rules associated to both
\be\label{spsoo}
\begin{tikzpicture}
\node (w1) at (-0.5,0.9) {$\sp(n_\alpha)_\theta^{(1)}$};
\begin{scope}[shift={(3,0)}]
\node (w2) at (-0.5,0.9) {$\so(2n_\beta+1)^{(1)}$};
\end{scope}
\draw (w1)--(w2);
\end{tikzpicture}
\ee
and
\be\label{soosp}
\begin{tikzpicture}
\node (w1) at (-0.5,0.9) {$\sp(n_\alpha)_\theta^{(1)}$};
\begin{scope}[shift={(3,0)}]
\node (w2) at (-0.5,0.9) {$\so(2n_\gamma+1)^{(1)}$};
\end{scope}
\draw (w1)--(w2);
\end{tikzpicture}
\ee
More explicitly, to obtain gluing rules for (\ref{soospsoo}), we can take any geometry with $1\le\nu\le2n_\alpha+8-n_\beta-n_\gamma$ for $\sp(n_\alpha)_\theta^{(1)}$, any geometry with $0\le\nu\le 2n_\beta-3-\Omega^{\beta\beta}-n_\alpha$ for $\so(2n_\beta+1)^{(1)}$, and any geometry with $0\le\nu\le 2n_\gamma-3-\Omega^{\gamma\gamma}-n_\alpha$ for $\so(2n_\gamma+1)^{(1)}$. The gluing rules for (\ref{spsoo}) are those listed above, and the gluing rules for (\ref{soosp}) are:
\bit
\item $f-x_{n_\beta+1}-x_{n_\beta+2}$ in $S_{0,\alpha}$ is glued to $f$ in $S_{0,\gamma}$.
\item $x_{n_\beta+i}-x_{n_\beta+i+1}$ in $S_{0,\alpha}$ is glued to $f$ in $S_{i,\gamma}$ for $i=1,\cdots,n_\gamma-1$.
\item $x_{n_{\beta+\gamma}},x_{n_{\beta+\gamma}}$ in $S_{0,\alpha}$ are glued to $x_1,y_1$ in $S_{n_\gamma,\gamma}$.
\item $x_{i+1}-x_{i},y_{i+1}-y_{i}$ in $S_{n_\gamma,\gamma}$ are glued to $f,f$ in $S_{i,\alpha}$ for $i=1,\cdots,n_\alpha-1$.
\item $f-x_{n_\alpha},f-y_{n_\alpha}$ in $S_{n_\gamma,\gamma}$ are glued to $f-x_1,x_1$ in $S_{n_\alpha,\alpha}$.
\eit
with the $x_1$ in $S_{n_\alpha,\alpha}$ being the same blowup as used in the gluing rules above for (\ref{spsoo}).

However, if we have a third neighbor $\so(2n_\delta+1)^{(1)}$ of $\sp(n_\alpha)_\theta^{(1)}$, then we must use a second blowup $x_2$ on $S_{n_\alpha,\alpha}$. As a consequence, we must choose a geometry with $2\le\nu\le 2n_\alpha+8-n_\beta+n_\gamma+n_\delta$ for $\sp(n_\alpha)_\theta^{(1)}$ to obtain the combined geometry for the configuration
\be
\begin{tikzpicture}
\node (w1) at (-0.5,0.9) {$\sp(n_\alpha)_\theta^{(1)}$};
\begin{scope}[shift={(3,0)}]
\node (w2) at (-0.5,0.9) {$\so(2n_\beta+1)^{(1)}$};
\end{scope}
\begin{scope}[shift={(-3,0)}]
\node (w3) at (-0.5,0.9) {$\so(2n_\gamma+1)^{(1)}$};
\end{scope}
\begin{scope}[shift={(0,1.5)}]
\node (w4) at (-0.5,0.9) {$\so(2n_\delta+1)^{(1)}$};
\end{scope}
\draw (w1)--(w2);
\draw  (w3) edge (w1);
\draw  (w4) edge (w1);
\end{tikzpicture}
\ee

\noindent\ubf{Gluing rules for \raisebox{-.25\height}{\begin{tikzpicture}
\node (w1) at (-0.5,0.9) {$\sp(n_\alpha)_\theta^{(1)}$};
\begin{scope}[shift={(3,0)}]
\node (w2) at (-0.5,0.9) {$\so(8)^{(1)}$};
\end{scope}
\draw [dashed] (w1)--(w2);
\end{tikzpicture}}}:
We can take any geometry with $0\le\nu\le 2n_\alpha+4$ for $\sp(n_\alpha)_\theta^{(1)}$, and any geometry with $0\le\nu\le 4-\Omega^{\beta\beta}-n_\alpha$ for $\so(8)^{(1)}$. The gluing rules are:
\bit
\item $f-x_1-x_2$ in $S_{0,\alpha}$ is glued to $f$ in $S_{0,\beta}$.
\item $x_1-x_{2}$ in $S_{0,\alpha}$ is glued to $f$ in $S_{3,\beta}$.
\item $x_2-x_{3}$ in $S_{0,\alpha}$ is glued to $f$ in $S_{2,\beta}$.
\item $x_3-x_{4}$ in $S_{0,\alpha}$ is glued to $f$ in $S_{1,\beta}$.
\item $x_{3},x_{4}$ in $S_{0,\alpha}$ are glued to $f-z_1,w_1$ in $S_{4,\beta}$.
\item $z_{i}-z_{i+1},w_{i+1}-w_{i}$ in $S_{4,\beta}$ are glued to $f,f$ in $S_{i,\alpha}$ for $i=1,\cdots,n_\alpha-1$.
\item $z_{n_\alpha}-w_{n_\alpha}$ in $S_{4,\beta}$ is glued to $f$ in $S_{n_\alpha,\alpha}$.
\eit
The theta angle is irrelevant as can be seen in the $\nu=1$ frame of $\sp(n_\alpha)_\theta^{(1)}$.

\noindent\ubf{Gluing rules for \raisebox{-.25\height}{\begin{tikzpicture}
\node (w1) at (-0.5,0.9) {$\sp(n_\alpha)_\theta^{(1)}$};
\begin{scope}[shift={(3,0)}]
\node (w2) at (-0.5,0.9) {$\so(7)^{(1)}$};
\end{scope}
\draw [dashed] (w1)--(w2);
\end{tikzpicture}}}:
We can take any geometry with $0\le\nu\le 2n_\alpha+4$ for $\sp(n_\alpha)_\theta^{(1)}$, and any geometry with $0\le\nu\le 8-2\Omega^{\beta\beta}-n_\alpha$ for $\so(7)^{(1)}$. The gluing rules are:
\bit
\item $f-x_1-x_2$ in $S_{0,\alpha}$ is glued to $f$ in $S_{0,\beta}$.
\item $x_2-x_{3}$ in $S_{0,\alpha}$ is glued to $f$ in $S_{2,\beta}$.
\item $x_1-x_2,x_3-x_{4}$ in $S_{0,\alpha}$ is glued to $f$ in $S_{3,\beta}$.
\item $x_{3},x_{4}$ in $S_{0,\alpha}$ are glued to $f-x_1,y_1$ in $S_{1,\beta}$.
\item $x_{i}-x_{i+1},y_{i+1}-y_{i}$ in $S_{1,\beta}$ are glued to $f,f$ in $S_{i,\alpha}$ for $i=1,\cdots,n_\alpha-1$.
\item $x_{n_\alpha}-y_{n_\alpha}$ in $S_{1,\beta}$ is glued to $f$ in $S_{n_\alpha,\alpha}$.
\eit
The theta angle is irrelevant as in the last case.

\noindent\ubf{Gluing rules for \raisebox{-.25\height}{\begin{tikzpicture}
\node (w1) at (-0.5,0.9) {$\sp(n_\alpha)_\theta^{(1)}$};
\begin{scope}[shift={(3,0)}]
\node (w2) at (-0.5,0.9) {$\fg_2^{(1)}$};
\end{scope}
\draw (w1)--(w2);
\end{tikzpicture}}}:
We can take any geometry with $1\le\nu\le 2n_\alpha+5$ for $\sp(n_\alpha)_\theta^{(1)}$, and any geometry with $0\le\nu\le 10-3\Omega^{\beta\beta}-n_\alpha$ for $\fg_2^{(1)}$. The gluing rules are:
\bit
\item $f-x_1-x_2$ in $S_{0,\alpha}$ is glued to $f$ in $S_{0,\beta}$.
\item $x_2-x_{3}$ in $S_{0,\alpha}$ is glued to $f$ in $S_{2,\beta}$.
\item $x_1-x_2,x_3,x_{3}$ in $S_{0,\alpha}$ are glued to $f,x_1,y_1$ in $S_{1,\beta}$.
\item $x_{i+1}-x_{i},y_{i+1}-y_{i}$ in $S_{1,\beta}$ are glued to $f,f$ in $S_{i,\alpha}$ for $i=1,\cdots,n_\alpha-1$.
\item $f-x_{n_\alpha},f-y_{n_\alpha}$ in $S_{1,\beta}$ are glued to $f-x_1,x_1$ in $S_{n_\alpha,\alpha}$.
\eit
The theta angle is irrelevant.

The blowup $x_1$ in $S_{n_\alpha,\alpha}$ can be repeated once more if there is another $\fg_2^{(1)}$ neighbor or an $\so(2n_\gamma+1)^{(1)}$ neighbor of $\sp(n_\alpha)_\theta^{(1)}$. That is, when we consider configurations of the form
\be
\begin{tikzpicture}
\node (w1) at (-0.5,0.9) {$\sp(n_\alpha)_\theta^{(1)}$};
\begin{scope}[shift={(3,0)}]
\node (w2) at (-0.5,0.9) {$\fg_2^{(1)}$};
\end{scope}
\begin{scope}[shift={(-3,0)}]
\node (w3) at (-0.5,0.9) {$\fg_2^{(1)}$};
\end{scope}
\draw (w1)--(w2);
\draw  (w3) -- (w1);
\end{tikzpicture}
\ee
or of the form
\be
\begin{tikzpicture}
\node (w1) at (-0.5,0.9) {$\sp(n_\alpha)_\theta^{(1)}$};
\begin{scope}[shift={(3,0)}]
\node (w2) at (-0.5,0.9) {$\fg_2^{(1)}$};
\end{scope}
\begin{scope}[shift={(-3.3,0)}]
\node (w3) at (-0.5,0.9) {$\so(2n_\gamma+1)^{(1)}$};
\end{scope}
\draw (w1)--(w2);
\draw  (w3) -- (w1);
\end{tikzpicture}
\ee
As before, if there is a third $\fg_2$ or $\so(2n_\delta+1)^{(1)}$ neighbor of $\sp(n_\alpha)_\theta^{(1)}$, then we must use another blowup $x_2$ on $S_{n_\alpha,\alpha}$ for the gluing rules corresponding to this neighbor.

\noindent\ubf{Gluing rules for \raisebox{-.25\height}{\begin{tikzpicture}
\node (w1) at (-0.5,0.9) {$\su(n_\alpha)^{(1)}$};
\begin{scope}[shift={(3,0)}]
\node (w2) at (-0.5,0.9) {$\su(n_\beta)^{(1)}$};
\end{scope}
\draw (w1)--(w2);
\end{tikzpicture}}}:
Here we allow $n_\alpha=\wh{n}_\alpha$ and $n_\alpha=\tilde 6$. We can take any geometry with $0\le\nu\le 2n_\alpha-n_\beta$ for $\su(n_\alpha)^{(1)}$, and any geometry with $0\le\nu\le 2n_\beta-n_\alpha$ for $\su(n_\beta)^{(1)}$. The gluing rules are:
\bit
\item $f-x_1,x_{n_\beta}$ in $S_{0,\alpha}$ are glued to $f-x_1,x_{n_\alpha}$ in $S_{0,\beta}$.
\item $x_i-x_{i+1}$ in $S_{0,\alpha}$ is glued to $f$ in $S_{i,\beta}$ for $i=1,\cdots,n_\beta-1$.
\item $x_i-x_{i+1}$ in $S_{0,\beta}$ is glued to $f$ in $S_{i,\alpha}$ for $i=1,\cdots,n_\alpha-1$.
\eit

\noindent\ubf{Gluing rules for \raisebox{-.25\height}{\begin{tikzpicture}
\node (w1) at (-0.5,0.9) {$\su(n_\alpha)^{(1)}$};
\begin{scope}[shift={(3.1,0)}]
\node (w2) at (-0.5,0.9) {$\so(2n_\beta)^{(1)}$};
\end{scope}
\node (w3) at (1,0.9) {\scriptsize{2}};
\draw (w1)--(w3);
\draw (w3)--(w2);
\end{tikzpicture}}}:
We can take any geometry with $n_\beta\le\nu\le 2n_\alpha-n_\beta$ for $\su(n_\alpha)^{(1)}$, and any geometry with $0\le\nu\le 2n_\beta-8-n_\alpha$ for $\so(2n_\beta)^{(1)}$. The gluing rules are:
\bit
\item $f-x_1-x_2,f-y_{1}-y_{2}$ in $S_{0,\alpha}$ are glued to $f,f$ in $S_{0,\beta}$.
\item $x_{i}-x_{i+1},y_{i}-y_{i+1}$ in $S_{0,\alpha}$ are glued to $f,f$ in $S_{i,\beta}$ for $i=1,\cdots,n_\beta-1$.
\item $x_{n_\beta-1},x_{n_\beta},y_{n_\beta-1},y_{n_\beta}$ in $S_{0,\alpha}$ are glued to $f-x_1,y_1,f-y_{n_\alpha},x_{n_\alpha}$ in $S_{n_\beta,\beta}$.
\item $x_{i}-x_{i+1},y_{i+1}-y_i$ in $S_{n_\beta,\beta}$ are glued to $f,f$ in $S_{i,\alpha}$ for $i=1,\cdots,n_\alpha-1$.
\eit

\noindent\ubf{Gluing rules for \raisebox{-.25\height}{\begin{tikzpicture}
\node (w1) at (-0.5,0.9) {$\su(n_\alpha)^{(1)}$};
\begin{scope}[shift={(3.4,0)}]
\node (w2) at (-0.5,0.9) {$\so(2n_\beta+1)^{(1)}$};
\end{scope}
\node (w3) at (1,0.9) {\scriptsize{2}};
\draw (w1)--(w3);
\draw (w3)--(w2);
\end{tikzpicture}}}:
We can take any geometry with $n_\beta\le\nu\le 2n_\alpha-n_\beta-1$ for $\su(n_\alpha)^{(1)}$, and any geometry with $0\le\nu\le 2n_\beta-7-n_\alpha$ for $\so(2n_\beta)^{(1)}$. The (non-geometric) gluing rules are:
\bit
\item $f-x_1-x_2,f-y_{1}-y_{2}$ in $S_{0,\alpha}$ are glued to $f,f$ in $S_{0,\beta}$.
\item $x_{i}-x_{i+1},y_{i}-y_{i+1}$ in $S_{0,\alpha}$ are glued to $f,f$ in $S_{i,\beta}$ for $i=1,\cdots,n_\beta-1$.
\item $x_{n_\beta}-x_{n_\beta+1},x_{n_\beta}-x_{n_\beta+1},y_{n_\beta},y_{n_\beta},x_{n_\beta+1},x_{n_\beta+1}$ in $S_{0,\alpha}$ are glued to $f,f,x_1,y_1,f-x_{n_\alpha},f-y_{n_\alpha}$ in $S_{n_\beta,\beta}$.
\item $x_{i+1}-x_{i},y_{i+1}-y_i$ in $S_{n_\beta,\beta}$ are glued to $f,f$ in $S_{i,\alpha}$ for $i=1,\cdots,n_\alpha-1$.
\eit

\noindent\ubf{Gluing rules for \raisebox{-.25\height}{\begin{tikzpicture}
\node (w1) at (-0.5,0.9) {$\su(2)^{(1)}$};
\begin{scope}[shift={(3,0)}]
\node (w2) at (-0.5,0.9) {$\so(7)^{(1)}$};
\end{scope}
\draw[dashed]  (w1)--(w2);
\end{tikzpicture}}}:
We must take the geometry with $\nu=0$ for $\su(2)^{(1)}$, and we can take any geometry with $0\le\nu\le 7-2\Omega^{\beta\beta}$ for $\so(7)^{(1)}$. The gluing rules are:
\bit
\item $f-x_1-x_2$ in $S_{0,\alpha}$ is glued to $f$ in $S_{0,\beta}$.
\item $x_2-x_{3}$ in $S_{0,\alpha}$ is glued to $f$ in $S_{2,\beta}$.
\item $x_1-x_2,x_3-x_{4}$ in $S_{0,\alpha}$ is glued to $f$ in $S_{3,\beta}$.
\item $x_{3},x_{4}$ in $S_{0,\alpha}$ are glued to $f-x_1,y_1$ in $S_{1,\beta}$.
\item $x_{1}-y_{1}$ in $S_{1,\beta}$ is glued to $f$ in $S_{1,\alpha}$.
\eit

\noindent\ubf{Gluing rules for \raisebox{-.25\height}{\begin{tikzpicture}
\node (w1) at (-0.5,0.9) {$\su(2)^{(1)}$};
\begin{scope}[shift={(3,0)}]
\node (w2) at (-0.5,0.9) {$\fg_2^{(1)}$};
\end{scope}
\draw  (w1)--(w2);
\end{tikzpicture}}}:
We must take the geometry with $\nu=1$ for $\su(2)^{(1)}$, and any geometry with $0\le\nu\le 9-3\Omega^{\beta\beta}$ for $\fg_2^{(1)}$. The gluing rules are:
\bit
\item $f-x_1-x_2$ in $S_{0,\alpha}$ is glued to $f$ in $S_{0,\beta}$.
\item $x_2-x_{3}$ in $S_{0,\alpha}$ is glued to $f$ in $S_{2,\beta}$.
\item $x_1-x_2,x_3,x_{3}$ in $S_{0,\alpha}$ are glued to $f,x_1,y_1$ in $S_{1,\beta}$.
\item $f-x_{1},f-y_{1}$ in $S_{1,\beta}$ are glued to $f-x_1,x_1$ in $S_{1,\alpha}$.
\eit

There is another possibility appearing in the twisted case that involves an undirected edge between two untwisted algebras. This possibility is 
\be
\begin{tikzpicture}
\node (v1) at (-0.5,0.4) {2};
\node at (-0.45,0.9) {$\su(n_\alpha)^{(1)}$};
\begin{scope}[shift={(2,0)}]
\node (v2) at (-0.5,0.4) {$2$};
\node at (-0.45,0.9) {$\su(n_\beta)^{(1)}$};
\end{scope}
\draw  (v1) -- (v2);
\draw (v1) .. controls (-1.5,-0.5) and (0.5,-0.5) .. (v1);
\end{tikzpicture}
\ee and it is displayed in Table \ref{FTR2}. The gluing rules for this case are the same as the gluing rules for 
\be
\begin{tikzpicture}
\node (w1) at (-0.5,0.9) {$\su(n_\alpha)^{(1)}$};
\begin{scope}[shift={(3,0)}]
\node (w2) at (-0.5,0.9) {$\su(n_\beta)^{(1)}$};
\end{scope}
\draw (w1)--(w2);
\end{tikzpicture}
\ee
presented above.

\subsubsection{Undirected edges between a twisted algebra and an untwisted algebra}\label{hell}
Now let us provide gluing rules for those cases in Table \ref{FTR2} in which both the nodes have non-trivial gauge algebras associated to them, such that at least one of the gauge algebras is twisted.

\noindent\ubf{Gluing rules for \raisebox{-.25\height}{\begin{tikzpicture}
\node (w1) at (-0.5,0.9) {$\sp(n_\alpha)_\theta^{(1)}$};
\begin{scope}[shift={(3,0)}]
\node (w2) at (-0.5,0.9) {$\so(2n_\beta)^{(2)}$};
\end{scope}
\draw (w1)--(w2);
\end{tikzpicture}}}:
Here we allow $2n_\beta=\wh{12}$. We can take any geometry with $1\le\nu\le 2n_\alpha+8-n_\beta$ for $\sp(n_\alpha)_\theta^{(1)}$, and any geometry with $0\le\nu\le 2n_\beta-4-\Omega^{\beta\beta}-n_\alpha$ for $\so(2n_\beta)^{(2)}$. The gluing rules are:
\bit
\item $f-x_1-x_2,x_1-x_2$ in $S_{0,\alpha}$ are glued to $f,f$ in $S_{0,\beta}$.
\item $x_i-x_{i+1}$ in $S_{0,\alpha}$ is glued to $f$ in $S_{i-1,\beta}$ for $i=2,\cdots,n_\beta-1$.
\item $x_{n_\beta},x_{n_\beta}$ in $S_{0,\alpha}$ are glued to $x_1,y_1$ in $S_{n_\beta-1,\beta}$.
\item $x_{i+1}-x_{i},y_{i+1}-y_i$ in $S_{n_\beta-1,\beta}$ are glued to $f,f$ in $S_{i,\alpha}$ for $i=1,\cdots,n_\alpha-1$.
\item $f-x_{n_\alpha},f-y_{n_\alpha}$ in $S_{n_\beta-1,\beta}$ are glued to $f-x_1,x_1$ in $S_{n_\alpha,\alpha}$.
\eit
The theta angle can be seen to be irrelevant by using the blowup $x_1$ on $S_{n_\alpha,\alpha}$. 

\ni The blowup $x_1$ in $S_{n_\alpha,\alpha}$ can be used in gluing rules corresponding to one more neighbor of the form $\so(2n_\gamma+1)^{(1)}$, $\fg_2^{(1)}$ or $\so(2n_\gamma)^{(2)}$ of $\sp(n_\alpha)_\theta^{(1)}$.

The fact that $n_\beta=2n_\alpha+8$ is not allowed manifests in the above gluing rules. The total number of blowups carried by $S_{0,\alpha}$ is at max $2n_\alpha+7$ but the gluing rules require the presence of $2n_\alpha+8$ blowups on $S_{0,\alpha}$. See the discussion around (\ref{bound1}) for an explanation of this restriction.

\noindent\ubf{Gluing rules for \raisebox{-.25\height}{\begin{tikzpicture}
\node (w1) at (-0.5,0.9) {$\su(n_\alpha)^{(1)}$};
\begin{scope}[shift={(3.1,0)}]
\node (w2) at (-0.5,0.9) {$\so(2n_\beta)^{(2)}$};
\end{scope}
\node (w3) at (1,0.9) {\scriptsize{2}};
\draw (w1)--(w3);
\draw (w3)--(w2);
\end{tikzpicture}}}:
We can take any geometry with $n_\beta-1\le\nu\le 2n_\alpha-n_\beta-1$ for $\su(n_\alpha)^{(1)}$, and any geometry with $0\le\nu\le 2n_\beta-8-n_\alpha$ for $\so(2n_\beta)^{(1)}$. The (non-geometric) gluing rules are:
\bit
\item $f-x_1-x_2,x_1-x_2,f-x_{1}-y_{1},x_1-y_1$ in $S_{0,\alpha}$ are glued to $f,f,f,f$ in $S_{0,\beta}$.
\item $x_{i+1}-x_{i+2},y_{i}-y_{i+1}$ in $S_{0,\alpha}$ are glued to $f,f$ in $S_{i,\beta}$ for $i=1,\cdots,n_\beta-2$.
\item $x_{n_\beta}-x_{n_\beta+1},x_{n_\beta}-x_{n_\beta+1},y_{n_\beta},y_{n_\beta},x_{n_\beta+1},x_{n_\beta+1}$ in $S_{0,\alpha}$ are glued to $f,f,x_1,y_1,f-x_{n_\alpha},f-y_{n_\alpha}$ in $S_{n_\beta,\beta}$.
\item $x_{i}-x_{i+1},y_{i+1}-y_i$ in $S_{n_\beta,\beta}$ are glued to $f,f$ in $S_{i,\alpha}$ for $i=1,\cdots,n_\alpha-1$.
\eit

\subsubsection{Directed edges}
Now we move onto gluing rules for edges listed in Table \ref{BTR2}.

\noindent\ubf{Gluing rules for \raisebox{-.25\height}{\begin{tikzpicture}
\node (w1) at (-0.5,0.9) {$\sp(n_\alpha)^{(1)}$};
\begin{scope}[shift={(3.1,0)}]
\node (w2) at (-0.5,0.9) {$\so(2n_\beta)^{(1)}$};
\end{scope}
\node (w3) at (1,0.9) {\scriptsize{2}};
\draw (w1)--(w3);
\draw[->] (w3)--(w2);
\end{tikzpicture}}}:
We can take any geometry with $0\le\nu\le 2n_\alpha+8-2n_\beta$ for $\sp(n_\alpha)^{(1)}$, and any geometry with $0\le\nu\le 2n_\beta-4-\Omega^{\beta\beta}-n_\alpha$ for $\so(2n_\beta)^{(1)}$. The gluing rules are:
\bit
\item $x_{n_\beta-1}-x_{n_\beta+1},x_{n_\beta}-x_{n_\beta+2}$ in $S_{0,\alpha}$ are glued to $f,f$ in $S_{0,\beta}$.
\item $x_{n_\beta-i}-x_{n_\beta-i+1},x_{n_\beta+i}-x_{n_\beta+i+1}$ in $S_{0,\alpha}$ are glued to $f,f$ in $S_{i,\beta}$ for $i=1,\cdots,n_\beta-1$.
\item $f-x_1-x_2$ in $S_{0,\alpha}$ is glued to $f$ in $S_{n_\beta,\beta}$. $x_{2n_\beta-1}$ in $S_{0,\alpha}$ is glued to $f-x_1$ in $S_{n_\beta,\beta}$. $x_{2n_\beta}$ in $S_{0,\alpha}$ is glued to $y_1$ in $S_{n_\beta,\beta}$.
\item $x_{i}-x_{i+1},y_{i+1}-y_i$ in $S_{n_\beta,\beta}$ are glued to $f,f$ in $S_{i,\alpha}$ for $i=1,\cdots,n_\alpha-1$.
\item $x_{n_\alpha}-y_{n_\alpha}$ in $S_{n_\beta,\beta}$ is glued to $f$ in $S_{n_\alpha,\alpha}$.
\eit
From this case onward, we are dropping the subscript $\theta$ on $\sp(n)^{(1)}$ whenever theta angle is not physically relevant. In such cases, the gluing rules will work uniformly for both values of $\theta$ and using arguments used earlier in the paper, the reader can easily check that the combined geometries descending from different values of theta angle are indeed isomorphic.

\noindent\ubf{Gluing rules for \raisebox{-.25\height}{\begin{tikzpicture}
\node (w1) at (-0.5,0.9) {$\sp(n_\alpha)^{(1)}$};
\begin{scope}[shift={(3.4,0)}]
\node (w2) at (-0.5,0.9) {$\so(2n_\beta+1)^{(1)}$};
\end{scope}
\node (w3) at (1,0.9) {\scriptsize{2}};
\draw (w1)--(w3);
\draw[->] (w3)--(w2);
\end{tikzpicture}}}:
We can take any geometry with $1\le\nu\le 2n_\alpha+7-2n_\beta$ for $\sp(n_\alpha)^{(1)}$, and any geometry with $0\le\nu\le 2n_\beta-3-\Omega^{\beta\beta}-n_\alpha$ for $\so(2n_\beta+1)^{(1)}$. The (non-geometric) gluing rules are:
\bit
\item $x_{n_\beta}-x_{n_\beta+2},x_{n_\beta+1}-x_{n_\beta+3}$ in $S_{0,\alpha}$ are glued to $f,f$ in $S_{0,\beta}$.
\item $x_{n_\beta-i+1}-x_{n_\beta-i+2},x_{n_\beta+i+1}-x_{n_\beta+i+2}$ in $S_{0,\alpha}$ are glued to $f,f$ in $S_{i,\beta}$ for $i=1,\cdots,n_\beta-1$.
\item $f-x_1-x_2,x_1-x_2$ in $S_{0,\alpha}$ are glued to $f,f$ in $S_{n_\beta,\beta}$. $x_{2n_\beta+1}$ in $S_{0,\alpha}$ is glued to $x_1$ in $S_{n_\beta,\beta}$. $x_{2n_\beta+1}$ in $S_{0,\alpha}$ is glued to $y_1$ in $S_{n_\beta,\beta}$.
\item $x_{i+1}-x_{i},y_{i+1}-y_i$ in $S_{n_\beta,\beta}$ are glued to $f,f$ in $S_{i,\alpha}$ for $i=1,\cdots,n_\alpha-1$.
\item $f-x_{n_\alpha}$ in $S_{n_\beta,\beta}$ is glued to $x_1$ in $S_{n_\alpha,\alpha}$. $f-y_{n_\alpha}$ in $S_{n_\beta,\beta}$ is glued to $f-x_1$ in $S_{n_\alpha,\alpha}$.
\eit
Notice that the blowup $x_1$ in $S_{n_\alpha,\alpha}$ can be used for gluing $\sp(n_\alpha)^{(1)}$ to one more neighbor, that is in configurations of the following form
\be
\begin{tikzpicture}
\node (w1) at (-0.5,0.9) {$\sp(n_\alpha)^{(1)}$};
\begin{scope}[shift={(3.4,0)}]
\node (w2) at (-0.5,0.9) {$\so(2n_\beta+1)^{(1)}$};
\end{scope}
\begin{scope}[shift={(-3,0)}]
\node (w4) at (-0.5,0.9) {$\so(2n_\gamma+1)^{(1)}$};
\end{scope}
\node (w3) at (1,0.9) {\scriptsize{2}};
\draw (w1)--(w3);
\draw[->] (w3)--(w2);
\draw  (w4) edge (w1);
\end{tikzpicture}
\ee
\be
\begin{tikzpicture}
\node (w1) at (-0.5,0.9) {$\sp(n_\alpha)^{(1)}$};
\begin{scope}[shift={(3.4,0)}]
\node (w2) at (-0.5,0.9) {$\so(2n_\beta+1)^{(1)}$};
\end{scope}
\begin{scope}[shift={(-3,0)}]
\node (w4) at (-0.5,0.9) {$\fg_2^{(1)}$};
\end{scope}
\node (w3) at (1,0.9) {\scriptsize{2}};
\draw (w1)--(w3);
\draw[->] (w3)--(w2);
\draw  (w4) edge (w1);
\end{tikzpicture}
\ee
\be
\begin{tikzpicture}
\node (w1) at (-0.5,0.9) {$\sp(n_\alpha)^{(1)}$};
\begin{scope}[shift={(3.4,0)}]
\node (w2) at (-0.5,0.9) {$\so(2n_\beta+1)^{(1)}$};
\end{scope}
\begin{scope}[shift={(-3,0)}]
\node (w4) at (-0.5,0.9) {$\so(2n_\gamma)^{(2)}$};
\end{scope}
\node (w3) at (1,0.9) {\scriptsize{2}};
\draw (w1)--(w3);
\draw[->] (w3)--(w2);
\draw  (w4) edge (w1);
\end{tikzpicture}
\ee
but cannot be used for gluing $\sp(n_\alpha)^{(1)}$ to two more neighbors.

\noindent\ubf{Gluing rules for \raisebox{-.25\height}{\begin{tikzpicture}
\node (w1) at (-0.5,0.9) {$\sp(n_\alpha)^{(1)}$};
\begin{scope}[shift={(3.1,0)}]
\node (w2) at (-0.5,0.9) {$\so(2n_\beta)^{(2)}$};
\end{scope}
\node (w3) at (1,0.9) {\scriptsize{2}};
\draw (w1)--(w3);
\draw[->] (w3)--(w2);
\end{tikzpicture}}}:
We can take any geometry with $1\le\nu\le 2n_\alpha+7-2n_\beta$ for $\sp(n_\alpha)^{(1)}$, and any geometry with $0\le\nu\le 2n_\beta-8-n_\alpha$ for $\so(2n_\beta)^{(2)}$. The (non-geometric) gluing rules are:
\bit
\item $x_{n_\beta}-x_{n_\beta+2},x_{n_\beta+1}-x_{n_\beta+3},x_{n_\beta}-x_{n_\beta+1},x_{n_\beta+2}-x_{n_\beta+3}$ in $S_{0,\alpha}$ are glued to $f,f,f,f$ in $S_{0,\beta}$.
\item $x_{n_\beta-i}-x_{n_\beta-i+1},x_{n_\beta+i+2}-x_{n_\beta+i+3}$ in $S_{0,\alpha}$ are glued to $f,f$ in $S_{i,\beta}$ for $i=1,\cdots,n_\beta-2$.
\item $f-x_1-x_2,x_1-x_2$ in $S_{0,\alpha}$ are glued to $f,f$ in $S_{n_\beta-1,\beta}$. $x_{2n_\beta+1}$ in $S_{0,\alpha}$ is glued to $x_1$ in $S_{n_\beta-1,\beta}$. $x_{2n_\beta+1}$ in $S_{0,\alpha}$ is glued to $y_1$ in $S_{n_\beta-1,\beta}$.
\item $x_{i+1}-x_{i},y_{i+1}-y_i$ in $S_{n_\beta-1,\beta}$ are glued to $f,f$ in $S_{i,\alpha}$ for $i=1,\cdots,n_\alpha-1$.
\item $f-x_{n_\alpha}$ in $S_{n_\beta-1,\beta}$ is glued to $x_1$ in $S_{n_\alpha,\alpha}$. $f-y_{n_\alpha}$ in $S_{n_\beta,\beta}$ is glued to $f-x_1$ in $S_{n_\alpha,\alpha}$.
\eit
The blowup $x_1$ in $S_{n_\alpha,\alpha}$ can be used to glue $\sp(n_\alpha)^{(1)}$ to exactly one more neighboring node connected to it by an undirected edge. The neighboring node can carry $\so(2n_\gamma+1)^{(1)}$, $\fg_2^{(1)}$ or $\so(2n_\gamma)^{(2)}$.

The fact that $n_\beta=n_\alpha+4$ is not allowed manifests in the above gluing rules. The total number of blowups carried by $S_{0,\alpha}$ is at max $2n_\alpha+7$ but the gluing rules require the presence of $2n_\alpha+9$ blowups on $S_{0,\alpha}$. See the discussion around (\ref{bound2}) for an explanation of this restriction.

\noindent\ubf{Gluing rules for \raisebox{-.25\height}{\begin{tikzpicture}
\node (w1) at (-0.5,0.9) {$\sp(n_\alpha)^{(1)}$};
\begin{scope}[shift={(2.9,0)}]
\node (w2) at (-0.5,0.9) {$\so(7)^{(1)}$};
\end{scope}
\node (w3) at (1,0.9) {\scriptsize{2}};
\draw[dashed] (w1)--(w3);
\draw[dashed,->] (w3)--(w2);
\end{tikzpicture}}}:
We can take any geometry with $0\le\nu\le 2n_\alpha$ for $\sp(n_\alpha)^{(1)}$, and any geometry with $0\le\nu\le 2-n_\alpha$ for $\so(7)^{(1)}$. The gluing rules are:
\bit
\item $x_{3}-x_{5},x_{4}-x_{6}$ in $S_{0,\alpha}$ are glued to $f,f$ in $S_{0,\beta}$.
\item $f-x_{1}-x_{2},x_{7},x_{8}$ in $S_{0,\alpha}$ are glued to $f,f-x_1,y_1$ in $S_{1,\beta}$.
\item $x_{2}-x_{3},x_{6}-x_{7}$ in $S_{0,\alpha}$ are glued to $f,f$ in $S_{2,\beta}$.
\item $x_{1}-x_{2},x_3-x_4,x_5-x_6,x_{7}-x_{8}$ in $S_{0,\alpha}$ are glued to $f,f,f,f$ in $S_{3,\beta}$.
\item $x_{i}-x_{i+1},y_{i+1}-y_i$ in $S_{1,\beta}$ are glued to $f,f$ in $S_{i,\alpha}$ for $i=1,\cdots,n_\alpha-1$.
\item $x_{n_\alpha}-y_{n_\alpha}$ in $S_{1,\beta}$ is glued to $f$ in $S_{n_\alpha,\alpha}$.
\eit

\noindent\ubf{Gluing rules for \raisebox{-.25\height}{\begin{tikzpicture}
\node (w1) at (-0.5,0.9) {$\sp(1)^{(1)}$};
\begin{scope}[shift={(2.7,0)}]
\node (w2) at (-0.5,0.9) {$\fg_2^{(1)}$};
\end{scope}
\node (w3) at (1,0.9) {\scriptsize{2}};
\draw (w1)--(w3);
\draw[->] (w3)--(w2);
\end{tikzpicture}}}:
We can take any geometry with $1\le\nu\le 3$ for $\sp(1)^{(1)}$. The (non-geometric) gluing rules are:
\bit
\item $x_{3}-x_{5},x_{4}-x_{6}$ in $S_{0,\alpha}$ are glued to $f,f$ in $S_{0,\beta}$.
\item $x_{2}-x_{3},x_{6}-x_{7}$ in $S_{0,\alpha}$ are glued to $f,f$ in $S_{2,\beta}$.
\item $f-x_1-x_2,x_{1}-x_{2},x_3-x_4,x_5-x_6,x_{7},x_{7}$ in $S_{0,\alpha}$ are glued to $f,f,f,f,x_1,y_1$ in $S_{1,\beta}$.
\item $f-x_{1},x_1$ in $S_{1,\alpha}$ are glued to $f-x_1,f-y_1$ in $S_{1,\beta}$.
\eit
Notice that the blowup $x_1$ in $S_{1,\alpha}$ can be used in gluing rules corresponding to exactly one more neighbor of $\sp(1)^{(1)}$ carrying algebra  $\so(2n_\gamma+1)^{(1)}$ or $\so(2n_\gamma)^{(2)}$.

\noindent\ubf{Gluing rules for \raisebox{-.25\height}{\begin{tikzpicture}
\node (w1) at (-0.5,0.9) {$\sp(n_\alpha)^{(1)}$};
\begin{scope}[shift={(3.1,0)}]
\node (w2) at (-0.5,0.9) {$\so(2n_\beta)^{(1)}$};
\end{scope}
\node (w3) at (1,0.9) {\scriptsize{3}};
\draw (w1)--(w3);
\draw[->] (w3)--(w2);
\end{tikzpicture}}}:
We can take any geometry with $0\le\nu\le 2n_\alpha+8-3n_\beta$ for $\sp(n_\alpha)^{(1)}$, and any geometry with $0\le\nu\le 2n_\beta-8-n_\alpha$ for $\so(2n_\beta)^{(1)}$. The gluing rules are:
\bit
\item $f-x_1-x_2,x_{2n_\beta-1}-x_{2n_\beta+1},x_{2n_\beta}-x_{2n_\beta+2}$ in $S_{0,\alpha}$ are glued to $f,f,f$ in $S_{0,\beta}$.
\item $x_i-x_{i+1},x_{2n_\beta-i}-x_{2n_\beta-i+1},x_{2n_\beta+i}-x_{2n_\beta+i+1}$ in $S_{0,\alpha}$ are glued to $f,f,f$ in $S_{i,\beta}$ for $i=1,\cdots,n_\beta-1$.
\item $x_{n_\beta-1}-x_{n_\beta+1},x_{n_\beta}-x_{n_\beta+2},x_{3n_\beta-1},x_{3n_\beta}$ in $S_{0,\alpha}$ are glued to $f,f,f-x_1,y_1$ in $S_{n_\beta,\beta}$.
\item $x_{i}-x_{i+1},y_{i+1}-y_i$ in $S_{n_\beta,\beta}$ are glued to $f,f$ in $S_{i,\alpha}$ for $i=1,\cdots,n_\alpha-1$.
\item $x_{n_\alpha}-y_{n_\alpha}$ in $S_{n_\beta,\beta}$ is glued to $f$ in $S_{n_\alpha,\alpha}$.
\eit

\noindent\ubf{Gluing rules for \raisebox{-.25\height}{\begin{tikzpicture}
\node (w1) at (-0.5,0.9) {$\sp(n_\alpha)^{(1)}$};
\begin{scope}[shift={(3.4,0)}]
\node (w2) at (-0.5,0.9) {$\so(2n_\beta+1)^{(1)}$};
\end{scope}
\node (w3) at (1,0.9) {\scriptsize{3}};
\draw (w1)--(w3);
\draw[->] (w3)--(w2);
\end{tikzpicture}}}:
We can take any geometry with $1\le\nu\le 2n_\alpha+7-3n_\beta$ for $\sp(n_\alpha)^{(1)}$, and any geometry with $0\le\nu\le 2n_\beta-7-n_\alpha$ for $\so(2n_\beta+1)^{(1)}$. The (non-geometric) gluing rules are:
\bit
\item $f-x_1-x_2,x_{2n_\beta}-x_{2n_\beta+2},x_{2n_\beta+1}-x_{2n_\beta+3}$ in $S_{0,\alpha}$ are glued to $f,f,f$ in $S_{0,\beta}$.
\item $x_i-x_{i+1},x_{2n_\beta-i+1}-x_{2n_\beta-i+2},x_{2n_\beta+i+1}-x_{2n_\beta+i+2}$ in $S_{0,\alpha}$ are glued to $f,f,f$ in $S_{i,\beta}$ for $i=1,\cdots,n_\beta-1$.
\item $x_{n_\beta}-x_{n_\beta+1},x_{n_\beta}-x_{n_\beta+1},x_{n_\beta+1}-x_{n_\beta+2},x_{n_\beta+1}-x_{n_\beta+2},x_{3n_\beta+1},x_{3n_\beta+1}$ in $S_{0,\alpha}$ are glued to $f,f,f,f,x_1,y_1$ in $S_{n_\beta,\beta}$.
\item $x_{i+1}-x_{i},y_{i+1}-y_i$ in $S_{n_\beta,\beta}$ are glued to $f,f$ in $S_{i,\alpha}$ for $i=1,\cdots,n_\alpha-1$.
\item $f-x_{n_\alpha},f-y_{n_\alpha}$ in $S_{n_\beta,\beta}$ are glued to $f-x_1,x_1$ in $S_{n_\alpha,\alpha}$.
\eit

\noindent\ubf{Gluing rules for \raisebox{-.25\height}{\begin{tikzpicture}
\node (w1) at (-0.5,0.9) {$\sp(n_\alpha)^{(1)}$};
\begin{scope}[shift={(3.1,0)}]
\node (w2) at (-0.5,0.9) {$\so(2n_\beta)^{(2)}$};
\end{scope}
\node (w3) at (1,0.9) {\scriptsize{3}};
\draw (w1)--(w3);
\draw[->] (w3)--(w2);
\end{tikzpicture}}}:
We can take any geometry with $1\le\nu\le 2n_\alpha+7-3n_\beta$ for $\sp(n_\alpha)^{(1)}$, and any geometry with $0\le\nu\le 2n_\beta-8-n_\alpha$ for $\so(2n_\beta)^{(2)}$. The (non-geometric) gluing rules are:
\bit
\item $f-x_1-x_2,x_1-x_2,x_{2n_\beta}-x_{2n_\beta+2},x_{2n_\beta}-x_{2n_\beta+1},x_{2n_\beta+1}-x_{2n_\beta+3},x_{2n_\beta+2}-x_{2n_\beta+3}$ in $S_{0,\alpha}$ are glued to $f,f,f,f,f,f$ in $S_{0,\beta}$.
\item $x_{i+1}-x_{i+2},x_{2n_\beta-i}-x_{2n_\beta-i+1},x_{2n_\beta+i+2}-x_{2n_\beta+i+3}$ in $S_{0,\alpha}$ are glued to $f,f,f$ in $S_{i,\beta}$ for $i=1,\cdots,n_\beta-2$.
\item $x_{n_\beta}-x_{n_\beta+1},x_{n_\beta}-x_{n_\beta+1},x_{n_\beta+1}-x_{n_\beta+2},x_{n_\beta+1}-x_{n_\beta+2},x_{3n_\beta+1},x_{3n_\beta+1}$ in $S_{0,\alpha}$ are glued to $f,f,f,f,x_1,y_1$ in $S_{n_\beta-1,\beta}$.
\item $x_{i+1}-x_{i},y_{i+1}-y_i$ in $S_{n_\beta-1,\beta}$ are glued to $f,f$ in $S_{i,\alpha}$ for $i=1,\cdots,n_\alpha-1$.
\item $f-x_{n_\alpha},f-y_{n_\alpha}$ in $S_{n_\beta-1,\beta}$ are glued to $f-x_1,x_1$ in $S_{n_\alpha,\alpha}$.
\eit
Again, the fact that $3n_\beta=2n_\alpha+8$ is not allowed manifests in the above gluing rules. The total number of blowups carried by $S_{0,\alpha}$ is at max $2n_\alpha+7$ but the gluing rules require the presence of $2n_\alpha+9$ blowups on $S_{0,\alpha}$. See the discussion around (\ref{bound3}) for an explanation of this restriction.

\noindent\ubf{Gluing rules for \raisebox{-.25\height}{\begin{tikzpicture}
\node (w1) at (-0.5,0.9) {$\su(n_\alpha)^{(1)}$};
\begin{scope}[shift={(3.1,0)}]
\node (w2) at (-0.5,0.9) {$\su(n_\beta)^{(1)}$};
\end{scope}
\node (w3) at (1,0.9) {\scriptsize{2}};
\draw (w1)--(w3);
\draw[->] (w3)--(w2);
\end{tikzpicture}}}:
We can take any geometry with $0\le\nu\le2n_\alpha-2n_\beta$ for $\su(n_\alpha)^{(1)}$, and any geometry with $0\le\nu\le2n_\beta-n_\alpha$ for $\su(n_\beta)^{(1)}$. The gluing rules are:
\bit
\item $f-x_1,x_{n_\beta}-x_{n_\beta+1},x_{2n_\beta}$ in $S_{0,\alpha}$ are glued to $f-x_1,f,x_{n_\alpha}$ in $S_{0,\beta}$.
\item $x_{i}-x_{i+1},x_{n_\beta+i}-x_{n_\beta+i+1}$ in $S_{0,\alpha}$ are glued to $f,f$ in $S_{i,\beta}$ for $i=1,\cdots,n_\beta-1$.
\item $x_{i}-x_{i+1}$ in $S_{0,\beta}$ is glued to $f$ in $S_{i,\alpha}$ for $i=1,\cdots,n_\alpha-1$.
\eit

\noindent\ubf{Gluing rules for \raisebox{-.25\height}{\begin{tikzpicture}
\node (w1) at (-0.5,0.9) {$\su(n_\alpha)^{(1)}$};
\begin{scope}[shift={(3.1,0)}]
\node (w2) at (-0.5,0.9) {$\su(n_\beta)^{(1)}$};
\end{scope}
\node (w3) at (1,0.9) {\scriptsize{3}};
\draw (w1)--(w3);
\draw[->] (w3)--(w2);
\end{tikzpicture}}}:
We can take any geometry with $0\le\nu\le2n_\alpha-3n_\beta$ for $\su(n_\alpha)^{(1)}$, and any geometry with $0\le\nu\le2n_\beta-n_\alpha$ for $\su(n_\beta)^{(1)}$. The gluing rules are:
\bit
\item $f-x_1,x_{n_\beta}-x_{n_\beta+1},x_{2n_\beta}-x_{2n_\beta+1},x_{3n_\beta}$ in $S_{0,\alpha}$ are glued to $f-x_1,f,f,x_{n_\alpha}$ in $S_{0,\beta}$.
\item $x_{i}-x_{i+1},x_{n_\beta+i}-x_{n_\beta+i+1},x_{2n_\beta+i}-x_{2n_\beta+i+1}$ in $S_{0,\alpha}$ are glued to $f,f,f$ in $S_{i,\beta}$ for $i=1,\cdots,n_\beta-1$.
\item $x_{i}-x_{i+1}$ in $S_{0,\beta}$ is glued to $f$ in $S_{i,\alpha}$ for $i=1,\cdots,n_\alpha-1$.
\eit

\noindent\ubf{Gluing rules for \raisebox{-.25\height}{\begin{tikzpicture}
\node (w1) at (-0.7,0.9) {$\su(2n_\alpha)^{(2)}$};
\begin{scope}[shift={(3.1,0)}]
\node (w2) at (-0.5,0.9) {$\su(n_\beta)^{(1)}$};
\end{scope}
\node (w3) at (1,0.9) {\scriptsize{2}};
\draw (w1)--(w3);
\draw[->] (w3)--(w2);
\end{tikzpicture}}}:
We can take any geometry with $0\le\nu\le2n_\beta-2n_\alpha$ for $\su(n_\beta)^{(1)}$. The gluing rules are:
\bit
\item $f-y_{n_\beta},x_{n_\beta},f-x_{1},y_{1}$ in $S_{0,\alpha}$ are glued to $x_{2n_\alpha-1},x_{2n_\alpha},f-x_2,f-x_1$ in $S_{0,\beta}$.
\item $x_{i}-x_{i+1},y_{i+1}-y_i$ in $S_{0,\alpha}$ are glued to $f,f$ in $S_{i,\beta}$ for $i=1,\cdots,n_\beta-1$.
\item $x_{i}-x_{i+1},x_{2n_\alpha-i}-x_{2n_\alpha-i+1}$ in $S_{0,\beta}$ are glued to $f,f$ in $S_{i,\alpha}$ for $i=1,\cdots,n_\alpha-1$.
\item $x_{n_\alpha}-x_{n_\alpha+1}$ in $S_{0,\beta}$ is glued to $f$ in $S_{n_\alpha,\alpha}$.
\eit

\noindent\ubf{Gluing rules for \raisebox{-.25\height}{\begin{tikzpicture}
\node (w1) at (-1,0.9) {$\su(2n_\alpha-1)^{(2)}$};
\begin{scope}[shift={(3.1,0)}]
\node (w2) at (-0.5,0.9) {$\su(n_\beta)^{(1)}$};
\end{scope}
\node (w3) at (1,0.9) {\scriptsize{2}};
\draw (w1)--(w3);
\draw[->] (w3)--(w2);
\end{tikzpicture}}}:
We can take any geometry with $1\le\nu\le2n_\beta-2n_\alpha+1$ for $\su(n_\beta)^{(1)}$. The (non-geometric) gluing rules are:
\bit
\item $y_{n_\beta},x_{n_\beta},f-x_{1},f-y_{1},f,f$ in $S_{0,\alpha}$ are glued to $x_{2n_\alpha-1},x_{2n_\alpha-1},y_1,f-x_1,x_1-x_2,f-x_2-y_1$ in $S_{0,\beta}$.
\item $x_{i}-x_{i+1},y_{i}-y_{i+1}$ in $S_{0,\alpha}$ are glued to $f,f$ in $S_{i,\beta}$ for $i=1,\cdots,n_\beta-1$.
\item $x_{i+1}-x_{i+2},x_{2n_\alpha-i-1}-x_{2n_\alpha-i}$ in $S_{0,\beta}$ are glued to $f,f$ in $S_{i,\alpha}$ for $i=1,\cdots,n_\alpha-2$.
\item $x_{n_\alpha}-x_{n_\alpha+1}$ in $S_{0,\beta}$ is glued to $f$ in $S_{n_\alpha-1,\alpha}$.
\eit

\noindent\ubf{Gluing rules for \raisebox{-.25\height}{\begin{tikzpicture}
\node (w1) at (-0.5,0.9) {$\fg_2^{(1)}$};
\begin{scope}[shift={(3,0)}]
\node (w2) at (-0.5,0.9) {$\su(2)^{(1)}$};
\end{scope}
\node (w3) at (0.9,0.9) {\scriptsize{2}};
\draw (w1)--(w3);
\draw[->] (w3)--(w2);
\end{tikzpicture}}}:
We can take any geometry with $1\le\nu\le 3$ for $\fg_2^{(1)}$, and we must use the geometry with $\nu=1$ for $\su(2)^{(1)}$. The (non-geometric) gluing rules are:
\bit
\item $f-x_1,y_1$ in $S_{0,\alpha}$ are glued to $f-x_2,f-x_1$ in $S_{0,\beta}$.
\item $x_{1}-y_1$ in $S_{0,\alpha}$ is glued to $f$ in $S_{1,\beta}$.
\item $x_2-x_3$ in $S_{0,\beta}$ is glued to $f$ in $S_{2,\alpha}$.
\item $x_1-x_2,x_{3},x_{3}$ in $S_{0,\beta}$ are glued to $f,x_1,y_1$ in $S_{1,\alpha}$.
\item $f-x_1,f-y_1$ in $S_{1,\alpha}$ are glued to $f-x_1,x_1$ in $S_{1,\beta}$.
\eit

\noindent\ubf{Gluing rules for \raisebox{-.25\height}{\begin{tikzpicture}
\node (w1) at (-0.5,0.9) {$\fg_2^{(1)}$};
\begin{scope}[shift={(3,0)}]
\node (w2) at (-0.5,0.9) {$\su(2)^{(1)}$};
\end{scope}
\node (w3) at (0.9,0.9) {\scriptsize{3}};
\draw (w1)--(w3);
\draw[->] (w3)--(w2);
\end{tikzpicture}}}:
We can take any geometry with $2\le\nu\le 3$ for $\fg_2^{(1)}$, and we must use the geometry with $\nu=1$ for $\su(2)^{(1)}$. The (non-geometric) gluing rules are:
\bit
\item $f-x_1,y_1,x_2-y_2$ in $S_{0,\alpha}$ are glued to $f-x_2,f-x_1,f$ in $S_{0,\beta}$.
\item $x_{1}-x_2,y_2-y_1$ in $S_{0,\alpha}$ are glued to $f,f$ in $S_{1,\beta}$.
\item $x_2-x_3$ in $S_{0,\beta}$ is glued to $f$ in $S_{2,\alpha}$.
\item $x_1-x_2,x_{3},x_{3}$ in $S_{0,\beta}$ are glued to $f,x_1,y_1$ in $S_{1,\alpha}$.
\item $f-x_1,f-y_1$ in $S_{1,\alpha}$ are glued to $f-x_1,x_1$ in $S_{1,\beta}$.
\eit

\noindent\ubf{Gluing rules for \raisebox{-.25\height}{\begin{tikzpicture}
\node (w1) at (-0.5,0.9) {$\so(7)^{(1)}$};
\begin{scope}[shift={(3,0)}]
\node (w2) at (-0.5,0.9) {$\sp(1)^{(1)}$};
\end{scope}
\node (w3) at (1,0.9) {\scriptsize{2}};
\draw[dashed] (w1)--(w3);
\draw[->,dashed] (w3)--(w2);
\end{tikzpicture}} and \raisebox{-.25\height}{\begin{tikzpicture}
\node (w1) at (-0.5,0.9) {$\so(7)^{(1)}$};
\begin{scope}[shift={(3,0)}]
\node (w2) at (-0.5,0.9) {$\su(2)^{(1)}$};
\end{scope}
\node (w3) at (1,0.9) {\scriptsize{2}};
\draw[dashed] (w1)--(w3);
\draw[->,dashed] (w3)--(w2);
\end{tikzpicture}}}:
We can take any geometry with $1\le\nu\le 7-2\Omega^{\alpha\alpha}$ for $\so(7)^{(1)}$, any geometry with $0\le\nu\le 6$ for $\sp(1)^{(1)}$, and we must use the geometry with $\nu=0$ for $\su(2)^{(1)}$. The gluing rules are:
\bit
\item $f-x_1,y_1$ in $S_{0,\alpha}$ are glued to $f-x_2,f-x_1$ in $S_{0,\beta}$.
\item $x_{1}-y_{1}$ in $S_{0,\alpha}$ is glued to $f$ in $S_{1,\beta}$.
\item $x_3,x_4$ in $S_{0,\beta}$ are glued to $f-x_1,y_1$ in $S_{1,\alpha}$.
\item $x_2-x_3$ in $S_{0,\beta}$ is glued to $f$ in $S_{2,\alpha}$.
\item $x_1-x_2,x_{3}-x_{4}$ in $S_{0,\beta}$ are glued to $f,f$ in $S_{3,\alpha}$.
\item $x_1-y_1$ in $S_{1,\alpha}$ is glued to $f$ in $S_{1,\beta}$.
\eit

\noindent\ubf{Gluing rules for \raisebox{-.25\height}{\begin{tikzpicture}
\node (w1) at (-0.5,0.9) {$\so(7)^{(1)}$};
\begin{scope}[shift={(3,0)}]
\node (w2) at (-0.5,0.9) {$\su(2)^{(1)}$};
\end{scope}
\node (w3) at (1,0.9) {\scriptsize{3}};
\draw[dashed] (w1)--(w3);
\draw[->,dashed] (w3)--(w2);
\end{tikzpicture}}}:
We can take any geometry with $2\le\nu\le 3$ for $\so(7)^{(1)}$, and we must use the geometry with $\nu=0$ for $\su(2)^{(1)}$. The gluing rules are:
\bit
\item $f-x_1,y_1,x_2-y_2$ in $S_{0,\alpha}$ are glued to $f-x_2,f-x_1,f$ in $S_{0,\beta}$.
\item $x_{1}-x_{2},y_2-y_1$ in $S_{0,\alpha}$ are glued to $f,f$ in $S_{1,\beta}$.
\item $x_3,x_4$ in $S_{0,\beta}$ are glued to $f-x_1,y_1$ in $S_{1,\alpha}$.
\item $x_2-x_3$ in $S_{0,\beta}$ is glued to $f$ in $S_{2,\alpha}$.
\item $x_1-x_2,x_{3}-x_{4}$ in $S_{0,\beta}$ are glued to $f,f$ in $S_{3,\alpha}$.
\item $x_1-y_1$ in $S_{1,\alpha}$ is glued to $f$ in $S_{1,\beta}$.
\eit

\noindent\ubf{Gluing rules for \raisebox{-.25\height}{\begin{tikzpicture}
\node (w1) at (-0.5,0.9) {$\so(8)^{(2)}$};
\begin{scope}[shift={(3,0)}]
\node (w2) at (-0.5,0.9) {$\sp(1)^{(1)}$};
\end{scope}
\node (w3) at (1,0.9) {\scriptsize{2}};
\draw[dashed] (w1)--(w3);
\draw[->] (w3)--(w2);
\end{tikzpicture}}}:
We can take any geometry with $0\le\nu\le 6$ for $\sp(1)^{(1)}$. The gluing rules are:
\bit
\item $f-x_1,y_1$ in $S_{1,\alpha}$ are glued to $x_3,x_4$ in $S_{0,\beta}$.
\item $x_{1}-y_{1}$ in $S_{1,\alpha}$ is glued to $f$ in $S_{1,\beta}$.
\item $f-x_1-x_4,f-x_2-x_3$ in $S_{0,\beta}$ are glued to $f,f$ in $S_{0,\alpha}$.
\item $x_2-x_3$ in $S_{0,\beta}$ is glued to $f$ in $S_{2,\alpha}$.
\item $x_1-x_2,x_{3}-x_{4}$ in $S_{0,\beta}$ are glued to $f,f$ in $S_{3,\alpha}$.
\eit

\noindent\ubf{Gluing rules for \raisebox{-.25\height}{\begin{tikzpicture}
\node (w1) at (-0.6,0.9) {$\so(2n_\alpha)^{(1)}$};
\begin{scope}[shift={(3.1,0)}]
\node (w2) at (-0.5,0.9) {$\sp(n_\beta)^{(1)}$};
\end{scope}
\node (w3) at (1,0.9) {\scriptsize{2}};
\draw (w1)--(w3);
\draw[->] (w3)--(w2);
\end{tikzpicture}}}:
We can take any geometry with $n_\beta\le\nu\le 2n_\alpha-4-\Omega^{\alpha\alpha}-n_\beta$ for $\so(2n_\alpha)^{(1)}$, and any geometry with $0\le\nu\le 2n_\beta+8-n_\alpha$ for $\sp(n_\beta)^{(1)}$. The gluing rules are:
\bit
\item $f-x_1,y_1$ in $S_{0,\alpha}$ are glued to $f-x_2,f-x_1$ in $S_{0,\beta}$.
\item $x_{i}-x_{i+1},y_{i+1}-y_i$ in $S_{0,\alpha}$ are glued to $f,f$ in $S_{i,\beta}$ for $i=1,\cdots,n_\beta-1$.
\item $x_{n_\beta}-y_{n_\beta}$ in $S_{0,\alpha}$ is glued to $f$ in $S_{n_\beta,\beta}$.
\item $x_{i}-x_{i+1}$ in $S_{0,\beta}$ is glued to $f$ in $S_{i,\alpha}$ for $i=1,\cdots,n_\alpha-1$.
\item $x_{n_\alpha-1},x_{n_\alpha}$ in $S_{0,\beta}$ are glued to $f-x_1,y_1$ in $S_{n_\alpha,\alpha}$.
\item $x_{i}-x_{i+1},y_{i+1}-y_i$ in $S_{n_\alpha,\alpha}$ are glued to $f,f$ in $S_{i,\beta}$ for $i=1,\cdots,n_\beta-1$.
\item $x_{n_\beta}-y_{n_\beta}$ in $S_{n_\alpha,\alpha}$ is glued to $f$ in $S_{n_\beta,\beta}$.
\eit

\noindent\ubf{Gluing rules for \raisebox{-.25\height}{\begin{tikzpicture}
\node (w1) at (-0.9,0.9) {$\so(2n_\alpha+1)^{(1)}$};
\begin{scope}[shift={(3.1,0)}]
\node (w2) at (-0.5,0.9) {$\sp(n_\beta)^{(1)}$};
\end{scope}
\node (w3) at (1,0.9) {\scriptsize{2}};
\draw (w1)--(w3);
\draw[->] (w3)--(w2);
\end{tikzpicture}}}:
We can take any geometry with $n_\beta\le\nu\le 2n_\alpha-3-\Omega^{\alpha\alpha}-n_\beta$ for $\so(2n_\alpha+1)^{(1)}$, and any geometry with $1\le\nu\le 2n_\beta+8-n_\alpha$ for $\sp(n_\beta)^{(1)}$. The gluing rules are:
\bit
\item $f-x_1,y_1$ in $S_{0,\alpha}$ are glued to $f-x_2,f-x_1$ in $S_{0,\beta}$.
\item $x_{i}-x_{i+1},y_{i+1}-y_i$ in $S_{0,\alpha}$ are glued to $f,f$ in $S_{i,\beta}$ for $i=1,\cdots,n_\beta-1$.
\item $x_{n_\beta}-y_{n_\beta}$ in $S_{0,\alpha}$ is glued to $f$ in $S_{n_\beta,\beta}$.
\item $x_{i}-x_{i+1}$ in $S_{0,\beta}$ is glued to $f$ in $S_{i,\alpha}$ for $i=1,\cdots,n_\alpha-1$.
\item $x_{n_\alpha},x_{n_\alpha}$ in $S_{0,\beta}$ are glued to $x_1,y_1$ in $S_{n_\alpha,\alpha}$.
\item $x_{i+1}-x_{i},y_{i+1}-y_i$ in $S_{n_\alpha,\alpha}$ are glued to $f,f$ in $S_{i,\beta}$ for $i=1,\cdots,n_\beta-1$.
\item $f-x_{n_\beta},f-y_{n_\beta}$ in $S_{n_\alpha,\alpha}$ are glued to $f-x_1,x_1$ in $S_{n_\beta,\beta}$.
\eit
The blowup $x_1$ in $S_{n_\beta,\beta}$ can be used to glue $\sp(n_\beta)^{(1)}$ to exactly one more neighboring node connected to it by an undirected edge. The neighboring node can carry $\so(2n_\gamma+1)^{(1)}$, $\fg_2^{(1)}$ or $\so(2n_\gamma)^{(2)}$.

\noindent\ubf{Gluing rules for \raisebox{-.25\height}{\begin{tikzpicture}
\node (w1) at (-0.7,0.9) {$\so(2n_\alpha)^{(2)}$};
\begin{scope}[shift={(3.1,0)}]
\node (w2) at (-0.5,0.9) {$\sp(n_\beta)^{(1)}$};
\end{scope}
\node (w3) at (1,0.9) {\scriptsize{2}};
\draw (w1)--(w3);
\draw[->] (w3)--(w2);
\end{tikzpicture}}}:
We can take any geometry with $n_\beta\le\nu\le 2n_\alpha-8-n_\beta$ for $\so(2n_\alpha)^{(2)}$, and any geometry with $1\le\nu\le 2n_\beta+8-n_\alpha$ for $\sp(n_\beta)^{(1)}$. The gluing rules are:
\bit
\item $f-x_1,y_1,f$ in $S_{0,\alpha}$ are glued to $f-x_2,f-x_1,x_1-x_2$ in $S_{0,\beta}$.
\item $x_{i}-x_{i+1},y_{i+1}-y_i$ in $S_{0,\alpha}$ are glued to $f,f$ in $S_{i,\beta}$ for $i=1,\cdots,n_\beta-1$.
\item $x_{n_\beta}-y_{n_\beta}$ in $S_{0,\alpha}$ is glued to $f$ in $S_{n_\beta,\beta}$.
\item $x_{i+1}-x_{i+2}$ in $S_{0,\beta}$ is glued to $f$ in $S_{i,\alpha}$ for $i=1,\cdots,n_\alpha-2$.
\item $x_{n_\alpha},x_{n_\alpha}$ in $S_{0,\beta}$ are glued to $x_1,y_1$ in $S_{n_\alpha-1,\alpha}$.
\item $x_{i+1}-x_{i},y_{i+1}-y_i$ in $S_{n_\alpha-1,\alpha}$ are glued to $f,f$ in $S_{i,\beta}$ for $i=1,\cdots,n_\beta-1$.
\item $f-x_{n_\beta},f-y_{n_\beta}$ in $S_{n_\alpha-1,\alpha}$ are glued to $f-x_1,x_1$ in $S_{n_\beta,\beta}$.
\eit
Again, the blowup $x_1$ in $S_{n_\beta,\beta}$ can be used to glue $\sp(n_\beta)^{(1)}$ to exactly one more neighboring node carrying $\so(2n_\gamma+1)^{(1)}$, $\fg_2^{(1)}$ or $\so(2n_\gamma)^{(2)}$.

\noindent\ubf{Gluing rules for \raisebox{-.25\height}{\begin{tikzpicture}
\node (w1) at (-0.6,0.9) {$\so(2n_\alpha)^{(1)}$};
\begin{scope}[shift={(3.1,0)}]
\node (w2) at (-0.5,0.9) {$\sp(n_\beta)^{(1)}$};
\end{scope}
\node (w3) at (1,0.9) {\scriptsize{3}};
\draw (w1)--(w3);
\draw[->] (w3)--(w2);
\end{tikzpicture}}}:
We can take any geometry with $2n_\beta\le\nu\le 2n_\alpha-8-n_\beta$ for $\so(2n_\alpha)^{(1)}$, and any geometry with $0\le\nu\le 2n_\beta+8-n_\alpha$ for $\sp(n_\beta)^{(1)}$. The gluing rules are:
\bit
\item $f-x_1,x_{2n_\beta}-y_{2n_\beta},y_1$ in $S_{0,\alpha}$ are glued to $f-x_2,f,f-x_1$ in $S_{0,\beta}$.
\item $x_{i}-x_{i+1},y_{i+1}-y_i,x_{2n_\beta-i}-x_{2n_\beta-i+1},y_{2n_\beta-i+1}-y_{2n_\beta-i}$ in $S_{0,\alpha}$ are glued to $f,f,f,f$ in $S_{i,\beta}$ for $i=1,\cdots,n_\beta-1$.
\item $x_{n_\beta}-x_{n_\beta+1},y_{n_\beta+1}-y_{n_\beta}$ in $S_{0,\alpha}$ are glued to $f,f$ in $S_{n_\beta,\beta}$.
\item $x_{i}-x_{i+1}$ in $S_{0,\beta}$ is glued to $f$ in $S_{i,\alpha}$ for $i=1,\cdots,n_\alpha-1$.
\item $x_{n_\alpha-1},x_{n_\alpha}$ in $S_{0,\beta}$ are glued to $f-x_1,y_1$ in $S_{n_\alpha,\alpha}$.
\item $x_{i}-x_{i+1},y_{i+1}-y_i$ in $S_{n_\alpha,\alpha}$ are glued to $f,f$ in $S_{i,\beta}$ for $i=1,\cdots,n_\beta-1$.
\item $x_{n_\beta}-y_{n_\beta}$ in $S_{n_\alpha,\alpha}$ is glued to $f$ in $S_{n_\beta,\beta}$.
\eit

\noindent\ubf{Gluing rules for \raisebox{-.25\height}{\begin{tikzpicture}
\node (w1) at (-0.9,0.9) {$\so(2n_\alpha+1)^{(1)}$};
\begin{scope}[shift={(3.1,0)}]
\node (w2) at (-0.5,0.9) {$\sp(n_\beta)^{(1)}$};
\end{scope}
\node (w3) at (1,0.9) {\scriptsize{3}};
\draw (w1)--(w3);
\draw[->] (w3)--(w2);
\end{tikzpicture}}}:
We can take any geometry with $2n_\beta\le\nu\le 2n_\alpha-7-n_\beta$ for $\so(2n_\alpha+1)^{(1)}$, and any geometry with $1\le\nu\le 2n_\beta+8-n_\alpha$ for $\sp(n_\beta)^{(1)}$. The (non-geometric) gluing rules are:
\bit
\item $f-x_1,x_{2n_\beta}-y_{2n_\beta},y_1$ in $S_{0,\alpha}$ are glued to $f-x_2,f,f-x_1$ in $S_{0,\beta}$.
\item $x_{i}-x_{i+1},y_{i+1}-y_i,x_{2n_\beta-i}-x_{2n_\beta-i+1},y_{2n_\beta-i+1}-y_{2n_\beta-i}$ in $S_{0,\alpha}$ are glued to $f,f,f,f$ in $S_{i,\beta}$ for $i=1,\cdots,n_\beta-1$.
\item $x_{n_\beta}-x_{n_\beta+1},y_{n_\beta+1}-y_{n_\beta}$ in $S_{0,\alpha}$ are glued to $f,f$ in $S_{n_\beta,\beta}$.
\item $x_{i}-x_{i+1}$ in $S_{0,\beta}$ is glued to $f$ in $S_{i,\alpha}$ for $i=1,\cdots,n_\alpha-1$.
\item $x_{n_\alpha},x_{n_\alpha}$ in $S_{0,\beta}$ are glued to $x_1,y_1$ in $S_{n_\alpha,\alpha}$.
\item $x_{i+1}-x_{i},y_{i+1}-y_i$ in $S_{n_\alpha,\alpha}$ are glued to $f,f$ in $S_{i,\beta}$ for $i=1,\cdots,n_\beta-1$.
\item $f-x_{n_\beta},f-y_{n_\beta}$ in $S_{n_\alpha,\alpha}$ are glued to $f-x_1,x_1$ in $S_{n_\beta,\beta}$.
\eit

\noindent\ubf{Gluing rules for \raisebox{-.25\height}{\begin{tikzpicture}
\node (w1) at (-0.7,0.9) {$\so(2n_\alpha)^{(2)}$};
\begin{scope}[shift={(3.1,0)}]
\node (w2) at (-0.5,0.9) {$\sp(n_\beta)^{(1)}$};
\end{scope}
\node (w3) at (1,0.9) {\scriptsize{3}};
\draw (w1)--(w3);
\draw[->] (w3)--(w2);
\end{tikzpicture}}}:
We can take any geometry with $2n_\beta\le\nu\le 2n_\alpha-8-n_\beta$ for $\so(2n_\alpha)^{(2)}$, and any geometry with $1\le\nu\le 2n_\beta+8-n_\alpha$ for $\sp(n_\beta)^{(1)}$. The (non-geometric) gluing rules are:
\bit
\item $f-x_1,x_{2n_\beta}-y_{2n_\beta},y_1,f$ in $S_{0,\alpha}$ are glued to $f-x_2,f,f-x_1,x_1-x_2$ in $S_{0,\beta}$.
\item $x_{i}-x_{i+1},y_{i+1}-y_i,x_{2n_\beta-i}-x_{2n_\beta-i+1},y_{2n_\beta-i+1}-y_{2n_\beta-i}$ in $S_{0,\alpha}$ are glued to $f,f,f,f$ in $S_{i,\beta}$ for $i=1,\cdots,n_\beta-1$.
\item $x_{n_\beta}-x_{n_\beta+1},y_{n_\beta+1}-y_{n_\beta}$ in $S_{0,\alpha}$ are glued to $f,f$ in $S_{n_\beta,\beta}$.
\item $x_{i+1}-x_{i+2}$ in $S_{0,\beta}$ is glued to $f$ in $S_{i,\alpha}$ for $i=1,\cdots,n_\alpha-2$.
\item $x_{n_\alpha},x_{n_\alpha}$ in $S_{0,\beta}$ are glued to $x_1,y_1$ in $S_{n_\alpha-1,\alpha}$.
\item $x_{i+1}-x_{i},y_{i+1}-y_i$ in $S_{n_\alpha-1,\alpha}$ are glued to $f,f$ in $S_{i,\beta}$ for $i=1,\cdots,n_\beta-1$.
\item $f-x_{n_\beta},f-y_{n_\beta}$ in $S_{n_\alpha-1,\alpha}$ are glued to $f-x_1,x_1$ in $S_{n_\beta,\beta}$.
\eit

\subsection{Gluing rules involving non-gauge-theoretic nodes}\label{ARng}
There are only three such nodes which are listed below
\be
\begin{tikzpicture}
\node at (-0.5,0.45) {1};
\node at (-0.45,0.9) {$\sp(0)^{(1)}_\theta$};
\end{tikzpicture}
\ee
\be
\begin{tikzpicture}
\node at (-0.5,0.45) {2};
\node at (-0.45,0.9) {$\su(1)^{(1)}$};
\end{tikzpicture}
\ee
\be
\begin{tikzpicture}
\node (v1) at (-0.5,0.4) {2};
\node at (-0.45,0.9) {$\su(1)^{(1)}$};
\draw (v1) .. controls (-1.5,-0.5) and (0.5,-0.5) .. (v1);
\end{tikzpicture}
\ee
The theta angle for $\sp(0)^{(1)}$ is physically irrelevant as long as there is no neighboring $\su(8)$.

First consider the edges shown as last two entries of Table \ref{R2P}. The gluing rules for these cases are as follows.

\noindent\ubf{Gluing rules for \raisebox{-.125\height}{\begin{tikzpicture}
\node (v1) at (-0.5,0.4) {$2$};
\node at (-0.45,0.9) {$\su(1)^{(1)}$};
\begin{scope}[shift={(2,0)}]
\node (v2) at (-0.5,0.4) {$1$};
\node at (-0.45,0.9) {$\sp(1)^{(1)}$};
\end{scope}
\draw  (v1) -- (v2);
\end{tikzpicture}} and \raisebox{-.125\height}{\begin{tikzpicture}
\node (v1) at (-0.5,0.4) {$2$};
\node at (-0.45,0.9) {$\su(1)^{(1)}$};
\begin{scope}[shift={(2,0)}]
\node (v2) at (-0.5,0.4) {$2$};
\node at (-0.45,0.9) {$\su(2)^{(1)}$};
\end{scope}
\draw  (v1) -- (v2);
\end{tikzpicture}}}:
We can choose any geometry with $1\le\nu\le10$ for $\sp(1)^{(1)}$ and any geometry with $1\le\nu\le4$ for $\su(2)^{(1)}$. The (non-geometric) gluing rules are:
\bit
\item $f-x-y$ in $S_{0,\alpha}$ is glued to $f$ in $S_{0,\beta}$.
\item $x,y$ in $S_{0,\alpha}$ are glued to $f-x_1,x_1$ in $S_{1,\beta}$.
\eit
As in cases discussed in last subsection, the blowup $x_1$ in $S_{1,\beta}$ can be used for gluing $\sp(1)^{(1)}$ or $\su(2)^{(1)}$ with another neighbor such that the gluing rules for $\sp(1)^{(1)}$ or $\su(2)^{(1)}$ with that neighbor allow a blowup on $S_{1,\beta}$ to be used for more than once.

\bigskip

\bigskip

The gluing rules for the edges shown in Table \ref{R2N} are as follows.

\noindent\ubf{Gluing rules for \raisebox{-.125\height}{\begin{tikzpicture}
\node (v1) at (-0.5,0.4) {$1$};
\node at (-0.45,0.9) {$\sp(0)^{(1)}$};
\begin{scope}[shift={(2,0)}]
\node (v2) at (-0.5,0.4) {$2$};
\node at (-0.45,0.9) {$\su(1)^{(1)}$};
\end{scope}
\draw  (v1) -- (v2);
\end{tikzpicture}}}:
\bit
\item $3l-\sum x_i$ in $S_{0,\alpha}$ is glued to $f$ in $S_{0,\beta}$.
\eit
See Appendix (\ref{grng}) for a derivation of the above gluing rules.

\noindent\ubf{Gluing rules for \raisebox{-.125\height}{\begin{tikzpicture}
\node (v1) at (-0.5,0.4) {$2$};
\node at (-0.45,0.9) {$\su(1)^{(1)}$};
\begin{scope}[shift={(2,0)}]
\node (v2) at (-0.5,0.4) {$2$};
\node at (-0.45,0.9) {$\su(1)^{(1)}$};
\end{scope}
\draw  (v1) -- (v2);
\end{tikzpicture}}}:
\bit
\item $f-x,x$ in $S_{0,\alpha}$ are glued to $f-x,x$ in $S_{0,\beta}$.
\eit
The blowups $x$ in $S_{0,\alpha}$ and $x$ in $S_{0,\beta}$ can be used for gluing to other $\su(1)^{(1)}$ neighbors. See Appendix (\ref{grng}) for a derivation of the above gluing rules.

\bigskip

\bigskip

Now consider the edges shown in the last entry of Table \ref{FTR2}:

\noindent\ubf{Gluing rules for \raisebox{-.325\height}{\begin{tikzpicture}
\node (v1) at (-0.5,0.4) {$2$};
\node at (-0.45,0.9) {$\su(2)^{(1)}$};
\begin{scope}[shift={(2,0)}]
\node (v2) at (-0.5,0.4) {$2$};
\node at (-0.45,0.9) {$\su(1)^{(1)}$};
\end{scope}
\draw  (v1) -- (v2);
\draw (v1) .. controls (-1.5,-0.5) and (0.5,-0.5) .. (v1);
\end{tikzpicture}}}:
\bit
\item $f-x_1,x_1$ in $S_{0,\alpha}$ are glued to $x,y$ in $S_{0,\beta}$.
\item $f$ in $S_{1,\alpha}$ is glued to $f-x-y$ in $S_{0,\beta}$.
\eit

\noindent\ubf{Gluing rules for \raisebox{-.325\height}{\begin{tikzpicture}
\node (v1) at (-0.5,0.4) {$2$};
\node at (-0.45,0.9) {$\su(1)^{(1)}$};
\begin{scope}[shift={(2,0)}]
\node (v2) at (-0.5,0.4) {$2$};
\node at (-0.45,0.9) {$\su(1)^{(1)}$};
\end{scope}
\draw  (v1) -- (v2);
\draw (v1) .. controls (-1.5,-0.5) and (0.5,-0.5) .. (v1);
\end{tikzpicture}}}:
\bit
\item $2h-x-2y,f-x$ in $S_{0,\alpha}$ are glued to $f-x,x$ in $S_{0,\beta}$.
\eit
The blowup $x$ in $S_{0,\beta}$ can be used for gluing to other $\su(1)^{(1)}$ neighbors. See Appendix (\ref{grng}) for a derivation of the above gluing rules. We remind the reader that this gluing rule involves the non-geometric node (\ref{nong}) and hence the above gluing rules should be viewed only as an algebraic description and not as a geometric description. See the discussion after equation (\ref{Gsu1l}) for more details.

\bigskip

\bigskip

Now consider the last entry of Table \ref{BTR2N}:

\noindent\ubf{Gluing rules for \raisebox{-.125\height}{\begin{tikzpicture}
\node (v1) at (-0.5,0.4) {$2$};
\node at (-0.45,0.9) {$\su(2)^{(1)}$};
\begin{scope}[shift={(2,0)}]
\node (v2) at (-0.5,0.4) {$2$};
\node at (-0.45,0.9) {$\su(1)^{(1)}$};
\end{scope}
\node (v3) at (0.5,0.4) {\tiny{$2$}};
\draw  (v1) edge (v3);
\draw  [->](v3) -- (v2);
\end{tikzpicture}}}:
We can use any geometry with $1\le\nu\le3$ for $\su(2)^{(1)}$. The gluing rules are:
\bit
\item $f-x_1,x_1$ in $S_{0,\alpha}$ are glued to $x,y$ in $S_{0,\beta}$.
\item $f-x_1,x_1$ in $S_{1,\alpha}$ are glued to $f-x,f-y$ in $S_{0,\beta}$.
\eit
The blowups $x_1$ in $S_{0,\alpha}$ and $x_1$ in $S_{1,\alpha}$ can also be used for gluing to other neighboring nodes of $\su(2)^{(1)}$ that carry some $\su(n)^{(1)}$.

\noindent\ubf{Gluing rules for \raisebox{-.125\height}{\begin{tikzpicture}
\node (v1) at (-0.5,0.4) {$2$};
\node at (-0.45,0.9) {$\su(2)^{(1)}$};
\begin{scope}[shift={(2,0)}]
\node (v2) at (-0.5,0.4) {$2$};
\node at (-0.45,0.9) {$\su(1)^{(1)}$};
\end{scope}
\node (v3) at (0.5,0.4) {\tiny{$3$}};
\draw  (v1) edge (v3);
\draw  [->](v3) -- (v2);
\end{tikzpicture}}}:
We can use any geometry with $1\le\nu\le3$ for $\su(2)^{(1)}$. The gluing rules are:
\bit
\item $f-x_1,x_1$ in $S_{0,\alpha}$ are glued to $x,y$ in $S_{0,\beta}$.
\item $f-x_1,x_1$ in $S_{1,\alpha}$ are glued to $2f-x,f-y$ in $S_{0,\beta}$.
\eit
The blowups $x_1$ in $S_{0,\alpha}$ and $x_1$ in $S_{1,\alpha}$ can also be used for gluing to other neighboring nodes of $\su(2)^{(1)}$ that carry some $\su(n)^{(1)}$.

\noindent\ubf{Gluing rules for \raisebox{-.125\height}{\begin{tikzpicture}
\node (v1) at (-0.5,0.4) {$2$};
\node at (-0.45,0.9) {$\su(1)^{(1)}$};
\begin{scope}[shift={(2,0)}]
\node (v2) at (-0.5,0.4) {$2$};
\node at (-0.45,0.9) {$\su(1)^{(1)}$};
\end{scope}
\node (v3) at (0.5,0.4) {\tiny{$2$}};
\draw  (v1) edge (v3);
\draw  [->](v3) -- (v2);
\end{tikzpicture}}}:
\bit
\item $f-x,x$ in $S_{0,\alpha}$ are glued to $2f-x,x$ in $S_{0,\beta}$.
\eit
(Note that the gluing rules proposed above are non-geometric.) The blowups $x$ in $S_{0,\alpha}$ and $x$ in $S_{0,\beta}$ can be used to further glue to other neighboring $\su(1)^{(1)}$. See Appendix (\ref{grng}) for a derivation of the above gluing rules.

\noindent\ubf{Gluing rules for \raisebox{-.125\height}{\begin{tikzpicture}
\node (v1) at (-0.5,0.4) {$2$};
\node at (-0.45,0.9) {$\su(1)^{(1)}$};
\begin{scope}[shift={(2,0)}]
\node (v2) at (-0.5,0.4) {$2$};
\node at (-0.45,0.9) {$\su(1)^{(1)}$};
\end{scope}
\node (v3) at (0.5,0.4) {\tiny{$3$}};
\draw  (v1) edge (v3);
\draw  [->](v3) -- (v2);
\end{tikzpicture}}}:
\bit
\item $f-x,x$ in $S_{0,\alpha}$ are glued to $3f-x,x$ in $S_{0,\beta}$.
\eit
(Note that the gluing rules proposed above are non-geometric.) The blowups $x$ in $S_{0,\alpha}$ and $x$ in $S_{0,\beta}$ can be used to further glue to other neighboring $\su(1)^{(1)}$. 

\subsubsection{$\sp(0)^{(1)}$ gluings: untwisted, without non-simply-laced}
At this point, we are only left with gluings of $\sp(0)^{(1)}$ to other nodes carrying non-trivial gauge algebras. In this case, we also have to provide gluing rules for two neighbors at a time. This is because the torus fiber for $dP_9$ is $3l-\sum x_i$ which involves all of the blowups. So all of the blowups must appear in the gluing rules associated to each edge. This is in stark contrast to the gluing rules for non-trivial algebras $\fg^{(q)}$ where (typically) the blowups used for gluing rules associated to different edges are different. Thus in the case of $\fg^{(q)}$, the gluing rules for different edges naturally decouple. However, in the case of $\sp(0)^{(1)}$, we have to provide gluing rules for multiple neighbors at a time and show explicitly that the curves inside $dP_9$ involved in gluing rules for different edges do not intersect. It turns out that in the context of $6d$ SCFTs, $\sp(0)^{(1)}$ can only have a maximum of two neighbors carrying non-trivial algebras.

In the case when all the neighbors are untwisted, $\sp(0)^{(1)}$ gluings were first studied in \cite{Bhardwaj:2018vuu}. For the completeness of our presentation, we reproduce their results in this subsection (providing enhanced explanations while we do so) before moving onto $\sp(0)^{(1)}$ gluings arising in the twisted case. Following \cite{Bhardwaj:2018vuu}, we will represent these $\sp(0)^{(1)}$ gluing rules in a graphical notation that we review as we review the results of \cite{Bhardwaj:2018vuu}.

To start with, let us consider the gluing rules for
\be
\begin{tikzpicture}
\begin{scope}[shift={(2,0)}]
\node (w2) at (-0.5,0.9) {$\sp(0)^{(1)}$};
\end{scope}
\begin{scope}[shift={(4,0)}]
\node (w3) at (-0.5,0.9) {$\fe_8^{(1)}$};
\end{scope}
\draw (w2)--(w3);
\end{tikzpicture}
\ee
which are displayed below
\be\label{ee8}
\begin{tikzpicture}
\node (v1) at (0,0) {\scriptsize{$x_8-x_9$}};
\node (v2) at (2,0) {\scriptsize{$x_7-x_8$}};
\node (v3) at (4,0) {\scriptsize{$x_6-x_7$}};
\node (v4) at (6,0) {\scriptsize{$x_5-x_6$}};
\node (v5) at (8,0) {\scriptsize{$x_4-x_5$}};
\node (v1_1) at (10,0) {\scriptsize{$x_1-x_4$}};
\node (v1_2) at (12,0) {\scriptsize{$x_2-x_1$}};
\node (v1_3) at (14,0) {\scriptsize{$x_3-x_2$}};
\node (v1_4) at (10,1) {\scriptsize{$l-x_1-x_2-x_3$}};
\draw  (v1) edge (v2);
\draw  (v2) edge (v3);
\draw  (v3) edge (v4);
\draw  (v4) edge (v5);
\draw  (v5) edge (v1_1);
\draw  (v1_1) edge (v1_2);
\draw  (v1_2) edge (v1_3);
\draw  (v1_1) edge (v1_4);
\begin{scope}[shift={(4.4,1.3)}]
\begin{scope}[shift={(2,0)}]
\node (w2) at (-0.5,0.9) {$\sp(0)^{(1)}$};
\end{scope}
\begin{scope}[shift={(4.4,0)}]
\node (w3) at (-0.5,0.9) {$\fe_8^{(1)}$};
\end{scope}
\draw (w2)--(w3);
\draw  (2.7,0.9) ellipse (2.3 and 0.7);
\end{scope}
\end{tikzpicture}
\ee
where each node denotes a curve in $dP_9$ whose genus is zero and self-intersection is $-2$. If there are $n$ edges between two nodes, it denotes that the two corresponding curves intersect in $n$ number of points. Each curve $C_a$ shown above is glued to the fiber $f$ of a Hirzebruch surface $S_{a}$ in the geometry associated to $\fe_8^{(1)}$. Which curve glues to the fiber of which $S_a$ can be figured out from the position of the curve in the graph above, because the graph takes the form of the corresponding Dynkin diagram, which in this case is $\fe_8^{(1)}$. Notice that 
\begin{align}
\sum_a d_a C_a=&(x_8-x_9)+2(x_7-x_8)+3(x_6-x_7)+4(x_5-x_6)+5(x_4-x_5)+6(x_1-x_4)\nn\\
&+4(x_2-x_1)+2(x_3-x_2)+3(l-x_1-x_2-x_3)\nn\\
=&\:\:3l-\sum x_i
\end{align}
and thus the torus fibers on both nodes are glued to each other, satisfying (\ref{tfgr}) for the untwisted case.

Now, we can use the above gluing rules to obtain gluing rules for regular maximal subalgebras of $\fe_8$ as follows. For example, to obtain the gluing rules for
\be\label{su2e7}
\begin{tikzpicture}
\node (w1) at (-0.5,0.9) {$\su(2)^{(1)}$};
\begin{scope}[shift={(3,0)}]
\node (w2) at (-0.5,0.9) {$\sp(0)^{(1)}$};
\end{scope}
\begin{scope}[shift={(5.5,0)}]
\node (w3) at (-0.5,0.9) {$\fe_7^{(1)}$};
\end{scope}
\draw (w1)--(w2);
\draw (w2)--(w3);
\end{tikzpicture}
\ee
we first delete the second curve from the left $x_7-x_8$ in (\ref{ee8}). After this deletion, the graph takes the form of Dynkin diagram for finite algebra $\su(2)\oplus\fe_7$. To obtain the gluing rules for (\ref{su2e7}), we simply need to add two extra $-2$ curves to the graph such that the finite Dynkin diagram of $\su(2)$ is converted to the affine Dynkin diagram of $\su(2)^{(1)}$ and similarly the finite Dynkin diagram of $\fe_7$ is converted to the affine Dynkin diagram of $\fe_7^{(1)}$. This is easy to do since we know that a weighted sum of the $-2$ curves participating in gluing to each affine Dynkin diagram must be $3l-\sum x_i$. This requirement uniquely fixes the extra $-2$ curves that need to be added. We thus obtain
\be\label{ee7su2}
\begin{tikzpicture}
\node (v1) at (6.9,-1.4) {\scriptsize{$x_8-x_9$}};
\node (v1_1) at (11.9,-1.4) {\scriptsize{$3l-x_1-x_2-x_3-x_4-x_5-x_6-x_7-2x_8$}};
\node (v2) at (16.25,0) {\scriptsize{$l-x_3-x_8-x_9$}};
\node (v3) at (4,0) {\scriptsize{$x_6-x_7$}};
\node (v4) at (6,0) {\scriptsize{$x_5-x_6$}};
\node (v5) at (8,0) {\scriptsize{$x_4-x_5$}};
\node (v1_1_1) at (10,0) {\scriptsize{$x_1-x_4$}};
\node (v1_2) at (12,0) {\scriptsize{$x_2-x_1$}};
\node (v1_3) at (14,0) {\scriptsize{$x_3-x_2$}};
\node (v1_4) at (10,1) {\scriptsize{$l-x_1-x_2-x_3$}};
\draw  (v3) edge (v4);
\draw  (v4) edge (v5);
\draw  (v5) edge (v1_1_1);
\draw  (v1_1_1) edge (v1_2);
\draw  (v1_2) edge (v1_3);
\draw  (v1_1_1) edge (v1_4);
\draw  (v1_3) edge (v2);
\draw[double]  (v1_1) -- (v1);
\begin{scope}[shift={(8.4,1.4)}]
\begin{scope}[shift={(4.4,0)}]
\node (w1) at (-0.5,0.9) {$\su(2)^{(1)}$};
\end{scope}
\begin{scope}[shift={(2,0)}]
\node (w2) at (-0.5,0.9) {$\sp(0)^{(1)}$};
\end{scope}
\begin{scope}[shift={(-0.4,0)}]
\node (w3) at (-0.5,0.9) {$\fe_7^{(1)}$};
\end{scope}
\draw (w2)--(w3);
\draw  (w1) edge (w2);
\draw  (1.6,0.9) ellipse (3.5 and 0.8);
\end{scope}
\end{tikzpicture}
\ee
as the gluing rules for (\ref{su2e7}). $l-x_3-x_8-x_9$ glues to the fiber of affine surface for $\fe_7^{(1)}$ and $x_8-x_9$ glues to the fiber of affine surface for $\su(2)^{(1)}$. Notice that the curves in each sub-Dynkin diagram sum up to $3l-\sum x_i$ if the sum is weighted by the Coxeter labels of the corresponding affine Dynkin diagram. Also notice that the curves forming the Dynkin diagram for $\fe_7^{(1)}$ do not intersect the curves forming the Dynkin diagram for $\su(2)^{(1)}$, which explicitly shows that the gluing rules for the two neighbors of $\sp(0)^{(1)}$ decouple from each other as required.

Incidentally, (\ref{ee7su2}) allows us to determined gluing rules for
\be
\begin{tikzpicture}
\begin{scope}[shift={(2,0)}]
\node (w2) at (-0.5,0.9) {$\sp(0)^{(1)}$};
\end{scope}
\begin{scope}[shift={(4.5,0)}]
\node (w3) at (-0.5,0.9) {$\fe_7^{(1)}$};
\end{scope}
\draw (w2)--(w3);
\end{tikzpicture}
\ee
and
\be
\begin{tikzpicture}
\begin{scope}[shift={(2,0)}]
\node (w2) at (-0.5,0.9) {$\sp(0)^{(1)}$};
\end{scope}
\begin{scope}[shift={(5,0)}]
\node (w3) at (-0.5,0.9) {$\su(2)^{(1)}$};
\end{scope}
\draw (w2)--(w3);
\end{tikzpicture}
\ee
without any other second neighbor for $\sp(0)^{(1)}$. This is done by only keeping the curves spanning the Dynkin diagram of $\fe_7^{(1)}$ or the Dynkin diagram of $\su(2)^{(1)}$, while omitting the rest of the curves from (\ref{ee7su2}). Thus, we obtain
\be\label{ee7}
\begin{tikzpicture}
\node (v2) at (16.25,0) {\scriptsize{$l-x_3-x_8-x_9$}};
\node (v3) at (4,0) {\scriptsize{$x_6-x_7$}};
\node (v4) at (6,0) {\scriptsize{$x_5-x_6$}};
\node (v5) at (8,0) {\scriptsize{$x_4-x_5$}};
\node (v1_1_1) at (10,0) {\scriptsize{$x_1-x_4$}};
\node (v1_2) at (12,0) {\scriptsize{$x_2-x_1$}};
\node (v1_3) at (14,0) {\scriptsize{$x_3-x_2$}};
\node (v1_4) at (10,1) {\scriptsize{$l-x_1-x_2-x_3$}};
\draw  (v3) edge (v4);
\draw  (v4) edge (v5);
\draw  (v5) edge (v1_1_1);
\draw  (v1_1_1) edge (v1_2);
\draw  (v1_2) edge (v1_3);
\draw  (v1_1_1) edge (v1_4);
\draw  (v1_3) edge (v2);
\begin{scope}[shift={(7.3,1.4)}]
\begin{scope}[shift={(2,0)}]
\node (w2) at (-0.5,0.9) {$\sp(0)^{(1)}$};
\end{scope}
\begin{scope}[shift={(4.4,0)}]
\node (w3) at (-0.5,0.9) {$\fe_7^{(1)}$};
\end{scope}
\draw (w2)--(w3);
\draw  (2.7,0.9) ellipse (2.3 and 0.7);
\end{scope}
\end{tikzpicture}
\ee
with the fiber in affine surface glued to $l-x_3-x_8-x_9$ and
\be
\begin{tikzpicture}
\node (v1) at (6.9,-1.4) {\scriptsize{$x_8-x_9$}};
\node (v1_1) at (11.9,-1.4) {\scriptsize{$3l-x_1-x_2-x_3-x_4-x_5-x_6-x_7-2x_8$}};
\draw[double]  (v1_1) -- (v1);
\begin{scope}[shift={(7.5,-0.8)}]
\begin{scope}[shift={(2,0)}]
\node (w2) at (-0.5,0.9) {$\sp(0)^{(1)}$};
\end{scope}
\begin{scope}[shift={(4.4,0)}]
\node (w3) at (-0.5,0.9) {$\su(2)^{(1)}$};
\end{scope}
\draw (w2)--(w3);
\draw  (2.7,0.9) ellipse (2.3 and 0.7);
\end{scope}
\end{tikzpicture}
\ee
with the fiber in affine surface glued to $x_8-x_9$.

Deleting other nodes from (\ref{ee8}), we can obtain the following gluing rules
\be\label{eso16}
\begin{tikzpicture}
\node (v1) at (0,0) {\scriptsize{$x_8-x_9$}};
\node (v2) at (2,0) {\scriptsize{$x_7-x_8$}};
\node (v3) at (4,0) {\scriptsize{$x_6-x_7$}};
\node (v4) at (6,0) {\scriptsize{$x_5-x_6$}};
\node (v5) at (8,0) {\scriptsize{$x_4-x_5$}};
\node (v1_1) at (10,0) {\scriptsize{$x_1-x_4$}};
\node (v1_2) at (12,0) {\scriptsize{$x_2-x_1$}};
\node (v1_3) at (2,1) {\scriptsize{$2l-x_1-x_2-x_4-x_5-x_6-x_7$}};
\node (v1_4) at (10,1) {\scriptsize{$l-x_1-x_2-x_3$}};
\draw  (v1) edge (v2);
\draw  (v2) edge (v3);
\draw  (v3) edge (v4);
\draw  (v4) edge (v5);
\draw  (v5) edge (v1_1);
\draw  (v1_1) edge (v1_2);
\draw  (v1_1) edge (v1_4);
\draw  (v1_3) edge (v2);
\begin{scope}[shift={(3.4,1.5)}]
\begin{scope}[shift={(2,0)}]
\node (w2) at (-0.5,0.9) {$\sp(0)^{(1)}$};
\end{scope}
\begin{scope}[shift={(4.4,0)}]
\node (w3) at (-0.5,0.9) {$\so(16)^{(1)}$};
\end{scope}
\draw (w2)--(w3);
\draw  (2.7,0.9) ellipse (2.3 and 0.7);
\end{scope}
\end{tikzpicture}
\ee
where $x_8-x_9$ glues to the fiber of affine surface for $\so(16)^{(1)}$.

\be
\begin{tikzpicture}
\node (v1) at (0,0) {\scriptsize{$x_8-x_9$}};
\node (v2) at (2,0) {\scriptsize{$x_7-x_8$}};
\node (v3) at (4,0) {\scriptsize{$x_6-x_7$}};
\node (v4) at (6,0) {\scriptsize{$x_5-x_6$}};
\node (v5) at (8,0) {\scriptsize{$x_4-x_5$}};
\node (v1_1) at (10,0) {\scriptsize{$x_1-x_4$}};
\node (v1_2) at (12,0) {\scriptsize{$x_2-x_1$}};
\node (v1_3) at (14,0) {\scriptsize{$x_3-x_2$}};
\node (v1_4) at (7,2) {\scriptsize{$3l-x_1-x_2-2x_3-x_4-x_5-x_6-x_7-x_8$}};
\draw  (v1) edge (v2);
\draw  (v2) edge (v3);
\draw  (v3) edge (v4);
\draw  (v4) edge (v5);
\draw  (v5) edge (v1_1);
\draw  (v1_1) edge (v1_2);
\draw  (v1_2) edge (v1_3);
\draw  (v1_4) edge (v1);
\draw  (v1_4) edge (v1_3);
\begin{scope}[shift={(4.2,2.7)}]
\begin{scope}[shift={(2,0)}]
\node (w2) at (-0.5,0.9) {$\sp(0)^{(1)}$};
\end{scope}
\begin{scope}[shift={(4.4,0)}]
\node (w3) at (-0.5,0.9) {$\su(9)^{(1)}$};
\end{scope}
\draw (w2)--(w3);
\draw  (2.7,0.9) ellipse (2.3 and 0.7);
\end{scope}
\end{tikzpicture}
\ee
where $x_8-x_9$ glues to the fiber of affine surface for $\su(9)^{(1)}$.

\be\label{ee6su3}
\begin{tikzpicture}
\node (v1) at (8,-2) {\scriptsize{$x_8-x_9$}};
\node (v1_1) at (12,-2) {\scriptsize{$x_7-x_8$}};
\node (v1_1_1) at (10,-1) {\scriptsize{$3l-x_1-x_2-x_3-x_4-x_5-x_6-2x_7-x_8$}};
\node (v2) at (10,2) {\scriptsize{$l-x_7-x_8-x_9$}};
\node (v4) at (6,0) {\scriptsize{$x_5-x_6$}};
\node (v5) at (8,0) {\scriptsize{$x_4-x_5$}};
\node (v1_1_1_1) at (10,0) {\scriptsize{$x_1-x_4$}};
\node (v1_2) at (12,0) {\scriptsize{$x_2-x_1$}};
\node (v1_3) at (14,0) {\scriptsize{$x_3-x_2$}};
\node (v1_4) at (10,1) {\scriptsize{$l-x_1-x_2-x_3$}};
\draw  (v4) edge (v5);
\draw  (v5) edge (v1_1_1_1);
\draw  (v1_1_1_1) edge (v1_2);
\draw  (v1_2) edge (v1_3);
\draw  (v1_1_1_1) edge (v1_4);
\draw  (v1_1_1) -- (v1_1);
\draw  (v1_4) edge (v2);
\draw  (v1) edge (v1_1_1);
\draw  (v1) edge (v1_1);
\begin{scope}[shift={(8.6,2.4)}]
\begin{scope}[shift={(4.4,0)}]
\node (w1) at (-0.5,0.9) {$\su(3)^{(1)}$};
\end{scope}
\begin{scope}[shift={(2,0)}]
\node (w2) at (-0.5,0.9) {$\sp(0)^{(1)}$};
\end{scope}
\begin{scope}[shift={(-0.4,0)}]
\node (w3) at (-0.5,0.9) {$\fe_6^{(1)}$};
\end{scope}
\draw (w2)--(w3);
\draw  (w1) edge (w2);
\draw  (1.6,0.9) ellipse (3.5 and 0.8);
\end{scope}
\end{tikzpicture}
\ee
where $l-x_7-x_8-x_9$ glues to the fiber of affine surface for $\fe_6^{(1)}$ and $x_8-x_9$ glues to the fiber of affine surface for $\su(3)^{(1)}$. Incidentally, this also allows us to obtain the following individual gluing rules
\be\label{ee6}
\begin{tikzpicture}
\node (v2) at (10,2) {\scriptsize{$l-x_7-x_8-x_9$}};
\node (v4) at (6,0) {\scriptsize{$x_5-x_6$}};
\node (v5) at (8,0) {\scriptsize{$x_4-x_5$}};
\node (v1_1_1_1) at (10,0) {\scriptsize{$x_1-x_4$}};
\node (v1_2) at (12,0) {\scriptsize{$x_2-x_1$}};
\node (v1_3) at (14,0) {\scriptsize{$x_3-x_2$}};
\node (v1_4) at (10,1) {\scriptsize{$l-x_1-x_2-x_3$}};
\draw  (v4) edge (v5);
\draw  (v5) edge (v1_1_1_1);
\draw  (v1_1_1_1) edge (v1_2);
\draw  (v1_2) edge (v1_3);
\draw  (v1_1_1_1) edge (v1_4);
\draw  (v1_4) edge (v2);
\begin{scope}[shift={(7.4,2.4)}]
\begin{scope}[shift={(2,0)}]
\node (w2) at (-0.5,0.9) {$\sp(0)^{(1)}$};
\end{scope}
\begin{scope}[shift={(4.4,0)}]
\node (w3) at (-0.5,0.9) {$\fe_6^{(1)}$};
\end{scope}
\draw (w2)--(w3);
\draw  (2.7,0.9) ellipse (2.3 and 0.7);
\end{scope}
\end{tikzpicture}
\ee
with the fiber in affine surface glued to $l-x_7-x_8-x_9$, and
\be\label{esu3}
\begin{tikzpicture}
\node (v1) at (8,-2) {\scriptsize{$x_8-x_9$}};
\node (v1_1) at (12,-2) {\scriptsize{$x_7-x_8$}};
\node (v1_1_1) at (10,-1) {\scriptsize{$3l-x_1-x_2-x_3-x_4-x_5-x_6-2x_7-x_8$}};
\draw  (v1_1_1) -- (v1_1);
\draw  (v1) edge (v1_1_1);
\draw  (v1) edge (v1_1);
\begin{scope}[shift={(7.3,-0.5)}]
\begin{scope}[shift={(2,0)}]
\node (w2) at (-0.5,0.9) {$\sp(0)^{(1)}$};
\end{scope}
\begin{scope}[shift={(4.4,0)}]
\node (w3) at (-0.5,0.9) {$\su(3)^{(1)}$};
\end{scope}
\draw (w2)--(w3);
\draw  (2.7,0.9) ellipse (2.3 and 0.7);
\end{scope}
\end{tikzpicture}
\ee
with the fiber in affine surface glued to $x_8-x_9$.

Now we can delete some nodes from the above set of gluing rules to obtain gluing rules for other algebras that arise as regular maximal subalgebras of the above algebras. For example, by deleting nodes from (\ref{eso16}), we can obtain the gluing rules for
\be
\begin{tikzpicture}
\node (w1) at (-0.5,0.9) {$\so(8)^{(1)}$};
\begin{scope}[shift={(3,0)}]
\node (w2) at (-0.5,0.9) {$\sp(0)^{(1)}$};
\end{scope}
\begin{scope}[shift={(6,0)}]
\node (w3) at (-0.5,0.9) {$\so(8)^{(1)}$};
\end{scope}
\draw (w1)--(w2);
\draw (w2)--(w3);
\end{tikzpicture}
\ee
since $\so(8)\oplus\so(8)$ is a regular maximal subalgebra of $\so(16)$. The gluing rules are
\be\label{eso8so8}
\begin{tikzpicture}
\node (v1) at (8,3.5) {\scriptsize{$x_8-x_9$}};
\node (v1_1) at (11,3.5) {\scriptsize{$x_7-x_8$}};
\node (v1_1_1) at (11,5) {\scriptsize{$2l-x_1-x_2-x_4-x_5-x_6-x_7$}};
\node (v2) at (11,-2) {\scriptsize{$2l-x_1-x_2-x_6-x_7-x_8-x_9$}};
\node (v4) at (14,3.5) {\scriptsize{$x_6-x_7$}};
\node (v5) at (8,-0.5) {\scriptsize{$x_4-x_5$}};
\node (v1_1_1_1) at (11,-0.5) {\scriptsize{$x_1-x_4$}};
\node (v1_2) at (14,-0.5) {\scriptsize{$x_2-x_1$}};
\node (v1_3) at (11,2) {\scriptsize{$l-x_3-x_6-x_7$}};
\node (v1_4) at (11,1) {\scriptsize{$l-x_1-x_2-x_3$}};
\draw  (v5) edge (v1_1_1_1);
\draw  (v1_1_1_1) edge (v1_2);
\draw  (v1_1_1_1) edge (v1_4);
\draw  (v1) edge (v1_1);
\draw  (v1_1) edge (v4);
\draw  (v1_1_1_1) edge (v2);
\draw  (v1_1_1) edge (v1_1);
\draw  (v1_1) edge (v1_3);
\begin{scope}[shift={(9.3,5.5)}]
\begin{scope}[shift={(4.4,0)}]
\node (w1) at (-0.5,0.9) {$\so(8)^{(1)}$};
\end{scope}
\begin{scope}[shift={(2,0)}]
\node (w2) at (-0.5,0.9) {$\sp(0)^{(1)}$};
\end{scope}
\begin{scope}[shift={(-0.4,0)}]
\node (w3) at (-0.5,0.9) {$\so(8)^{(1)}$};
\end{scope}
\draw (w2)--(w3);
\draw  (w1) edge (w2);
\draw  (1.6,0.9) ellipse (3.5 and 0.8);
\end{scope}
\end{tikzpicture}
\ee
where the fibers in affine surfaces glue to $x_8-x_9$ and $2l-x_1-x_2-x_6-x_7-x_8-x_9$. Tha bove gluing rules imply that the gluing rules for a single $\so(8)^{(1)}$ are obtained by amputating one of the $\so(8)^{(1)}$ sub-graph from (\ref{eso8so8}).
\be\label{eso81}
\begin{tikzpicture}
\node (v1) at (8,3.5) {\scriptsize{$x_8-x_9$}};
\node (v1_1) at (11,3.5) {\scriptsize{$x_7-x_8$}};
\node (v1_1_1) at (11,5) {\scriptsize{$2l-x_1-x_2-x_4-x_5-x_6-x_7$}};
\node (v4) at (14,3.5) {\scriptsize{$x_6-x_7$}};
\node (v1_3) at (11,2) {\scriptsize{$l-x_3-x_6-x_7$}};
\draw  (v1) edge (v1_1);
\draw  (v1_1) edge (v4);
\draw  (v1_1_1) edge (v1_1);
\draw  (v1_1) edge (v1_3);
\begin{scope}[shift={(8.2,5.5)}]
\begin{scope}[shift={(2,0)}]
\node (w2) at (-0.5,0.9) {$\sp(0)^{(1)}$};
\end{scope}
\begin{scope}[shift={(4.4,0)}]
\node (w3) at (-0.5,0.9) {$\so(8)^{(1)}$};
\end{scope}
\draw (w2)--(w3);
\draw  (2.7,0.9) ellipse (2.3 and 0.7);
\end{scope}
\end{tikzpicture}
\ee
with the fiber in affine surface glued to $x_8-x_9$. The reader might wonder what happens if amputate the other $\so(8)^{(1)}$ sub-graph from (\ref{eso8so8}) to obtain the gluing rules as
\be\label{eso82}
\begin{tikzpicture}
\node (v2) at (11,-2) {\scriptsize{$2l-x_1-x_2-x_6-x_7-x_8-x_9$}};
\node (v5) at (8,-0.5) {\scriptsize{$x_4-x_5$}};
\node (v1_1_1_1) at (11,-0.5) {\scriptsize{$x_1-x_4$}};
\node (v1_2) at (14,-0.5) {\scriptsize{$x_2-x_1$}};
\node (v1_4) at (11,1) {\scriptsize{$l-x_1-x_2-x_3$}};
\draw  (v5) edge (v1_1_1_1);
\draw  (v1_1_1_1) edge (v1_2);
\draw  (v1_1_1_1) edge (v1_4);
\draw  (v1_1_1_1) edge (v2);
\end{tikzpicture}
\ee
It turns out that (\ref{eso81}) and (\ref{eso82}) are related by an automorphism of $dP_9$. To see this, let's first relabel the blowups as
\begin{align}
x_1&\lra x_7\\
x_2&\lra x_6\\
x_3&\lra x_5\\
x_4&\lra x_8
\end{align}
so that (\ref{eso81}) is converted to
\be \label{eso83}
\begin{tikzpicture}
\node (v1) at (8,3.5) {\scriptsize{$x_4-x_9$}};
\node (v1_1) at (11,3.5) {\scriptsize{$x_1-x_4$}};
\node (v1_1_1) at (11,5) {\scriptsize{$2l-x_1-x_2-x_3-x_6-x_7-x_8$}};
\node (v4) at (14,3.5) {\scriptsize{$x_2-x_1$}};
\node (v1_3) at (11,2) {\scriptsize{$l-x_1-x_2-x_5$}};
\draw  (v1) edge (v1_1);
\draw  (v1_1) edge (v4);
\draw  (v1_1_1) edge (v1_1);
\draw  (v1_1) edge (v1_3);
\end{tikzpicture}
\ee
Now we perform two basic automorphisms of $dP_9$. The basic automorphisms are described in Appendix \ref{DP} and involve a choice of three blowups. For the first basic automorphism we choose the blowups $x_1$, $x_2$ and $x_4$, and after performing this operation the gluing rules (\ref{eso83}) are transformed to
\be\label{eso84}
\begin{tikzpicture}
\node (v1) at (8,3.5) {\scriptsize{$l-x_1-x_2-x_9$}};
\node (v1_1) at (11,3.5) {\scriptsize{$x_1-x_4$}};
\node (v1_1_1) at (11,5) {\scriptsize{$2l-x_1-x_2-x_3-x_6-x_7-x_8$}};
\node (v4) at (14,3.5) {\scriptsize{$x_2-x_1$}};
\node (v1_3) at (11,2) {\scriptsize{$x_4-x_5$}};
\draw  (v1) edge (v1_1);
\draw  (v1_1) edge (v4);
\draw  (v1_1_1) edge (v1_1);
\draw  (v1_1) edge (v1_3);
\end{tikzpicture}
\ee
For the second basic automorphism we choose $x_6$, $x_7$ and $x_8$ thus transforming (\ref{eso84}) to
\be\label{eso85}
\begin{tikzpicture}
\node (v1) at (6.5,3.5) {\scriptsize{$2l-x_1-x_2-x_6-x_7-x_8-x_9$}};
\node (v1_1) at (11,3.5) {\scriptsize{$x_1-x_4$}};
\node (v1_1_1) at (11,5) {\scriptsize{$l-x_1-x_2-x_3$}};
\node (v4) at (14,3.5) {\scriptsize{$x_2-x_1$}};
\node (v1_3) at (11,2) {\scriptsize{$x_4-x_5$}};
\draw  (v1) edge (v1_1);
\draw  (v1_1) edge (v4);
\draw  (v1_1_1) edge (v1_1);
\draw  (v1_1) edge (v1_3);
\end{tikzpicture}
\ee
which precisely matches (\ref{eso82}), thus demonstrating that (\ref{eso81}) and (\ref{eso82}) are isomorphic gluing rules.

This will hold true in general in what follows. Whenever we will find two seemingly different gluing rules, they will always turn out to be related by an automorphism, except for two cases. These two cases are the gluing rules for $\su(8)^{(1)}$ and $\su(8)^{(2)}$, where we find two possible gluing rules in each case. The two possibilities correspond to different choices of theta angle for $\sp(0)$ in the $6d$ theory.

Let us collect all of the remaining gluing rules below
\be

\ee
where the fibers in affine surfaces glue to $x_8-x_9$ and $2l-x_2-x_3-x_6-x_7-x_8-x_9$.
\be
%
\ee
where the fiber in affine surface glues to $x_8-x_9$.
\be
%
\ee
where the fibers in affine surfaces glue to $x_8-x_9$ and $l-x_7-x_8-x_9$.
\be
%
\ee
where the fibers in affine surfaces glue to $x_8-x_9$ and $2l-x_1-x_2-x_4-x_5-x_8-x_9$.
\be
%
\ee
where the fibers in affine surfaces glue to $x_8-x_9$ and $2l-x_2-x_3-x_6-x_7-x_8-x_9$.
\be
%
\ee
where the fibers in affine surfaces glue to $x_8-x_9$ and $2l-x_2-x_3-x_6-x_7-x_8-x_9$.
\be\label{eso12}
%
\ee
where the fiber in affine surface glues to $2l-x_1-x_2-x_4-x_5-x_8-x_9$.
\be
%
\ee
where the fiber in affine surface glues to $x_8-x_9$.
\be
%
\ee
where the fibers in affine surfaces glue to $x_8-x_9$ and $2l-x_2-x_3-x_6-x_7-x_8-x_9$.
\be\label{eso10}
%
\ee
where the fiber in affine surface glues to $2l-x_2-x_3-x_6-x_7-x_8-x_9$.
\be
%
\ee
where the fiber in affine surface glues to $x_8-x_9$.
\be
%
\ee
where the fibers in affine surfaces glue to $x_8-x_9$ and $3l-x_1-x_2-2x_3-x_4-x_5-x_7-x_8-x_9$.
\be
%
\ee
where the fibers in affine surfaces glue to $x_8-x_9$ and $3l-x_1-x_2-2x_3-x_4-x_5-x_7-x_8-x_9$.
\be
%
\ee
where the fibers in affine surfaces glue to $x_8-x_9$ and $3l-x_1-x_2-2x_3-x_4-x_5-x_7-x_8-x_9$.
\be
%
\ee
where the fiber in affine surface glues to $x_8-x_9$.

Finally, we come to the gluing rules for $\su(8)^{(1)}$ for which we have two versions depending on the choice of theta angle for $\sp(0)$. The adjoint of $\fe_8$ decomposes into the adjoint plus an irreducible spinor of $\so(16)$. In our study, this spinor corresponds to the node of $\so(16)$ Dynkin diagram whose corresponding fiber is glued to $x_2-x_1$ in (\ref{eso16}). This is visible from the gluing rules (\ref{ee8}) for $\fe_8^{(1)}$ since the extra particles in adjoint of $\fe_8$ come from the curve $x_3-x_2$ which indeed transform in the spinor of $\so(16)$ associated to $x_2-x_1$ since $x_3-x_2$ intersects $x_2-x_1$.

Now, to obtain the gluing rules for $\su(8)^{(1)}$, we delete $2l-x_1-x_2-x_4-x_5-x_6-x_7$ from (\ref{eso16}), and we have the choice to either delete $l-x_1-x_2-x_3$ or $x_2-x_1$. This latter choice leads to another choice of spinor of $\so(16)$. If we delete $x_2-x_1$, then this matches the previous choice of spinor we had, and leads to the gluing rules for $\theta=0$. If we delete $l-x_1-x_2-x_3$, then this does not match the previous choice of spinor we had, and leads to the gluing rules for $\theta=\pi$. In the latter case, $\su(8)$ gauges the spinor of $\so(16)$ in the adjoint of $\fe_8$, and in the former case it does not. Thus the latter case has less global symmetry compared to former. We refer the reader to \cite{Mekareeya:2017jgc} for more details. The two gluing rules are thus as follows:
\be
\begin{tikzpicture}
\node (v1) at (0,0) {\scriptsize{$x_8-x_9$}};
\node (v2) at (2,0) {\scriptsize{$x_7-x_8$}};
\node (v3) at (4,0) {\scriptsize{$x_6-x_7$}};
\node (v4) at (6,0) {\scriptsize{$x_5-x_6$}};
\node (v5) at (8,0) {\scriptsize{$x_4-x_5$}};
\node (v1_1) at (10,0) {\scriptsize{$x_1-x_4$}};
\node (v1_2) at (12,0) {\scriptsize{$x_2-x_1$}};
\node (v1_4) at (6,2) {\scriptsize{$3l-x_1-2x_2-x_3-x_4-x_5-x_6-x_7-x_8$}};
\draw  (v1) edge (v2);
\draw  (v2) edge (v3);
\draw  (v3) edge (v4);
\draw  (v4) edge (v5);
\draw  (v5) edge (v1_1);
\draw  (v1_1) edge (v1_2);
\draw  (v1_4) edge (v1);
\draw  (v1_4) edge (v1_2);
\begin{scope}[shift={(3.3,2.6)}]
\begin{scope}[shift={(2,0)}]
\node (w2) at (-0.5,0.9) {$\sp(0)_\pi^{(1)}$};
\end{scope}
\begin{scope}[shift={(4.4,0)}]
\node (w3) at (-0.5,0.9) {$\su(8)^{(1)}$};
\end{scope}
\draw (w2)--(w3);
\draw  (2.7,0.9) ellipse (2.3 and 0.7);
\end{scope}
\end{tikzpicture}
\ee
\be
\begin{tikzpicture}
\node (v1) at (0,0) {\scriptsize{$x_8-x_9$}};
\node (v2) at (2,0) {\scriptsize{$x_7-x_8$}};
\node (v3) at (4,0) {\scriptsize{$x_6-x_7$}};
\node (v4) at (6,0) {\scriptsize{$x_5-x_6$}};
\node (v5) at (8,0) {\scriptsize{$x_4-x_5$}};
\node (v1_1) at (10,0) {\scriptsize{$x_1-x_4$}};
\node (v1_2) at (12.4,0) {\scriptsize{$l-x_1-x_2-x_3$}};
\node (v1_4) at (6,2) {\scriptsize{$2l-x_1-x_4-x_5-x_6-x_7-x_8$}};
\draw  (v1) edge (v2);
\draw  (v2) edge (v3);
\draw  (v3) edge (v4);
\draw  (v4) edge (v5);
\draw  (v5) edge (v1_1);
\draw  (v1_1) edge (v1_2);
\draw  (v1_4) edge (v1);
\draw  (v1_4) edge (v1_2);
\begin{scope}[shift={(3.3,2.6)}]
\begin{scope}[shift={(2,0)}]
\node (w2) at (-0.5,0.9) {$\sp(0)_0^{(1)}$};
\end{scope}
\begin{scope}[shift={(4.4,0)}]
\node (w3) at (-0.5,0.9) {$\su(8)^{(1)}$};
\end{scope}
\draw (w2)--(w3);
\draw  (2.7,0.9) ellipse (2.3 and 0.7);
\end{scope}
\end{tikzpicture}
\ee
In both the cases, the fiber in affine surface glues to $x_8-x_9$.

\subsubsection{$\sp(0)^{(1)}$ gluings: untwisted, with non-simply-laced}

Until now, we have only considered simply laced subalgebras of $\fe_8$. To generalize our gluing rules to non-simply laced subalgebras of $\fe_8$, we use the folding of Dynkin diagrams. The Dynkin diagrams for untwisted affine non-simply laced algebras can be produced by folding the Dynkin diagrams for untwisted affine simply laced algebras. The foldings relevant in our analysis are:
\begin{align}
\so(2n)^{(1)}&\to\so(2n-1)^{(1)}\\
\fe_6^{(1)}&\to\ff_4^{(1)}\\
\so(8)^{(1)}&\to\so(7)^{(1)}\to\fg_2^{(1)}
\end{align}
For example, to obtain the gluing rules for
\be
\begin{tikzpicture}
\begin{scope}[shift={(2,0)}]
\node (w2) at (-0.5,0.9) {$\sp(0)^{(1)}$};
\end{scope}
\begin{scope}[shift={(5,0)}]
\node (w3) at (-0.5,0.9) {$\so(15)^{(1)}$};
\end{scope}
\draw (w2)--(w3);
\end{tikzpicture}
\ee
we simply fold the graph (\ref{eso16}) to obtain
\be\label{eso15}
\begin{tikzpicture}
\node (v1) at (0,0) {\scriptsize{$x_8-x_9$}};
\node (v2) at (2,0) {\scriptsize{$x_7-x_8$}};
\node (v3) at (4,0) {\scriptsize{$x_6-x_7$}};
\node (v4) at (6,0) {\scriptsize{$x_5-x_6$}};
\node (v5) at (8,0) {\scriptsize{$x_4-x_5$}};
\node (v1_1) at (10,0) {\scriptsize{$x_1-x_4$}};
\node (v1_2) at (13,0) {\scriptsize{$x_2-x_1$,\:\:$l-x_1-x_2-x_3$}};
\node (v1_3) at (2,1) {\scriptsize{$2l-x_1-x_2-x_4-x_5-x_6-x_7$}};
\draw  (v1) edge (v2);
\draw  (v2) edge (v3);
\draw  (v3) edge (v4);
\draw  (v4) edge (v5);
\draw  (v5) edge (v1_1);
\draw[double]  (v1_1) -- (v1_2);
\draw  (v1_3) edge (v2);
\begin{scope}[shift={(3.3,1.5)}]
\begin{scope}[shift={(2,0)}]
\node (w2) at (-0.5,0.9) {$\sp(0)^{(1)}$};
\end{scope}
\begin{scope}[shift={(4.4,0)}]
\node (w3) at (-0.5,0.9) {$\so(15)^{(1)}$};
\end{scope}
\draw (w2)--(w3);
\draw  (2.7,0.9) ellipse (2.3 and 0.7);
\end{scope}
\end{tikzpicture}
\ee
where the fiber in affine surface glues to $x_8-x_9$ and the rightmost node denotes two $-2$ curves $x_2-x_1$ and $l-x_1-x_2-x_3$. Both of these curves glue to a copy of the fiber of the corresponding surface in the geometry for $\so(15)^{(1)}$. We can check that the weighted sum of fibers equals $3l-\sum x_i$.

Since we can now have multiple gluing curves associated to the gluing of $dP_9$ to some other surface, we have to make sure that all of the gluing curves are on an equal footing. More precisely, we have to make sure that the condition (\ref{CYP}) is satisfied, which translates to the following condition. Let $S_a$ be the different surfaces $dP_9$ is glued to, and let $C^i_a$ be the different gluing curves in $dP_9$ for the gluing to $S_a$. The \emph{total} gluing curve for the gluing to $S_a$ is
\be
C_a:=\sum_i C^i_a
\ee
Then (\ref{CYP}) translates to the condition that
\be\label{CYE}
C^i_a\cdot C_b = C^j_a\cdot C_b
\ee
for all $i,j,a,b$. It can be easily verified that (\ref{eso15}) satisfies this condition. This condition (\ref{CYE}) will be an important consistency condition in what follows and the reader can verify that all of the geometries that follow satisfy (\ref{CYE}).

By folding other gluing rules presented above, we can obtain the following gluing rules
\be

\ee
where the fibers in affine surfaces glue to $x_8-x_9$ and $2l-x_2-x_3-x_6-x_7-x_8-x_9$.
\be\label{eso13}
%
\ee
where the fiber in affine surface glues to $x_8-x_9$.
\be
%
\ee
where the fibers in affine surfaces glue to $x_8-x_9$ and $l-x_7-x_8-x_9$.
\be
%
\ee
where the fibers in affine surfaces glue to $x_8-x_9$ and $2l-x_1-x_2-x_6-x_7-x_8-x_9$.
\be\label{eso7so7}
%
\ee
where the fibers in affine surfaces glue to $x_8-x_9$ and $2l-x_1-x_2-x_6-x_7-x_8-x_9$.
\be
%
\ee
where the fibers in affine surfaces glue to $x_8-x_9$ and $2l-x_1-x_2-x_6-x_7-x_8-x_9$.
\be
%
\ee
where the fibers in affine surfaces glue to $x_8-x_9$ and $2l-x_1-x_2-x_4-x_5-x_8-x_9$.
\be
%
\ee
where the fiber in affine surface glues to $2l-x_1-x_2-x_4-x_5-x_8-x_9$.
\be
%
\ee
where the fibers in affine surfaces glue to $x_8-x_9$ and $l-x_7-x_8-x_9$.
\be
%
\ee
where the fibers in affine surfaces glue to $x_8-x_9$ and $3l-x_1-2x_2-x_3-x_4-x_6-x_7-x_8-x_9$.
\be
%
\ee
where the fiber in affine surface glues to $2l-x_2-x_3-x_6-x_7-x_8-x_9$.
\be\label{ef4}
%
\ee
where the fiber in affine surface glues to $l-x_7-x_8-x_9$.
\be
%
\ee
where the fibers in affine surfaces glue to $x_8-x_9$ and $3l-2x_1-x_2-x_3-x_4-x_6-x_7-x_8-x_9$.
\be
%
\ee
where the fiber in affine surface glues to $2l-x_2-x3-x_6-x_7-x_8-x_9$.
\be
%
\ee
where the fibers in affine surfaces glue to $x_8-x_9$ and $3l-x_1-x_2-x_3-2x_4-x_6-x_7-x_8-x_9$.
\be\label{efg2}
%
\ee
where the fiber in affine surface glues to $x_8-x_9$.

The above cases do not completely exhaust all the possible non-simply laced subalgebras of $\fe_8$. Some of these subalgebras cannot be thought of as foldings of simply laced subalgebras of $\fe_8$. One such example is $\ff_4\oplus\fg_2$. Notice that unfolding $\ff_4^{(1)}\oplus\fg_2^{(1)}$ leads to $\fe_6^{(1)}\oplus\so(8)^{(1)}$, but $\fe_6\oplus\so(8)$ is not a subalgebra of $\fe_8$. To obtain the gluing rules for this example, we find a collection of curves giving rise to $\fg_2^{(1)}$ not intersecting (\ref{ef4}) and satisfying (\ref{CYE}):
\be\label{ef4g2}
\begin{tikzpicture}
\node (v2) at (5.5,6.5) {\scriptsize{$l-x_7-x_8-x_9$}};
\node (v1_1_1_1) at (11,6.5) {\scriptsize{$x_1-x_4$}};
\node (v1_2) at (13.5,6.5) {\scriptsize{$x_2-x_1$,\:\:$x_4-x_5$}};
\node (v1_3) at (16.5,6.5) {\scriptsize{$x_3-x_2$,\:\:$x_5-x_6$}};
\node (v1_4) at (8.5,6.5) {\scriptsize{$l-x_1-x_2-x_3$}};
\draw [double] (v1_1_1_1) -- (v1_2);
\draw[double]  (v1_2) -- (v1_3);
\draw  (v1_1_1_1) edge (v1_4);
\draw  (v1_4) edge (v2);
\node (v1) at (6,5.3) {\scriptsize{$x_8-x_9$}};
\node (v1_1) at (8,5.3) {\scriptsize{$x_7-x_8$}};
\node (v4_1) at (13,5.3) {\scriptsize{$l-x_1-x_4-x_7$,\:\:$l-x_2-x_5-x_7$,\:\:$l-x_3-x_6-x_7$}};
\draw  (v1) edge (v1_1);
\draw (8.58,5.3) -- (9.54,5.3);
\draw (8.58,5.26) -- (9.54,5.26);
\draw (8.58,5.34) -- (9.54,5.34);
\begin{scope}[shift={(9.4,7.5)}]
\begin{scope}[shift={(4.4,0)}]
\node (w1) at (-0.5,0.9) {$\fg_2^{(1)}$};
\end{scope}
\begin{scope}[shift={(2,0)}]
\node (w2) at (-0.5,0.9) {$\sp(0)^{(1)}$};
\end{scope}
\begin{scope}[shift={(-0.4,0)}]
\node (w3) at (-0.5,0.9) {$\ff_4^{(1)}$};
\end{scope}
\draw (w2)--(w3);
\draw  (w1) edge (w2);
\draw  (1.6,0.9) ellipse (3.5 and 0.8);
\end{scope}
\end{tikzpicture}
\ee
where the fibers in affine surfaces glue to $x_8-x_9$ and $l-x_7-x_8-x_9$. Notice that even though, by the virtue of (\ref{CYE}), the total gluing curves see different component gluing curves equally, the different components do not. For example, even though the gluing curve $x_3-x_2$ has different intersections with the gluing curves $l-x_2-x_5-x_7$ and $l-x_1-x_4-x_7$, the total gluing curve $(x_3-x_2)+(x_5-x_6)$ equal intersections with the two gluing curves $l-x_2-x_5-x_7$ and $l-x_1-x_4-x_7$, as required by (\ref{CYE}). Similar remarks apply to many of the gluing rules that follow.

To obtain the gluing rules for $\so(9)\oplus\so(7)$, we start from (\ref{eso7so7}) and extend the chains for one of the $\so(7)$:
\be
\begin{tikzpicture}
\node (v1) at (8,3.5) {\scriptsize{$x_8-x_9$}};
\node (v1_1) at (11,3.5) {\scriptsize{$x_7-x_8$}};
\node (v1_1_1) at (11,5) {\scriptsize{$2l-x_1-x_2-x_4-x_5-x_6-x_7$}};
\node (v2) at (11.8,2.5) {\scriptsize{$2l-x_1-x_2-x_6-x_7-x_8-x_9$}};
\node (v4) at (14,3.5) {\scriptsize{$x_6-x_7$}};
\node (v5) at (8.8,1) {\scriptsize{$x_4-x_5$}};
\node (v1_1_1_1) at (11.8,1) {\scriptsize{$x_1-x_4$}};
\node (v1_2) at (15.8,1) {\scriptsize{$x_2-x_1$,\:\:$l-x_1-x_2-x_3$}};
\node (v1_3) at (17.5,3.5) {\scriptsize{$x_2-x_6$,\:\:$l-x_2-x_3-x_6$}};
\draw  (v5) edge (v1_1_1_1);
\draw[double]  (v1_1_1_1) -- (v1_2);
\draw  (v1) edge (v1_1);
\draw  (v1_1) edge (v4);
\draw  (v1_1_1_1) edge (v2);
\draw  (v1_1_1) edge (v1_1);
\draw[double]  (v4) -- (v1_3);
\begin{scope}[shift={(11.2,5.8)}]
\begin{scope}[shift={(4.4,0)}]
\node (w1) at (-0.5,0.9) {$\so(7)^{(1)}$};
\end{scope}
\begin{scope}[shift={(2,0)}]
\node (w2) at (-0.5,0.9) {$\sp(0)^{(1)}$};
\end{scope}
\begin{scope}[shift={(-0.4,0)}]
\node (w3) at (-0.5,0.9) {$\so(9)^{(1)}$};
\end{scope}
\draw (w2)--(w3);
\draw  (w1) edge (w2);
\draw  (1.6,0.9) ellipse (3.5 and 0.8);
\end{scope}
\end{tikzpicture}
\ee
where the fibers in affine surfaces glue to $x_8-x_9$ and $2l-x_1-x_2-x_6-x_7-x_8-x_9$.

By folding $\so(7)^{(1)}$ we can obtain $\fg_2^{(1)}$, so folding the above gluing rules we obtain the following gluing rules
\be
\begin{tikzpicture}
\node (v1) at (8,3.5) {\scriptsize{$x_8-x_9$}};
\node (v1_1) at (11,3.5) {\scriptsize{$x_7-x_8$}};
\node (v1_1_1) at (11,5) {\scriptsize{$2l-x_1-x_2-x_4-x_5-x_6-x_7$}};
\node (v4) at (14,3.5) {\scriptsize{$x_6-x_7$}};
\node (v5) at (9.2,2.1) {\scriptsize{$2l-x_1-x_2-x_6-x_7-x_8-x_9$}};
\node (v1_1_1_1) at (13.4,2.1) {\scriptsize{$x_1-x_4$}};
\node (v1_2) at (17.9,2.1) {\scriptsize{$x_2-x_1$,\:\:$l-x_1-x_2-x_3$,\:\:$x_4-x_5$}};
\node (v1_3) at (17.5,3.5) {\scriptsize{$x_2-x_6$,\:\:$l-x_2-x_3-x_6$}};
\draw  (v5) edge (v1_1_1_1);
\draw  (v1) edge (v1_1);
\draw  (v1_1) edge (v4);
\draw  (v1_1_1) edge (v1_1);
\draw[double]  (v4) -- (v1_3);
\begin{scope}[shift={(11.3,5.7)}]
\begin{scope}[shift={(4.4,0)}]
\node (w1) at (-0.5,0.9) {$\fg_2^{(1)}$};
\end{scope}
\begin{scope}[shift={(2,0)}]
\node (w2) at (-0.5,0.9) {$\sp(0)^{(1)}$};
\end{scope}
\begin{scope}[shift={(-0.4,0)}]
\node (w3) at (-0.5,0.9) {$\so(9)^{(1)}$};
\end{scope}
\draw (w2)--(w3);
\draw  (w1) edge (w2);
\draw  (1.6,0.9) ellipse (3.5 and 0.8);
\end{scope}
\draw (13.95,2.15) -- (15.4,2.15) (15.4,2.1) -- (13.95,2.1) (13.95,2.05) -- (15.4,2.05);
\end{tikzpicture}
\ee
where the fibers in affine surfaces glue to $x_8-x_9$ and $2l-x_1-x_2-x_6-x_7-x_8-x_9$.

\subsubsection{$\sp(0)^{(1)}$ gluings: twisted algebras, undirected edges}
Now we provide gluing rules for the cases involving twisted gauge algebras and undirected edges, that is gluing rules of the form 
\be
\begin{tikzpicture}
\node (w1) at (-0.5,0.9) {$\fg_\alpha^{(q_\alpha)}$};
\begin{scope}[shift={(2,0)}]
\node (w2) at (-0.5,0.9) {$\sp(0)^{(1)}$};
\end{scope}
\begin{scope}[shift={(4,0)}]
\node (w3) at (-0.5,0.9) {$\fg_\gamma^{(q_\gamma)}$};
\end{scope}
\draw (w1)--(w2);
\draw (w2)--(w3);
\end{tikzpicture}
\ee
Most of these gluing rules can be understood as foldings of gluing rules of the form
\be
\begin{tikzpicture}
\node (w1) at (-0.5,0.9) {$\fg_\alpha^{(1)}$};
\begin{scope}[shift={(2,0)}]
\node (w2) at (-0.5,0.9) {$\sp(0)^{(1)}$};
\end{scope}
\begin{scope}[shift={(4,0)}]
\node (w3) at (-0.5,0.9) {$\fg_\gamma^{(1)}$};
\end{scope}
\draw (w1)--(w2);
\draw (w2)--(w3);
\end{tikzpicture}
\ee
provided above. The relevant foldings are
\begin{align}
\so(4n)^{(1)}&\to\su(2n)^{(2)}\to\su(2n-1)^{(2)}\\
\so(7)^{(1)}&\to\su(4)^{(2)}\to\su(3)^{(2)}\\
\so(2n+1)^{(1)}&\to\so(2n)^{(2)}\\
\fg_2^{(1)}&\to\su(3)^{(2)}\\
\fe_7^{(1)}&\to\fe_6^{(2)}\\
\ff_4^{(1)}&\to\so(8)^{(3)}
\end{align}
For example, for $\so(14)^{(2)}$, we fold (\ref{eso15}) to obtain
\be\label{eso14t}
\begin{tikzpicture}
\node (v1) at (0,0) {\scriptsize{$x_8-x_9,$}};
\node (v2) at (2,0) {\scriptsize{$x_7-x_8$}};
\node (v3) at (4,0) {\scriptsize{$x_6-x_7$}};
\node (v4) at (6,0) {\scriptsize{$x_5-x_6$}};
\node (v5) at (8,0) {\scriptsize{$x_4-x_5$}};
\node (v1_1) at (10,0) {\scriptsize{$x_1-x_4$}};
\node (v1_2) at (12,0) {\scriptsize{$x_2-x_1$,}};
\draw[double]  (v1) -- (v2);
\draw  (v2) edge (v3);
\draw  (v3) edge (v4);
\draw  (v4) edge (v5);
\draw  (v5) edge (v1_1);
\draw[double]  (v1_1) -- (v1_2);
\node at (-0.2,-0.4) {\scriptsize$2l-x_1-x_2-x_4$};
\node at (12,-0.4) {\scriptsize$l-x_1-x_2-x_3$};
\node at (-0.4,-0.7) {\scriptsize{$-x_5-x_6-x_7$}};
\begin{scope}[shift={(3.2,0.8)}]
\begin{scope}[shift={(2,0)}]
\node (w2) at (-0.5,0.9) {$\sp(0)^{(1)}$};
\end{scope}
\begin{scope}[shift={(4.4,0)}]
\node (w3) at (-0.5,0.9) {$\so(14)^{(2)}$};
\end{scope}
\draw (w2)--(w3);
\draw  (2.7,0.9) ellipse (2.3 and 0.7);
\end{scope}
\end{tikzpicture}
\ee
where two copies of fibers in affine surface glue to $x_8-x_9$, $2l-x_1-x_2-x_4-x_5-x_6-x_7$. Let $d_a$ be the dual Coxeter labels for $\so(14)^{(2)}$ and $f_a$ be the fibers in the Hirzebruch surfaces corresponding to $\so(14)^{(2)}$. Then,
\begin{align}
2d_af_a=&(x_8-x_9)+(2l-x_1-x_2-x_4-x_5-x_6-x_7)+2(x_7-x_8)+2(x_6-x_7)+2(x_5-x_6)\nn\\
&+2(x_4-x_5)+2(x_1-x_4)+(x_2-x_1)+(l-x_1-x_2-x_3)\nn\\
=&3l-\sum x_i
\end{align}
Thus, (\ref{tfgr}) holds true in this case. Same holds true for all the following examples in this subsection, as the reader can verify.

To obtain other $\so(2n)^{(2)}$ of lower rank, we add the curves lying in the middle of the chain in (\ref{eso14t}). Adding $x_4-x_5$ to $x_1`-x_4$, we obtain the gluing rules for $\so(12)^{(2)}$:
\be

\ee
where $x_8-x_9$, $2l-x_1-x_2-x_4-x_5-x_6-x_7$ glue to fibers in affine surface.

Continuing in this fashion, we obtain
\be
%
\ee
where $x_8-x_9$, $2l-x_1-x_2-x_4-x_5-x_6-x_7$ glue to fibers in affine surface.
\be
%
\ee
where $x_8-x_9$, $2l-x_1-x_2-x_4-x_5-x_6-x_7$ glue to fibers in affine surface.
\be
%
\ee
where $x_8-x_9$, $2l-x_1-x_2-x_4-x_5-x_6-x_7$ glue to fibers in affine surface.

By folding (\ref{eso16}), we obtain the following two gluing rules
\be\label{esu8t0}
%
\ee
where $x_8-x_9$, $x_2-x_1$ glue to fibers in the affine surface.
\be
%
\ee
where $x_8-x_9$, $l-x_1-x_2-x_3$ glue to fibers in the affine surface.

Combining $x_6-x_7$, $x_5-x_6$ and $x_4-x_5$ in (\ref{esu8t0}), we obtain the gluing rules for $\su(6)^{(2)}$:
\be
%
\ee
where $x_8-x_9$, $x_2-x_1$ glue to fibers in the affine surface.

Folding (\ref{esu8t0}), we obtain
\be
%
\ee
where $x_8-x_9$, $x_2-x_1$, $l-x_1-x_2-x_3$ and$2l-x_1-x_2-x_4-x_5-x_6-x_7$ glue to four copies of fiber in the affine surface.

By adding the curves in the previous configuration, we obtain the following two:
\be
%
\ee
where $x_8-x_9$, $x_2-x_1$, $l-x_1-x_2-x_3$ and $2l-x_1-x_2-x_4-x_5-x_6-x_7$ glue to four copies of fiber in the affine surface.
\be
%
\ee
where $x_8-x_9$, $x_2-x_1$, $l-x_1-x_2-x_3$ and $2l-x_1-x_2-x_4-x_5-x_6-x_7$ glue to four copies of fiber in the affine surface.

Folding (\ref{ef4}), we obtain
\be
%
\ee
where $x_3-x_2$, $x_5-x_6$ and $l-x_7-x_8-x_9$ glue to three copies of fiber in the affine surface.

By folding (\ref{ee7}) and (\ref{ee7su2}) we obtain:
\be\label{ee6t}
%
\ee
where $l-x_3-x_8-x_9$ and $x_6-x_7$ glue to two copies of fiber in the affine surface.
\be
%
\ee
where $x_8-x_9$, $l-x_3-x_8-x_9$ and $x_6-x_7$ glue to fibers inside corresponding affine surfaces.

In a similar fashion, by folding other configurations and sometimes adding some of the curves in them, we can obtain the following configurations:
\be
%
\ee
where $x_8-x_9$, $2l-x_2-x_3-x_6-x_7-x_8-x_9$ and $x_3-x_2$ glue to fibers inside corresponding affine surfaces.
\be
%
\ee
where $x_8-x_9$, $2l-x_2-x_3-x_6-x_7-x_8-x_9$ and $x_3-x_2$ glue to fibers inside corresponding affine surfaces.
\be
%
\ee
where $x_8-x_9$, $2l-x_1-x_2-x_6-x_7-x_8-x_9$ and $l-x_1-x_2-x_3$ glue to fibers inside corresponding affine surfaces.
\be
%
\ee
where $x_8-x_9$, $2l-x_1-x_2-x_6-x_7-x_8-x_9$, $x_2-x_1$, $x_4-x_5$ and $l-x_1-x_2-x_3$ glue to fibers inside corresponding affine surfaces.
\be
%
\ee
where $x_8-x_9$, $2l-x_1-x_2-x_6-x_7-x_8-x_9$ and $l-x_1-x_2-x_3$ glue to fibers inside corresponding affine surfaces.
\be
%
\ee
where $x_8-x_9$, $2l-x_1-x_2-x_6-x_7-x_8-x_9$, $x_2-x_1$, $x_4-x_5$ and $l-x_1-x_2-x_3$ glue to fibers inside corresponding affine surfaces.
\be
%
\ee
where $x_8-x_9$, $2l-x_1-x_2-x_6-x_7-x_8-x_9$ and $l-x_1-x_2-x_3$ glue to fibers inside corresponding affine surfaces.
\be
%
\ee
where $x_8-x_9$, $2l-x_1-x_2-x_6-x_7-x_8-x_9$, $x_2-x_1$, $x_4-x_5$ and $l-x_1-x_2-x_3$ glue to fibers inside corresponding affine surfaces.
\be
%
\ee
where $x_8-x_9$ and $2l-x_1-x_2-x_4-x_5-x_8-x_9$, $x_6-x_7$ glue to fibers inside corresponding affine surfaces.
\be
%
\ee
where $x_8-x_9$, $2l-x_1-x_2-x_4-x_5-x_6-x_7$ and $x_5-x_9$ glue to fibers inside corresponding affine surfaces.
\be
%
\ee
where $x_8-x_9$, $2l-x_1-x_2-x_4-x_5-x_6-x_7$ and $2l-x_1-x_2-x_6-x_7-x_8-x_9$ glue to fibers inside corresponding affine surfaces.
\be
%
\ee
where $x_8-x_9$, $2l-x_1-x_2-x_4-x_5-x_6-x_7$ and $2l-x_1-x_2-x_6-x_7-x_8-x_9$ glue to fibers inside corresponding affine surfaces.
\be
%
\ee
where $x_8-x_9$, $2l-x_1-x_2-x_4-x_5-x_6-x_7$, $x_4-x_5$ and $2l-x_1-x_2-x_6-x_7-x_8-x_9$ glue to fibers inside corresponding affine surfaces.
\be
%
\ee
where $2l-x_1-x_2-x_4-x_5-x_6-x_7$, $x_8-x_9$, $x_4-x_5$, $x_2-x_1$, $l-x_1-x_2-x_3$ and $2l-x_1-x_2-x_6-x_7-x_8-x_9$ glue to fibers inside corresponding affine surfaces.
\be
%
\ee
where $2l-x_1-x_2-x_4-x_5-x_6-x_7$, $x_8-x_9$, $x_4-x_5$, $x_2-x_1$, $l-x_1-x_2-x_3$ and $2l-x_1-x_2-x_6-x_7-x_8-x_9$ glue to fibers inside corresponding affine surfaces.
\be
%
\ee
where $2l-x_1-x_2-x_4-x_5-x_6-x_7$, $x_8-x_9$, $x_2-x_6$, $l-x_2-x_3-x_6$, $x_4-x_5$, $x_2-x_1$, $l-x_1-x_2-x_3$ and $2l-x_1-x_2-x_6-x_7-x_8-x_9$ glue to fibers inside corresponding affine surfaces.
\be
%
\ee
where $x_8-x_9$, $l-x_1-x_4-x_7$, $l-x_2-x_5-x_7$, $l-x_3-x_6-x_7$ and $l-x_7-x_8-x_9$ glue to fibers inside corresponding affine surfaces.
\be
%
\ee
where $x_8-x_9$, $l-x_1-x_4-x_7$, $l-x_2-x_5-x_7$, $l-x_3-x_6-x_7$, $x_3-x_2$, $x_5-x_6$, and $l-x_7-x_8-x_9$ glue to fibers inside corresponding affine surfaces.
\be
%
\ee
where $x_8-x_9$, $x_3-x_2$, $x_5-x_6$, and $l-x_7-x_8-x_9$ glue to fibers inside corresponding affine surfaces.
\be
%
\ee
where $x_8-x_9$, $x_3-x_2$, $x_5-x_6$, and $l-x_7-x_8-x_9$ glue to fibers inside corresponding affine surfaces.
\be
%
\ee
where $x_8-x_9$, $x_3-x_2$ and $2l-x_2-x_3-x_6-x_7-x_8-x_9$ glue to fibers inside corresponding affine surfaces.
\be
%
\ee
where $x_8-x_9$, $x_3-x_2$, $x_1-x_4$, $l-x_2-x_3-x_5$ and $2l-x_2-x_3-x_6-x_7-x_8-x_9$ glue to fibers inside corresponding affine surfaces.
\be
%
\ee
where $x_8-x_9$, $x_3-x_2$, $x_1-x_4$, $l-x_2-x_3-x_5$ and $2l-x_2-x_3-x_6-x_7-x_8-x_9$ glue to fibers inside corresponding affine surfaces.
\be
%
\ee
where $x_8-x_9$, $x_4-x_5$ and $2l-x_1-x_2-x_6-x_7-x_8-x_9$ glue to fibers inside corresponding affine surfaces.
\be
%
\ee
where $x_8-x_9$, $x_2-x_1$, $x_4-x_5$, $l-x_1-x_2-x_3$ and $2l-x_1-x_2-x_6-x_7-x_8-x_9$ glue to fibers inside corresponding affine surfaces.

Now, we are left with some possibilities that do not arise as foldings. For example, the unfolding of $\fe_6^{(2)}\oplus\su(3)^{(1)}$ is $\fe_7^{(1)}\oplus\su(3)^{(1)}$ which cannot be embedded into $\fe_8^{(1)}$. To obtain the gluing rules for this case, we notice that folding of (\ref{ee7}) has zero mutual intersection with (\ref{esu3}).
\be
%
\ee
where $x_8-x_9$, $x_6-x_7$, $l-x_3-x_6-x_7$, $2l-x_1-x_2-x_4-x_5-x_6-x_7$, $x_6-x_7$ and $l-x_7-x_8-x_9$ glue to fibers inside corresponding affine surfaces.

In a similar fashion, by folding, adding curves or by guessing a correct configuration of curves, we can obtain all the other following gluing rules:
\be
%
\ee
where $x_8-x_9$, $l-x_1-x_4-x_5$, $l-x_1-x_3-x_6$ and $l-x_1-x_2-x_7$ glue to four copies of fiber in the affine surface.
\be
%
\ee
where $x_8-x_9$,  $2l-x_1-x_2-x_4-x_5-x_6-x_7$, $x_4-x_5$ and $2l-x_1-x_2-x_6-x_7-x_8-x_9$ glue to fibers inside corresponding affine surfaces.
\be
%
\ee
where $x_8-x_9$,  $2l-x_1-x_2-x_4-x_5-x_6-x_7$, $x_4-x_5$ and $2l-x_1-x_2-x_6-x_7-x_8-x_9$ glue to fibers inside corresponding affine surfaces.
\be
%
\ee
where $x_8-x_9$, $x_5-x_6$, $l-x_3-x_5-x_6$, $2l-x_1-x_2-x_4-x_5-x_6-x_7$, $x_4-x_5$ and $2l-x_1-x_2-x_6-x_7-x_8-x_9$ glue to fibers inside corresponding affine surfaces.
\be
%
\ee
where $x_8-x_9$, $l-x_1-x_4-x_7$, $l-x_2-x_5-x_7$, $l-x_3-x_6-x_7$, $x_5-x_9$ and $l-x_2-x_7-x_8$ glue to fibers inside corresponding affine surfaces.
\be
%
\ee
where $x_8-x_9$, $l-x_1-x_4-x_7$, $l-x_2-x_5-x_7$, $l-x_3-x_6-x_7$, $x_5-x_9$, $l-x_2-x_3-x_6$, $l-x_1-x_2-x_4$ and $l-x_2-x_7-x_8$ glue to fibers inside corresponding affine surfaces.
\be
%
\ee
where $x_8-x_9$, $x_5-x_6$, $x_2-x_3$, $x_6-x_9$, $x_5-x8$ and $x_4-x_7$ glue to fibers inside corresponding affine surfaces.

\subsubsection{$\sp(0)^{(1)}$ gluings: Directed edges}\label{DE}
Finally we consider cases in which one or both the neighbors of $\sp(0)^{(1)}$ are connected to it via directed edges. Our main constraint comes from (\ref{tfgr}) which states that the torus fibers must be glued appropriately. Let us define $C_{0,\alpha}$ be a $-2$ curve in $dP_9$ which glues to the affine surface for $\fg_\alpha^{(q_\alpha)}$ in the gluing rule associated to an undirected edge, that is gluing rule for
\be
\begin{tikzpicture}
\begin{scope}[shift={(3,0)}]
\node (w2) at (-0.5,0.9) {$\sp(0)^{(1)}$};
\end{scope}
\begin{scope}[shift={(6,0)}]
\node (w3) at (-0.5,0.9) {$\fg_\alpha^{(q_\alpha)}$};
\end{scope}
\draw  (w2) edge (w3);
\end{tikzpicture}
\ee
If $q_\alpha=1$, then there is a unique $C_{0,\alpha}$. If $q_\alpha>1$, then there can be multiple such $-2$ curves. In this case, we pick the curve containing the blowup $x_9$ as $C_{0,\alpha}$. This uniquely fixes the $-2$ curve $C_{0,\alpha}$. The reason for the prominence of the blowup $x_9$ in this definition is that the KK mass $\frac{1}{R}$ enters into the volume of $x_9$, and the volume of any other curve in $dP_9$ that does not involve $x_9$ is independent of $\frac{1}{R}$. We refer the reader to \cite{Bhardwaj:2018vuu} for more details.

To obtain the gluing rules for
\be
\begin{tikzpicture}
\node (w1) at (-0.5,0.9) {$\fg_\alpha^{(q_\alpha)}$};
\begin{scope}[shift={(3,0)}]
\node (w2) at (-0.5,0.9) {$\sp(0)^{(1)}$};
\end{scope}
\begin{scope}[shift={(6,0)}]
\node (w3) at (-0.5,0.9) {$\fg_\gamma^{(q_\gamma)}$};
\end{scope}
\node (w4) at (4.2,0.9) {\scriptsize{$e_{\gamma}$}};
\draw (w1)--(w2);
\draw[<-]  (w2) edge (w4);
\draw  (w4) edge (w3);
\end{tikzpicture}
\ee
we start from the gluing rules for
\be
\begin{tikzpicture}
\node (w1) at (-0.5,0.9) {$\fg_\alpha^{(q_\alpha)}$};
\begin{scope}[shift={(3,0)}]
\node (w2) at (-0.5,0.9) {$\sp(0)^{(1)}$};
\end{scope}
\begin{scope}[shift={(6,0)}]
\node (w3) at (-0.5,0.9) {$\fg_\gamma^{(q_\gamma)}$};
\end{scope}
\draw (w1)--(w2);
\draw (w2)--(w3);
\end{tikzpicture}
\ee
and simply replace the curve $C_{0,\gamma}$ in $dP_9$ by the curve $C_{0,\gamma}+e_\gamma\left(3l-\sum x_i\right)$. Similarly, to obtain the gluing rules for
\be
\begin{tikzpicture}
\node (w1) at (-0.5,0.9) {$\fg_\alpha^{(q_\alpha)}$};
\begin{scope}[shift={(3,0)}]
\node (w2) at (-0.5,0.9) {$\sp(0)^{(1)}$};
\end{scope}
\begin{scope}[shift={(6,0)}]
\node (w3) at (-0.5,0.9) {$\fg_\gamma^{(q_\gamma)}$};
\end{scope}
\node (w4) at (4.2,0.9) {\scriptsize{$e_{\gamma}$}};
\node (w5) at (0.9,0.9) {\scriptsize{$e_{\alpha}$}};
\draw[<-]  (w2) edge (w4);
\draw  (w4) edge (w3);
\draw  (w1) edge (w5);
\draw[->]  (w5) edge (w2);
\end{tikzpicture}
\ee
we start from the gluing rules for
\be
\begin{tikzpicture}
\node (w1) at (-0.5,0.9) {$\fg_\alpha^{(q_\alpha)}$};
\begin{scope}[shift={(3,0)}]
\node (w2) at (-0.5,0.9) {$\sp(0)^{(1)}$};
\end{scope}
\begin{scope}[shift={(6,0)}]
\node (w3) at (-0.5,0.9) {$\fg_\gamma^{(q_\gamma)}$};
\end{scope}
\draw (w1)--(w2);
\draw (w2)--(w3);
\end{tikzpicture}
\ee
and simply replace the curves $C_{0,\gamma}$ and $C_{0,\alpha}$ in $dP_9$ by the curves $C_{0,\gamma}+e_\gamma\left(3l-\sum x_i\right)$ and $C_{0,\alpha}+e_\alpha\left(3l-\sum x_i\right)$ respectively. It is trivial to see that this replacement satisfies (\ref{tfgr}).

Now we only need to consider gluing rules of the form
\be\label{ede}
\begin{tikzpicture}
\begin{scope}[shift={(3,0)}]
\node (w2) at (-0.5,0.9) {$\sp(0)^{(1)}$};
\end{scope}
\begin{scope}[shift={(6,0)}]
\node (w3) at (-0.5,0.9) {$\fg_\gamma^{(q_\gamma)}$};
\end{scope}
\node (w4) at (4.2,0.9) {\scriptsize{$e_{\gamma}$}};
\draw  (w2) edge (w4);
\draw[->]  (w4) edge (w3);
\end{tikzpicture}
\ee
since in the context of $6d$ SCFTs, it is not possible for any other node to attach to $\sp(0)^{(1)}$ in (\ref{ede}).

We first work out the following gluing rules by hand:
\be
\begin{tikzpicture}
\node (v1) at (8,3.5) {\scriptsize{$x_8-x_9,$}};
\node (v1_1) at (11,3.5) {\scriptsize{$x_7-x_8,$}};
\node (v1_1_1) at (11,5) {\scriptsize{$2l-x_1-x_2-x_4-x_5-x_6-x_7$}};
\node (v4) at (14,3.5) {\scriptsize{$x_4-x_5,$}};
\node (v1_3) at (11,2) {\scriptsize{$x_4-x_6,$}};
\draw [double] (v1) -- (v1_1);
\draw [double] (v1_1) -- (v4);
\draw [double] (v1_1_1) -- (v1_1);
\begin{scope}[shift={(7,5.8)}]
\begin{scope}[shift={(3,0)}]
\node (w2) at (-0.5,0.9) {$\sp(0)^{(1)}$};
\end{scope}
\begin{scope}[shift={(6,0)}]
\node (w3) at (-0.5,0.9) {$\so(8)^{(1)}$};
\end{scope}
\node (w4) at (4.1,0.9) {\scriptsize{$2$}};
\draw  (w2) edge (w4);
\draw[->]  (w4) edge (w3);
\draw  (4,0.9) ellipse (2.5 and 0.7);
\end{scope}
\node at (8,3.2) {\scriptsize{$x_2-x_1$}};
\node (v2) at (11,3.2) {\scriptsize{$x_1-x_4$}};
\draw [double] (v2) -- (v1_3);
\node at (11,1.7) {\scriptsize{$x_5-x_7$}};
\node at (14,3.2) {\scriptsize{$x_6-x_7$}};
\node at (11,5.3) {\scriptsize{$l-x_1-x_2-x_3,$}};
\end{tikzpicture}
\ee
where $x_8-x_9$, $x_2-x_1$ glue to two copies of fiber in the affine surface. Indeed we can check that twice the torus fiber for $\so(8)^{(1)}$ is glued to $3l-\sum x_i$.

By folding the above gluing rules, we obtain:
\be\label{eso7d}
\begin{tikzpicture}
\node (v1) at (8,3.5) {\scriptsize{$x_4-x_6,$}};
\node (v1_1) at (11,3.5) {\scriptsize{$x_7-x_8,$}};
\node (v1_1_1) at (11,5) {\scriptsize{$2l-x_1-x_2-x_4-x_5-x_6-x_7$}};
\node (v4) at (14,3.5) {\scriptsize{$x_4-x_5,$}};
\node (v1_3) at (14,2.9) {\scriptsize{$x_2-x_1,$}};
\draw [double] (v1) -- (v1_1);
\draw [double] (v1_1_1) -- (v1_1);
\begin{scope}[shift={(7,5.8)}]
\begin{scope}[shift={(3,0)}]
\node (w2) at (-0.5,0.9) {$\sp(0)^{(1)}$};
\end{scope}
\begin{scope}[shift={(6,0)}]
\node (w3) at (-0.5,0.9) {$\so(7)^{(1)}$};
\end{scope}
\node (w4) at (4.1,0.9) {\scriptsize{$2$}};
\draw  (w2) edge (w4);
\draw[->]  (w4) edge (w3);
\draw  (4,0.9) ellipse (2.5 and 0.7);
\end{scope}
\node at (8,3.2) {\scriptsize{$x_5-x_7$}};
\node (v2) at (11,3.2) {\scriptsize{$x_1-x_4$}};
\node at (14,2.6) {\scriptsize{$x_8-x_9$}};
\node at (14,3.2) {\scriptsize{$x_6-x_7,$}};
\node at (11,5.3) {\scriptsize{$l-x_1-x_2-x_3,$}};
\draw (11.55,3.55) -- (13.4,3.55) (13.4,3.5) -- (11.55,3.5) (11.55,3.45) -- (13.4,3.45) (13.4,3.4) -- (11.55,3.4);
\end{tikzpicture}
\ee
where $x_8-x_9$, $x_2-x_1$, $x_6-x_7$, $x_4-x_5$ glue to four copies of fiber in the affine surface.

Treating $\su(3)^{(1)}$ as a subalgebra of $\so(7)^{(1)}$, we can obtain the following gluing rules
\be
\begin{tikzpicture}
\node (v1_1) at (11,3.5) {\scriptsize{$x_7-x_8,$}};
\node (v1_1_1) at (11,5) {\scriptsize{$2l-x_1-x_2-x_4-x_5-x_6-x_7$}};
\node (v4) at (14,3.5) {\scriptsize{$x_2-x_7,$}};
\draw [double] (v1_1_1) -- (v1_1);
\begin{scope}[shift={(7.6,5.9)}]
\begin{scope}[shift={(3,0)}]
\node (w2) at (-0.5,0.9) {$\sp(0)^{(1)}$};
\end{scope}
\begin{scope}[shift={(6,0)}]
\node (w3) at (-0.5,0.9) {$\su(3)^{(1)}$};
\end{scope}
\node (w4) at (4.1,0.9) {\scriptsize{$2$}};
\draw  (w2) edge (w4);
\draw[->]  (w4) edge (w3);
\draw  (4,0.9) ellipse (2.5 and 0.7);
\end{scope}
\node (v2) at (11,3.2) {\scriptsize{$x_1-x_4$}};
\node at (14,3.2) {\scriptsize{$x_4-x_9$}};
\node at (11,5.3) {\scriptsize{$l-x_1-x_2-x_3,$}};
\draw [double] (v1_1) -- (v4);
\draw [double] (v1_1_1) -- (v4);
\end{tikzpicture}
\ee
where $x_4-x_9$, $x_2-x_7$ glue to two copies of fiber in the affine surface.

Finally, folding (\ref{eso7d}), we obtain
\be
\begin{tikzpicture}
\node (v1) at (14,2.3) {\scriptsize{$x_4-x_6,$}};
\node (v1_1) at (11,3.5) {\scriptsize{$x_7-x_8,$}};
\node (v1_1_1) at (14,1.4) {\scriptsize{$2l-x_1-x_2-x_4-x_5-x_6-x_7$}};
\node (v4) at (14,3.5) {\scriptsize{$x_4-x_5,$}};
\node (v1_3) at (14,2.9) {\scriptsize{$x_2-x_1,$}};
\begin{scope}[shift={(9.3,4.1)}]
\begin{scope}[shift={(3,0)}]
\node (w2) at (-0.5,0.9) {$\sp(0)^{(1)}$};
\end{scope}
\begin{scope}[shift={(6,0)}]
\node (w3) at (-0.5,0.9) {$\su(3)^{(2)}$};
\end{scope}
\node (w4) at (4.1,0.9) {\scriptsize{$2$}};
\draw  (w2) edge (w4);
\draw[->]  (w4) edge (w3);
\draw  (4,0.9) ellipse (2.5 and 0.7);
\end{scope}
\node at (14,2) {\scriptsize{$x_5-x_7$}};
\node (v2) at (11,3.2) {\scriptsize{$x_1-x_4$}};
\node at (14,2.6) {\scriptsize{$x_8-x_9,$}};
\node at (14,3.2) {\scriptsize{$x_6-x_7,$}};
\node at (14,1.7) {\scriptsize{$l-x_1-x_2-x_3,$}};
\draw (11.55,3.55) -- (13.4,3.55) (13.4,3.5) -- (11.55,3.5) (11.55,3.45) -- (13.4,3.45) (13.4,3.4) -- (11.55,3.4);
\draw (11.55,3.35) -- (13.4,3.35) (13.4,3.3) -- (11.55,3.3) (11.55,3.25) -- (13.4,3.25) (13.4,3.2) -- (11.55,3.2);
\end{tikzpicture}
\ee
where $x_8-x_9$, $x_2-x_1$, $x_6-x_7$, $x_4-x_5$, $x_4-x_6$, $x_5-x_7$, $l-x_1-x_2-x_3$, $2l-x_1-x_2-x_4-x_5-x_6-x_7$ glue to eight copies of fiber in the affine surface.

\section{Conclusions and future directions}\label{conc}
In this paper, we have associated a genus-one fibered Calabi-Yau threefold to every $5d$ KK theory, except a few cases for which we provide an algebraic description mimicking the properties of genus-one fibered Calabi-Yau threefolds. Compactifying M-theory on the threefold constructs the KK theory on its Coulomb branch. The threefold is presented as a local neighborhood of a collection of surfaces intersecting with each other. We explicitly identify all the surfaces and their intersections for every KK theory. Such a description of the threefold allows an easy determination of the set of all compact holomorphic curves (known as the Mori cone) inside the threefold along with their intersection numbers with other cycles in the threefold. The Mori cone encodes crucial non-perturbative data needed to perform RG flows on the KK theory which lead to $5d$ SCFTs. For the cases without a completely geometric description we propose an analog of Mori cone using which one can perform RG flows on these outlying KK theories as well.

According to a conjecture (see \cite{Jefferson:2018irk,Bhardwaj:2018yhy,Bhardwaj:2018vuu}) for which substantial evidence was provided in \cite{Jefferson:2018irk}, all the $5d$ SCFTs sit at the end points of such RG flows emanating from $5d$ KK theories. Thus, this work can be viewed as providing a preliminary step towards an explicit classification of $5d$ SCFTs. In principle, the Coulomb branch data of all $5d$ SCFTs is encoded in the properties of Calabi-Yau threefolds presented in this paper (see Section \ref{GKK}). Explicitly, such RG flows are performed by performing sequences of flops and blowdowns on the Calabi-Yau threefolds associated to $5d$ KK theories. See \cite{Jefferson:2018irk,Bhardwaj:2018yhy,Bhardwaj:2018vuu} for a general discussion and \cite{Bhardwaj:2019jtr} for the explicit classification of $5d$ SCFTs up to rank three using the results of this paper. Extending the classification to higher ranks, perhaps using a computer program, would be of significant interest.

The Calabi-Yau threefold associated to a $5d$ KK theory is determined by combining the data of the prepotential of the KK theory with certain geometric consistency conditions. We provide a concrete proposal for the computation of this prepotential based on the definition of the $5d$ KK theory in terms of a $6d$ SCFT on a circle and twisted by a discrete global symmetry around the circle. See Section \ref{PPKK} for more details.

Along the way, we provide a graphical classification scheme for $5d$ KK theories which mimics the graphical classification scheme used to classify $6d$ SCFTs. In fact the graphs associated to $5d$ KK theories generalize the graphs associated to $6d$ SCFTs just as Dynkin graphs associated to general Lie algebras generalize the Dynkin graphs associated to simply laced Lie algebras. We provide a full list of all the possible vertices and edges that can appear in graphs associated to $5d$ KK theories. See Section \ref{SKK} for more details. We leave an explicit classification of $5d$ KK theories to a future work. Such a classification can be performed in a straightforward fashion starting from the explicit classification of $6d$ SCFTs presented in \cite{Heckman:2015bfa,Bhardwaj:2019hhd} and applying the folding operations discussed in Section \ref{SKK}.

A noteworthy point deserving a special mention is that our work applies uniformly to all $6d$ SCFTs irrespective of whether they are constructed in the frozen phase of F-theory or in the unfrozen phase of F-theory. In other words, the dictionary relating M-theory and $5d$ KK theories applies uniformly to all $5d$ KK theories irrespective of the F-theory origin of the associated $6d$ SCFT. This is in stark contrast with the case of $6d$ SCFTs for which the dictionary relating F-theory and the resulting $6d$ theory is modified depending on the presence (called the frozen phase) or absence (called the unfrozen phase) of O$7^+$ planes in the base of the elliptic Calabi-Yau threefold used for compactification of F-theory. See \cite{Bhardwaj:2018jgp} for more details.

In the future, it will be interesting to use the geometries presented in this paper to derive $5d$ gauge theory descriptions associated to $6d$ SCFTs compactified on a circle (possibly with a twist). This can be done by performing local S-dualities on the geometries associated to $5d$ KK theories. See the recent work \cite{Bhardwaj:2019ngx} for more details on the methodology.

\section*{Acknowledgements}
The authors thank Sheldon Katz, Sung Soo Kim, and Gabi Zafrir for valuable discussions, as well as the members of the Simons Center for Geometry and Physics for their hospitality during the 2018 and 2019 Simons Summer Workshop(s) while part of this work was being completed. HK also thanks the members of Harvard University for their hospitality while this work was being completed. The research of CV is supported in part by the NSF grant PHY-1719924 and by a grant from the Simons Foundation (602883, CV). The research of LB and HCT is supported by the NSF grant PHY-1719924, and HCT also acknowledges partial support from the Greek America Foundation through the Mavroyannis Scholarship. The research of HK is supported by the POSCO Science Fellowship of POSCO TJ Park Foundation and the National Research Foundation of Korea (NRF) Grant 2018R1D1A1B07042934. The research of PJ is supported by the U.S. Department of Energy, Office of Science, Office of High Energy Physics of U.S. Department of Energy under grant Contract Numbers DE-SC0012567 and DE-SC0019127.

\appendix
\section{Geometric background}\label{back}
In this section, we recall some background useful for this paper.  We refer the reader to Section 2 of \cite{Bhardwaj:2018vuu} for a more detailed background on various points discussed below in this appendix.
\subsection{Hirzebruch surfaces}\label{Hirz}
A Hirzebruch surface is a $\P^1$ fibration over $\P^1$. We denote a Hirzebruch surface with a degree $-n$ fibration as $\bF_n$. We refer to the fiber $\P^1$ as $f$ and the base $\P^1$ as $e$. Their intersection numbers are
\begin{align}
e^2&=-n\\
f^2&=0\\
e\cdot f&=1
\end{align}
Another very important curve in $\bF_n$ is
\be
h:=e+nf
\ee
whose genus is zero and intersection numbers are
\begin{align}
h^2&=n\\
h\cdot e&=0\\
h\cdot f&=1
\end{align}
Note that $e=h$ for $\bF_0$. The set of holomorphic curves, often referred to as \emph{Mori cone}, for $\bF_n$ with $n\ge0$ is generated by $e$ and $f$. For $\bF_n$ with $n\le0$, the Mori cone is generated by $h$ and $f$.

The canonical class $K$ of $\bF_n$ is an antiholomorphic curve which can be determined by the virtue of adjunction formula which states that for a surface $S$ and a curve $C$ inside $S$, the canonical class $K_S$ of $S$ satisfies
\be\label{adjunct}
\left(K_S+C\right)\cdot C=2g(C)-2
\ee
where $g(C)$ is the genus of $C$. Demanding that $K$ satisfies (\ref{adjunct}) for $e,f$ determines it to be
\be
K=-(e+h+2f)
\ee
from which we can compute that
\be
K^2=8
\ee

Notice that $\bF_n$ and $\bF_{-n}$ are isomorphic to each other via the map
\begin{align}
e&\lra h\\
f&\lra f\\
h&\lra e
\end{align}
Thus, we will restrict our attention to Hirzebruch surfaces with $n\ge0$ in what follows. However, at various points in the main body of the paper we find it useful to include Hirzebruch surfaces with negative degrees since they allow us to express answers in a more uniform way.

We also deal with surfaces which arise by performing $b$ number of blowups on $\bF_n$. The blowups will often be non-generic. We can obtain different surfaces by performing $b$ blowups in different fashions on $\bF_n$. In this paper, we refer to all the different surfaces arising via $b$ blowups of $\bF_n$ as $\bF_n^b$. The curves inside $\bF_n^b$ can be described by adding the curves $x_i$ with $i=1,\cdots,b$ which are the exceptional divisors created by the blowups. We will use the convention that the total transforms\footnote{If $B:\tilde S\to S$ is a blowup of a surface $S$, then the total transform of a curve $C$ in $S$ is the curve $B^{-1}(C)$ in $\tilde S$.} of the curves $e$, $f$ and $h$ are denoted by the same names $e$, $f$ and $h$ in $\bF_n^b$. Thus, the intersection numbers between $e$, $f$ and $h$ are those mentioned above, and their intersections with $x_i$ are
\begin{align}
x_i\cdot x_j&=-\delta_{ij}\\
e\cdot x_i&=0\\
f\cdot x_i&=0\\
h\cdot x_i&=0
\end{align}
The blowup procedure creates curves that can be written as
\be\label{K2}
\alpha e+\beta f-\sum \gamma_i x_i
\ee
with $\alpha,\beta,\gamma_i\ge0$. The important point is that the blowups $x_i$ can appear with negative sign.

Again, using the adjunction formula (\ref{adjunct}) we can find the canonical class $K$ for $\bF_n^b$ to be
\be
K=-(e+h+2f)+\sum x_i
\ee
from which we compute
\be
K^2=8-b
\ee

An important isomorphism exists between $\bF_0^1$ and $\bF_1^1$ with the blowup on both surfaces being performed at a generic point. In fact, a single blowup of $\bF_0$ is always generic. The map from $\bF_1^1$ to $\bF_0^1$ is
\begin{align}
e&\to e-x\\
f-x&\to x\\
x&\to f-x
\end{align}
It is easy to see that the above isomorphism only works when the blowups are generic. For, the non-generic one point blowup of $\bF_1$ contains the curve $e-x$, which would be sent to $e-f$ inside $\bF^1_0$. But $e-f$ is not a holomorphic curve in $\bF^1_0$. The above isomorphism is responsible for the equivalence of geometries corresponding to
\be
\begin{tikzpicture}
\node at (-0.5,0.45) {1};
\node at (-0.45,0.9) {$\sp(n)_{(n+1)\pi}^{(1)}$};
\end{tikzpicture}
\ee
and
\be
\begin{tikzpicture}
\node at (-0.5,0.45) {1};
\node at (-0.45,0.9) {$\sp(n)_{n\pi}^{(1)}$};
\end{tikzpicture}
\ee
whenever the theta angle is physically irrelevant. In the situations where theta angle is physically relevant, the above isomorphism is broken by the presence of neighboring surfaces.

To differentiate between the different surfaces $\bF_n^b$ for fixed $n$ and $b$, we have to track the data of their Mori cone. One important point is that the gluing curves inside the surfaces must be the generators of Mori cone. In the paper, we find many instances in which a surface $\bF_n^b$ appearing in different contexts carries different kinds of gluing curves, thus demonstrating that the two $\bF_n^b$ are different surfaces. For example, the geometry with $\nu=0$ for 
\be
\begin{tikzpicture}
\node at (-0.5,0.45) {2};
\node at (-0.45,0.9) {$\su(n+4)^{(1)}$};
\end{tikzpicture}
\ee
and the geometry with $\nu=0$ for
\be
\begin{tikzpicture}
\node at (-0.5,0.45) {1};
\node at (-0.45,0.9) {$\sp(n)_{(n+1)\pi}^{(1)}$};
\end{tikzpicture}
\ee
both contain a surface $\bF_0^{2n+8}$ with different gluing curves $e-\sum x_i$ and $2e+f-\sum x_i$ respectively. Thus the $\bF_0^{2n+8}$ appearing in the two theories are different blowups of $\bF_0$.

The final point we want to address is that $\bF_2$ and $\bF_0$ are same up to decoupled states. This can be seen by noticing that the Mori cone of latter embeds into the Mori cone of former. This embedding $\bF_0\to\bF_2$ is
\begin{align}
e&\to e+f\\
f&\to f
\end{align}
This means that $\bF_2$ equals $\bF_0$ plus some decoupled states. Decoupling these states corresponds to performing a complex structure deformation $\bF_2\to\bF_0$. When $\bF_0$ and $\bF_2$ carry blowups, this conclusion might be changed or unchanged depending on how the blowups are done. See the discussion after (\ref{sp1ab2}) for an example where this conclusion still holds true even in the presence of blowups.

\subsection{del Pezzo surfaces}\label{DP}
The discussion of del Pezzo surfaces starts with the discussion of complex projective plane $\P^2$ which contains a single curve $l$ whose genus is zero and intersection number is
\be
l^2=1
\ee
(\ref{adjunct}) determines the canonical class to be
\be
K=-3l
\ee
from which we compute
\be
K^2=9
\ee

Performing $n$ blowups on $\P^2$ at generic locations leads to the del Pezzo surface $dP_n$. It can be described in terms of curve $l$ and $x_i$ with intersection numbers
\begin{align}
x_i\cdot x_j&=-\delta_{ij}\\
l\cdot x_i&=0
\end{align}
Again, the blowups create new holomorphic curves which can be written as
\be
\alpha l-\sum \gamma_i x_i
\ee
with $\alpha,\gamma_i\ge0$. In the paper, we abuse the notation and call a non-generic $n$ point blowup of $\P^2$ as $dP_n$ too. The canonical class for $dP_n$ is
\be
K=-3l+\sum x_i
\ee
with
\be
K^2=9-n
\ee

del Pezzo surfaces and Hirzebruch surfaces are related to each other by virtue of an isomorphism $dP_1\to\bF_1$ which acts as
\begin{align}
x&\to e\\
l-x&\to f\\
l&\to h
\end{align}
A one point blowup of $\P^2$ is always generic and thus there is a unique $dP_1$ which appears in the above isomorphism.

A special example of del Pezzo surfaces for us in this paper will be $dP_9$ which is the geometry associated to
\be
\begin{tikzpicture}
\node at (-0.5,0.45) {1};
\node at (-0.45,0.9) {$\sp(0)^{(1)}$};
\end{tikzpicture}
\ee
The curve
\be\label{tf}
F=3l-\sum x_i
\ee
has the properties that
\be
F^2=0
\ee
and
\be
K\cdot F=0
\ee
Thus, $F$ is a fiber of genus one, or in other words a torus fiber inside $dP_9$.

$dP_n$ for $n\ge3$ admits the following basic automorphism. We first choose three distinct blowups $x_i$, $x_j$ and $x_k$, and then implement
\begin{align}
x_i&\to l-x_j-x_k\\
x_j&\to l-x_i-x_k\\
x_k&\to l-x_i-x_j\\
l&\to 2l-x_i-x_j-x_k
\end{align}
Combining this automorphism with permutations of blowups, we can obtain more general automorphisms of $dP_n$ (with $n\ge3$) which can be decomposed as a sequence comprising of above mentioned basic automorphisms and permutations of blowups. Notice that for $dP_9$, any such automorphism leaves the torus fiber (\ref{tf}) invariant.

\subsection{Arithmetic genus for curves in a self-glued surface}\label{genus}
When a surface has no self-gluings, then the arithmetic genus\footnote{Throughout this paper, we never use the geometric genus. Whenever the word ``genus'' appears in this paper, it always refers to arithmetic genus.} of curves living inside the surface can be computed using the adjunction formula (\ref{adjunct}). 

However, when the surface has self-gluings, the genus of the curve is modified. For example, consider gluing the exceptional curves $x$ and $y$ in a generic two point blowup of $\bF_1$. The curve $h-x-y$ (which is a rational curve before gluing) looks like an elliptic fiber with nodal singularity after the gluing, so its arithmetic genus should be one instead of zero, which is what would be suggested by (\ref{adjunct}). This example suggests that the intersection numbers of a curve $C$ with the curves $C_1$ and $C_2$ participating in a self-gluing should be used to modify (\ref{adjunct}) in order to obtain the correct arithmetic genus. However, not all such intersection numbers participate in such a modification. To see this, consider the curve $f-x$ in the above example. This curve remains rational even after gluing. Thus, even though it intersects $x$, its genus is correctly captured by (\ref{adjunct}).

The examples of $h-x-y$ and $f-x$ above suggest that the genus of a curve $C$ should only be modified whenever an intersection with $C_1$ has a partner intersection with $C_2$. Thus our proposal for the computation of genus of an arbitrary curve $C$ is as follows: Let $n_1$ and $n_2$ be the intersections of $C$ with $C_1$ and $C_2$ respectively, and let $n=\text{min}(n_1,n_2)$. Then, our proposal for computation of genus is
\be\label{gformula}
2g(C)-2=(K_S+C)\cdot C+2n
\ee

(\ref{gformula}) allows certain curves to have a non-negative genus even though they did not have a non-negative genus before self-gluing. For example, consider
\bit
\item A surface $\bF_m^2$ with $x$ glued to $y$. The curve $e-x-2y$ has $g=0$ according to (\ref{gformula}) while it has $g=-1$ according to (\ref{adjunct}) which is the formula we would use in the absence of self-gluing. $e-x-2y$ appears as a gluing curve in some of our geometries, for example (\ref{GsueF}), (\ref{GsuoF}), (\ref{GsueS}) and (\ref{GsuoS}).
\item A surface $\bF_0^2$ with $e-x$ glued to $e-y$. The curve $2f-x$ has $g=0$ according to (\ref{gformula}) while it has $g=-1$ according to (\ref{adjunct}). $2f-x$ appears as a gluing curve in the gluing rules for
\be
\begin{tikzpicture}
\node (v1) at (-0.5,0.4) {$2$};
\node at (-0.45,0.9) {$\su(1)^{(1)}$};
\begin{scope}[shift={(2,0)}]
\node (v2) at (-0.5,0.4) {$2$};
\node at (-0.45,0.9) {$\su(1)^{(1)}$};
\end{scope}
\node (v3) at (0.5,0.4) {\tiny{$2$}};
\draw  (v1) edge (v3);
\draw  [->](v3) -- (v2);
\end{tikzpicture}
\ee
\eit

\section{Exceptional cases}\label{Exc}
In this Appendix we study some of the exceptional cases where the methods used in the paper are not applicable in a straightforward manner.
\subsection{Geometries for non-gauge theoretic nodes}\label{ng}
The following non-gauge theoretic nodes arise in our analysis
\be\label{ngsp}
\begin{tikzpicture}
\node at (-0.5,0.45) {1};
\node at (-0.45,0.9) {$\sp(0)^{(1)}$};
\end{tikzpicture}
\ee
\be\label{ngsu}
\begin{tikzpicture}
\node at (-0.5,0.45) {2};
\node at (-0.45,0.9) {$\su(1)^{(1)}$};
\end{tikzpicture}
\ee
\be\label{ngsul}
\begin{tikzpicture}
\node (v1) at (-0.5,0.4) {2};
\node at (-0.45,0.9) {$\su(1)^{(1)}$};
\draw (v1) .. controls (-1.5,-0.5) and (0.5,-0.5) .. (v1);
\end{tikzpicture}
\ee
According to our proposal the prepotential $6\tilde\cF$ for each case must be zero. So the geometry cannot be directly guessed from the prepotential. One can try to take corresponding limits of the geometries for the following gauge theoretic nodes
\be
\begin{tikzpicture}
\node at (-0.5,0.45) {1};
\node at (-0.45,0.9) {$\sp(n)^{(1)}$};
\end{tikzpicture}
\ee
\be\label{gsu}
\begin{tikzpicture}
\node at (-0.5,0.45) {2};
\node at (-0.45,0.9) {$\su(n)^{(1)}$};
\end{tikzpicture}
\ee
\be
\begin{tikzpicture}
\node (v1) at (-0.5,0.4) {2};
\node at (-0.45,0.9) {$\su(n)^{(1)}$};
\draw (v1) .. controls (-1.5,-0.5) and (0.5,-0.5) .. (v1);
\end{tikzpicture}
\ee
But this procedure is unreliable. For example, taking the limit of the geometry (\ref{Gsuo2}) would suggest that there should exist a phase of (\ref{ngsu}) governed by the geometry
\be
\begin{tikzpicture} [scale=1.9]
\node (v2) at (-2.2,-0.5) {$\mathbf{0^{1+1}_{0}}$};
\draw (v2) .. controls (-1.4,0) and (-1.4,-1) .. (v2);
\node at (-1.9,-0.2) {\scriptsize{$e$-$x$-$y$}};
\node at (-1.9,-0.8) {\scriptsize{$e$}};
\end{tikzpicture}
\ee
However, even though the self-gluing here satisfies the Calabi-Yau condition (\ref{CY}), it does not satisfy the condition (\ref{CYP}). So, this is not a consistent geometry, and there should be no such phase for (\ref{ngsu}).

Fortunately, a gauge theory description of the KK theories (\ref{ngsp}), (\ref{ngsu}) and (\ref{ngsul}) is known, which allows us to reliably compute the corresponding geometries. In terms of the language used throughout this paper, this gauge theory description is a ``non-canonical'' gauge theory description of these KK theories, since it does not correspond to the $6d$ gauge theory description on the tensor branch of the corresponding $6d$ SCFT.

To start with, it is known that (\ref{ngsp}) can be described by the gauge theory $\su(2)$ with eight fundamental hypers. We can compute the prepotential  via
\be
6\mathcal{F}=\frac{1}{2}\left(\sum_{r}|r \cdot \phi|^3- \sum_f \sum_{w(\cR_f)}|w(\cR_f) \cdot \phi + m_f|^3\right)
\ee
and convert it into a geometry as described in Section \ref{GF}. When all mass parameters are turned off, we obtain the geometry
\be\label{sp05d}
\begin{tikzpicture} [scale=1.9]
\node (v2) at (-2.2,-0.5) {$\mathbf{0^{8}_{1}}$};
\end{tikzpicture}
\ee
which equals $dP_9$. See the discussion that follows (\ref{Gsp0}).

Next, it is known that (\ref{ngsu}) can be described by the gauge theory $\sp(1)$ with an adjoint hyper and $\theta=0$. Moreover, it is known that upon integrating out the adjoint matter of $\sp(n)$, the theta angle remains unchanged. We know that the geometry corresponding to pure $\sp(1)$ with $\theta=0$ is
\be\label{sp1p}
\begin{tikzpicture} [scale=1.9]
\node (v2) at (-2.2,-0.5) {$\mathbf{0_{0}}$};
\end{tikzpicture}
\ee
where we adopt the convention that $f$ is the W-boson of $\sp(1)$ and $e$ is an instanton. So, we just have to integrate the adjoint matter into (\ref{sp1p}) to figure out the geometry for (\ref{ngsu}). We can write the weights of the adjoint as $w_1=(2)$, $w_2=(0)$ and $w_3=(-2)$ in terms of their Dynkin coefficient. When mass parameter for adjoint is very large, then according to the discussion in Section \ref{GF}, we should be able to find a $-1$ curve $C$ living inside a non-compact surface $N$ such that $C$ intersects $S_0=\bF_0$ transversely at two points. We can consistently choose the gluing curve for $N$ inside $S_0$ to be $f$ since $N\cdot f$ must be zero as the mass of the W-boson must be independent of the mass parameter associated to $N$ which is the mass parameter associated to adjoint hyper. As we bring the mass of adjoint to zero, $C$ undergoes a flop transition. If a $-1$ curve living outside a surface $S$ intersects $S$ at two points transversely, then flopping the $-1$ curve leads to the emergence of self-gluing on the surface $S$. Thus, the geometry for (\ref{ngsu}) is
\be\label{sp1ab}
\begin{tikzpicture} [scale=1.9]
\node (v2) at (-2.2,-0.5) {$\mathbf{0^{1+1}_{0}}$};
\draw (v2) .. controls (-1.4,0) and (-1.4,-1) .. (v2);
\node at (-1.9,-0.2) {\scriptsize{$x$}};
\node at (-1.9,-0.8) {\scriptsize{$y$}};
\end{tikzpicture}
\ee
with the gluing curve to $N$ being the genus one curve $f-x-y$. We can write the geometry in an isomorphic way by first exchanging $e$ with $f$, which keeps the description (\ref{sp1ab}) while changing the gluing curve to $N$ as $e-x-y$. Now we perform the isomorphism $\bF^2_0\to\bF^2_2$ such that
\begin{align}
e-x-y&\to e\\
f-x&\to x\\
f-y&\to y\\
x&\to f-x\\
y&\to f-y
\end{align}
which changes (\ref{sp1ab}) to
\be\label{sp1ab2}
\begin{tikzpicture} [scale=1.9]
\node (v2) at (-2.2,-0.5) {$\mathbf{0^{1+1}_{2}}$};
\draw (v2) .. controls (-1.4,0) and (-1.4,-1) .. (v2);
\node at (-1.9,-0.2) {\scriptsize{$f$-$x$}};
\node at (-1.9,-0.8) {\scriptsize{$f$-$y$}};
\end{tikzpicture}
\ee
with the gluing curve to $N$ being $e$. As discussed at the end of Appendix \ref{Hirz}, this geometry gives rise to some decoupled states which can be decoupled by doing a complex structure deformation to
\be
\begin{tikzpicture} [scale=1.9]
\node (v2) at (-2.2,-0.5) {$\mathbf{0^{1+1}_{0}}$};
\draw (v2) .. controls (-1.4,0) and (-1.4,-1) .. (v2);
\node at (-1.9,-0.2) {\scriptsize{$f$-$x$}};
\node at (-1.9,-0.8) {\scriptsize{$f$-$y$}};
\end{tikzpicture}
\ee
Performing an exchange of $e$ and $f$ again leads to the geometry
\be\label{Gsu1G}
\begin{tikzpicture} [scale=1.9]
\node (v2) at (-2.2,-0.5) {$\mathbf{0^{1+1}_{0}}$};
\draw (v2) .. controls (-1.4,0) and (-1.4,-1) .. (v2);
\node at (-1.9,-0.2) {\scriptsize{$e$-$x$}};
\node at (-1.9,-0.8) {\scriptsize{$e$-$y$}};
\end{tikzpicture}
\ee
which is what is displayed in (\ref{Gsu1}) because the fiber $f$ becomes an elliptic fiber in this frame (with a nodal singularity). This is as we would expect from the fact that (\ref{ngsu}) arises from an untwisted unfrozen $6d$ SCFT and hence it must be possible to feed the geometry (\ref{Gsu1G}) into F-theory, which requires the presence of an elliptic fibration. The gluing curves for the non-compact surface responsible for mass parameter of adjoint are $x$ and $y$ in this frame.

Finally, it is known that (\ref{ngsul}) can be described by the gauge theory $\sp(1)$ with an adjoint hyper and $\theta=\pi$. The geometry corresponding to pure $\sp(1)$ with $\theta=\pi$ is
\be
\begin{tikzpicture} [scale=1.9]
\node (v2) at (-2.2,-0.5) {$\mathbf{0_{1}}$};
\end{tikzpicture}
\ee
In a similar fashion as above, integrating in the adjoint leads to
\be\label{Gsu1lG}
\begin{tikzpicture} [scale=1.9]
\node (v2) at (-2.2,-0.5) {$\mathbf{0^{1+1}_{1}}$};
\draw (v2) .. controls (-1.4,0) and (-1.4,-1) .. (v2);
\node at (-1.9,-0.2) {\scriptsize{$x$}};
\node at (-1.9,-0.8) {\scriptsize{$y$}};
\end{tikzpicture}
\ee
which is indeed the ``geometry'' presented in (\ref{Gsu1l}). We write the word geometry in quotation marks because it is only to be understood as an algebraic description mimicking the properties of the geometric description available for other KK theories. See the discussion after equation (\ref{Gsu1l}) for more details.

\subsection{Gluing rules between non-gauge theoretic nodes}\label{grng}
As we combine non-gauge theoretic nodes via edges, the prepotential $6\tilde\cF$ still remains zero. Thus, another method to compute the gluing rules presented in the main body of this paper is desirable. The goal of this section is to provide this alternative derivation.

\noindent\ubf{Gluing rules for \raisebox{-.125\height}{\begin{tikzpicture}
\node (v1) at (-0.5,0.4) {$2$};
\node at (-0.45,0.9) {$\su(1)^{(1)}$};
\begin{scope}[shift={(2,0)}]
\node (v2) at (-0.5,0.4) {$1$};
\node at (-0.45,0.9) {$\sp(0)^{(1)}$};
\end{scope}
\draw  (v1) -- (v2);
\end{tikzpicture}}}: It is known that this KK theory is equivalent to a $5d$ $\sp(2)$ gauge theory with eight fundamentals and an antisymmetric. The theta angle for $\sp(2)$ is irrelevant due to the presence of fundamentals. So we can start with geometry corresponding to any theta angle for pure $\sp(2)$ and then integrate in the matter. The geometry with theta angle zero is
\be\label{sp2io}
\begin{tikzpicture} [scale=1.9]
\node (v1) at (-4.25,-0.5) {$\mathbf{1_{6}}$};
\node (v2) at (-2.6,-0.5) {$\mathbf{2_1}$};
\draw  (v1) edge (v2);
\node at (-4,-0.4) {\scriptsize{$e$}};
\node at (-2.9,-0.4) {\scriptsize{$2h$}};
\end{tikzpicture}
\ee
where we have labeled the surfaces according to the labeling of the corresponding simple co-roots of $\sp(2)$. Notice that this is different than a similar labeling of the surfaces in terms of simple co-roots of affine algebras used in the main body of the text. The weights for fundamental are
\begin{gather*}
(1,0)^+\\
(-1,1)^+\\
(1,-1)^+\\
(-1,0)^+
\end{gather*}
where we have arranged the weights in a spindle shape according to their level and the superscripts on top of the weights denotes the sign of virtual volume of the weights in the totally integrated out phase (\ref{sp2io}). The last weight $(-1,0)$ can be recognized as a $-1$ curve living in a non-compact surface and intersecting $S_1$ once. Since there are eight fundamentals, there are eight copies of the above weight system. Making the virtual volume of $(-1,0)$ negative for all eight copies leads to the phase
\be\label{sp2ii1}
\begin{tikzpicture} [scale=1.9]
\node (v1) at (-4.25,-0.5) {$\mathbf{1^8_{6}}$};
\node (v2) at (-2.6,-0.5) {$\mathbf{2_1}$};
\draw  (v1) edge (v2);
\node at (-4,-0.4) {\scriptsize{$e$}};
\node at (-2.9,-0.4) {\scriptsize{$2h$}};
\end{tikzpicture}
\ee
The weight system in this phase can be written as eight copies of
\begin{gather*}
(1,0)^+\\
(-1,1)^+\\
(1,-1)^+\\
(-1,0)^-
\end{gather*}
The blowups $x_i$ correspond to eight copies of the weight $(-1,0)$. Indeed, the volume of $x_i$ is $\phi_1$ which is negative of the virtual volume of the weight $(-1,0)$ in this phase. The other weights are obtained by adding the fibers $f_i$ of the two surfaces $S_i$. For example, $f_1-x_i$ are eight copies of the weight $(1,-1)$ and indeed $\text{vol}(f_1-x_i)=\phi_1-\phi_2$ which matches the virtual volume of $(1,-1)$. Now making the virtual volume of all the eight copies of the weight $(1,-1)$ negative corresponds to flopping the curves $f_1-x_i$ in (\ref{sp2ii1}) where $f_1$ is the fiber of $S_1$. The resulting geometry is
\be\label{sp2ii}
\begin{tikzpicture} [scale=1.9]
\node (v1) at (-4.25,-0.5) {$\mathbf{1_{2}}$};
\node (v2) at (-2.6,-0.5) {$\mathbf{2^8_1}$};
\draw  (v1) edge (v2);
\node at (-4,-0.4) {\scriptsize{$h$}};
\node at (-3.1,-0.4) {\scriptsize{$2h$-$\sum x_i$}};
\end{tikzpicture}
\ee
with the weight system being eight copies of
\begin{gather*}
(1,0)^+\\
(-1,1)^+\\
(1,-1)^-\\
(-1,0)^-
\end{gather*}
The curves $x_i$ in the phase (\ref{sp2ii}) correspond to eight copies of the weight $(1,-1)$. Notice that we can take mass parameter for all eight fundamentals to be zero in this phase since weights which are negatives of each other have virtual volumes of opposite signs. Thus, we have completely integrated in the eight fundamentals. Now we move onto the integration of antisymmetric.

The weight system for antisymmetric of $\sp(2)$ in phase (\ref{sp2ii}) is
\begin{gather*}
(0,1)^+\\
(2,-1)^+\\
(0,0)^+\\
(-2,1)^+\\
(0,-1)^+
\end{gather*}
Flipping the sign for $(0,-1)$, we obtain
\be
\begin{tikzpicture} [scale=1.9]
\node (v1) at (-4.25,-0.5) {$\mathbf{1_{2}}$};
\node (v2) at (-2.4,-0.5) {$\mathbf{2^{8+1}_1}$};
\draw  (v1) edge (v2);
\node at (-4,-0.4) {\scriptsize{$e$}};
\node at (-3.1,-0.4) {\scriptsize{$2h$-$\sum x_i$}};
\end{tikzpicture}
\ee
with the ninth blowup $y$ on $S_2$ not participating in the gluing curve for $S_1$ inside $S_2$. Now, flipping the sign for $(-2,1)$ corresponds to flopping $f_2-y$. Since it intersects the gluing curve $2h-\sum x_i$ twice, this results in a self-gluing on $S_1$
\be
\begin{tikzpicture} [scale=1.9]
\node (v1) at (-4.25,-0.5) {$\mathbf{1^{1+1}_{2}}$};
\node (v2) at (-2.3,-0.5) {$\mathbf{2^{8}_1}$};
\draw  (v1) edge (v2);
\node at (-3.7,-0.4) {\scriptsize{$h$-$x$-$y$}};
\node at (-2.9,-0.4) {\scriptsize{$2h$+$f$-$\sum x_i$}};
\draw (v1) .. controls (-5,-0.1) and (-5,-0.9) .. (v1);
\node at (-4.7,-0.2) {\scriptsize{$x$}};
\node at (-4.7,-0.8) {\scriptsize{$y$}};
\end{tikzpicture}
\ee
The reader can check that both (\ref{CYP}) and (\ref{CY}) are satisfied here. The weight system of antisymmetric corresponding to this phase is
\begin{gather*}
(0,1)^+\\
(2,-1)^+\\
(0,0)^+\\
(-2,1)^-\\
(0,-1)^-
\end{gather*}
with $x\sim y$ being identified with the weight $(-2,1)$. After performing an isomorphism on $S_1$ can be rewritten as
\be\label{sp2II}
\begin{tikzpicture} [scale=1.9]
\node (v1) at (-4.25,-0.5) {$\mathbf{1^{1+1}_{0}}$};
\node (v2) at (-2.3,-0.5) {$\mathbf{2^{8}_1}$};
\draw  (v1) edge (v2);
\node at (-3.8,-0.4) {\scriptsize{$f$}};
\node at (-2.9,-0.4) {\scriptsize{$2h$+$f$-$\sum x_i$}};
\draw (v1) .. controls (-5,-0.1) and (-5,-0.9) .. (v1);
\node at (-4.7,-0.2) {\scriptsize{$e$-$x$}};
\node at (-4.7,-0.8) {\scriptsize{$e$-$y$}};
\end{tikzpicture}
\ee
leading to the same gluing rules as those presented in the main text.

\noindent\ubf{Gluing rules for \raisebox{-.125\height}{\begin{tikzpicture}
\node (v1) at (-0.5,0.4) {$2$};
\node at (-0.45,0.9) {$\su(1)^{(1)}$};
\begin{scope}[shift={(2,0)}]
\node (v2) at (-0.5,0.4) {$2$};
\node at (-0.45,0.9) {$\su(1)^{(1)}$};
\end{scope}
\draw  (v1) -- (v2);
\end{tikzpicture}}}: It is known that this KK theory is equivalent to a $5d$ $\su(3)$ gauge theory with an adjoint and Chern-Simons level zero. The geometry for $\su(3)$ with CS level zero is
\be
\begin{tikzpicture} [scale=1.9]
\node (v1) at (-4.25,-0.5) {$\mathbf{1_{1}}$};
\node (v2) at (-2.9,-0.5) {$\mathbf{2^{}_1}$};
\draw  (v1) edge (v2);
\node at (-4,-0.4) {\scriptsize{$e$}};
\node at (-3.2,-0.4) {\scriptsize{$e$}};
\end{tikzpicture}
\ee
The weight system for adjoint in this phase is
\begin{gather*}
(1,1)^+\\
(-1,2)^+~(2,-1)^+\\
(0,0)^+~(0,0)^+\\
(1,-2)^+~(-2,1)^+\\
(-1,-1)^+
\end{gather*}
The weight $(-1,-1)$ can be identified with a $-1$ curve living in a non-compact surface and intersecting both $S_1$ and $S_2$ at one point each. Flipping the sign of this weight leads to the appearance of a blowup on both $S_1$ and $S_2$
\be
\begin{tikzpicture} [scale=1.9]
\node (v1) at (-4.4,-0.5) {$\mathbf{1^1_{1}}$};
\node (v2) at (-2.8,-0.5) {$\mathbf{2^{1}_1}$};
\node at (-4.1,-0.4) {\scriptsize{$e$,$x$}};
\node at (-3.1,-0.4) {\scriptsize{$e$,$x$}};
\node (v3) at (-3.6,-0.5) {\scriptsize{2}};
\draw  (v1) edge (v3);
\draw  (v3) edge (v2);
\end{tikzpicture}
\ee
Notice that both the blowups are glued to each other. This can be understood as a consequence of the fact that they both correspond to the same weight i.e. $(-1,-1)^-$, but since there is a single such weight, these two curves must be identified with each other. In this flop frame, the weight system is
\begin{gather*}
(1,1)^+\\
(-1,2)^+~(2,-1)^+\\
(0,0)^+~(0,0)^+\\
(1,-2)^+~(-2,1)^+\\
(-1,-1)^-
\end{gather*}
and the curves corresponding $(-1,2)^+$ and $(-2,1)^+$ can be identified as $(f-x)_{S_1}$ and $(f-x)_{S_2}$ respectively. Flopping both of these, flips the sign of both the weights $(-1,2)$ and $(-2,1)$ and leads to the geometry
\be
\begin{tikzpicture} [scale=1.9]
\node (v1) at (-4.5,-0.5) {$\mathbf{1^{1+1}_{0}}$};
\node (v2) at (-2.1,-0.5) {$\mathbf{2^{1+1}_0}$};
\node at (-3.9,-0.4) {\scriptsize{$e$-$y$, $f$-$x$}};
\node at (-2.7,-0.4) {\scriptsize{$e$-$y$, $f$-$x$}};
\draw (v1) .. controls (-5.2,-0.1) and (-5.2,-0.9) .. (v1);
\draw (v2) .. controls (-1.4,-0.1) and (-1.4,-0.9) .. (v2);
\node at (-4.9,-0.2) {\scriptsize{$x$}};
\node at (-4.9,-0.8) {\scriptsize{$y$}};
\node at (-1.7,-0.2) {\scriptsize{$x$}};
\node at (-1.7,-0.8) {\scriptsize{$y$}};
\node (v3) at (-3.3,-0.5) {\scriptsize{2}};
\draw  (v1) edge (v3);
\draw  (v3) edge (v2);
\end{tikzpicture}
\ee
which after performing an isomorphism of both the surfaces can be written as
\be
\begin{tikzpicture} [scale=1.9]
\node (v1) at (-4.5,-0.5) {$\mathbf{1^{1+1}_{0}}$};
\node (v2) at (-2.1,-0.5) {$\mathbf{2^{1+1}_0}$};
\node at (-4,-0.4) {\scriptsize{$f$-$x$,$x$}};
\node at (-2.6,-0.4) {\scriptsize{$f$-$x$,$x$}};
\draw (v1) .. controls (-5.2,-0.1) and (-5.2,-0.9) .. (v1);
\draw (v2) .. controls (-1.4,-0.1) and (-1.4,-0.9) .. (v2);
\node at (-4.9,-0.2) {\scriptsize{$e$-$x$}};
\node at (-4.9,-0.8) {\scriptsize{$e$-$y$}};
\node at (-1.7,-0.2) {\scriptsize{$e$-$x$}};
\node at (-1.7,-0.8) {\scriptsize{$e$-$y$}};
\node (v3) at (-3.3,-0.5) {\scriptsize{2}};
\draw  (v1) edge (v3);
\draw  (v3) edge (v2);
\end{tikzpicture}
\ee
leading to the same gluing rules as those presented in the main text.

\noindent\ubf{Gluing rules for \raisebox{-.125\height}{\begin{tikzpicture}
\node (v1) at (-0.5,0.4) {$2$};
\node at (-0.45,0.9) {$\su(1)^{(1)}$};
\begin{scope}[shift={(2,0)}]
\node (v2) at (-0.5,0.4) {$2$};
\node at (-0.45,0.9) {$\su(1)^{(1)}$};
\end{scope}
\node (v3) at (0.5,0.4) {\tiny{$2$}};
\draw  (v1) edge (v3);
\draw  [->](v3) -- (v2);
\end{tikzpicture}}}: It is known that this KK theory is equivalent to a $5d$ $\sp(2)$ gauge theory with an adjoint and theta angle zero. The geometry for pure $\sp(2)$ with zero theta angle is known to be
\be
\begin{tikzpicture} [scale=1.9]
\node (v1) at (-4.25,-0.5) {$\mathbf{1_{6}}$};
\node (v2) at (-2.9,-0.5) {$\mathbf{2^{}_1}$};
\draw  (v1) edge (v2);
\node at (-4,-0.4) {\scriptsize{$e$}};
\node at (-3.2,-0.4) {\scriptsize{$2h$}};
\end{tikzpicture}
\ee
The weight system for adjoint in this phase is
\begin{gather*}
(2,0)^+\\
(0,1)^+\\
(-2,2)^+~(2,-1)^+\\
(0,0)^+~(0,0)^+\\
(2,-2)^+~(-2,1)^+\\
(0,-1)^+\\
(-2,0)^+
\end{gather*}
Flipping the sign for $(-2,0)$ leads to the geometry
\be
\begin{tikzpicture} [scale=1.9]
\node (v1) at (-4.25,-0.5) {$\mathbf{1^{1+1}_{6}}$};
\node (v2) at (-2.8,-0.5) {$\mathbf{2_1}$};
\draw  (v1) edge (v2);
\node at (-3.9,-0.4) {\scriptsize{$e$}};
\node at (-3.1,-0.4) {\scriptsize{$2h$}};
\draw (v1) .. controls (-5,-0.1) and (-5,-0.9) .. (v1);
\node at (-4.7,-0.2) {\scriptsize{$x$}};
\node at (-4.7,-0.8) {\scriptsize{$y$}};
\end{tikzpicture}
\ee
In this phase, the weight $(0,-1)^+$ can be identified with curves $f_1-x$ and $f_1-y$, along with a $-1$ curve $z$ living in a non-compact surface and intersecting $S_2$ at one point. $z$ is glued to $f_1-x$ but not to $f_1-y$. Since if it glues also to $f_1-y$, then it would mean that $f_1-x$ is glued to $f_1-y$ resulting in another self-gluing of $S_1$, namely $f_1-x\sim f_1-y$. After this self-gluing, the volumes of $f_1-x$ and $f_1-y$ will be $\phi_1-\phi_2$ leading to a contradiction with our starting step that their volume is $-\phi_2$. 

Now, to flip the sign of the weight $(0,-1)$, we have to flop $f_1-x\sim z$ which automatically flops $f_1-y$ since its volume is same. The flop of $f_1-x$ creates a new blowup on $S_1$ that we call $x'$. Similarly, the flop of $f_1-y$ creates a new blowup on $S_1$ that we call $y'$. Moreover the flop of $z$ creates a blowup on $S_2$ that we call $z'$.

After the flop $S_1=\bF_4^2$ with $f_1-x'$ glued to $f_1-y'$ and $S_2=\bF_2^1$. The total gluing curve for $S_2$ in $S_1$ is $e_1+x'+y'$, and the total gluing curve for $S_1$ in $S_2$ is $2h$. The gluing $f_1-x\sim z$ transforms into the gluing $x'\sim z'$ in the new frame. Thus, the total gluing curve splits into two gluing curves:
\begin{align}
e_1+y'&\sim 2h-z'\\
x'&\sim z'
\end{align}
The reader can check that the curves involved on both sides in both of these gluings have same genus, and moreover (\ref{CYP}) and (\ref{CY}) are satisfied for both gluings. Notice that if we would have tried to split the total gluing curve into three gluing curves $e_1,x',y'$ glued respectively to $2h-2z',z',z'$, we would have run into two problems. First is the same problem that we noted before the flop was performed, that this would imply a second self gluing $x'\sim y'$ of $S_1$ and the weight system won't match with the system of curves in the geometry anymore. Second, the genus of $2h_2-2z'$ is $-1$ and the genus of $e_1$ is $+1$, so the first gluing curve wouldn't make sense.

Thus at this step of the integration process, the geometry is
\be
\begin{tikzpicture} [scale=1.9]
\node (v1) at (-4.5,-0.5) {$\mathbf{1^{1+1}_{4}}$};
\node (v2) at (-2.1,-0.5) {$\mathbf{2^{1}_1}$};
\node at (-3.9,-0.4) {\scriptsize{$e$+$y$, $x$}};
\node at (-2.6,-0.4) {\scriptsize{$2h$-$z$, $z$}};
\draw (v1) .. controls (-5.2,-0.1) and (-5.2,-0.9) .. (v1);
\node at (-4.9,-0.2) {\scriptsize{$f$-$x$}};
\node at (-4.9,-0.8) {\scriptsize{$f$-$y$}};
\node (v3) at (-3.3,-0.5) {\scriptsize{2}};
\draw  (v1) edge (v3);
\draw  (v3) edge (v2);
\end{tikzpicture}
\ee
where we have dropped the primes on the blowups. The corresponding weight system is
\begin{gather*}
(2,0)^+\\
(0,1)^+\\
(-2,2)^+~(2,-1)^+\\
(0,0)^+~(0,0)^+\\
(2,-2)^+~(-2,1)^+\\
(0,-1)^-\\
(-2,0)^-
\end{gather*}
By performing an isomorphism, we can write the geometry as
\be
\begin{tikzpicture} [scale=1.9]
\node (v1) at (-4.7,-0.5) {$\mathbf{1^{1+1}_{2}}$};
\node (v2) at (-1.7,-0.5) {$\mathbf{2^{1}_1}$};
\node at (-3.9,-0.4) {\scriptsize{$e$+$f$-$x$-$2y$, $f$-$x$}};
\node at (-2.2,-0.4) {\scriptsize{$2h$-$z$, $z$}};
\draw (v1) .. controls (-5.4,-0.1) and (-5.4,-0.9) .. (v1);
\node at (-5.1,-0.2) {\scriptsize{$x$}};
\node at (-5.1,-0.8) {\scriptsize{$y$}};
\node (v3) at (-3.1,-0.5) {\scriptsize{2}};
\draw  (v1) edge (v3);
\draw  (v3) edge (v2);
\end{tikzpicture}
\ee
The weight $(2,-2)^+$ corresponds to the curve $x\sim y$, and the weight $(-2,1)^+$ corresponds to the curve $f_2-z$. Upon flopping them, we obtain the geometry with adjoint matter completely integrated in
\be\label{sp2AII}
\begin{tikzpicture} [scale=1.9]
\node (v1) at (-4.5,-0.5) {$\mathbf{1^{1+1}_{2}}$};
\node (v2) at (-1.5,-0.5) {$\mathbf{2^{1+1}_0}$};
\node at (-3.8,-0.4) {\scriptsize{$e$+$f$-$y$, $f$-$x$}};
\node at (-2.3,-0.4) {\scriptsize{$2e$+$f$-$x$-$2y$, $f$-$x$}};
\draw (v1) .. controls (-5.2,-0.1) and (-5.2,-0.9) .. (v1);
\draw (v2) .. controls (-0.8,-0.1) and (-0.8,-0.9) .. (v2);
\node at (-4.9,-0.2) {\scriptsize{$x$}};
\node at (-4.9,-0.8) {\scriptsize{$y$}};
\node at (-1.1,-0.2) {\scriptsize{$x$}};
\node at (-1.1,-0.8) {\scriptsize{$y$}};
\node (v3) at (-3.1,-0.5) {\scriptsize{2}};
\draw  (v1) edge (v3);
\draw  (v3) edge (v2);
\end{tikzpicture}
\ee
After an isomorphism, we obtain
\be
\begin{tikzpicture} [scale=1.9]
\node (v1) at (-4.5,-0.5) {$\mathbf{1^{1+1}_{0}}$};
\node (v2) at (-2.1,-0.5) {$\mathbf{2^{1+1}_0}$};
\node at (-4,-0.4) {\scriptsize{$f$-$x$, $x$}};
\node at (-2.7,-0.4) {\scriptsize{$2f$-$x$, $x$}};
\draw (v1) .. controls (-5.2,-0.1) and (-5.2,-0.9) .. (v1);
\draw (v2) .. controls (-1.4,-0.1) and (-1.4,-0.9) .. (v2);
\node at (-4.9,-0.2) {\scriptsize{$e$-$x$}};
\node at (-4.9,-0.8) {\scriptsize{$e$-$y$}};
\node at (-1.7,-0.2) {\scriptsize{$e$-$x$}};
\node at (-1.7,-0.8) {\scriptsize{$e$-$y$}};
\node (v3) at (-3.3,-0.5) {\scriptsize{2}};
\draw  (v1) edge (v3);
\draw  (v3) edge (v2);
\end{tikzpicture}
\ee
which shows that gluing rules are precisely those quoted in the main text.

\noindent\ubf{Gluing rules for \raisebox{-.325\height}{\begin{tikzpicture}
\node (v1) at (-0.5,0.4) {$2$};
\node at (-0.45,0.9) {$\su(1)^{(1)}$};
\begin{scope}[shift={(2,0)}]
\node (v2) at (-0.5,0.4) {$2$};
\node at (-0.45,0.9) {$\su(1)^{(1)}$};
\end{scope}
\draw  (v1) -- (v2);
\draw (v2) .. controls (0.5,-0.5) and (2.5,-0.5) .. (v2);
\end{tikzpicture}}}: It is known that this KK theory is equivalent to a $5d$ $\sp(2)$ gauge theory with an adjoint and theta angle $\pi$. Thus, the analysis for this case is similar to that of the last case which was
\be
\begin{tikzpicture}
\node (v1) at (-0.5,0.4) {$2$};
\node at (-0.45,0.9) {$\su(1)^{(1)}$};
\begin{scope}[shift={(2,0)}]
\node (v2) at (-0.5,0.4) {$2$};
\node at (-0.45,0.9) {$\su(1)^{(1)}$};
\end{scope}
\node (v3) at (0.5,0.4) {\tiny{$2$}};
\draw  (v1) edge (v3);
\draw  [->](v3) -- (v2);
\end{tikzpicture}
\ee
since only the theta angle is different for these two cases. Following similar steps as above, the final ``geometry''\footnote{We remind the reader that it should only be viewed as an algebraic description since the KK theory involves the non-geometric node.} analogous to (\ref{sp2AII}) is found to be
\be
\begin{tikzpicture} [scale=1.9]
\node (v1) at (-4.5,-0.5) {$\mathbf{1^{1+1}_{2}}$};
\node (v2) at (-1.5,-0.5) {$\mathbf{2^{1+1}_1}$};
\node at (-3.8,-0.4) {\scriptsize{$e$+$f$-$y$, $f$-$x$}};
\node at (-2.3,-0.4) {\scriptsize{$2h$-$x$-$2y$, $f$-$x$}};
\draw (v1) .. controls (-5.2,-0.1) and (-5.2,-0.9) .. (v1);
\draw (v2) .. controls (-0.8,-0.1) and (-0.8,-0.9) .. (v2);
\node at (-4.9,-0.2) {\scriptsize{$x$}};
\node at (-4.9,-0.8) {\scriptsize{$y$}};
\node at (-1.1,-0.2) {\scriptsize{$x$}};
\node at (-1.1,-0.8) {\scriptsize{$y$}};
\node (v3) at (-3.1,-0.5) {\scriptsize{2}};
\draw  (v1) edge (v3);
\draw  (v3) edge (v2);
\end{tikzpicture}
\ee
which after an isomorphism becomes
\be
\begin{tikzpicture} [scale=1.9]
\node (v1) at (-4.5,-0.5) {$\mathbf{1^{1+1}_{0}}$};
\node (v2) at (-1.7,-0.5) {$\mathbf{2^{1+1}_1}$};
\node at (-3.9,-0.4) {\scriptsize{$f$-$x$, $x$}};
\node at (-2.5,-0.4) {\scriptsize{$2h$-$x$-$2y$, $f$-$x$}};
\draw (v1) .. controls (-5.2,-0.1) and (-5.2,-0.9) .. (v1);
\draw (v2) .. controls (-1,-0.1) and (-1,-0.9) .. (v2);
\node at (-4.9,-0.2) {\scriptsize{$e$-$x$}};
\node at (-4.9,-0.8) {\scriptsize{$e$-$y$}};
\node at (-1.3,-0.2) {\scriptsize{$x$}};
\node at (-1.3,-0.8) {\scriptsize{$y$}};
\node (v3) at (-3.2,-0.5) {\scriptsize{2}};
\draw  (v1) edge (v3);
\draw  (v3) edge (v2);
\end{tikzpicture}
\ee
which matches the gluing rules claimed in the text.

\subsection{Theta angle for $\sp(n)$}\label{theta}
Notice that there are two inequivalent geometries which give rise to a $5d$ pure $\sp(n)$ gauge theory:
\be\label{spnt0}
\begin{tikzpicture} [scale=1.9]
\node (v10) at (-0.5,-2) {$\mathbf{1_{2n+2}}$};
\node (v9) at (0.4,-2) {$\cdots$};
\node (v8) at (1.5,-2) {$\mathbf{(n-2)_{8}}$};
\node (v5) at (4.7,-2) {$\mathbf{n_0}$};
\node (v7) at (3.05,-2) {$\mathbf{(n-1)_{6}}$};
\draw  (v7) edge (v8);
\draw  (v8) edge (v9);
\draw  (v9) edge (v10);
\node at (-0.1,-1.9) {\scriptsize{$e$}};
\node at (4.3,-1.9) {\scriptsize{$2e$+$f$}};
\node at (3.6,-1.9) {\scriptsize{$e$}};
\node at (2.5,-1.9) {\scriptsize{$h$}};
\node at (1,-1.9) {\scriptsize{$h$}};
\node at (2,-1.9) {\scriptsize{$e$}};
\draw  (v5) edge (v7);
\end{tikzpicture}
\ee
and
\be\label{spnt1}
\begin{tikzpicture} [scale=1.9]
\node (v10) at (-0.5,-2) {$\mathbf{1_{2n+2}}$};
\node (v9) at (0.4,-2) {$\cdots$};
\node (v8) at (1.5,-2) {$\mathbf{(n-2)_{8}}$};
\node (v5) at (4.7,-2) {$\mathbf{n_1}$};
\node (v7) at (3.05,-2) {$\mathbf{(n-1)_{6}}$};
\draw  (v7) edge (v8);
\draw  (v8) edge (v9);
\draw  (v9) edge (v10);
\node at (-0.1,-1.9) {\scriptsize{$e$}};
\node at (4.4,-1.9) {\scriptsize{$2h$}};
\node at (3.6,-1.9) {\scriptsize{$e$}};
\node at (2.5,-1.9) {\scriptsize{$h$}};
\node at (1,-1.9) {\scriptsize{$h$}};
\node at (2,-1.9) {\scriptsize{$e$}};
\draw  (v5) edge (v7);
\end{tikzpicture}
\ee
These two geometries correspond to two different possible values of theta angle. The only difference between (\ref{spnt0}) and (\ref{spnt1}) is whether $S_n=\bF_0$ or $S_n=\bF_1$. It is well-known that (see for instance \cite{Jefferson:2018irk}) for $\sp(1)$, $\theta=0$ has $S_1=\bF_0$ and $\theta=\pi$ has $S_1=\bF_1$, while for $\sp(2)$, $\theta=0$ has $S_2=\bF_1$ and $\theta=\pi$ has $S_2=\bF_0$. 

We claim that for higher $n$, the same pattern continues to hold and the theta angle corresponding to $\bF_0$ (or $\bF_1$) changes by $\pi~(\text{mod}~ 2\pi)$ every time one increases the rank $n$ by one unit. To see this, one can start from the statement \cite{Tachikawa:2011ch} that the KK theory
\be\label{spKK}
\begin{tikzpicture}
\node (v1) at (-0.5,0.4) {$2$};
\node at (-0.45,0.9) {$\su(1)^{(1)}$};
\begin{scope}[shift={(2,0)}]
\node (v2) at (-0.5,0.4) {$2$};
\node at (-0.45,0.9) {$\su(1)^{(1)}$};
\end{scope}
\begin{scope}[shift={(-2,0)}]
\node (v3) at (-0.5,0.4) {$2$};
\node at (-0.45,0.9) {$\su(1)^{(1)}$};
\end{scope}
\begin{scope}[shift={(-3.5,0)}]
\node (v4) at (-0.5,0.4) {$\cdots$};
\end{scope}
\begin{scope}[shift={(-5,0)}]
\node (v5) at (-0.5,0.4) {$2$};
\node at (-0.45,0.9) {$\su(1)^{(1)}$};
\end{scope}
\draw  (v1) -- (v2);
\draw (v2) .. controls (0.5,-0.5) and (2.5,-0.5) .. (v2);
\draw  (v1) edge (v3);
\draw  (v3) edge (v4);
\draw  (v4) edge (v5);
\end{tikzpicture}
\ee
with a total of $n$ nodes is equivalent to a $5d$ $\sp(n)$ gauge theory with an adjoint hyper and $\theta=\pi$. We can build the geometry corresponding to (\ref{spKK}) by using the data presented in this paper and derived in Appendix (\ref{grng}). Now the key point is that integrating out the adjoint matter does not change the theta angle. So, we can simply integrate out the adjoint matter from the geometry corresponding to (\ref{spKK}) to land on to pure $\sp(n)$ theory with $\theta=\pi$. This process is inverse of the process of integrating in of matter discussed in Appendices (\ref{ng}) and (\ref{grng}) and corresponds to making the virtual volumes of all the weights of adjoint of $\sp(n)$ to have the same sign. Once this is done, it is found that the geometry for $\theta=\pi$ is (\ref{spnt0}) whenever $n$ is even, and the geometry (\ref{spnt1}) whenever $n$ is odd. From this we conclude that the geometry (\ref{spnt0}) corresponds to $\theta=\theta_0$ and the geometry (\ref{spnt1}) corresponds to $\theta=\theta_1$ where
\begin{gather}
\theta_1=n\pi~(\text{mod}~2\pi)\\
\theta_0=\theta_1+\pi~(\text{mod}~2\pi)
\end{gather}

\section{A concrete non-trivial check of our proposal}\label{concrete}
We devote this section to a concrete and non-trivial check of our proposal. It is known that \cite{Hayashi:2015vhy} the KK theory
\be\label{su2B}

\ee
is equivalent to the $5d$ gauge theory with gauge algebra $\su(2)\oplus\su(4)$ with a hyper transforming in $\F\otimes\Asym$. More precisely, the gauge-theoretic phase diagram for the $\su(2)\oplus\su(4)$ embeds into the phase diagram for the KK theory (\ref{su2B}). In this section we will show this explicitly.

Let us start with the geometry assigned to (\ref{su2B}) in the paper with $\nu$ chosen to be zero for both $\su(2)^{(1)}$:
\be
%
\ee
where the surfaces $S_0$ and $S_1$ correspond to the left $\su(2)^{(1)}$ in (\ref{su2B}), and the surfaces $S'_0$ and $S'_1$ correspond to the right $\su(2)^{(1)}$ in (\ref{su2B}). As visible in the above diagram, $x_4$ in $S_0$ is glued to $x_2$ in $S'_0$. Flopping this curve, we obtain
\be
%
\ee
Now flopping $f-x$ in $S_1$ which is glued to $x_1$ in $S'_0$, we obtain
\be
%
\ee
which after performing an isomorphism on $S_0$ can be written as
\be
%
\ee
Now, flopping the $e$ curves inside $S_0$ and $S_1$ (which are glued to each other), we obtain
\be
%
\ee
where a surface without a subscript denotes that the surface is a del Pezzo surface rather than a Hirzebruch surface. That is, $S_0=dP_4$ and $S_1=\P^2=dP_0$. Let us use the blowup $x_4$ on $S_0$ to write $S_0$ in terms of the Hirzebruch surface $\bF_1$
\be
%
\ee
Flopping $x_3$ in $S_0$ glued to $f-x_1$ in $S'_1$ gives rise to
\be
%
\ee
We use $x$ in $S_1$ to write $S_1$ in terms of Hirzebruch surface $\bF_1$
\be
%
\ee
Performing the automorphism on $S'_0$ that exchanges $e$ and $f$, we obtain
\be
%
\ee
Now let us write $S'_1$ as a del Pezzo surface. This rewrites the $e$ curve as a blowup which we denote by $w$
\be
%
\ee
We can now perform a basic automorphism (of del Pezzo surfaces) on $S'_1$ involving the three blowups $x_1$, $x_2$ and $y_1$ to obtain
\be
%
\ee
Converting $S'_1$ back into $\bF_1$ using the blowup $y_2$, we obtain
\be\label{su2so6}
%
\ee
This is the final form of the geometry that we wanted to obtain. 

It is clear that $S_0$, $S'_0$ and $S_1$ describe an $\su(4)$ and $S'_1$ describes an $\su(2)$ in (\ref{su2so6}). This can be checked by intersecting the fibers of the corresponding Hirzebruch surfaces with these surfaces. The intersection matrix yields the Cartan matrix for $\su(4)\oplus\su(2)$. Now, let us show that the configuration of blowups indeed describes $\Asym\otimes\F$ of $\su(4)\oplus\su(2)$. For this we relabel the surfaces as
\begin{align}
S_0&\to S_1\\
S'_0&\to S_2\\
S_1&\to S_3
\end{align}
thus rewriting the geometry as
\be\label{su2so6re}
\begin{tikzpicture} [scale=1.9]
\node (v1) at (-0.5,0.9) {$\mathbf{1_{2}^{}}$};
\node (v2) at (2.1,0.9) {$\mathbf{2^{}_{0}}$};
\node (v10) at (-0.5,-0.9) {$\mathbf{3_{2}^{}}$};
\node (v6) at (2.1,-0.9) {$\mathbf{1_{1}^{'3+1+1+1}}$};
\node[rotate=0] at (-0.1,1) {\scriptsize{$e$}};
\node at (0.25,0.6) {\scriptsize{$f$, $f$}};
\node at (1.7,1) {\scriptsize{$e$}};
\node (v4) at (2.1,0) {\scriptsize{2}};
\draw  (v2) edge (v4);
\draw  (v4) edge (v6);
\node (v7) at (1.05,-0.2) {\scriptsize{2}};
\draw  (v1) edge (v7);
\draw  (v7) edge (v6);
\node[rotate=0] at (2.4,0.6) {\scriptsize{$f$, $f$}};
\node at (2.5,-0.5) {\scriptsize{$w$-$z$, $y$-$x_3$}};
\node at (-0.1,-1.1) {\scriptsize{$f$, $f$}};
\node at (1.1,-1.1) {\scriptsize{$x_2$-$w$, $f$-$x_1$-$y$}};
\node at (1.7,-0.3) {\scriptsize{$f$-$w$-$x_2$,}};
\draw  (v10) edge (v2);
\node at (-0.3,-0.6) {\scriptsize{$e$}};
\node at (1.6,0.7) {\scriptsize{$e$}};
\draw  (v1) edge (v2);
\node (v3) at (0.4,-0.9) {\scriptsize{2}};
\draw  (v10) edge (v3);
\draw  (v3) edge (v6);
\node at (1.8,-0.5) {\scriptsize{$x_1$-$y$}};
\end{tikzpicture}
\ee
The weight system for $\Asym\otimes\F$ can be written as
\begin{gather*}
(0,1,0|1)\\
(1,-1,1|1)~(0,1,0|-1)\\
(-1,0,1|1)~(1,0,-1|1)~(1,-1,1|-1)\\
(-1,1,-1|1)~(-1,0,1|-1)~(1,0,-1|-1)\\
(0,-1,0|1)~(-1,1,-1|-1)\\
(0,-1,0|-1)
\end{gather*}
where the three entries on the left hand side of slash denote the weights with respect to $\su(4)$ comprised by $S_1$, $S_2$ and $S_3$, and the entry on the right hand side of slash denotes the weight with respect to $\su(2)$ comprised by $S'_1$.

\ni From the geometry (\ref{su2so6re}) we see that the holomorphic curves
\begin{align}
\text{vol}(x_1)&=(1,0,-1|1)\\
\text{vol}(x_2)&=(-1,0,1|1)\\
\text{vol}(x_3)&=(0,-1,0|1)\\
\text{vol}(y)&=(-1,1,-1|1)\\
\text{vol}(f-z)&=(0,1,0|1)\\
\text{vol}(f-w)&=(1,-1,1|1)
\end{align}
match weights of the form $(x,y,z|1)$, and the antiholomorphic curves $x_1-f,x_2-f,x_3-f,y-f,-z,-w$ match weights of the form $(x,y,z|-1)$, where $f$ denotes the fiber of Hirzebruch surface $S'_1=\bF_1^6$. Thus we have reproduced the full weight system for $\Asym\otimes\F$, justifying our claim. More precisely, the geometry (\ref{su2so6re}) describes the $\su(4)\oplus\su(2)$ gauge theory in the gauge-theoretic phase given by the following virtual volumes
\begin{gather*}
(0,1,0|1)^+\\
(1,-1,1|1)^+~(0,1,0|-1)^-\\
(-1,0,1|1)^+~(1,0,-1|1)^+~(1,-1,1|-1)^-\\
(-1,1,-1|1)^+~(-1,0,1|-1)^-~(1,0,-1|-1)^-\\
(0,-1,0|1)^+~(-1,1,-1|-1)^-\\
(0,-1,0|-1)^-
\end{gather*}

\section{Comparisons with known cases in the literature}\label{comparison}
In this section we provide a comparison with some $5d$ KK theories known in the literature via other methods. In particular, we show that the geometries we obtain for these $5d$ KK theories allow us to see the $5d$ gauge theory descriptions of these $5d$ KK theories that have been proposed in the literature.

\subsection{Untwisted}
Let us start with an example of untwisted compactification. It has been proposed \cite{Hayashi:2015fsa} that
\be
\begin{tikzpicture}
\node at (-0.5,0.45) {1};
\node at (-0.45,0.9) {$\sp(n)^{(1)}$};
\end{tikzpicture}
\ee
can be described by the $5d$ gauge theory having gauge algebra $\su(n+2)$ with $2n+8$ hypers in fundamental. To see this consider the $\nu=1$ phase of (\ref{spt1})
\be
\begin{tikzpicture} [scale=1.9]
\node (v1) at (-2.5,-2) {$\mathbf{0_1^{2n+7}}$};
\node (v10) at (-0.5,-2) {$\mathbf{1_{2n+1}}$};
\node (v9) at (0.4,-2) {$\cdots$};
\node (v8) at (1.5,-2) {$\mathbf{(n-2)_{7}}$};
\node (v5) at (5.1,-2) {$\mathbf{n^1_1}$};
\node (v7) at (3.25,-2) {$\mathbf{(n-1)_{5}}$};
\draw  (v7) edge (v8);
\draw  (v8) edge (v9);
\draw  (v9) edge (v10);
\draw  (v10) edge (v1);
\node at (-0.1,-1.9) {\scriptsize{$e$}};
\node at (-0.9,-1.9) {\scriptsize{$h$}};
\node at (4.7,-1.9) {\scriptsize{$2h$-$x$}};
\node at (3.8,-1.9) {\scriptsize{$e$}};
\node at (2.7,-1.9) {\scriptsize{$h$}};
\node at (1,-1.9) {\scriptsize{$h$}};
\node at (-1.8,-1.9) {\scriptsize{$2h$- $\sum x_i$}};
\node at (2,-1.9) {\scriptsize{$e$}};
\draw  (v5) edge (v7);
\end{tikzpicture}
\ee
which after an isomorphism can be written as
\be
\begin{tikzpicture} [scale=1.9]
\node (v1) at (-2.5,-2) {$\mathbf{0_{2n+3}^{2n+7}}$};
\node (v10) at (-0.5,-2) {$\mathbf{1_{2n+1}}$};
\node (v9) at (0.4,-2) {$\cdots$};
\node (v8) at (1.5,-2) {$\mathbf{(n-2)_{7}}$};
\node (v5) at (5.1,-2) {$\mathbf{n^1_1}$};
\node (v7) at (3.25,-2) {$\mathbf{(n-1)_{5}}$};
\draw  (v7) edge (v8);
\draw  (v8) edge (v9);
\draw  (v9) edge (v10);
\draw  (v10) edge (v1);
\node at (-0.1,-1.9) {\scriptsize{$e$}};
\node at (-0.9,-1.9) {\scriptsize{$h$}};
\node at (4.6,-1.9) {\scriptsize{$e$+$2f$-$x$}};
\node at (3.8,-1.9) {\scriptsize{$e$}};
\node at (2.7,-1.9) {\scriptsize{$h$}};
\node at (1,-1.9) {\scriptsize{$h$}};
\node at (-2,-1.9) {\scriptsize{$e$}};
\node at (2,-1.9) {\scriptsize{$e$}};
\draw  (v5) edge (v7);
\end{tikzpicture}
\ee
Now flopping the blowup sitting on $S_n$ back to $S_0$, we obtain
\be
\begin{tikzpicture} [scale=1.9]
\node (v1) at (-2.5,-2) {$\mathbf{0_{2n+4}^{2n+8}}$};
\node (v10) at (-0.5,-2) {$\mathbf{1_{2n+2}}$};
\node (v9) at (0.4,-2) {$\cdots$};
\node (v8) at (1.5,-2) {$\mathbf{(n-2)_{8}}$};
\node (v5) at (5.1,-2) {$\mathbf{n_0}$};
\node (v7) at (3.25,-2) {$\mathbf{(n-1)_{6}}$};
\draw  (v7) edge (v8);
\draw  (v8) edge (v9);
\draw  (v9) edge (v10);
\draw  (v10) edge (v1);
\node at (-0.1,-1.9) {\scriptsize{$e$}};
\node at (-0.9,-1.9) {\scriptsize{$h$}};
\node at (4.7,-1.9) {\scriptsize{$e$+2$f$}};
\node at (3.8,-1.9) {\scriptsize{$e$}};
\node at (2.7,-1.9) {\scriptsize{$h$}};
\node at (1,-1.9) {\scriptsize{$h$}};
\node at (-2,-1.9) {\scriptsize{$e$}};
\node at (2,-1.9) {\scriptsize{$e$}};
\draw  (v5) edge (v7);
\end{tikzpicture}
\ee
where we can see that the associated Cartan matrix is that of $\su(n+2)$ and the $2n+8$ blowups sitting on $S_0$ can be identified with the fundamentals. This identification is done by noticing that the volume for a blowup matches the absolute value of virtual volume of a weight for the fundamental of $\su(n+2)$.

\subsection{Twisted}
Now, let us consider an example when we twist by an outer automorphism. It has been proposed in \cite{Hayashi:2015vhy} that
\be
\begin{tikzpicture}
\node at (-0.5,0.45) {2};
\node at (-0.45,0.9) {$\su(n)^{(2)}$};
\end{tikzpicture}
\ee
can be described by $5d$ gauge theory with gauge algebra $\so(n+2)$ and $n$ fundamental hypers. First let us consider the case when $n=2m$. In this case the geometry is displayed in (\ref{suet}). Flopping all the $y_i$, we obtain
\be
\begin{tikzpicture}[scale=1.9]
\node (v2) at (-2.5,-0.5) {$\mathbf{m_1}$};
\node (v3) at (-1,-0.5) {$\mathbf{(m-1)_6}$};
\node (v4) at (0.35,-0.5) {$\mathbf{\cdots}$};
\node (v5) at (1.65,-0.5) {$\mathbf{2_{2m}}$};
\node (v6) at (3.5,-1.55) {$\mathbf{0^{2m}_{2m+2}}$};
\draw  (v2) edge (v3);
\draw  (v3) edge (v4);
\draw  (v4) edge (v5);
\draw  (v5) edge (v6);
\node at (3.8,-0.2) {\scriptsize{$f$-$x_i$}};
\node at (3.8,-1.2) {\scriptsize{$f$-$ x_i$}};
\node at (-2.1,-0.4) {\scriptsize{$2h$}};
\node at (-1.6,-0.4) {\scriptsize{$e$}};
\node at (-0.3,-0.4) {\scriptsize{$h$}};
\node at (2,-0.9) {\scriptsize{$h$}};
\node at (2.9,-1.4) {\scriptsize{$e$}};
\node at (1.3,-0.4) {\scriptsize{$e$}};
\node (v7) at (3.5,-0.65) {\scriptsize{$2m$}};
\node (v1) at (3.5,0.2) {$\mathbf{1^{2m}_{2m+2}}$};
\draw  (v5) edge (v1);
\draw  (v1) edge (v7);
\draw  (v6) edge (v7);
\node at (2,-0.2) {\scriptsize{$h$}};
\node at (2.9,0.1) {\scriptsize{$e$}};
\end{tikzpicture}
\ee
Now flopping all the $f-x_i$, we obtain
\be
\begin{tikzpicture}[scale=1.9]
\node (v2) at (-2.5,-0.5) {$\mathbf{m_1}$};
\node (v3) at (-1,-0.5) {$\mathbf{(m-1)_6}$};
\node (v4) at (0.35,-0.5) {$\mathbf{\cdots}$};
\node (v5) at (1.65,-0.5) {$\mathbf{2^{2m}_{2m}}$};
\node (v6) at (3.5,-1.55) {$\mathbf{0^{}_{2}}$};
\draw  (v2) edge (v3);
\draw  (v3) edge (v4);
\draw  (v4) edge (v5);
\draw  (v5) edge (v6);
\node at (-2.1,-0.4) {\scriptsize{$2h$}};
\node at (-1.6,-0.4) {\scriptsize{$e$}};
\node at (-0.3,-0.4) {\scriptsize{$h$}};
\node at (2,-0.9) {\scriptsize{$h$-$\sum x_i$}};
\node at (3.1,-1.5) {\scriptsize{$e$}};
\node at (1.3,-0.4) {\scriptsize{$e$}};
\node (v1) at (3.5,0.2) {$\mathbf{1^{}_{2}}$};
\draw  (v5) edge (v1);
\node at (2,-0.2) {\scriptsize{$h$-$\sum x_i$}};
\node at (3.1,0.2) {\scriptsize{$e$}};
\end{tikzpicture}
\ee
Now we can carry the $2m$ blowups onto $S_m$ to obtain the geometry
\be
\begin{tikzpicture}[scale=1.9]
\node (v2) at (-2.5,-0.5) {$\mathbf{m^{2m}_1}$};
\node (v3) at (-0.4,-0.5) {$\mathbf{(m-1)_{2m-6}}$};
\node (v4) at (0.9,-0.5) {$\mathbf{\cdots}$};
\node (v5) at (1.65,-0.5) {$\mathbf{2^{}_{0}}$};
\node (v6) at (3.3,-1.55) {$\mathbf{0^{}_{2}}$};
\draw  (v2) edge (v3);
\draw  (v3) edge (v4);
\draw  (v4) edge (v5);
\draw  (v5) edge (v6);
\node at (-1.9,-0.4) {\scriptsize{$2h$-$\sum x_i$}};
\node at (-1.2,-0.4) {\scriptsize{$h$}};
\node at (0.3,-0.4) {\scriptsize{$e$}};
\node at (2,-0.9) {\scriptsize{$e$}};
\node at (2.9,-1.5) {\scriptsize{$e$}};
\node at (1.4,-0.4) {\scriptsize{$h$}};
\node (v1) at (3.3,0.2) {$\mathbf{1^{}_{2}}$};
\draw  (v5) edge (v1);
\node at (2,-0.2) {\scriptsize{$e$}};
\node at (2.9,0.2) {\scriptsize{$e$}};
\end{tikzpicture}
\ee
which after an isomorphism on $S_m$ can be rewritten as
\be
\begin{tikzpicture}[scale=1.9]
\node (v2) at (-2.5,-0.5) {$\mathbf{m^{2m}_{2m-4}}$};
\node (v3) at (-0.4,-0.5) {$\mathbf{(m-1)_{2m-6}}$};
\node (v4) at (0.9,-0.5) {$\mathbf{\cdots}$};
\node (v5) at (1.65,-0.5) {$\mathbf{2^{}_{0}}$};
\node (v6) at (3.3,-1.55) {$\mathbf{0^{}_{2}}$};
\draw  (v2) edge (v3);
\draw  (v3) edge (v4);
\draw  (v4) edge (v5);
\draw  (v5) edge (v6);
\node at (-1.9,-0.4) {\scriptsize{$e$}};
\node at (-1.2,-0.4) {\scriptsize{$h$}};
\node at (0.3,-0.4) {\scriptsize{$e$}};
\node at (2,-0.9) {\scriptsize{$e$}};
\node at (2.9,-1.5) {\scriptsize{$e$}};
\node at (1.4,-0.4) {\scriptsize{$h$}};
\node (v1) at (3.3,0.2) {$\mathbf{1^{}_{2}}$};
\draw  (v5) edge (v1);
\node at (2,-0.2) {\scriptsize{$e$}};
\node at (2.9,0.2) {\scriptsize{$e$}};
\end{tikzpicture}
\ee
The Cartan matrix associated to this geometry is indeed that for $\so(2m+2)$ and the $2m$ blowups can be identified as $2m$ hypers in fundamental of $\so(2m+2)$.

Similarly, the geometry for $n=2m+1$ is given in (\ref{suot}). Flopping $x_i\sim y_i$ living on $S_0$, we obtain
\be
\begin{tikzpicture} [scale=1.9]
\node (v2) at (-1.6,-0.5) {$\mathbf{ m_1}$};
\node (v3) at (-0.3,-0.5) {$\mathbf{(m-1)_6}$};
\node (v4) at (0.8,-0.5) {$\cdots$};
\node (v5) at (1.8,-0.5) {$\mathbf{ 1^{2m+1}_{2m +2}}$};
\node (v6) at (3.8,-0.5) {$\mathbf{0_{6}}$};
\draw  (v2) edge (v3);
\draw  (v3) edge (v4);
\draw  (v4) edge (v5);
\draw  (v5) edge (v6);
\node at (-1.2,-0.4) {\scriptsize{$2h$}};
\node at (-0.8,-0.4) {\scriptsize{$e$}};
\node at (0.3,-0.4) {\scriptsize{$h$}};
\node at (2.5,-0.4) {\scriptsize{$2h$-2$\sum x_i$}};
\node at (3.4,-0.4) {\scriptsize{$e$}};
\node at (1.3,-0.4) {\scriptsize{$e$}};
\end{tikzpicture}
\ee
After performing an isomorphism we can write the above geometry as
\be
\begin{tikzpicture} [scale=1.9]
\node (v2) at (-1.8,-0.5) {$\mathbf{ m_1}$};
\node (v3) at (-0.3,-0.5) {$\mathbf{(m-1)_6}$};
\node (v4) at (0.8,-0.5) {$\cdots$};
\node (v5) at (2,-0.5) {$\mathbf{ 1^{2m+1}_{1}}$};
\node (v6) at (3.6,-0.5) {$\mathbf{0_{6}}$};
\draw  (v2) edge (v3);
\draw  (v3) edge (v4);
\draw  (v4) edge (v5);
\draw  (v5) edge (v6);
\node at (-1.4,-0.4) {\scriptsize{$2h$}};
\node at (-0.8,-0.4) {\scriptsize{$e$}};
\node at (0.3,-0.4) {\scriptsize{$h$}};
\node at (2.5,-0.4) {\scriptsize{$2h$}};
\node at (3.3,-0.4) {\scriptsize{$e$}};
\node at (1.4,-0.4) {\scriptsize{$e$-$\sum x_i$}};
\end{tikzpicture}
\ee
Now moving the blowups onto $S_m$ we obtain
\be
\begin{tikzpicture} [scale=1.9]
\node (v2) at (-2,-0.5) {$\mathbf{ m^{2m+1}_1}$};
\node (v3) at (0,-0.5) {$\mathbf{(m-1)_{2m-5}}$};
\node (v4) at (1.2,-0.5) {$\cdots$};
\node (v5) at (2,-0.5) {$\mathbf{ 1_{1}}$};
\node (v6) at (3.6,-0.5) {$\mathbf{0_{6}}$};
\draw  (v2) edge (v3);
\draw  (v3) edge (v4);
\draw  (v4) edge (v5);
\draw  (v5) edge (v6);
\node at (-1.3,-0.4) {\scriptsize{$2h$-$\sum x_i$}};
\node at (-0.7,-0.4) {\scriptsize{$e$}};
\node at (0.7,-0.4) {\scriptsize{$h$}};
\node at (2.5,-0.4) {\scriptsize{$2h$}};
\node at (3.3,-0.4) {\scriptsize{$e$}};
\node at (1.7,-0.4) {\scriptsize{$e$}};
\end{tikzpicture}
\ee
which can be rewritten as 
\be
\begin{tikzpicture} [scale=1.9]
\node (v2) at (-2,-0.5) {$\mathbf{ m^{2m+1}_{2m-3}}$};
\node (v3) at (0,-0.5) {$\mathbf{(m-1)_{2m-5}}$};
\node (v4) at (1.2,-0.5) {$\cdots$};
\node (v5) at (2,-0.5) {$\mathbf{ 1_{1}}$};
\node (v6) at (3.6,-0.5) {$\mathbf{0_{6}}$};
\draw  (v2) edge (v3);
\draw  (v3) edge (v4);
\draw  (v4) edge (v5);
\draw  (v5) edge (v6);
\node at (-1.4,-0.4) {\scriptsize{$e$}};
\node at (-0.8,-0.4) {\scriptsize{$h$}};
\node at (0.7,-0.4) {\scriptsize{$e$}};
\node at (2.5,-0.4) {\scriptsize{$2h$}};
\node at (3.3,-0.4) {\scriptsize{$e$}};
\node at (1.7,-0.4) {\scriptsize{$e$}};
\end{tikzpicture}
\ee
which precisely describes $\so(2m+3)$ with $2m+1$ hypers in fundamental of $\so(2m+3)$.

\section{Instructions for using the attached Mathematica notebook}\label{mathematica}
	A Mathematica notebook is included as an ancillary file with the arXiv submission of this paper. The use of this notebook requires installation of the Mathematica package LieArt.nb which can be found online at. 	In particular, the notebook provides the evaluation of two functions Geometry5dKK and SignsKK. The former can be used to compute the shifted prepotential $6\tilde\cF$ (defined in Section \ref{Sshift}) for $5d$ KK theories whose associated graph contains either one or two nodes; see Tables \ref{TR1}--\ref{R2N} and Tables \ref{KR1}--\ref{BTR2N}. The latter function can be used for the evaluation of all possible signs associated to different phases of the above prepotential.

	The Mathematica notebook is built around the use of the function

\color{blue}

	\begin{Verbatim}[frame=single,fillcolor=\color{yellow}]
 			Geometry5dKK[...]
	\end{Verbatim} 

	\color{black}
The above function outputs a graphical representation of the shifted prepotential $6\tilde\cF$ associated to the input $5d$ KK theory. The graphical output is naturally organized in the form of triple intersection numbers for the associated geometry. See Section \ref{tripre} for the map between triple intersection numbers and the shifted prepotential.
	\newline

	\noindent \underline{\textbf{Input}}
	\\
Let us now describe possible inputs for the function Geometry5dKK:

	\begin{itemize}
	\item For a single node 
	\be
	\begin{tikzpicture}
\node at (-0.5,0.45) {$k$};
\node at (-0.45,0.9) {$\fg^{(q)}$};
\end{tikzpicture}
\ee
the first input is the number $k$ as shown below
	\color{blue}
	\begin{Verbatim}[frame=single,fillcolor=\color{yellow}]
			Geometry5dKK[{k,...}]
		\end{Verbatim}
		\color{black}
		
		\item For two nodes $\alpha$ and $\beta$, the first input is the matrix $\Omega=\begin{pmatrix}\Omega_S^{\alpha\alpha} & \Omega_S^{\alpha\beta}\\ \Omega_S^{\beta\alpha} & \Omega_S^{\beta\beta}  \end{pmatrix}$:
\color{blue}
		\begin{Verbatim}[frame=single,fillcolor=\color{yellow}]
			Geometry5dKK[{Ω,...}]
		\end{Verbatim} 
		\color{black}
See Section \ref{DS2} for the definition of $\Omega_S^{\alpha\beta}$ etc.
	\item When there is a single node, the second and final input captures the data of $\fg^{(q)}$. When there are two nodes, the second input captures the data of $\fg_\alpha^{(q_\alpha)}$, and the third and final input captures the data of $\fg_\beta^{(q_\beta)}$. The data of an affine algebra is captured by dividing it into the ``algebra part'' and the ``twist part''. For example, the algebra part of $\fg^{(q)}$ is $\fg$ which is a finite Lie algebra, and the twist part of $\fg^{(q)}$ is $q$. The algebra part can be inserted in LieArt format. For example, A-type can be inserted as
	\color{blue}
	\begin{verbatim}
	A1, A2, ..., An
	\end{verbatim}
\color{black}
	
	B-type can be inserted as 
	\color{blue}
	\begin{verbatim}
	B2, B3, ..., Bn
	\end{verbatim} 
\color{black}
	
	C-type can be inserted as
	\color{blue}
	\begin{verbatim}
	C2, C3, ..., Cn
	\end{verbatim}
\color{black}
	
	D-type can be inserted as
	\color{blue}
	\begin{verbatim}
	D3, D4, ..., Dn\end{verbatim}
\color{black}
	
	E-type can be inserted as
	\color{blue}
	\begin{verbatim}       
	E6, E7, E8\end{verbatim}
	\color{black}
         And other types can be inserted as \color{blue}
	\begin{verbatim}
	G2, F4
	\end{verbatim}
	\color{black}
	
	The twist part can be inserted as
	\color{blue}
	\begin{verbatim}
	U, T2, T3
	\end{verbatim}
	\color{black}
	where U means  `untwisted' (corresponding to $ q=1$), T2 means  `$\mathbb{Z}_2 $ twisted
	 (corresponding to $ q=2$) and T3 means   `$\mathbb{Z}_3 $ twisted' (corresponding to $ q=3$).		
	\end{itemize}

\noindent The full input thus is as follows:

	\begin{itemize}
		\item For a single node, the following format is used:
			\color{blue}
	\begin{Verbatim}[frame=single,fillcolor=\color{yellow}]
		Geometry5dKK[{k,{Algebra,Twist}}]
		\end{Verbatim}
			\color{black}
		For example,
				\color{blue}
	\begin{Verbatim}[frame=single,fillcolor=\color{yellow}]
		Geometry5dKK[{2,{A4,T2}}]
		\end{Verbatim}
		\color{black}

		\item For two nodes, the format is:
		\color{blue}
	\begin{Verbatim}[frame=single,fillcolor=\color{yellow}]
	Geometry5dKK[{Ω,{Algebra1,Twist1},{Algebra2,Twist2}}]
		\end{Verbatim}
		\color{black}
For example,		
		\color{blue}
	\begin{Verbatim}[frame=single,fillcolor=\color{yellow}]
	Geometry5dKK[{Ω,{C3,U},{D6,T2}}]
		\end{Verbatim}
		\color{black}

		In order to consider trivial gauge algebras of type $\mathfrak{su}(1)$, $\mathfrak{sp}(0)$, one needs to insert a zero in the place of the algebra and twist input: that is we perform the replacement $\{Algebra,Twist\}\rightarrow 0$.
		For example, if $\fg_\alpha$ is trivial, but $\fg_\beta$ is not, then the input takes the form
		\color{blue}
	\begin{Verbatim}[frame=single,fillcolor=\color{yellow}]
	Geometry5dKK[{Ω,0,{Algebra2,Twist2}}]
		\end{Verbatim}
	\end{itemize}
	\color{black}
	
\noindent Some of the nodes contain extra decorations. Such nodes can be inserted by using extra identifiers as follows:	

	\begin{itemize}
		\item \raisebox{-.125\height}{ \begin{tikzpicture}
			\node at (-0.5,0.45) {$1$};
			\node at (-0.45,0.9) {$\mathfrak{su}(n)^{(1)}$};
			\end{tikzpicture}}  vs.  \raisebox{-.125\height}{ \begin{tikzpicture}
			\node at (-0.5,0.45) {$1$};
			\node at (-0.45,0.9) {$\mathfrak{su}(\wh{n})^{(1)}$};
			\end{tikzpicture}}  
		\newline
		
		To incorporate the second case, we replace $Twist$ with  $\{Twist,Frozen\}$.
		For example,
			\color{blue}
	\begin{Verbatim}[frame=single,fillcolor=\color{yellow}]
		Geometry5dKK[{1,{A8,U}}]
		\end{Verbatim}
			\color{black}
		becomes
			\color{blue}
	\begin{Verbatim}[frame=single,fillcolor=\color{yellow}]
		Geometry5dKK[{1,{A8,{U,Frozen}}}]
		\end{Verbatim}
			\color{black}
		\item   \raisebox{-.125\height}{ \begin{tikzpicture}
			\node at (-0.5,0.45) {$1$};
			\node at (-0.45,0.9) {$\mathfrak{su}(6)^{(1)}$};
			\end{tikzpicture}}  vs.   \raisebox{-.125\height}{ \begin{tikzpicture}
			\node at (-0.5,0.45) {$1$};
			\node at (-0.45,0.9) {$\mathfrak{su}(\tilde{6})^{(1)}$};
			\end{tikzpicture}} 
		\newline
	
		To incorporate the second case, we replace $Twist$ with $\{Twist,Three \}$, so that
		
	\color{blue}
\begin{Verbatim}[frame=single,fillcolor=\color{yellow}]
		Geometry5dKK[{1,{A5,U}}]
		\end{Verbatim}
		\color{black}
	becomes
			\color{blue}
\begin{Verbatim}[frame=single,fillcolor=\color{yellow}]
		Geometry5dKK[{1,{A5,{U,Three}}}]
		\end{Verbatim}
		\color{black}
			\item   \raisebox{-.125\height}{ \begin{tikzpicture}
			\node at (-0.5,0.45) {$2$};
			\node at (-0.45,0.9) {$\mathfrak{su}(n)^{(1)}$};
			\end{tikzpicture}}  vs  \raisebox{-.4\height}{\begin{tikzpicture}
			\node (v1) at (-0.5,0.4) {2};
			\node at (-0.45,0.9) {$\su(n)^{(1)}$};
			\draw (v1) .. controls (-1.5,-0.5) and (0.5,-0.5) .. (v1);
			\end{tikzpicture}} 
		\newline

		To incorporate the second case, we replace $Twist$ with $\{Twist,Loop\}$, so that
		
		\color{blue}
		\begin{Verbatim}[frame=single,fillcolor=\color{yellow}]
		Geometry5dKK[{2,{A5,U}}]
		\end{Verbatim}
		\color{black}
	becomes
		\color{blue}
		\begin{Verbatim}[frame=single,fillcolor=\color{yellow}]
		Geometry5dKK[{2,{A5,{U,Loop}}}]
		\end{Verbatim}
			\color{black}
		\item \raisebox{-.125\height}{ \begin{tikzpicture}
			\node at (-0.5,0.45) {$k$};
			\node at (-0.45,0.9) {$\mathfrak{so}(12)^{(q)}$};
			\end{tikzpicture}} vs  \raisebox{-.125\height}{ \begin{tikzpicture}
			\node at (-0.5,0.45) {$1$};
			\node at (-0.45,0.9) {$\mathfrak{so}(\widehat{12})^{(q)}$};
			\end{tikzpicture}}
		\newline
		
		To incorporate the second case, we replace $Twist$ with $\{Twist,Cospinor \}$, so that
		\color{blue}
	\begin{Verbatim}[frame=single,fillcolor=\color{yellow}]
		Geometry5dKK[{1,{D6,U}}]
		\end{Verbatim}
			\color{black}
			becomes
				\color{blue}
	\begin{Verbatim}[frame=single,fillcolor=\color{yellow}]
		Geometry5dKK[{1,{D6,{U,Cospinor}}}]
		\end{Verbatim}
			\color{black}
			\item \raisebox{-.125\height}{ 
				\begin{tikzpicture}
				\node (v1) at (-0.5,0.45) {3};
				\node at (-0.45,0.9) {$\mathfrak{so}(8)^{(2)}$};
				\begin{scope}[shift={(2,0)}]
				\node (v2) at (-0.5,0.45) {$1$};
				\node at (-0.45,0.9) {$\mathfrak{sp}(1)^{(1)}$};
				\end{scope}
				\node (v3) at (0.5,0.45) {\tiny{ $2$}};
				\draw[dashed]  (v1) edge (v3);
				\draw[->]  (v3) edge (v2);
				\end{tikzpicture}} 
			
			To incorporate this case we use the usual input without any extra identifiers.
				\color{blue}
		\begin{Verbatim}[frame=single,fillcolor=\color{yellow}]
		Geometry5dKK[{Ω,{D4,T2},{A1,U}}]
		\end{Verbatim}

				\color{black}
		\item \raisebox{-.125\height}{ \begin{tikzpicture}
		\node (v1) at (-0.5,0.45) {1};
		\node at (-0.45,0.9) {$\mathfrak{sp}(n_i)^{(1)}$};
		\begin{scope}[shift={(2,0)}]
		\node (v2) at (-0.5,0.45) {$k$};
		\node at (-0.45,0.9) {$\mathfrak{so}(7)^{(1)}$};
		\end{scope}
		\draw  (v1) edge (v2);
		\end{tikzpicture}} vs. \raisebox{-.125\height}{ \begin{tikzpicture}
		\node (v1) at (-0.5,0.45) {1};
		\node at (-0.45,0.9) {$\mathfrak{sp}(n_i)^{(1)}$};
		\begin{scope}[shift={(2,0)}]
		\node (v2) at (-0.5,0.45) {$k$};
		\node at (-0.45,0.9) {$\mathfrak{so}(7)^{(1)}$};
		\end{scope}
		\draw[dashed]  (v1) edge (v2);
		\end{tikzpicture}} and
		\newline
		\raisebox{-.125\height}{ \begin{tikzpicture}
		\node (v1) at (-0.5,0.45) {1};
		\node at (-0.45,0.9) {$\mathfrak{sp}(n_i)^{(1)}$};
		\begin{scope}[shift={(2,0)}]
		\node (v2) at (-0.5,0.45) {$k$};
		\node at (-0.4,0.9) {$\mathfrak{so}(8)^{(q)}$};
		\end{scope}
		\draw  (v1) edge (v2);
		\end{tikzpicture}} vs. 	\raisebox{-.125\height}{ \begin{tikzpicture}
		\node (v1) at (-0.5,0.45) {1};
		\node at (-0.45,0.9) {$\mathfrak{sp}(n_i)^{(1)}$};
		\begin{scope}[shift={(2,0)}]
		\node (v2) at (-0.5,0.45) {$k$};
		\node at (-0.4,0.9) {$\mathfrak{so}(8)^{(q)}$};
		\end{scope}
		\draw[dashed]  (v1) edge (v2);
		\end{tikzpicture}}
		  	
	To incorporate these cases, we replace $Twist$ with $\{Twist,S  \}$.
	For example, one would use the following formats:
	\color{blue}
\begin{Verbatim}[frame=single,fillcolor=\color{yellow}]
	Geometry5dKK[{Ω,{C2,U},{B3,{U,S}}}]
	\end{Verbatim}
			\color{black}
	and 
			\color{blue}
	\begin{Verbatim}[frame=single,fillcolor=\color{yellow}]
	Geometry5dKK[{Ω,{C2,U},{D4,{U,S}}}]
	\end{Verbatim}
		\end{itemize}
		\color{black}
\noindent	\underline{\textbf{Choice of Phase}}
	\newline
	
	\noindent For each input, the output (i.e. the prepotential) depends on a particular choice of gauge-theoretic phase for the theory. The different gauge-theoretic phases correspond to different choices of signs for the virtual volumes of the weights of the representations associated to the matter content for the input KK theory. See Sections \ref{PPKK} and \ref{flops} along with Appendix \ref{Exc} for more details.
	
	After the input is inserted, the notebook will request as additional input the signs of virtual volumes for all the weights corresponding to matter hypermultiplets. 
A pop-up window appears containing the information needed to make a consistent choice of signs.  
	For example, consider
	 \raisebox{-.125\height}{$\begin{tikzpicture}
			\node at (-0.5,0.45) {$1$};
			\node at (-0.45,0.9) {$\mathfrak{su}(5)^{(1)}$};
			\end{tikzpicture}$}.
	\begin{figure}
		\begin{center}
		\includegraphics[scale=0.45]{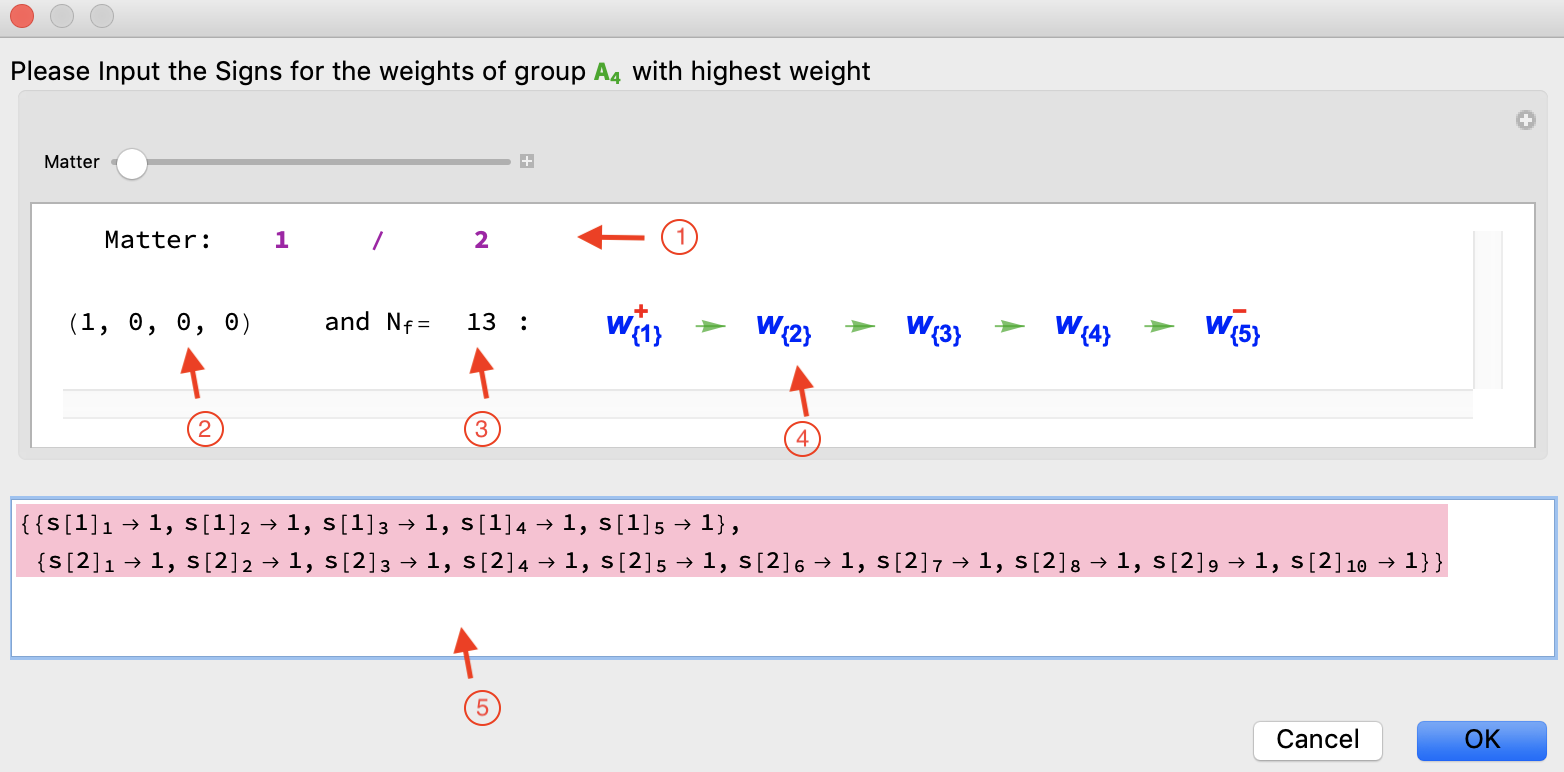}
		\caption{An illustration of the various features of the initial (sign input) pop-up window of the function Geometry5dKK. The various aspects, numbered 1 through 5, are explained in the body of this appendix.}
		\label{fig:fig1}
		\end{center}
	\end{figure}
 After inputting the correct data associated to this theory, a window appears as depicted in Figure \ref{fig:fig1}. The information indicated in the window can be understood as follows: 
\newline
		
\begin{figure}[]
	\hspace{1.5 cm}
	\includegraphics[scale=0.45]{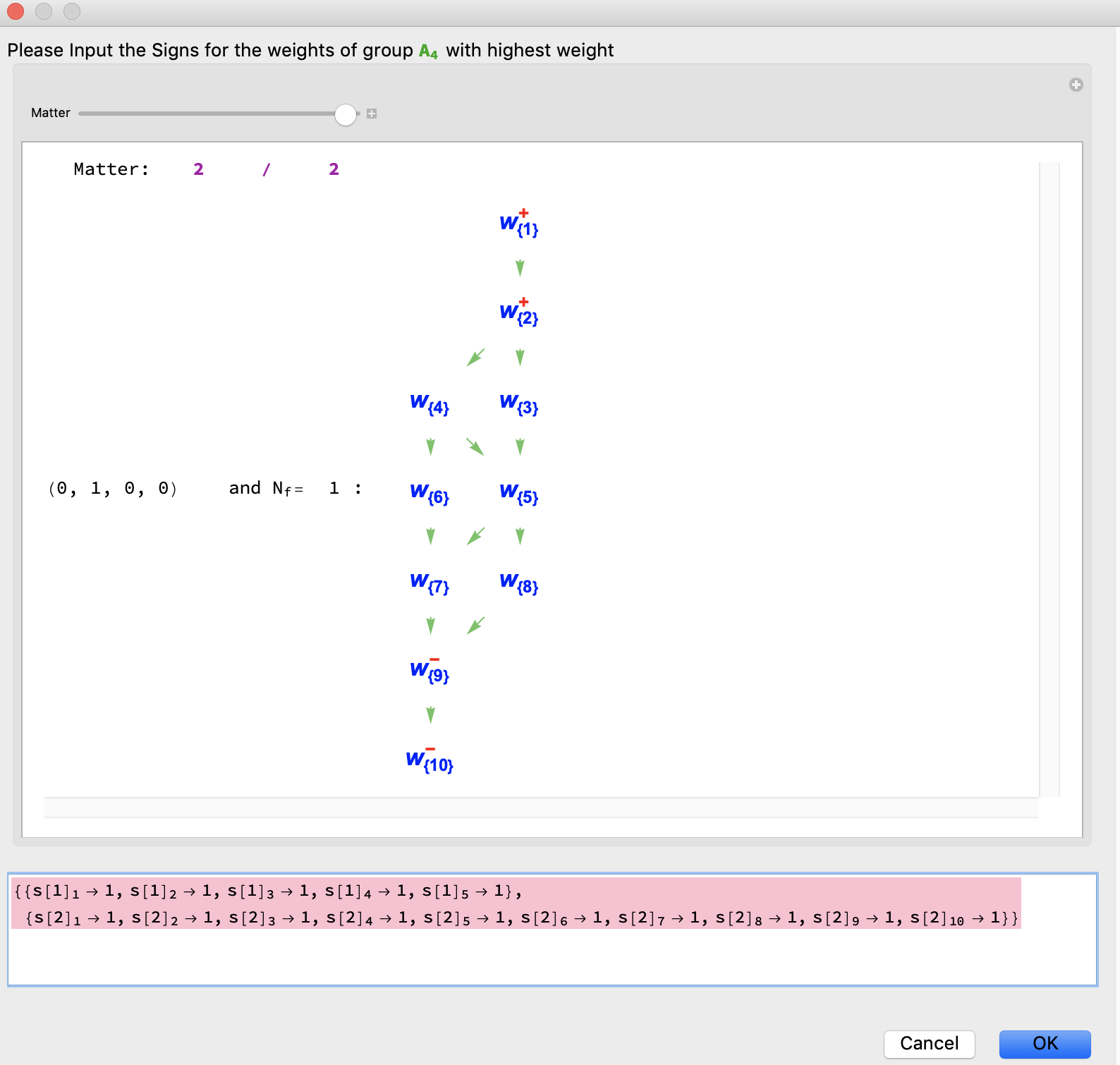}
		\caption{The slider moves between different representations; in the example depicted above, the slider moves from the first to the second representation.}
	\label{fig:fig2}
\end{figure}

\noindent \raisebox{-.125\height}{	\begin{tikzpicture}
	\node at (0,0) {\color{red}{$1$}};
	\draw[red]  (0,0) circle (0.2cm);
	\end{tikzpicture}} This labels the difference choices of irreducible representaitons of the invariant subalgebra (under the twist) in which the hypers of the canonical 5d gauge theory associated to the KK theory transform. In this particular case we have two distinct representations, namely the fundamental and the antisymmetric representations of $\su(5)$, as can be seen from Table \ref{TR1}. The slider on top can be used to slide between the two irreps. For example in Figure \ref{fig:fig1}, we see data associated to fundamental representation and in \ref{fig:fig2} we see the data associated to antisymmetric representation.
\newline
	
\noindent 

\noindent \raisebox{-.125\height}{	\begin{tikzpicture}
		\node at (0,0) {\color{red}{$2$}};
		\draw[red]  (0,0) circle (0.2cm);
		\end{tikzpicture}} This indicates the highest weight of the representation.
	\newline
		
	\noindent \raisebox{-.125\height}{	\begin{tikzpicture}
		\node at (0,0) {\color{red}{$3$}};
		\draw[red]  (0,0) circle (0.2cm);
		\end{tikzpicture}} Here, $N_f$ represents the number full hypermultiplets transforming in the given representation. In Figure \ref{fig:fig1} there are 13 hypermultiplets transforming in the fundamental representation, while in Figure \ref{fig:fig2} there is one hypermultiplet transforming in the antisymmetric representation.
	\newline

\noindent		\raisebox{-.125\height}{	\begin{tikzpicture}
		\node at (0,0) {\color{red}{$4$}};
		\draw[red]  (0,0) circle (0.2cm);
		\end{tikzpicture}} shows the Hasse diagram of the weight system of the  representation. The Hasse diagram is a graphical representation of the partial  order of the weight system. Recall, that given a highest weight $w_1$ one can construct the entire weight system by subtracting positive simple roots, $w_i=w_{i-1}-n_i \alpha_i$ ($\alpha_i$ denote the simple roots). For example, the fundamental representation of $\mathfrak{su}(5)$, which is comprised of weights $w_{i=1,\dots,5}$, is characterized by the partial order $w_1 \geq w_2 \geq \cdots \geq w_5$, where $w_i \geq w_j$ means that $w_i - w_j = n_i \alpha_i$ where $n_i \geq 0$. This information is important when determining the possible choices of signs for the virtual volumes of weights lying in this weight system. For example, if we choose $w_3$ to be have a positive virtual volume, then $w_2$  needs to also have a positive virtual volume since $w_2 \geq w_3$ according to the Hasse diagram.
		
	   The red superscript indicates whether a weight is positive or negative. A positive (resp. negative) weight is defined as the positive (resp. negative) linear combination of simple roots. When no mass parameters are turned on, then the signs of virtual volumes for positive and negative weights are fixed to be positive and negative, respectively (assuming the dual of the irreducible Weyl chamber is defined as the region in which the virtual volumes of all positive simple roots are positive.) The signs of the rest of the weights are undetermined by the signs of simple roots and hence can be chosen freely as long as the ordering described by the Hasse diagram is satisfied. When mass parameters are turned on, then it is possible for positive weights to have negative virtual volume and negative weights to have positive virtual volume, for some values of the mass parameters. For a generic choice of mass parameters, the only constraint for any of the signs of the weights is that the ordering provided by the Hasse diagram is respected.
	   \newline

		\noindent \raisebox{-.125\height}{	\begin{tikzpicture}
		\node at (0,0) {\color{red}{$5$}};
		\draw[red]  (0,0) circle (0.2cm);
		\end{tikzpicture}} This is the area in which a choice of signs should be specified.A default input is given where all the signs are positive, that is ``+1''. The notation $ s[i]_j$ is explained as follows:	 $i$ labels each different representation (in this case, $i$ runs over two two representations) and $j$ labels the different of weights (in this case, for the fundamental representation, $j$ runs from $1$ to $5$, while for the antisymmetric representation, $j$ runs from $1$ to $10$).  For example, based on the Hasse diagram presented in Figure \ref{fig:fig1} and assuming we do not turn on any mass parameters, we can make a list of all the allowed choices of signs for the fundamental representation of $\mathfrak{su}(5)$:
	\begin{align}
	\begin{split}
&s(1)_1\to 1,s(1)_2\to 1,s(1)_3\to 1,s(1)_4\to 1,s(1)_5\to -1\\
&	s(1)_1\to 1,s(1)_2\to 1,s(1)_3\to 1,s(1)_4\to -1,s(1)_5\to -1\\
&	s(1)_1\to 1,s(1)_2\to 1,s(1)_3\to -1,s(1)_4\to -1,s(1)_5\to -1\\
&s(1)_1\to 1,s(1)_2\to -1,s(1)_3\to -1,s(1)_4\to -1,s(1)_5\to -1.
	\end{split}
\end{align}
	If we choose to turn mass parameters on then we can also have the following sign choices:
	\begin{align}
	\begin{split}
&s(1)_1\to 1,s(1)_2\to 1,s(1)_3\to 1,s(1)_4\to 1,s(1)_5\to 1\\
&	s(1)_1\to -1,s(1)_2\to -1,s(1)_3\to -1,s(1)_4\to -1,s(1)_5\to -1.
	\end{split}
	\end{align}

In the case of two nodes, the code first asks for the signs of the weights associated to the first algebra. The pop-up window is exactly as discussed above, with the sole difference being that the notation for the signs is modified to $ s[i]_{j,1}$, where in addition to the subscripts $i,j$ that respectively label the different representations and weights, there is another subscript $1$ that indicates the representation is charged under the first algebra. After the signs associated to the representations of the first algebra have been specified, a second window appears requesting the signs associated to the second algebra. The format is identical, with the distinction that the signs are denoted by $ s[i]_{j,2}$, with the subscript $2$ labeling the second algebra. Finally, a third window appears requesting signs for the weights of tensor product representations charged under both the first and second algebras.

		For example consider 		\begin{tikzpicture}
		\node (v1) at (-0.5,0.45) {2};
		\node at (-0.45,0.9) {$\mathfrak{su}(2)$};
		\begin{scope}[shift={(1.5,0)}]
		\node (v2) at (-0.5,0.45) {$2$};
		\node at (-0.45,0.9) {$\mathfrak{su}(2)$};
		\end{scope}
		\draw  (v1) edge (v2);
		\end{tikzpicture} ,
		for which the input is:
		\newline
			\color{blue}
	\begin{Verbatim}[frame=single,fillcolor=\color{yellow}]
		Geometry5dKK[{{{2,-1},{-1,2}},{A1,U},{A1,U}}]
	\end{Verbatim}
		\color{black}
		
\begin{figure}[h!]
		\hspace{1.5 cm}
		\includegraphics[scale=0.45]{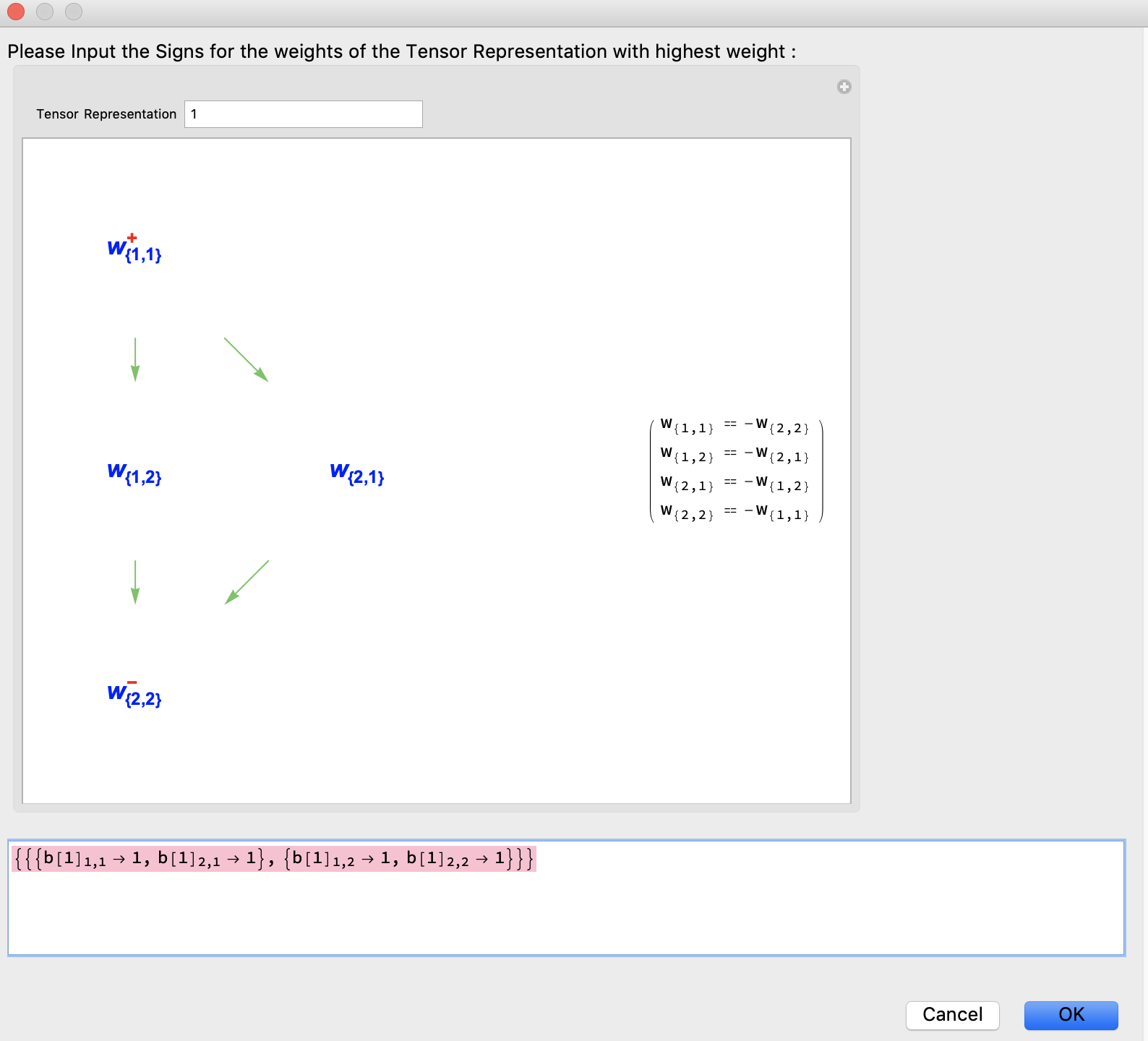}
		\caption{ Signs for the tensor product representation.}
		\label{fig:fig3}
	\end{figure}

An example of the third window is displayed in Figure \ref{fig:fig3}.
In this case, on the upper left side of the window instead of a slider one can find the number of hypermultiplets transforming in a mixed representation. In Figure  \ref{fig:fig3} there is one such hypermultiplet, but in other cases there can be a half-integer number of hypermultiplets. This information is necessary to determine a consistent choice of signs, since for example mass parameters cannot be switched on for half-hypermultiplets. The Hasse diagram in this case is that of the tensor product representation $R_1 \otimes R_2$, where $R_1=R_2=\mathbf{2}$ of $\mathfrak{su}(2)$.
Let $v_i$ denote the weights associated to the first $\mathfrak{su}(2)$ and let $\omega_i$ denote the weights associated to the second $\mathfrak{su}(2)$.
The weight system of the tensor product of these two representations is
\begin{align}
w_{\{i,j\}}=v_i \oplus \omega_j.
\end{align}
The Hasse diagram of this weight system can now be determined based on the ordering of the weights $v_i$ and $\omega_j$. For example,
\begin{align}
w_{\{1,1\}}=v_1+ \omega_1 \geq v_2+ \omega_1 =w_{\{2,1\}} 
\end{align}
The  Hasse diagram and the number of hypermultiplets is enough to determine a consistent choice of signs. The signs follow a similar notation as above, namely
\begin{equation}
b[1]_{i,j}
	\end{equation}
	where the bracketed `$1$' indicates that there is only one mixed representation and the subscripts $i,j$ are the same as the subscripts for  $w_{\{i,j\}}$, referring to weights of the first and second algebras respectively.
\newline

	\noindent \underline{\textbf{Allowed signs for the representations}}
\\ \newline
	\noindent As mentioned above the choice of signs depends on the Hasse diagrams, the values of mass parameters, and on which combinations of representations are chosen. The function
	\color{blue}
	\begin{Verbatim}[frame=single,fillcolor=\color{yellow}]
      			SignsKK[]
	\end{Verbatim}
		\color{black}
		determines all the possible allowed signs for each hypermultiplet of a specific theory. A word of caution: the computational cost of this function increases very quickly with the dimensions of the representations.

	The input of for this function is of the same format described in the previous section:
		\color{blue}
	\begin{Verbatim}[frame=single,fillcolor=\color{yellow}]
   		  SignsKK[{k,{Algebra,Twist}}]  
	\end{Verbatim}
	 	\color{black}
	 	OR
	 			\color{blue}
	\begin{Verbatim}[frame=single,fillcolor=\color{yellow}]
	SignsKK[{Ω,{Algebra1,Twist1},{Algebra2,Twist2}}]
	\end{Verbatim}
		\color{black}
	The output of this function is the appropriate number of hypermultiplets and the type of representation, together with the Hasse diagrams of the weight systems. As described above, the Hasse diagram includes superscripts indicating whether a weight is positive, negative, or indeterminate sign. In the absence of mass parameters the only signs that need to be determined are those of the indeterminate weights. Note that zero weights have superscript `$0$'.  
	The output, namely all consistent gauge-theoretic phases of the theory, is presented both as a collection of Hasse diagrams and as a list of sign choices. The Hasse diagrams for the allowed signs includes superscripts indicating when the signs are taken to be positive (blue) or negative (red). This function is useful for determining all allowed phases and corresponding sign choices when computing the geometry.
	
It is important to note that in some cases the signs associated to different hypermultiplets are not independent. For example, consider 

\begin{align}
\label{eqn:branch}
\raisebox{-.4\height}{\begin{tikzpicture}
	 	\draw (-2.9,0)--(-0.7,0);
	 	\draw (-2.9,0)--(-2.9,0.7);
	 	\draw (-2.9,0.7)--(-0.7,0.7);
	 	\draw (-0.7,0)--(-0.7,0.7);
	 	\node at (-1.85,0.9) {$N_f$};
	 	\node at (-1.85,0.4) {\tiny{$2n_\alpha+8-\frac{n_\beta-1}{2}$}};
	 	\node (v1) at (0.5,0.4) {1};
	 	\node at (0.55,0.9) {$\sp(n_\alpha)^{(1)}$};
	 	\begin{scope}[shift={(2,0)}]
	 	\node (v2) at (0.6,0.4) {$k$};
	 	\node at (0.65,0.9) {$\so(n_\beta)^{(2)}$};	
	 	\end{scope}
	 	\node at (1.52,0) {\tiny{  $\frac{1}{2}( n_\alpha \otimes  (n_\beta -1)) $}};
	 	\draw  (v1) edge (v2);
	 	\draw (3.7,0)--(5.7,0);
	 	\draw (3.7,0)--(3.7,0.7);
	 	\draw (3.7,0.7)--(5.7,0.7);
	 	\draw (5.7,0)--(5.7,0.7);
	 	\node at (4.7,0.4) {\tiny{$n_\beta-8-\frac{n_\alpha}{2}$}};
	 	\node at (4.7,0.9) {$N_f$};
	 	\end{tikzpicture}}
\end{align}
	
	 where the extra labels indicate the number of hypermultiplets included in the theory. In particular, note that there are $2 n_\alpha+8-\frac{n_\beta}{2}$ full hypermultiplets of $\sp(n_\alpha)^{(1)}$ and one half-hyper in a mixed representation. This half-hypermultiplet comes from the branching of the bifundamental $n_\alpha \otimes n_\beta \rightarrow n_\alpha \otimes ( (n_\beta -1)\oplus 1)  $
	 after performing the twist of $\so(n_\beta)^{(2)}$, which leaves invariant the algebra $\so(n_\beta-1)$.  This implies that the signs associated to the half-hypermultiplet are not independent but rather depend on the signs chosen for the bifundamental representation. In this case the function SignsKK returns all possible sign choices consistent with these branching rules.
	 
\begin{figure}[htbp]
		\begin{center}
		\includegraphics[scale=0.4]{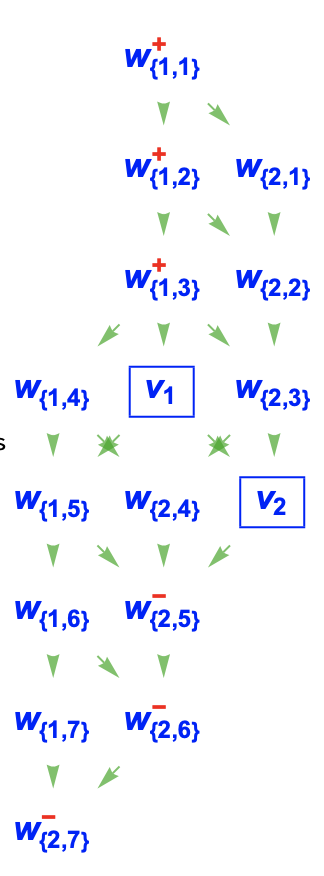}
		\caption{Hasse diagram for the case $n_\alpha = 1, k=3, n_\beta = 4$ of the theory displayed in (\ref{eqn:branch}). Note that $w_{\{i,j\}} $ are the weights of the bifundamental and $v_1,v_2$ are the weights of the half-hypermultiplet. }
		\label{fig:fig51}
		\end{center}
	\end{figure}

\begin{sidewaysfigure}[htbp]
	\begin{center}
	\includegraphics[scale=0.4]{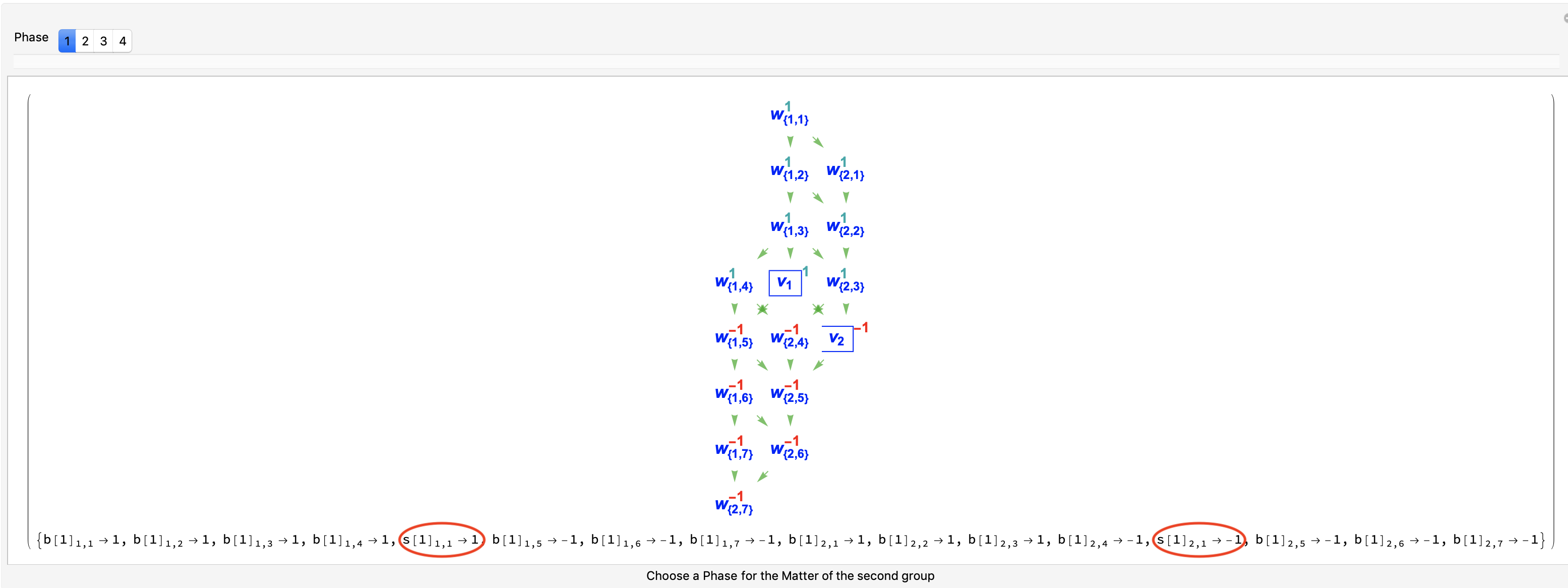}
	\caption{Signs choices for the case $n_\alpha = 1, k=3, n_\beta = 4$ of the theory displayed in (\ref{eqn:branch}). The list provides the signs for both the bifundamental and the half hypermultiplet. The signs with the red circles represent those of the half-hypermultiplet. }
	\label{fig:fig52}
	\end{center}
\end{sidewaysfigure}

	 For example, consider $n_\alpha=1$ , $k=3$ and $n_\beta=4$. The Hasse diagram for the bifundamental combined with the half-hypermultiplet of $\sp(1)$ is displayed in Figure \ref{fig:fig51}.
The possible sign choices are displayed in Figure \ref{fig:fig52}.

	\noindent \underline{\textbf{Output}}
	\newline
		\begin{sidewaysfigure}[htbp]
		\begin{center}
		\includegraphics[scale=0.4]{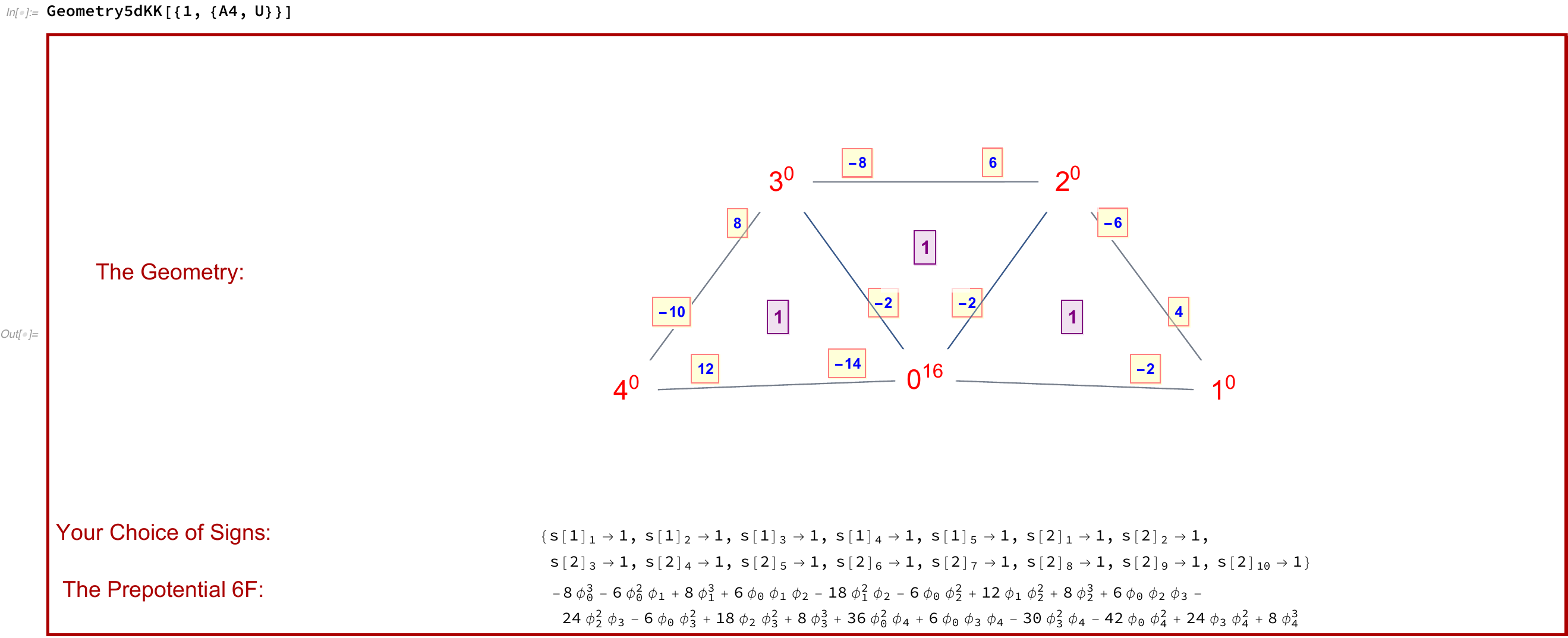}
		\caption{This is the result of the evaluation appearing in Figure \ref{fig:fig1}.}
		\label{fig:fig4}
		\end{center}
	\end{sidewaysfigure}
	
	\clearpage
	\noindent Once the signs have been specified in Geometry5dKK, the following output is returned (see an example shown in Figure \ref{fig:fig4}), and is comprised of the following elements:
	\begin{enumerate}
		\item The triple intersection numbers for the corresponding geometry are presented in a graphical form similar to the graphs presented in Section \ref{GKK} of this paper. The vertices of the graph are surfaces and edges between the vertices indicate the intersections between the corresponding surfaces. The superscript on a vertex $i$ denotes $8-S_i^3$. If the superscript is zero, then it is not displayed. Every edge carries two yellow boxes at either ends. Consider an edge going between vertices $i$ and $j$. The number in the yellow box near the vertex $i$ denotes the triple intersection number $S_iS_j^2$, and the number in the yellow box near the vertex $j$ denotes the triple intersection number $S_i^2S_j$. If the number carried by some yellow box is zero, then that box is not displayed. There is a purple box placed in the middle of every face formed by three edges joining three vertices, say $i$, $j$ and $k$. The number in the purple box denotes the triple intersection number $S_iS_jS_k$. If the number carried by purple box is zero, then it is not displayed.
		\item The choice of signs made by the user.
		\item The the shifted prepotential $6\tilde\cF$.
		In the case of a KK theory with a single node,  $\phi_0$ is the Coulomb branch parameter associated to the affine node of the Dynkin diagram and $\phi_i$ with $i=1,...Rank[Algebra]$ are the Coulomb branch parameters associated to the finite part of the diagram. 
		In the case of a KK theory with two nodes, $\phi_{0,1},\phi_{i,1}$ are the Coulomb branch parameters associated to the first (affine) algebra and $\phi_{0,2},\phi_{i,2}$ are the Coulomb branch parameters associated to the second (affine) algebra.
	\end{enumerate}

\bibliographystyle{ytphys}
\let\bbb\bibitem\def\bibitem{\itemsep4pt\bbb}
\bibliography{ref}

\end{document}